\documentstyle[12pt,dina4p,epsfig]{article}
\title{\bf On Stern-Gerlach forces allowed by special relativity and
the special case of the classical spinning particle of
Derbenev-Kondratenko}
\author{K. Heinemann\thanks{heineman@mail.desy.de} \\
 Deutsches Elektronen-Synchrotron  DESY ,\ Hamburg  }
\begin{document}
\newcounter{INDEX}
\setcounter{section}{0}
\setcounter{subsection}{0}
\setcounter{equation}{0}
\renewcommand
{\theequation}{\mbox{\thesection .\arabic{equation}\alph{INDEX}}}
\def\wtimes{\wedge}
\maketitle
%\footnote{e-mail address: t00hei@dhhdesy3}
%
\begin{abstract}
This work is devoted to an examination of Stern-Gerlach forces
consistent with special relativity and is motivated by recent interest
in the relativistic Stern-Gerlach force acting on polarized protons in
high-energy particle accelerators. The equations for the orbital and
spin motion of a classical charged particle with arbitrary intrinsic
magnetic dipole moment in an external electromagnetic field are
considered and by imposing the constraints of special relativity
and restricting to first order in spin (= first order $\hbar$) a
well-defined class of spin-orbit systems is obtained. All these systems
can be treated on an equal footing including such prominent cases as
those considered by Frenkel and by Good. 
\par The Frenkel case is
considered in great detail because I show that this system is identical
with the one introduced by Derbenev and Kondratenko for studying spin
motion in accelerators. In particular I prove that the spin-orbit system
of Derbenev and Kondratenko is (nonmanifestly) $\rm{Poincar\acute{e}}$
covariant and identify the transformation properties of this system
under the $\rm{Poincar\acute{e}}$ group. The Derbenev-Kondratenko
Hamiltonian was originally proposed as a way to combine relativistic
spin precession and the Lorentz force. The aforementioned findings now
demonstrate that the Derbenev-Kondratenko Hamiltonian also provides a
legitimate framework for handling the relativistic Stern-Gerlach force.
\par Numerical examples based on the Frenkel and Good cases for the HERA
proton ring and electromagnetic traps are provided.
\end{abstract}
\newpage
\tableofcontents
\vspace{1cm}
%
%\newpage
\section*{Introduction}
\addcontentsline{toc}{section}{Introduction}
\setcounter{equation}{0}
\setcounter{subsection}{0}
\setcounter{section}{0}
\vspace{1cm}
One of the most economical descriptions of the motion of polarized
beams of spin 1/2 particles in high-energy accelerators is provided by
the semiclassical Hamiltonian given by Derbenev and Kondratenko
\cite{DK73}, which in the language of the Dirac equation can be
written in the\\ form
\footnote{I choose units where the vacuum light velocity is $c=1$.}:
\begin{eqnarray}
&& H_{{}_{M,op}} \equiv H_{{}_{M,op,orb}} + H_{{}_{M,op,spin}} \; ,
\label{0.1}
\end{eqnarray}
with
\begin{eqnarray}
&& H_{{}_{M,op,orb}} = \beta\cdot\sqrt{\vec{\pi}_{{}_{M,op}}^\dagger\cdot
\vec{\pi}_{{}_{M,op}}+m^2} + e\cdot \phi_{{}_{M,op}} \; ,
\nonumber\\
&& H_{{}_{M,op,spin}} = \vec{\sigma}_{{}_{op}}^{\,\dagger}\cdot
                       \vec{W}_{{}_{M,op}}  \; ,
\label{0.2}
\end{eqnarray}
where
\begin{eqnarray}
 && \vec{\pi}_{{}_{M,op}} = \vec{p}_{{}_{M,op}} -
e\cdot\vec{A}_{{}_{M,op}}
  \;  {\equiv \;\; canonical\;\;momentum\;\;vector} \; ,
\nonumber\\
&& \vec{W}_{{}_{M,op}} = -\frac{e}{2\cdot m}\cdot\biggl( \lbrack
\frac{m}{J_{{}_{M,op}}}+\frac{g-2}{2} \rbrack \cdot\beta\cdot
\vec{B}_{{}_{M,op}} \nonumber\\&&\qquad
-\frac{g-2}{2}\cdot\frac{1}{J_{{}_{M,op}}\cdot(J_{{}_{M,op}}+m)}\cdot
\beta\cdot\vec{\pi}_{{}_{M,op}}^\dagger\cdot
\vec{B}_{{}_{M,op}}\cdot\vec{\pi}_{{}_{M,op}}\nonumber\\
&&\qquad
\qquad -\lbrack \frac{g}{2\cdot J_{{}_{M,op}}} -
\frac{1}{J_{{}_{M,op}}+m}\rbrack\cdot
(\vec{\pi}_{{}_{M,op}}\wtimes\vec{E}_{{}_{M,op}})\biggr)
 + {\rm hermitian\;conjugate} \; ,
\end{eqnarray}
with
\begin{eqnarray}
 && J_{{}_{M,op}} =
\sqrt{\vec{\pi}_{{}_{M,op}}^\dagger\cdot \vec{\pi}_{{}_{M,op}}+m^2} \; .
\label{0.3}
\end{eqnarray}
Furthermore I define
\begin{eqnarray}
 && \vec{v}_{{}_{M,op}} \equiv \frac{i}{\hbar}\cdot
(H_{{}_{M,op}}\cdot\vec{r}_{{}_{M,op}} - \vec{r}_{{}_{M,op}}\cdot
H_{{}_{M,op}} ) \; , \nonumber\\
                    &&     \gamma_{{}_{M,op}} \equiv
(1-\vec{v}_{{}_{M,op}}^\dagger\cdot \vec{v}_{{}_{M,op}})^{-1/2} \; ,
\nonumber\\
&& m\cdot  \gamma_{{}_{M,op}}  \cdot \vec{v}_{{}_{M,op}}
  \;  {\equiv \;\; kinetic\;\;momentum\;\;vector\;\;\equiv\;\;mechanical\;
momentum\;vector} \; . \nonumber\\
\end{eqnarray}
Here $\phi_{{}_{M,op}},\vec{A}_{{}_{M,op}},
\vec{E}_{{}_{M,op}},\vec{B}_{{}_{M,op}}$ represent the
electromagnetic field, $e,m$ denote charge and (nonvanishing) rest mass
of the particle whereas $g$ denotes the gyromagnetic factor $\lbrack a
\rbrack$.
The operators $\beta,\frac{2}{\hbar}\cdot\vec{\sigma}_{{}_{op}}$ are Dirac 
matrices. This semiclassical Hamiltonian can be obtained from the Dirac
Hamiltonian (modified by the Pauli term) by a certain relativistic 
generalization of the Foldy-Wouthuysen transformation $\lbrack b \rbrack$ 
in which terms of second and higher order in the spin are dropped. The orbital
variables $\vec{r}_{{}_{M,op}},\vec{p}_{{}_{M,op}}$ and the spin
variable $\vec{\sigma}_{{}_{op}}$ 
%are operators which for this
%generalized Foldy-Wouthuysen representation are the same as those for
%the nonrelativistic Pauli equation,
%\footnote{See also \cite{DS72,Hei}.}
%so that they 
obey the usual commutation relations.
\footnote{Specifically $\vec{r}_{{}_{M,op}},\vec{p}_{{}_{M,op}}$ obey
the canonical commutation relations whereas
$2\cdot\vec{\sigma}_{{}_{op}}/\hbar$ obeys the same commutation
relations as the Pauli matrices.}
The operator $\vec{r}_{{}_{M,op}}$ is often called the
operator of `canonical mean position', which
explains the presence of the subscript `M'. Note also that for
vanishing  electromagnetic field $\vec{r}_{{}_{M,op}}$ is 
the position operator of Newton and Wigner \cite{NW49}. 
%Because
%(even for vanishing field) the eigenstates of the Dirac position
%operator $\vec{r}_{{}_{D,op}}$
%\footnote{i.e. the position operator appearing in the Dirac
%representation of the Dirac equation \cite{Fol62}}
%are spread over a region of the order of the particle Compton
%wavelength in the configuration space of $\vec{r}_{{}_{M,op}}$,
%the canonical mean position describes a particle which is not pointlike 
%\cite{Fol62}.
%\footnote{However it is localizable in the sense of \cite{NW49,Wi62}.}
The operator $\vec{\sigma}_{{}_{op}}$ defines the spin in a frame at
rest w.r.t. the canonical mean position. The spin part 
$H_{{}_{M,op,spin}}$ of the
semiclassical Hamiltonian is in the form of a Stern-Gerlach (SG) energy
term and the orbital part $H_{{}_{M,op,orb}}$ resembles the standard
orbital Hamiltonian. With this semiclassical Hamiltonian the Heisenberg
equations of motion for the orbital operators comprise the Lorentz force
and SG like terms. The Heisenberg equation of motion for the spin
operator is equivalent to the Thomas-Bargmann-Michel-Telegdi equation
(Thomas-BMT equation) \cite{BMT59,Tho27}. This semiclassical Hamiltonian
appears in the calculation of natural radiative polarization of
electrons (the Sokolov-Ternov effect) \cite{DK73,Jac76,ST64} in inhomogeneous
fields. However, the interpretation of this Hamiltonian in terms of
classical variables is also of great utility and has been used to
describe the SG force for relativistic particles as well as to construct
the action-angle variables for classical spin-orbit motion
\cite{BHR94a,BHR94b,DK73,Der90a,Der90b,Yok87}.
 Since from the outset the
semiclassical Hamiltonian is to be applied in a context where $\hbar^2$
can be neglected, the classical version contains the same information
without additional approximation. So it would be an unnecessary
complication to continue to use the quantum version -- one can just view
the latter as a catalyst.
Thus in the remainder of this paper I will use the classical form.
\par My interest in relativistic SG forces stems from the
suggestion that they could be used to separate spin states in storage
rings and thereby provide polarized (anti-) proton beams
\cite{CPP95,NR87}.
Clearly, since the SG force is very small, one must be careful to
include {\it all} terms in calculations. This is especially true in
storage ring physics where there are not only strong transverse magnetic
field gradients but also strong longitudinal gradients as well as large
high frequency electric fields. One way to include all terms
automatically is to use a formulation based on a Hamiltonian and of
course one immediately thinks of the Derbenev-Kondratenko (DK) Hamiltonian. A
formulation based on a Hamiltonian is also symplectic.
The Hamiltonian approach has already been developed in
\cite{BHR94a,BHR94b}.
\par I have opened this paper with a review of the DK approach to spin and
orbital motion but there is a large literature on classical,
$\rm{Poincar\acute{e}}$ covariant
equations of motion in the presence of SG forces going back as far as
the paper of Frenkel of 1926 \cite{Fre26}. See the reviews in
\cite{BT80,Nyb62,Pla66b,Roh72,TVW80}. So, naturally,
given the prominence of the DK Hamiltonian in the
handling of spin effects in storage rings, one would like to know how it
is related to other formalisms and whether it leads to
$\rm{Poincar\acute{e}}$ covariant equations of motion. One would also
like to know what general form the covariant equations of
spin-orbit motion can take and to study the numerical
implications of various choices.
\par Although I also treat some other topics
the main burden of this paper is then the following:
\begin{itemize}
\item{I prove that the equations of motion derived from the
DK Hamiltonian are, after a transformation of the coordinates, identical
to those parts of the Frenkel equations \cite{Fre26} obtained after
dropping second and higher order terms in the spin, and are therefore
$\rm{Poincar\acute{e}}$ covariant.
To the knowledge of the author the present proof is the first
and it occupies most of the present work.
\footnote{For the special case of the first order SG force a proof was
effectively contained in  \cite{DS70}.}
                                          The key to the proof is the
realization that the position variable of the DK
equations is not the spatial part of a space-time position but that
this can be remedied if one first transforms it into the position
variable  given by Pryce which does not have this disadvantage.
\footnote{I learnt this `trick' from \cite{DS70} who applied it to the
special case of the first order SG force.}
This leads  to the   conclusion that the
approach of \cite{DK73} is consistent with special relativity $\lbrack c
\rbrack$. Thus the treatments in \cite{BHR94a,BHR94b,Der90a,Der90b} are
also relativistic.}
\item{I then demonstrate, at first order in spin,
                         how to construct more general classical
$\rm{Poincar\acute{e}}$ covariant equations of spin-orbit motion and
find  that there is an unlimited number of possibilities allowed by
the kinematic constraints with each possibility being characterized by
the choice of five  parameters.
This unified approach enables one to classify equations appearing in the
literature by identifying their characteristic parameters.
In particular, besides the Frenkel equations, three other
cases are considered in more detail: the Good-Nyborg-Rafanelli (=GNR)
equations \cite{Goo62,Nyb64,Raf70}
(which is also the approach chosen in \cite{CPP95,NR87}), a case dealt
with in \cite{CoM94}, a case dealt with in \cite{Cos94}, 
and the simple case of vanishing SG force.}
This technique could be extended to higher order in spin and the number
of characteristic parameters would then increase.
\item{ Having achieved my main aims I then investigate in more detail
another relationship between the DK equations and
some of the other spin-orbit systems and show that
some spin-orbit systems can be related by transformations. In
particular I relate the Frenkel equations to the GNR equations.}
\item{I obtain numerical estimates of the SG force for simple spin and
field configurations in the HERA proton storage ring \cite{Bri95} and in
traps and compare the expectations from various equations.}
\end{itemize}
\par  Even if one has a basis for choosing among the plethora of
spin-orbit systems available, for example by experiment or by
appealing to the Dirac equation, one must still
decide which  variables to use for representing the particle
position. The natural choice is the Pryce position variable since it
transforms like the spatial part of the space-time position.
On the other hand, in  Newton-Wigner coordinates the DK equations are
in canonical form.
As we  shall see later, Newton-Wigner coordinates can lead to equations
of motion very different from those for the Pryce coordinates.
However, the Newton-Wigner  coordinate
differs by less than the Compton wave length from the Pryce
coordinate, so that the relationship between the Frenkel
and DK equations is not just mathematical but
expresses the effective physical identity of the DK particle
and the Frenkel particle.
\par The paper is organized as follows..
In sections 1 and 2 the spin-orbit system defined by the
DK equations is considered.
In sections 3 and 4 I consider the spin-orbit system defined by
the Frenkel equations and I show that it is the same system as in
sections 1 and 2.
The central part of the work is section 5 where the most general form
for the spin-orbit systems under study is defined (equations (5.5)) and
their main properties are explored.
In sections 6,7,8,9 I simply expose some further properties and
give numerical examples. In particular in section 8 I consider
the SG force in magnetic fields typical for HERA-p and in section 9
the SG force in electromagnetic traps.
In section 10 I introduce the transformations
between different spin-orbit systems, mentioned above.
\par Throughout this work the time evolution of the spin is always that of
\cite{BMT59,Tho27} and the orbital motion is determined by the Lorentz
force plus the SG force. The SG force consists of two parts, a part
linear in the electromagnetic field vectors (the `first order SG
force') and a part quadratic in these fields (the `second order SG
force').
\setcounter{equation}{0}
\section{The DK equations}
\subsection{The DK Hamiltonian}
The classical DK Hamiltonian $\lbrack d \rbrack$ is
obtained from the semiclassical version (0.1) by replacing the operators
by classical canonical variables and the commutators by the following
Poisson bracket relations
\footnote{The variables $\vec{r}_{{}_M},\vec{p}_{{}_M},\vec{\sigma}$
together with the
Poisson bracket $\lbrace \;,\;\rbrace_{{}_M}$
define a Poisson algebra for the DK Hamiltonian.}
$\lbrack e \rbrack$:
\begin{eqnarray}
&& \lbrace r_{{}_{M,j}},p_{{}_{M,k}}\rbrace_{{}_M} = \delta_{jk} \; ,\qquad
 \lbrace r_{{}_{M,j}},r_{{}_{M,k}}\rbrace_{{}_M} = 
 \lbrace p_{{}_{M,j}},p_{{}_{M,k}}\rbrace_{{}_M} =
   \lbrace r_{{}_{M,j}} , \sigma_k\rbrace_{{}_M} =
    \lbrace p_{{}_{M,j}} , \sigma_k\rbrace_{{}_M} = 0 \; , 
\nonumber\\
&& \lbrace \sigma_j , \sigma_k\rbrace_{{}_M} =
            \sum_{m=1}^3 \varepsilon_{jkm}\cdot \sigma_m
                                             \; .
\qquad (j,k=1,2,3)
\label{1.1}
\end{eqnarray}
Note that $\vec{r}_{{}_M},t$ are the position and   time variables
used in the everyday business of accelerator physics if calculations are
made in `cartesian coordinates'.
%The subscript `M' on the Poisson
%bracket indicates that we are to take partial derivatives with respect
%to `M' variables.
%\par The spin vector is constrained by $\lbrack f \rbrack$:
%
%\begin{eqnarray}
% && \vec{\sigma}^{\,\dagger}\cdot\vec{\sigma} = \hbar^2/4 \; .
%\label{1.2}
%\end{eqnarray}
%
The electromagnetic field is characterized either by the
potentials $\phi_{{}_M},\vec{A}_{{}_M}$ or the field vectors
$\vec{E}_{{}_M},\vec{B}_{{}_M}$. Thus one has $\lbrack f \rbrack$:
\begin{eqnarray}
&& \vec{B}_{{}_M} = \vec{\nabla}_{{}_M} \wtimes
                                          \vec{A}_{{}_M} \; , \qquad
  \vec{E}_{{}_M} = - \vec{\nabla}_{{}_M}\phi_{{}_M} -
                  \frac{\partial \vec{A}_{{}_M}}{\partial t} \; .
\label{1.2}
\end{eqnarray}
The functions $\phi_{{}_M},\vec{A}_{{}_M},
\vec{E}_{{}_M},\vec{B}_{{}_M}$ depend on $\vec{r}_{{}_M},t$. This also
fixes the meaning of the partial derivative $\partial/\partial t$ in
(1.2).
The corresponding Maxwell equations are  given   in section 2.
On introducing the quantities $\lbrack g \rbrack$:
\begin{eqnarray}
 && \vec{\pi}_{{}_M} \equiv \vec{p}_{{}_M} - e\cdot\vec{A}_{{}_M} \; ,
\nonumber\\
 && J_{{}_M} \equiv \sqrt{\vec{\pi}_{{}_M}^\dagger\cdot
                                            \vec{\pi}_{{}_M}+m^2} \; ,
\nonumber\\
 && \vec{W}_{{}_M} \equiv -\frac{e}{m}\cdot\biggl(
 \lbrack \frac{m}{J_{{}_M}}+\frac{g-2}{2} \rbrack \cdot\vec{B}_{{}_M}
 -\frac{g-2}{2}\cdot\frac{1}{J_{{}_M}\cdot(J_{{}_M}+m)}\cdot
  \vec{\pi}_{{}_M}^\dagger\cdot\vec{B}_{{}_M}\cdot\vec{\pi}_{{}_M}
\nonumber\\ &&\qquad
-\lbrack \frac{g}{2\cdot J_{{}_M}} -\frac{1}{J_{{}_M}+m}\rbrack\cdot
 (\vec{\pi}_{{}_M}\wtimes\vec{E}_{{}_M}) \biggr)    \; ,
\label{1.3}
\end{eqnarray}
the classical version of the DK Hamiltonian reads as
\cite{DK73}:
\begin{eqnarray}
 && H_{{}_M}(\vec{r}_{{}_M},\vec{p}_{{}_M},\vec{\sigma},t)
  \equiv   J_{{}_M} + e\cdot\phi_{{}_M}+
  \vec{\sigma}^{\,\dagger}\cdot \vec{W}_{{}_M}  \; .
\label{1.4}
\end{eqnarray}
To maintain consistency with the semiclassical nature of the
DK Hamiltonian, throughout this paper all terms of second
and higher order in the spin are
neglected. This is the underlying approximation used in this paper. In
particular one then always obtains the spin motion given by \cite{BMT59,Tho27}.
\footnote{This approximation is standard also in applications to
accelerator physics. For the meaning of a power series expansion in
spin, see also \cite{Pla66b}.}
\subsection{The DK equations}
With this Hamiltonian the equations of motion for the orbital and spin
variables - the DK equations - can be obtained using
(1.1) and are:
\footnote{In this paper the total time derivative is denoted by the
prime $'$.}
\setcounter{INDEX}{1}
\begin{eqnarray}
&&\vec{r}^{\;{}_{'}}_{{}_M}=
                          \lbrace\vec{r}_{{}_M},H_{{}_M}\rbrace_{{}_M} \; ,
\label{1.5a}
\\
\addtocounter{equation}{-1}
\addtocounter{INDEX}{1}
&&\vec{p}^{\;{}_{'}}_{{}_M}=
                        \lbrace \vec{p}_{{}_M}, H_{{}_M}
                                               \rbrace_{{}_M}  =
    \frac{e}{J_{{}_M}}\cdot(\vec{\pi}_{{}_M}\wtimes\vec{B}_{{}_M})
   -e \cdot\vec{\nabla}_{{}_M}\phi_{{}_M} +
    \frac{e}{J_{{}_M}}\cdot
   (\vec{\pi}_{{}_M}^\dagger\cdot\vec{\nabla}_{{}_M}) \vec{A}_{{}_M}
  - \underbrace{
   \vec{\nabla}_{{}_M} (\vec{\sigma}^{\,\dagger}\cdot \vec{W}_{{}_M})
                       }_{{\rm SG\;term}} \; ,
\nonumber\\ &&\qquad
\label{1.5b}
\\
\addtocounter{equation}{-1}
\addtocounter{INDEX}{1}
&&\vec{\sigma}\ ' =\lbrace\vec{\sigma}, H_{{}_M} \rbrace_{{}_M}
                  =\vec{W}_{{}_M}\wtimes\vec{\sigma} \; ,
\label{1.5c}
\end{eqnarray}
\setcounter{INDEX}{0}
\par\noindent
where the SG terms in the orbital equations have been explicitly
identified.
The covariance of (1.5) under the
$\rm{Poincar\acute{e}}$ group will be demonstrated in
section 5.
Note that (1.5c) is equivalent to the Thomas-BMT equation
\cite{BMT59,Tho27}.
\par In the following it will be useful to define:
\setcounter{INDEX}{1}
\begin{eqnarray}
 && \vec{v}_{{}_M} \equiv
  \vec{r}^{\;{}_{'}}_{{}_M} \; ,
\\
\addtocounter{equation}{-1}
\addtocounter{INDEX}{1}
\label{1.6a}
&&\gamma_{{}_M}\equiv
         (1-\vec{v}_{{}_M}^{\,\dagger}\cdot\vec{v}_{{}_M})^{-1/2} \; ,
\\
\addtocounter{equation}{-1}
\addtocounter{INDEX}{1}
\label{1.6b}
 && \vec{\Omega}_{{}_M}\equiv
                        -\frac{e}{m}\cdot\biggl(
 \lbrack \frac{1}{\gamma_{{}_M}}+
  \frac{g-2}{2} \rbrack \cdot\vec{B}_{{}_M}
 -\frac{g-2}{2}\cdot\frac{\gamma_{{}_M}}{\gamma_{{}_M}+1}\cdot
  \vec{v}_{{}_M}^{\,\dagger}\cdot\vec{B}_{{}_M}\cdot\vec{v}_{{}_M}
 \nonumber\\ &&\qquad
-\lbrack \frac{g}{2} -\frac{\gamma_{{}_M}}{\gamma_{{}_M}+1}\rbrack\cdot
 (\vec{v}_{{}_M}\wtimes\vec{E}_{{}_M}) \biggr)    \; .
\label{1.6c}
\end{eqnarray}
\setcounter{INDEX}{0}
By neglecting the spin terms in (1.5-6) one gets the familiar
relations:
\begin{eqnarray*}
 && m\cdot (\gamma_{{}_M}\cdot\vec{v}_{{}_M})'  =
    e\cdot(\vec{v}_{{}_M}\wtimes\vec{B}_{{}_M})+
                                            e\cdot \vec{E}_{{}_M} \; ,
\qquad \gamma_{{}_M}'  =
    \frac{e}{m}\cdot\vec{v}_{{}_M}^{\,\dagger}\cdot\vec{E}_{{}_M} \; .
\end{eqnarray*}
With (1.6) I find that (1.5c) reads as:
\begin{eqnarray}
&& \vec{\sigma}\ ' =  \vec{\Omega}_{{}_M}\wtimes\vec{\sigma} \; .
\label{1.7}
\end{eqnarray}
This is the usual form of the Thomas-BMT equation \cite{BMT59,Tho27}. The
transformation properties of $\vec{r}_{{}_M},t,\vec{v}_{{}_M},
\vec{\sigma}$ under the $\rm{Poincar\acute{e}}$ group are discussed in detail
in section 5.
\subsection*{1.3}
By (1.1),(1.4),(1.5a) I have:
\footnote{This equation is derived in Appendix B, see: (B.3-4).}
\begin{eqnarray}
 && \vec{v}_{{}_M} =
    \lbrace \vec{r}_{{}_M}, H_{{}_M} \rbrace_{{}_M}
 =\frac{\vec{\pi}_{{}_M}}{J_{{}_M}}
 +  \underbrace{
    \frac{e}{J_{{}_M}^3} \cdot\vec{\sigma}^{\,\dagger}\cdot
   \vec{B}_{{}_M}\cdot\vec{\pi}_{{}_M}
 +\frac{e}{m\cdot J_{{}_M}\cdot(J_{{}_M}+m)^2}\cdot
                \vec{\sigma}^{\,\dagger}
 \cdot (\vec{\pi}_{{}_M}\wtimes\vec{E}_{{}_M})\cdot\vec{\pi}_{{}_M}
                       }_{{\rm SG\;terms}}
\nonumber\\ &&
 -  \underbrace{
  \frac{e}{m\cdot(J_{{}_M}+m)}\cdot(\vec{E}_{{}_M}\wtimes\vec{\sigma})
 + \frac{e\cdot g}{2\cdot m\cdot J_{{}_M}}\cdot
           (\vec{E}_{{}_M}\wtimes\vec{\sigma})
 - \frac{e\cdot g}{2\cdot m\cdot J_{{}_M}^3}\cdot
             \vec{\sigma}^{\,\dagger}
                                                                \cdot
    (\vec{\pi}_{{}_M}\wtimes\vec{E}_{{}_M})\cdot\vec{\pi}_{{}_M}
                       }_{{\rm SG\;terms}}
\nonumber\\ &&\qquad
+\frac{g-2}{2}\cdot \biggl(
    \underbrace{
 \frac{e}{m}\cdot\frac{1}{J_{{}_M}\cdot(J_{{}_M}+m)}\cdot
  \vec{\pi}_{{}_M}^{\,\dagger}\cdot\vec{\sigma}
\cdot\vec{B}_{{}_M}
                       }_{{\rm SG\;term}}
\nonumber\\ &&
 -  \underbrace{
   \frac{e}{m}\cdot
  \frac{m+2\cdot J_{{}_M}}{J_{{}_M}^3\cdot(J_{{}_M}+m)^2}\cdot
\vec{\pi}_{{}_M}^{\,\dagger}\cdot\vec{\sigma}\cdot\vec{\pi}_{{}_M}^\dagger\cdot
                           \vec{B}_{{}_M}\cdot \vec{\pi}_{{}_M}
+ \frac{e}{m}\cdot\frac{1}{J_{{}_M}\cdot(J_{{}_M}+m)}\cdot
\vec{\pi}_{{}_M}^\dagger\cdot\vec{B}_{{}_M}\cdot
                                                \vec{\sigma} \biggr)
                       }_{{\rm SG\;terms}} \; .
\label{1.8}
\end{eqnarray}
Thus for the `M' variables      the canonical momentum
is different from the kinetic momentum, i.e.
\begin{eqnarray*}
&& \vec{\pi}_{{}_M} \neq m\cdot \gamma_{{}_M}\cdot \vec{v}_{{}_M} \; ,
\end{eqnarray*}
which says that in the `M' variables one has zitterbewegung.
This effect disappears with the  electromagnetic field.
\footnote{The term `zitterbewegung' was introduced in \cite{Sch30} and in its
original sense it only applies to a free quantum mechanical particle.
Thus my term deals with the classical analogue and even applies in
the presence of electromagnetic fields. For more details, see
\cite{Cor68,Fol62}.}
\par The spin vector is constrained by:
\begin{eqnarray}
 && \vec{\sigma}^{\,\dagger}\cdot\vec{\sigma} = \hbar^2/4 \; .
\label{1.9}
\end{eqnarray}
Because (1.9) is of second order in spin it plays no role in this paper. It
is only applied in sections 8 and 9 for numerical calculations,
where it gives the spin vector its correct length. Hence (1.9) acts as a
numerical constraint to be inserted if the formulae are numerically evaluated.
Note also that (1.9) is conserved under (1.5c).
\par My next task will be to rewrite equations (1.5) in a form which
facilitates comparison with other formalisms.
\setcounter{equation}{0}
\section{Reexpressing the DK equations in terms of
auxiliary variables}
\subsection*{2.1}
In this section I replace the variables $\vec{r}_{{}_M},
\vec{v}_{{}_M}, \vec{\sigma}$ by new variables $\vec{r}_{{}_P},
\vec{v}_{{}_P},\vec{s}$ which will later serve as the building blocks of
the $\rm{Poincar\acute{e}}$ covariant formulae to be derived.
\par I define
\begin{eqnarray}
 && \vec{r}_{{}_P} \equiv
% \vec{r}_{{}_M}+\frac{1}{m}\cdot\frac{1}{J_{{}_M}+m}
% \cdot (\vec{\sigma}\wtimes\vec{\pi}_{{}_M})=
    \vec{r}_{{}_M} + \frac{1}{m}\cdot
            \frac{\gamma_{{}_M}}{\gamma_{{}_M}+1}
\cdot (\vec{\sigma}\wtimes\vec{v}_{{}_M}) \; .
\label{2.1}
\end{eqnarray}
The quantity $\vec{r}_{{}_P}$ is the position
variable of Pryce \cite{Pry49}. The corresponding operator 
$\vec{r}_{{}_{P,op}}$ describes (as $\vec{r}_{{}_{M,op}}$ does) a particle 
which is not pointlike from the point of view of the Dirac position operator. 
For the explicit form of these operators in the Dirac
representation, see \cite{Hei}. 
For the special case of the first order
SG force, see for example: \cite{DS70,DS72}. Note that
$\vec{r}_{{}_M}$ and $\vec{r}_{{}_P}$ differ by less than the Compton
wave length. In contrast to $\vec{r}_{{}_M}$, the variable
$\vec{r}_{{}_P}$ transforms under $\rm{Poincar\acute{e}}$
transformations as the spatial part of the space-time position (see
section 3).
\footnote{The quantum mechanical proof of this covariance property,
based on the Dirac equation (plus Pauli term), deals with the operator
$\vec{r}_{{}_{P,op}}$ and is given for the special case of
the first order SG force in \cite{DS72}. For the general proof, see
\cite{Hei}.}
In the present paper, where I work classically, I do not rely on the
Dirac equation but instead derive the
$\rm{Poincar\acute{e}}$ covariance of the
DK equations from the covariance property of
$\vec{r}_{{}_P}$.
\par The components of $\vec{r}_{{}_P}$ are not canonical, i.e.
\begin{eqnarray}
&&\lbrace r_{{}_{P,j}}, r_{{}_{P,k}}\rbrace_{{}_M} =
%     \frac{1}{J_{{}_M}^2}\cdot\sum_{l=1}^3\varepsilon_{jkl}\cdot
% ( s_l + \frac{1}{m^2}\cdot \pi_{{}_{M,l}} \cdot
%  \vec{s}^{\,\dagger}\cdot\vec{\pi}_{{}_M} )
%\nonumber\\
%&=&
     \frac{1}{m^2} \cdot \sum_{l=1}^3\varepsilon_{jkl}\cdot
   (\frac{1}{\gamma_{{}_M}^2}\cdot  s_l +
                            v_{{}_{M,l}} \cdot
  \vec{v}_{{}_M}^{\,\dagger}\cdot\vec{s}
)    \neq 0 \; .
                                    \qquad (j,k=1,2,3)
\label{2.2}
\end{eqnarray}
But the complementary virtues of my two position variables are now
clear; the variable $\vec{r}_{{}_P}$ is useful for studying covariance
of the equations of motion (as seen below) whereas $\vec{r}_{{}_M}$ is
useful for symplectic calculus (calculating spin-orbit transport maps)
because the Poisson brackets for $\vec{r}_{{}_M}$ are canonical (see
(1.1)).
\par The velocity vector corresponding to $\vec{r}_{{}_P}$ reads as:
\setcounter{INDEX}{1}
\begin{eqnarray}
 && \vec{v}_{{}_P} \equiv
  \vec{r}^{\;{}_{'}}_{{}_P}=
   \vec{v}_{{}_M} +
    \biggl(   \frac{1}{m}\cdot\frac{\gamma_{{}_M}}{\gamma_{{}_M}+1}
\cdot (\vec{\sigma}\wtimes\vec{v}_{{}_M}) \biggr)' \; ,
\label{2.3a}
\end{eqnarray}
\addtocounter{equation}{-1}
\addtocounter{INDEX}{1}
\par\noindent
and I define:
\begin{eqnarray}
 && \gamma_{{}_P} \equiv
  (1-\vec{v}_{{}_P}^\dagger\cdot\vec{v}_{{}_P})^{-1/2} \; .
\label{2.3b}
\end{eqnarray}
\setcounter{INDEX}{0}
\par\noindent
Furthermore I define:
\begin{eqnarray}
&& \vec{s} \equiv
   \gamma_{{}_M} \cdot \vec{\sigma} -
         \frac{\gamma_{{}_M}^2}{\gamma_{{}_M}+1} \cdot
\vec{v}_{{}_M}^{\,\dagger}\cdot\vec{\sigma}
\cdot\vec{v}_{{}_M} \; ,
\label{2.4}
\end{eqnarray}
from which follows:
\begin{eqnarray}
 &&  \vec{\sigma} =
     \frac{1}{\gamma_{{}_P}}\cdot
     \vec{s} + \frac{\gamma_{{}_P}}{\gamma_{{}_P}+1}
   \cdot
  \vec{v}_{{}_P}^{\,\dagger}\cdot\vec{s}
\cdot\vec{v}_{{}_P}\; ,
\nonumber\\
&&
\vec{v}_{{}_P}^{\,\dagger}\cdot\vec{\sigma}
=
  \vec{v}_{{}_P}^{\,\dagger}\cdot\vec{s}
 \; ,
\nonumber\\
&&\lbrace s_j, s_k\rbrace_{{}_M}
%  = \sum_{l=1}^3\varepsilon_{jkl}\cdot
% ( s_l + \frac{1}{m^2}\cdot \pi_{{}_{M,l}} \cdot
%  \vec{s}^{\,\dagger}\cdot\vec{\pi}_{{}_M} )
%  =   \sum_{l=1}^3\varepsilon_{jkl}\cdot
%   (   s_l + \gamma_{{}_M}^2\cdot v_{{}_{M,l}} \cdot
%  \vec{s}^{\,\dagger}\cdot\vec{v}_{{}_M} )
%\nonumber\\&&\qquad
  =   \sum_{l=1}^3\varepsilon_{jkl}\cdot
   (   s_l + \gamma_{{}_P}^2\cdot v_{{}_{P,l}} \cdot
 \vec{v}_{{}_P}^{\,\dagger}\cdot\vec{s}
)   \; . \qquad
                                           (j,k=1,2,3)
\label{2.5}
\end{eqnarray}
One can therefore express the evolutions of $\vec{r}_{{}_M}(t),
\vec{v}_{{}_M}(t),\vec{\sigma}(t)$ in terms of $\vec{r}_{{}_P}(t),
\vec{v}_{{}_P}(t), \vec{s}(t)$.
The field   equations for the  `M' fields are
\begin{eqnarray}
&& 0 = \vec{\nabla}_{{}_M} \wtimes \vec{B}_{{}_M}-
                  \frac{\partial \vec{E}_{{}_M}}{\partial t} \; , \qquad
   0 = \vec{\nabla}_{{}_M} \wtimes \vec{E}_{{}_M} +
                  \frac{\partial \vec{B}_{{}_M}}{\partial t} \; , \qquad
   0 = \vec{\nabla}_{{}_M}^\dagger\cdot \vec{E}_{{}_M} \; , \qquad
   0 = \vec{\nabla}_{{}_M}^\dagger\cdot \vec{B}_{{}_M} \; ,
\nonumber\\&&
\label{2.6}
\end{eqnarray}
which are the vacuum   Maxwell equations.
On introducing the abbreviations
\begin{eqnarray}
&&\vec{E}_{{}_P} \equiv  \vec{E}_{{}_M}(\vec{r}_{{}_P},t) \; , \qquad
  \vec{B}_{{}_P} \equiv  \vec{B}_{{}_M}(\vec{r}_{{}_P},t) \; ,
\label{2.7}
\end{eqnarray}
one then obtains $\lbrack h \rbrack$:
\begin{eqnarray}
&& 0 = \vec{\nabla}_{{}_P} \wtimes \vec{B}_{{}_P}-
                  \frac{\partial \vec{E}_{{}_P}}{\partial t} \; , \qquad
   0 = \vec{\nabla}_{{}_P} \wtimes \vec{E}_{{}_P} +
                  \frac{\partial \vec{B}_{{}_P}}{\partial t} \; , \qquad
   0 = \vec{\nabla}_{{}_P}^\dagger\cdot \vec{E}_{{}_P} \; , \qquad
   0 = \vec{\nabla}_{{}_P}^\dagger\cdot \vec{B}_{{}_P} \; ,
\nonumber\\
\label{2.8}
\end{eqnarray}
which also are the vacuum   Maxwell equations.
The transformation properties of $\vec{r}_{{}_P},t,\vec{v}_{{}_P},
\vec{s},\vec{B}_{{}_P},\vec{E}_{{}_P}$ under the
$\rm{Poincar\acute{e}}$ group are discussed in detail in section 5.
\subsection*{2.2}
In Appendix A it is shown that the DK equations (1.5) lead by (2.4) to:
\begin{eqnarray}
 && \vec{s}\ ' =  \frac{e}{m}\cdot \lbrack
 \frac{g-2}{2}\cdot\gamma_{{}_P}\cdot
 \vec{v}_{{}_P}^{\,\dagger}\cdot\vec{s}
\cdot
    (\vec{E}_{{}_P}+\vec{v}_{{}_P}\wtimes\vec{B}_{{}_P})
+\frac{1}{\gamma_{{}_P}}\cdot
 \vec{v}_{{}_P}^\dagger\cdot\vec{E}_{{}_P}\cdot\vec{s}
\nonumber\\&&
-\frac{g-2}{2}\cdot\gamma_{{}_P}\cdot
 \vec{v}_{{}_P}^{\,\dagger}\cdot\vec{s}
\cdot
        \vec{v}_{{}_P}^\dagger\cdot\vec{E}_{{}_P}\cdot \vec{v}_{{}_P}
+\frac{g}{2}\cdot\frac{1}{\gamma_{{}_P}}\cdot
            (\vec{s}\wtimes\vec{B}_{{}_P})
-\frac{g}{2}\cdot\frac{1}{\gamma_{{}_P}}\cdot
\vec{s}^{\,\dagger}\cdot\vec{E}_{{}_P}\cdot \vec{v}_{{}_P} \rbrack \; .
\label{2.9}
\end{eqnarray}
\subsection*{2.3}
Moreover in Appendix B it is shown that the DK equations
(1.5) lead by (2.3) to:
\begin{eqnarray}
&& m\cdot (\gamma_{{}_P}\cdot\vec{v}_{{}_P})' =
    e\cdot (\vec{v}_{{}_P}\wtimes\vec{B}_{{}_P}) + e\cdot\vec{E}_{{}_P} +
\frac{e\cdot g}{2\cdot m\cdot\gamma_{{}_P}}\cdot \vec{\nabla}_{{}_P}
                                                               \biggl(
\vec{s}^{\,\dagger}\cdot\vec{B}_{{}_P}-\vec{E}_{{}_P}^\dagger
         \cdot(\vec{s}\wtimes\vec{v}_{{}_P}) \biggr)
\nonumber\\&&\qquad
+\frac{e\cdot \gamma_{{}_P}}{2\cdot m}\cdot \lbrack
 2\cdot\vec{s}^{\,\dagger}\cdot\vec{B}_{{}_P}'\cdot\vec{v}_{{}_P}
-g\cdot(\vec{s}\wtimes\vec{v}_{{}_P})^\dagger\cdot
  \vec{E}_{{}_P}'\cdot\vec{v}_{{}_P}
+(g-2)\cdot(\vec{E}_{{}_P}'\wtimes\vec{s})
\nonumber\\&&\qquad
+(g-2)\cdot\vec{v}_{{}_P}^\dagger\cdot\vec{s}\cdot\vec{B}_{{}_P}'
-(g-2)\cdot\vec{v}_{{}_P}^\dagger\cdot\vec{E}_{{}_P}'
     \cdot(\vec{v}_{{}_P}\wtimes\vec{s})\rbrack
\nonumber\\&&\qquad
+\frac{e^2}{4\cdot m^2}\cdot\lbrack
   -(g-2)^2\cdot
   \vec{v}_{{}_P}^\dagger\cdot\vec{B}_{{}_P}\cdot
                                      (\vec{B}_{{}_P} \wtimes\vec{s})
   +(g-2)^2\cdot
   \vec{B}_{{}_P}^\dagger\cdot\vec{B}_{{}_P}\cdot
                                 (\vec{v}_{{}_P} \wtimes\vec{s})
\nonumber\\&&\qquad
   -(g-2)^2\cdot
   \vec{s}^{\,\dagger}\cdot\vec{E}_{{}_P}
\cdot\vec{B}_{{}_P}
   +(-g^2\cdot \vec{v}_{{}_P}^\dagger\cdot\vec{v}_{{}_P}
   +2\cdot g\cdot \vec{v}_{{}_P}^\dagger\cdot\vec{v}_{{}_P} +4)\cdot
   \vec{s}^{\,\dagger}\cdot\vec{B}_{{}_P}\cdot\vec{E}_{{}_P}
\nonumber\\&&\qquad
   -(g-2)^2\cdot
   \vec{v}_{{}_P}^\dagger\cdot\vec{E}_{{}_P}\cdot
                                      (\vec{E}_{{}_P} \wtimes\vec{s})
   +(g-2)^2\cdot
       ( \vec{v}_{{}_P}\wtimes\vec{B}_{{}_P})^\dagger
 \cdot\vec{E}_{{}_P} \cdot(\vec{v}_{{}_P}\wtimes\vec{s})
\nonumber\\&&\qquad
   +(g-2)^2\cdot
       \vec{E}_{{}_P}^\dagger\cdot\vec{E}_{{}_P} \cdot
                                      (\vec{v}_{{}_P}\wtimes\vec{s})
   -(g^2-4\cdot g)\cdot
   \vec{s}^{\,\dagger}\cdot\vec{B}_{{}_P}\cdot
                               (\vec{v}_{{}_P} \wtimes\vec{B}_{{}_P})
\nonumber\\&&\qquad
   +(g-2)\cdot g\cdot
   \vec{E}_{{}_P}^\dagger\cdot\vec{B}_{{}_P}\cdot \vec{s}
   -(g-2)\cdot g\cdot
   \vec{v}_{{}_P}^\dagger\cdot\vec{E}_{{}_P}\cdot
          \vec{v}_{{}_P}^\dagger\cdot\vec{B}_{{}_P}\cdot\vec{s}
\nonumber\\&&\qquad
   +(g-2)\cdot g\cdot
   \vec{v}_{{}_P}^\dagger\cdot\vec{E}_{{}_P}\cdot
           \vec{s}^{\,\dagger}\cdot\vec{B}_{{}_P}\cdot\vec{v}_{{}_P}
   +(g-2)\cdot g\cdot
   \vec{v}_{{}_P}^\dagger\cdot\vec{s}\cdot
            \vec{v}_{{}_P}^\dagger\cdot\vec{B}_{{}_P}\cdot\vec{E}_{{}_P}
\nonumber\\&&\qquad
   -(g^2-4\cdot g)\cdot
  \vec{E}_{{}_P}^\dagger\cdot( \vec{v}_{{}_P}\wtimes
                                             \vec{s})\cdot\vec{E}_{{}_P}
   +2\cdot g\cdot
\vec{E}_{{}_P}^\dagger\cdot( \vec{v}_{{}_P}\wtimes
      \vec{s})\cdot(\vec{v}_{{}_P}\wtimes\vec{B}_{{}_P})
\rbrack   \; .
\label{2.10}
\end{eqnarray}
%
%Note that the $\vec{E}_{{}_P}'$ and $\vec{B}_{{}_P}'$ factors
%only   appear in SG terms, i.e. in leading order in
%spin.
%
\subsection*{2.4}
On collecting (2.3a),(2.9-10) one finds   that the DK
equations (1.5) are equivalent to:
\setcounter{INDEX}{1}
\begin{eqnarray}
 && \vec{r}^{\;{}_{'}}_{{}_P}
                              =  \vec{v}_{{}_P} \; ,
\\
\addtocounter{equation}{-1}
\addtocounter{INDEX}{1}
\label{2.11a}
 && m\cdot (\gamma_{{}_P}\cdot\vec{v}_{{}_P})' =
    e\cdot (\vec{v}_{{}_P}\wtimes\vec{B}_{{}_P}) + e\cdot\vec{E}_{{}_P} +
\frac{e\cdot g}{2\cdot m\cdot\gamma_{{}_P}}\cdot \vec{\nabla}_{{}_P}
                                                               \biggl(
\vec{s}^{\,\dagger}\cdot\vec{B}_{{}_P}-\vec{E}_{{}_P}^\dagger
         \cdot(\vec{s}\wtimes\vec{v}_{{}_P}) \biggr)
\nonumber\\&&\qquad
+\frac{e\cdot \gamma_{{}_P}}{2\cdot m}\cdot \lbrack
 2\cdot\vec{s}^{\,\dagger}\cdot\vec{B}_{{}_P}'\cdot\vec{v}_{{}_P}
-g\cdot(\vec{s}\wtimes\vec{v}_{{}_P})^\dagger\cdot
  \vec{E}_{{}_P}'\cdot\vec{v}_{{}_P}
+(g-2)\cdot(\vec{E}_{{}_P}'\wtimes\vec{s})
\nonumber\\&&\qquad
+(g-2)\cdot\vec{v}_{{}_P}^\dagger\cdot\vec{s}\cdot\vec{B}_{{}_P}'
-(g-2)\cdot\vec{v}_{{}_P}^\dagger\cdot\vec{E}_{{}_P}'
     \cdot(\vec{v}_{{}_P}\wtimes\vec{s})\rbrack
\nonumber\\&&\qquad
+\frac{e^2}{4\cdot m^2}\cdot\lbrack
   -(g-2)^2\cdot
   \vec{v}_{{}_P}^\dagger\cdot\vec{B}_{{}_P}\cdot
                                      (\vec{B}_{{}_P} \wtimes\vec{s})
   +(g-2)^2\cdot
   \vec{B}_{{}_P}^\dagger\cdot\vec{B}_{{}_P}\cdot
                                 (\vec{v}_{{}_P} \wtimes\vec{s})
\nonumber\\&&\qquad
   -(g-2)^2\cdot
   \vec{s}^{\,\dagger}\cdot\vec{E}_{{}_P}
\cdot\vec{B}_{{}_P}
   +(-g^2\cdot \vec{v}_{{}_P}^\dagger\cdot\vec{v}_{{}_P}
   +2\cdot g\cdot \vec{v}_{{}_P}^\dagger\cdot\vec{v}_{{}_P} +4)\cdot
   \vec{s}^{\,\dagger}\cdot\vec{B}_{{}_P}\cdot\vec{E}_{{}_P}
\nonumber\\&&\qquad
   -(g-2)^2\cdot
   \vec{v}_{{}_P}^\dagger\cdot\vec{E}_{{}_P}\cdot
                                      (\vec{E}_{{}_P} \wtimes\vec{s})
   +(g-2)^2\cdot
       ( \vec{v}_{{}_P}\wtimes\vec{B}_{{}_P})^\dagger
 \cdot\vec{E}_{{}_P} \cdot(\vec{v}_{{}_P}\wtimes\vec{s})
\nonumber\\&&\qquad
   +(g-2)^2\cdot
       \vec{E}_{{}_P}^\dagger\cdot\vec{E}_{{}_P} \cdot
                                      (\vec{v}_{{}_P}\wtimes\vec{s})
   -(g^2-4\cdot g)\cdot
   \vec{s}^{\,\dagger}\cdot\vec{B}_{{}_P}\cdot
                               (\vec{v}_{{}_P} \wtimes\vec{B}_{{}_P})
\nonumber\\&&\qquad
   +(g-2)\cdot g\cdot
   \vec{E}_{{}_P}^\dagger\cdot\vec{B}_{{}_P}\cdot \vec{s}
   -(g-2)\cdot g\cdot
   \vec{v}_{{}_P}^\dagger\cdot\vec{E}_{{}_P}\cdot
          \vec{v}_{{}_P}^\dagger\cdot\vec{B}_{{}_P}\cdot\vec{s}
\nonumber\\&&\qquad
   +(g-2)\cdot g\cdot
   \vec{v}_{{}_P}^\dagger\cdot\vec{E}_{{}_P}\cdot
           \vec{s}^{\,\dagger}\cdot\vec{B}_{{}_P}\cdot\vec{v}_{{}_P}
   +(g-2)\cdot g\cdot
   \vec{v}_{{}_P}^\dagger\cdot\vec{s}\cdot
            \vec{v}_{{}_P}^\dagger\cdot\vec{B}_{{}_P}\cdot\vec{E}_{{}_P}
\nonumber\\&&\qquad
   -(g^2-4\cdot g)\cdot
  \vec{E}_{{}_P}^\dagger\cdot( \vec{v}_{{}_P}\wtimes
                                             \vec{s})\cdot\vec{E}_{{}_P}
   +2\cdot g\cdot
\vec{E}_{{}_P}^\dagger\cdot( \vec{v}_{{}_P}\wtimes
      \vec{s})\cdot(\vec{v}_{{}_P}\wtimes\vec{B}_{{}_P})
\rbrack   \; ,
\\
\addtocounter{equation}{-1}
\addtocounter{INDEX}{1}
\label{2.11b}
 && \vec{s}\ ' =
      \frac{e}{m}\cdot \lbrack
 \frac{g-2}{2}\cdot\gamma_{{}_P}\cdot
 \vec{v}_{{}_P}^{\,\dagger}\cdot\vec{s}
\cdot
    (\vec{E}_{{}_P}+\vec{v}_{{}_P}\wtimes\vec{B}_{{}_P})
+\frac{1}{\gamma_{{}_P}}\cdot
 \vec{v}_{{}_P}^\dagger\cdot\vec{E}_{{}_P}\cdot\vec{s}
\nonumber\\&&
-\frac{g-2}{2}\cdot\gamma_{{}_P}\cdot
 \vec{v}_{{}_P}^{\,\dagger}\cdot\vec{s}
\cdot
        \vec{v}_{{}_P}^\dagger\cdot\vec{E}_{{}_P}\cdot \vec{v}_{{}_P}
+\frac{g}{2}\cdot\frac{1}{\gamma_{{}_P}}\cdot
            (\vec{s}\wtimes\vec{B}_{{}_P})
-\frac{g}{2}\cdot\frac{1}{\gamma_{{}_P}}\cdot
\vec{s}^{\,\dagger}\cdot\vec{E}_{{}_P}\cdot \vec{v}_{{}_P} \rbrack \; .
\label{2.11c}
\end{eqnarray}
\setcounter{INDEX}{0}
\par\noindent
The covariance of (2.11) under the $\rm{Poincar\acute{e}}$ group will be
demonstrated in section 5. By neglecting the spin terms in (2.11b) one gets
the familiar relations:
\begin{eqnarray}
 && m\cdot (\gamma_{{}_P}\cdot\vec{v}_{{}_P})'  =
    e\cdot(\vec{v}_{{}_P}\wtimes\vec{B}_{{}_P})+
                                            e\cdot \vec{E}_{{}_P} \; ,
\qquad \gamma_{{}_P}'  =
    \frac{e}{m}\cdot\vec{v}_{{}_P}^\dagger\cdot\vec{E}_{{}_P} \; .
\label{2.12}
\end{eqnarray}
\setcounter{equation}{0}
\section{The equivalence between the DK equations and
the Frenkel equations}
\subsection*{3.1}
In this section the DK equations are shown to be
equivalent to the Frenkel equations.
\par To study the covariance properties of $\vec{r}_{{}_P},
\vec{v}_{{}_P},\vec s,t$ I first construct the variables
$X^{{}^P},U^{{}^P}$ where the components
$X^{{}^P}_{\mu}$ are defined by $\lbrack i \rbrack$:
\begin{eqnarray}
&& X^{{}^P}_{\mu} = (\vec{r}_{{}_P}^{\,\dagger},i\cdot t)_{\mu} \;,
 \qquad (\mu =1,...,4)
\label{3.1}
\end{eqnarray}
and represent the space-time position. The components
$U^{{}^P}_{\mu}$ denote the corresponding 4-velocity, i.e.
\footnote{In this paper the upper dot symbol denotes the total proper
time derivative.}
\begin{eqnarray}
&& U^{{}^P}_{\mu} = \dot{X}^{{}^P}_{\mu}\equiv
   \frac{d}{d\tau} \; X^{{}^P}_{\mu} =
\gamma_{{}_P}\cdot \frac{d}{dt}\; X^{{}^P}_{\mu} =
(\gamma_{{}_P}\cdot\vec{v}_{{}_P}^\dagger,i\cdot
                                          \gamma_{{}_P})_{\mu} \; ,
 \qquad (\mu =1,...,4)
\label{3.2}
\end{eqnarray}
where $\tau$ denotes the proper time.
\par The utility   of the spin vector $\vec{s}$ becomes
apparent by encoding it
in a spin tensor $S^{{}^P}$. To achieve this I first introduce the `dipole
moment tensor' $\lbrack j \rbrack$:
\begin{eqnarray}
 && M^{{}^P}_{\mu\nu} = \left( \begin{array}{cccc}
      0 & \mu_{{}_{P,3}} & -\mu_{{}_{P,2}} & i\cdot\varepsilon_{{}_{P,1}}   \\
     -\mu_{{}_{P,3}} & 0 & \mu_{{}_{P,1}} & i\cdot \varepsilon_{{}_{P,2}}   \\
      \mu_{{}_{P,2}} & -\mu_{{}_{P,1}} & 0 & i\cdot \varepsilon_{{}_{P,3}}   \\
      -i\cdot \varepsilon_{{}_{P,1}}   &
      -i\cdot \varepsilon_{{}_{P,2}}   &
      -i\cdot \varepsilon_{{}_{P,3}}   &  0
                \end{array}
         \right)_{\mu\nu}
                \; ,
 \qquad (\mu,\nu =1,...,4)
\label{3.3}
\end{eqnarray}
which by definition transforms as a tensor of rank 2.
One calls $\vec{\mu}_{{}_P}$ the `magnetic dipole moment' and
$\vec{\varepsilon}_{{}_P}$ the `electric dipole moment'
\cite{Nyb64}. The magnetic dipole moment resp.
the electric dipole moment in the rest frame is denoted by
$\vec{\mu}_{{}_R}$ resp. $\vec{\varepsilon}_{{}_R}$ and by definition it
is the intrinsic magnetic dipole moment resp. intrinsic electric dipole
moment of the particle.
The tensor $M^{{}^P}$ appears
frequently in the literature on the relativistic SG force and it occurs
already in the 1926 paper of Frenkel \cite{Fre26}.
Note that $M^{{}^P}$ is also used in the theory of relativistic fluids
resp. composite particles.
\footnote{See for example the textbook treatments in \cite{DS72,Moe72,Syn58}.}
\par Although $M^{{}^P}$ is especially useful for treating particles with
combined intrinsic magnetic and intrinsic electric dipole moments, such
particles are not treated
in this paper. In this study I only consider particles without intrinsic
electric
dipole moment. Denoting the dipole moment tensor in the rest frame by
$M^{{}^R}$, I then have:
\begin{eqnarray*}
 && M^{{}^R}_{\mu\nu} = \left( \begin{array}{cccc}
      0 & \mu_{{}_{R,3}} & -\mu_{{}_{R,2}} & 0    \\
     -\mu_{{}_{R,3}} & 0 & \mu_{{}_{R,1}} &  0    \\
      \mu_{{}_{R,2}} & -\mu_{{}_{R,1}} & 0 & 0    \\
       0  &
       0  &
       0  &  0
                \end{array}
         \right)_{\mu\nu}
                \; .
 \qquad (\mu,\nu =1,...,4)
\end{eqnarray*}
Bearing in mind that for a charged particle the intrinsic magnetic dipole 
moment is related to the rest frame spin by:
\begin{eqnarray}
&&\vec{\mu}_{{}_R} \equiv \frac{e\cdot g}{2\cdot m}\cdot\vec{\sigma} \; ,
\label{3.4}
\end{eqnarray}
one obtains:
\begin{eqnarray*}
 && M^{{}^R}_{\mu\nu} = \frac{e\cdot g}{2\cdot m}\cdot
\left( \begin{array}{cccc}
      0 & \sigma_3 & -\sigma_2 & 0    \\
     -\sigma_3 & 0 & \sigma_1 &  0    \\
      \sigma_2 & -\sigma_1 & 0 & 0    \\
       0  &
       0  &
       0  &  0
                \end{array}
         \right)_{\mu\nu}
                \; .
 \qquad (\mu,\nu =1,...,4)
\end{eqnarray*}
One now  defines the spin tensor $S^{{}^P}$ as that tensor, which in
the rest frame reads as:
\begin{eqnarray}
  S^{{}^R}_{\mu\nu} &=&
 \left( \begin{array}{cccc}
      0 & \sigma_3 & -\sigma_2 & 0    \\
     -\sigma_3 & 0 & \sigma_1 &  0    \\
      \sigma_2 & -\sigma_1 & 0 & 0    \\
       0  &
       0  &
       0  &  0
                \end{array}
         \right)_{\mu\nu}
                \; .
 \qquad (\mu,\nu =1,...,4)
\label{3.5}
\end{eqnarray}
Thus the spin tensor $S^{{}^P}$ is given by:
\begin{eqnarray*}
&& M^{{}^P} = \frac{e\cdot g}{2\cdot m}\cdot S^{{}^P}  \; .
\end{eqnarray*}
By transforming back from the rest frame
(see subsection 7.2) one obtains:
\footnote{The corresponding operator $S^{{}^{P,op}}$ is obtained in 
\cite{FG61a,HW63}.}
\begin{eqnarray}
 && S^{{}^P}_{\mu\nu} = \left( \begin{array}{cccc}
      0 & s_3 & -s_2 &-i\cdot q_1     \\
     -s_3 & 0 & s_1 &-i\cdot q_2     \\
      s_2 & -s_1 & 0 &-i\cdot q_3     \\
      i\cdot q_1   &
      i\cdot q_2   &
      i\cdot q_3   &  0
                \end{array}
         \right)_{\mu\nu}
                \; , \qquad
  \vec{q} = \vec{s} \wtimes \vec{v}_{{}_P} \; ,
 \qquad (\mu,\nu =1,...,4)
\label{3.6}
\end{eqnarray}
so that one has
\begin{eqnarray}
&&\vec{\mu}_{{}_P} \equiv \frac{e\cdot g}{2\cdot m}\cdot\vec{s} \; , \qquad
  \vec{\varepsilon}_{{}_P} \equiv \frac{e\cdot g}{2\cdot m}\cdot
(\vec{v}_{{}_P}\wtimes \vec s) \; .
\label{3.7}
\end{eqnarray}
For a particle with vanishing intrinsic electric dipole moment
equations (3.2),(3.6) enshrine the kinematic
constraints $\lbrack k \rbrack$:
\setcounter{INDEX}{1}
\begin{eqnarray}
 && U^{{}^P}_{\mu}\cdot  U^{{}^P}_{\mu} = -1 \; ,
\\
\addtocounter{equation}{-1}
\addtocounter{INDEX}{1}
\label{3.8a}
 && S^{{}^P}_{\mu\nu}\cdot U^{{}^P}_{\nu} = 0 \; .
 \qquad (\mu =1,...,4)
%\\
%\addtocounter{equation}{-1}
%\addtocounter{INDEX}{1}
\label{3.8b}
% && \sqrt{S^{{}^P}_{\mu\nu}\cdot S^{{}^P}_{\mu\nu}} =
%    \sqrt{2\cdot\vec{\sigma}^{\,\dagger}\cdot\vec{\sigma}} =
%  \hbar/\sqrt{2} \; .
%\label{3.8c}
\end{eqnarray}
\setcounter{INDEX}{0}
\par\noindent
With (3.6) I have encoded the spin vector $\vec s$ in the spin tensor
$S^{{}^P}$.
For the consistency of the rank 2 tensor property of $S^{{}^P}$ with the
transformation properties of the spin vector $\vec{s}$ and of
$\vec{v}_{{}_P}$, see subsections 5.4 and 5.5.
\par I also introduce the rank two antisymmetric tensor field describing
the electromagnetic field and whose components are defined by
\begin{eqnarray}
 && F^{{}^P}_{\mu\nu} \equiv \left( \begin{array}{cccc}
      0 & B_{{}_{P,3}} & -B_{{}_{P,2}} & -i\cdot E_{{}_{P,1}}   \\
-B_{{}_{P,3}} & 0 & B_{{}_{P,1}} & -i\cdot E_{{}_{P,2}}     \\
 B_{{}_{P,2}} & -B_{{}_{P,1}} & 0 & -i\cdot E_{{}_{P,3}}     \\
      i\cdot E_{{}_{P,1}}   &
      i\cdot E_{{}_{P,2}}   &
      i\cdot E_{{}_{P,3}}   &  0
                \end{array}
         \right)_{\mu\nu} \; ,
 \qquad (\mu,\nu =1,...,4)
\label{3.9}
\end{eqnarray}
i.e.
\begin{eqnarray*}
&& F^{{}^P} \leftrightarrow
     (\vec{B}_{{}_P},
-i\cdot\vec{E}_{{}_P})    \; .
\end{eqnarray*}
Then     $\lbrack l \rbrack$ with the definition:
\begin{eqnarray}
&&  \partial^{{}^P}_{\mu} \equiv
    (\frac{\partial}{\partial X^{{}^P}_1} \ , \
     \frac{\partial}{\partial X^{{}^P}_2} \ , \
     \frac{\partial}{\partial X^{{}^P}_3} \ , \
     \frac{\partial}{\partial X^{{}^P}_4})_{\mu} \; ,
 \qquad (\mu =1,...,4)
\label{3.10}
\end{eqnarray}
the vacuum Maxwell equations (2.8) read as:
\begin{eqnarray}
&& 0 = \partial^{{}^P}_{\rho} F^{{}^P}_{\mu\nu} +
    \partial^{{}^P}_{\mu} F^{{}^P}_{\nu\rho} +
    \partial^{{}^P}_{\nu} F^{{}^P}_{\rho\mu} \; , \qquad
0 = \partial^{{}^P}_{\mu} F^{{}^P}_{\mu\nu} \; .
 \qquad (\mu,\nu,\rho =1,...,4)
\label{3.11}
\end{eqnarray}
\subsection*{3.2}
I can now reexpress the evolution (2.11) of $\vec{r}_{{}_P}(t),
\vec{v}_{{}_P}(t), \vec{s}(t)$ in terms of
$X^{{}^P}(\tau),U^{{}^P}(\tau), S^{{}^P}(\tau)$. In
subsection 5.2 it is shown that equations (2.11) are
equivalent to:
\setcounter{INDEX}{1}
\begin{eqnarray}
 && \dot{X}^{{}^P}_{\mu} = U^{{}^P}_{\mu} \; ,
\\
\addtocounter{equation}{-1}
\addtocounter{INDEX}{1}
\label{3.12a}
&&\dot{U}^{{}^P}_{\mu}=
     \frac{e}{m}\cdot F^{{}^P}_{\mu\nu}\cdot U^{{}^P}_{\nu}
  -   \frac{e\cdot g}{4\cdot m^2}\cdot\biggl(
      S^{{}^P}_{\nu\omega}\cdot \partial^{{}^P}_{\mu} F^{{}^P}_{\omega\nu}
+U^{{}^P}_{\mu}\cdot S^{{}^P}_{\nu\omega}\cdot U^{{}^P}_{\lambda}\cdot
                                        \partial^{{}^P}_{\lambda}
                            F^{{}^P}_{\omega\nu}  \biggr)
\nonumber\\&&\qquad
  +   \frac{e\cdot(g-2)}{2\cdot m^2}\cdot
              S^{{}^P}_{\mu\nu}\cdot U^{{}^P}_{\omega}\cdot
 U^{{}^P}_{\lambda}\cdot\partial^{{}^P}_{\lambda} F^{{}^P}_{\nu\omega}
  -  \frac{e^2\cdot(g-2)^2}{4\cdot m^3}\cdot
S^{{}^P}_{\mu\nu}\cdot F^{{}^P}_{\nu\omega}\cdot
F^{{}^P}_{\omega\rho}\cdot U^{{}^P}_{\rho}
\nonumber\\&&\qquad
  +  \frac{e^2\cdot(g-2)\cdot g}{4\cdot m^3}\cdot
F^{{}^P}_{\mu\nu}\cdot S^{{}^P}_{\nu\omega}\cdot
        F^{{}^P}_{\omega\rho}\cdot U^{{}^P}_{\rho}
  -   \frac{e^2\cdot g}{4\cdot m^3}\cdot
              F^{{}^P}_{\mu\nu}\cdot U^{{}^P}_{\nu}\cdot
              F^{{}^P}_{\lambda\omega}\cdot S^{{}^P}_{\omega\lambda} \; ,
\\
\addtocounter{equation}{-1}
\addtocounter{INDEX}{1}
\label{3.12b}
 && \dot{S}^{{}^P}_{\mu\nu} =  \frac{e\cdot g}{2\cdot m}\cdot\biggl(
                   F^{{}^P}_{\mu\omega}\cdot S^{{}^P}_{\omega\nu}-
                   S^{{}^P}_{\mu\omega}\cdot F^{{}^P}_{\omega\nu} \biggr)
\nonumber\\&&
  -   \frac{e\cdot (g-2)}{2\cdot m}\cdot\biggl(
S^{{}^P}_{\mu\omega}\cdot F^{{}^P}_{\omega\lambda}\cdot
                              U^{{}^P}_{\lambda}\cdot U^{{}^P}_{\nu}-
S^{{}^P}_{\nu\omega}\cdot F^{{}^P}_{\omega\lambda}\cdot
   U^{{}^P}_{\lambda}\cdot U^{{}^P}_{\mu}
                                           \biggr) \; .
 \qquad (\mu,\nu =1,...,4) \qquad
\label{3.12c}
\end{eqnarray}
\setcounter{INDEX}{0}
\par\noindent
These are the equations given by Frenkel in 1926 \cite{Fre26}.
\footnote{Equations (3.12) are equivalent to equations (13a),(14),(21),
(21a-b) in \cite{Fre26}. They were rederived by many authors. See also
the reviews in \cite{BT80,Nyb62,Pla66b,Roh72,TVW80}.}
They respect the kinematic constraints (3.8) and I can therefore
conclude that:
\begin{itemize}
\item The DK equations are equivalent to the Frenkel
equations.
\end{itemize}
In this derivation it was essential that the  electromagnetic
field obeys the vacuum Maxwell equations.
The covariance of (3.12) under the
$\rm{Poincar\acute{e}}$ group will be demonstrated in
section 5. Note also that (3.12c) is equivalent to the BMT equation
\cite{BMT59,Nyb64}. The term
\begin{eqnarray*}
 &&  \frac{e\cdot g}{2\cdot m}\cdot\biggl(
                   F^{{}^P}_{\mu\omega}\cdot S^{{}^P}_{\omega\nu}-
                   S^{{}^P}_{\mu\omega}\cdot F^{{}^P}_{\omega\nu}
-  S^{{}^P}_{\mu\omega}\cdot F^{{}^P}_{\omega\lambda}\cdot
                              U^{{}^P}_{\lambda}\cdot U^{{}^P}_{\nu} +
S^{{}^P}_{\nu\omega}\cdot F^{{}^P}_{\omega\lambda}\cdot
   U^{{}^P}_{\lambda}\cdot U^{{}^P}_{\mu}  \biggr)
\end{eqnarray*}
is that part of the rhs of (3.12c) which is independent of the forces
acting on the orbital motion. The remaining term
\begin{eqnarray*}
&& \frac{e}{m}\cdot\biggl(
S^{{}^P}_{\mu\omega}\cdot F^{{}^P}_{\omega\lambda}\cdot
                              U^{{}^P}_{\lambda}\cdot U^{{}^P}_{\nu}-
S^{{}^P}_{\nu\omega}\cdot F^{{}^P}_{\omega\lambda}\cdot
   U^{{}^P}_{\lambda}\cdot U^{{}^P}_{\mu}
                                           \biggr)
\end{eqnarray*}
represents the Thomas precession \cite{Tho27}.
\par It follows
      from the normalization of the spin vector that:
\begin{eqnarray}
 && S^{{}^P}_{\mu\nu}\cdot S^{{}^P}_{\mu\nu} =
  \hbar^2/2 \; .
\label{3.13}
\end{eqnarray}
As in
   (1.9) this equation is of second order in spin so that it plays no role
in this paper. Note also that (3.13) is conserved under (3.12c).
\setcounter{equation}{0}
\section{Rederiving the Frenkel equations in terms of a Hamiltonian which is
a Poincare scalar}
\subsection*{4.1}
The reader will perhaps be interested to learn that one can rederive
the Frenkel equations from   a Hamiltonian
with the proper time as independent variable.
\par One begins by noting that $X^{{}^P}, P^{{}^P}$ and $S^{{}^P}$
obey the following Poisson
bracket relations \cite{Cor68}:
\footnote{In the quantum mechanical analogue to the system (4.1)
the position operator correponding to $\vec{r}_{{}_P}$ must be different from 
$\vec{r}_{{}_{P,op}}$ because this system involves (unlike the system for
the `M' variables) negative energy states. See for example \cite{JM63,Cor68}.}
\begin{eqnarray}
&& \lbrace X^{{}^P}_{\mu} ,
     P^{{}^P}_{\nu} \rbrace_{{}_P} =  \delta_{\mu\nu} \; ,
\nonumber\\
 && \lbrace S^{{}^P}_{\mu\nu},  S^{{}^P}_{\lambda\omega}  \rbrace_{{}_P} =
            S^{{}^P}_{\mu\lambda}\cdot\delta_{\nu\omega}
       +    S^{{}^P}_{\nu\omega}\cdot\delta_{\mu\lambda}
       -    S^{{}^P}_{\mu\omega}\cdot\delta_{\nu\lambda}
       -    S^{{}^P}_{\nu\lambda}\cdot\delta_{\mu\omega} \; ,
\nonumber\\
 && \lbrace X^{{}^P}_{\mu} , X^{{}^P}_{\nu} \rbrace_{{}_P}
      = \lbrace X^{{}^P}_{\mu} , S^{{}^P}_{\nu\lambda} \rbrace_{{}_P}
      = \lbrace P^{{}^P}_{\mu} , P^{{}^P}_{\nu} \rbrace_{{}_P}
      = \lbrace P^{{}^P}_{\mu} , S^{{}^P}_{\nu\lambda} \rbrace_{{}_P} = 0 \; .
 \;\; (\mu,\nu,\lambda,\omega =1,...,4) \qquad
\label{4.1}
\end{eqnarray}
Then I introduce the abbreviations
\begin{eqnarray}
 && M^{{}^P} \equiv  m + \frac{e\cdot g}{4\cdot m}\cdot
              S^{{}^P}_{\nu\omega}\cdot F^{{}^P}_{\omega\nu} \; ,
\nonumber\\
 && A^{{}^P}_{\mu} \equiv
       (\vec{A}_{{}_P}^\dagger,i\cdot \phi_{{}_P})_{\mu} \; ,
\nonumber\\
 && \Pi^{{}^P}_{\mu} \equiv P^{{}^P}_{\mu} - e\cdot A^{{}^P}_{\mu}
\equiv (\vec{\pi}_{{}_P}^\dagger, \Pi^{{}^P}_4)_{\mu} \; ,
 \qquad (\mu =1,...,4)
\label{4.2}
\end{eqnarray}
where $\vec{A}_{{}_P}(\vec{r}_{{}_P},t),\phi_{{}_P}(\vec{r}_{{}_P},t)$
are the potentials of the  electromagnetic field so that:
\begin{eqnarray}
&& \vec{B}_{{}_P} = \vec{\nabla}_{{}_P} \wtimes
                                          \vec{A}_{{}_P} \; , \qquad
  \vec{E}_{{}_P} = - \vec{\nabla}_{{}_P}\phi_{{}_P} -
                  \frac{\partial \vec{A}_{{}_P}}{\partial t} \; .
\label{4.3}
\end{eqnarray}
i.e:
\begin{eqnarray}
 && F^{{}^P}_{\mu\nu} =  \partial^{{}^P}_{\mu} A^{{}^P}_{\nu} -
                         \partial^{{}^P}_{\nu} A^{{}^P}_{\mu} \; .
 \qquad (\mu,\nu =1,...,4)
\label{4.4}
\end{eqnarray}
The Hamiltonian is  $\lbrack m \rbrack$:
\begin{eqnarray}
 && H^{{}^P} = (\frac{1}{m}-\frac{M^{{}^P}}{2\cdot m^2})\cdot
 \Pi^{{}^P}_{\mu}\cdot \Pi^{{}^P}_{\mu}  +  \frac{1}{2}\cdot M^{{}^P}
   + \frac{e\cdot(g-2)}{2\cdot m^3}\cdot
     \Pi^{{}^P}_{\mu}\cdot S^{{}^P}_{\mu\nu}\cdot F^{{}^P}_{\nu\omega}\cdot
     \Pi^{{}^P}_{\omega}  \; ,
\label{4.5}
\end{eqnarray}
and  the corresponding equations of motion are:
\begin{eqnarray}
 && \dot{X}^{{}^P}_{\mu}  =
   \lbrace X^{{}^P}_{\mu},  H^{{}^P} \rbrace_{{}_P} \; ,
\nonumber\\
 && \dot{U}^{{}^P}_{\mu}  =
   \lbrace U^{{}^P}_{\mu},  H^{{}^P} \rbrace_{{}_P}
=  \lbrace \lbrace X^{{}^P}_{\mu},  H^{{}^P} \rbrace_{{}_P}  ,
                                  H^{{}^P} \rbrace_{{}_P} \; ,
\nonumber\\
 && \dot{S}^{{}^P}_{\mu\nu} = \lbrace S^{{}^P}_{\mu\nu},  H^{{}^P}
\rbrace_{{}_P} \; .
 \qquad (\mu,\nu =1,...,4)
\label{4.6}
\end{eqnarray}
By evaluating the Poisson brackets one finds that these are the
Frenkel equations (3.12).
\par Because $S^{{}^P}$ is a tensor of rank 2 and $F^{{}^P}$ is a tensor field
of rank 2 it follows  by (4.2) that $M^{{}^P}$ is a scalar field.
Moreover, because $P^{{}^P}$ is a 4-vector and $A^{{}^P}$ a 4-vector
field, one finds   that all three terms of the Hamiltonian (4.5)
are scalar fields.
\subsection*{4.2}
With (4.5) one gets:
\begin{eqnarray}
&&\dot{\Pi}^{{}^P}_{\mu} =
 \lbrace \Pi^{{}^P}_{\mu},  H^{{}^P} \rbrace_{{}_P}
 =         \frac{e}{m}\cdot F^{{}^P}_{\mu\nu}\cdot \Pi^{{}^P}_{\nu}
- \frac{e\cdot g}{4\cdot m}\cdot S^{{}^P}_{\nu\omega}\cdot
     \partial^{{}^P}_{\mu} F^{{}^P}_{\omega\nu}
  +  \frac{e^2\cdot(g-2)}{2\cdot m^3}\cdot
F^{{}^P}_{\mu\nu}\cdot S^{{}^P}_{\nu\omega}\cdot
        F^{{}^P}_{\omega\rho}\cdot \Pi^{{}^P}_{\rho}
\nonumber\\&&  \qquad
  +   \frac{e^2\cdot g}{4\cdot m^3}\cdot
              F^{{}^P}_{\mu\nu}\cdot \Pi^{{}^P}_{\nu}\cdot
              F^{{}^P}_{\lambda\omega}\cdot S^{{}^P}_{\omega\lambda} \; ,
 \qquad (\mu =1,...,4) \qquad
\label{4.7}
\end{eqnarray}
and:
\begin{eqnarray}
 && U^{{}^P}_{\mu}  = (\frac{2}{m}-\frac{M^{{}^P}}{m^2})\cdot
 \Pi^{{}^P}_{\mu} + \frac{e\cdot(g-2)}{2\cdot m^3}\cdot
   S^{{}^P}_{\mu\nu}\cdot F^{{}^P}_{\nu\omega}\cdot
        \Pi^{{}^P}_{\omega}   \; . \qquad (\mu =1,...,4)
\label{4.8}
\end{eqnarray}
With (4.8) one can write (4.7) in the elegant form:
\begin{eqnarray}
&&\dot{\Pi}^{{}^P}_{\mu} = e\cdot F^{{}^P}_{\mu\nu}\cdot
 U^{{}^P}_{\nu}  -  \frac{e\cdot g}{4\cdot m}\cdot
 S^{{}^P}_{\nu\omega}\cdot \partial^{{}^P}_{\mu} F^{{}^P}_{\omega\nu} \; ,
 \qquad (\mu =1,...,4)
\label{4.9}
\end{eqnarray}
which will be useful in subsection 7.3.
\par Using (4.8) the constraints (3.8), expressed in terms of
the variables $X^{{}^P},P^{{}^P}$ and $S^{{}^P}$, read as:
\begin{eqnarray*}
&&\Pi^{{}^P}_{\mu}\cdot\Pi^{{}^P}_{\mu} = m^2 - 2\cdot m\cdot M^{{}^P}
 \; , \nonumber\\
 && S^{{}^P}_{\mu\nu}\cdot \Pi^{{}^P}_{\nu} = 0 \; ,
 \qquad (\mu =1,...,4) \nonumber\\
% && \sqrt{S^{{}^P}_{\mu\nu}\cdot S^{{}^P}_{\mu\nu}} =
%  \hbar/\sqrt{2} \; ,
\end{eqnarray*}
so that the Hamiltonian $H^{{}^P}$ vanishes, if the constraints are
taken into account. One thus has a constrained Hamiltonian system.
The constraints are to be taken into account only in the final results
(e.g. the equations of motion).
\subsection*{4.3}
In the `P' variables the canonical momentum is
$\vec{\pi}_{{}_P}$ and using (4.2),(4.8) and Appendix D I get:
\begin{eqnarray*}
 && m\cdot \gamma_{{}_P}\cdot\vec{v}_{{}_P}
 =  \vec{\pi}_{{}_P}  +\frac{e}{m^2}\cdot
 \vec{s}^{\,\dagger}\cdot\vec{B}_{{}_P}\cdot \vec{\pi}_{{}_P}
 -\frac{e\cdot g}{2\cdot m^3\cdot\gamma_{{}_P}}\cdot
   \vec{E}_{{}_P}^\dagger \cdot(\vec{s}\wtimes\vec{\pi}_{{}_P})
                                       \cdot \vec{\pi}_{{}_P}
\nonumber\\ &&\qquad
 +\frac{e\cdot (g-2)}{2\cdot m^2}\cdot
 \vec{\pi}_{{}_P}^{\,\dagger}\cdot\vec{s}
\cdot\vec{B}_{{}_P}
 +\frac{e\cdot (g-2)}{2\cdot m}\cdot \gamma_{{}_P}\cdot
  (\vec{E}_{{}_P}\wtimes\vec{s}) \; .
\end{eqnarray*}
I thus observe for the `P' variables that the canonical momentum
is different from the kinetic momentum, i.e.
\begin{eqnarray*}
&& \vec{\pi}_{{}_P} \neq m\cdot \gamma_{{}_P}\cdot \vec{v}_{{}_P} \; .
\end{eqnarray*}
As in case of the `M' variables (see the end of section 1), one has
zitterbewegung for the `P' variables. This effects disappears with the
electromagnetic field.
\footnote{One can modify the Hamiltonian (4.5) in a way such that
zitterbewegung arises even for the free particle. Then higher orders of the
spin are important \cite{Cor68,Pla66b}.}
\setcounter{equation}{0}
\section{Relating the Frenkel equations to other approaches. The
Poincare covariance}
\subsection*{5.1}
The Frenkel equations are just a special case allowed by the
kinematic constraints (3.8) and in fact various   forms for the SG
forces are possible even if one requires, as I do, that the spin
equation is equivalent to the BMT equation and that the
electromagnetic field obeys the vacuum Maxwell equations.
\footnote{An early work about the nonuniqueness of the relativistic SG force
is: \cite{Moe49}.}
In fact the constraints (3.8) allow the following generalization of
(3.12):
\setcounter{INDEX}{1}
\begin{eqnarray}
 && \dot{X}^{{}^P}_{\mu} = U^{{}^P}_{\mu} \; ,
\\
\addtocounter{equation}{-1}
\addtocounter{INDEX}{1}
\label{5.1a}
&&\dot{U}^{{}^P}_{\mu}= \frac{e}{m}\cdot
 F^{{}^P}_{\mu\nu}\cdot U^{{}^P}_{\nu}
  +   Y^{{}^P}_{\mu} \; ,
\\
\addtocounter{equation}{-1}
\addtocounter{INDEX}{1}
\label{5.1b}
 && \dot{S}^{{}^P}_{\mu\nu} =  \frac{e\cdot g}{2\cdot m}\cdot\biggl(
                   F^{{}^P}_{\mu\omega}\cdot S^{{}^P}_{\omega\nu}-
                   S^{{}^P}_{\mu\omega}\cdot F^{{}^P}_{\omega\nu} \biggr)
\nonumber\\&&
  -   \frac{e\cdot (g-2)}{2\cdot m}\cdot\biggl(
S^{{}^P}_{\mu\omega}\cdot F^{{}^P}_{\omega\lambda}\cdot U^{{}^P}_{\lambda}\cdot
                                            U^{{}^P}_{\nu}-
S^{{}^P}_{\nu\omega}\cdot F^{{}^P}_{\omega\lambda}\cdot U^{{}^P}_{\lambda}\cdot
                                            U^{{}^P}_{\mu}
                                           \biggr) \; ,
 \qquad (\mu,\nu =1,...,4)  \qquad
\label{5.1c}
\end{eqnarray}
\setcounter{INDEX}{0}
\par\noindent
where the 4-vector $Y^{{}^P}$ collects the SG force
terms.
%I require that $Y^{{}^P}$ depends only on
%$m,U^{{}^P}, e\cdot
%F^{{}^P},S^{{}^P},\partial^{{}^P}$ and on an arbitrary
%number of dimensionless constants $c_1,c_2,...$ and that the dependence
%on $S^{{}^P}$ be linear. If one only allows the algebraic tensor
%operations of addition, direct product and contraction then by
%dimensional analysis the dependence on $F^{{}^P}$ resp.
%$\partial^{{}^P}$ is at most quadratic resp. linear and the
%constraints (3.8) lead to the following most general ansatz:
I require that  $Y^{{}^P}$ only depends on the following
                   dimensional quantities:                              
$m,\, U^{{}^P},\, e\cdot F^{{}^P},\, S^{{}^P},\, \partial^{{}^P}$.
%and on an arbitrary
%number of dimensionless constants $c_1,c_2,...$.
If one assumes that the dependence on
$U^{{}^P}, e\cdot F^{{}^P},S^{{}^P},\partial^{{}^P}$
is polynomial and in particular of first order in
$S^{{}^P}$, then by dimensional analysis the dependence on $F^{{}^P}$ resp.
$\partial^{{}^P}$ is at most quadratic resp. linear and the
constraints (3.8) lead to the following most general ansatz
$\lbrack n \rbrack$:
\begin{eqnarray}
&& Y^{{}^P}_{\mu} =
   -  \frac{e\cdot c_2}{4\cdot m^2}\cdot\biggl(
S^{{}^P}_{\nu\omega}\cdot \partial^{{}^P}_{\mu} F^{{}^P}_{\omega\nu}
+U^{{}^P}_{\mu}\cdot S^{{}^P}_{\nu\omega}\cdot U^{{}^P}_{\lambda}
                                       \cdot\partial^{{}^P}_{\lambda}
                            F^{{}^P}_{\omega\nu}  \biggr)
\nonumber\\&&
  +   \frac{e\cdot(c_2-c_1-2)}{2\cdot m^2}\cdot
S^{{}^P}_{\mu\nu}\cdot U^{{}^P}_{\omega}\cdot
       U^{{}^P}_{\lambda}\cdot\partial^{{}^P}_{\lambda}
                            F^{{}^P}_{\nu\omega}
  +   \frac{e\cdot c_6}{2\cdot m^2}\cdot\biggl(
S^{{}^P}_{\nu\omega}\cdot \partial^{{}^P}_{\omega} F^{{}^P}_{\mu\nu}
+U^{{}^P}_{\mu}\cdot S^{{}^P}_{\lambda\omega}\cdot
    U^{{}^P}_{\nu}\cdot\partial^{{}^P}_{\omega}
                            F^{{}^P}_{\nu\lambda}  \biggr)
\nonumber\\&&\qquad
  +   \frac{e\cdot c_7}{2\cdot m^2}\cdot
              S^{{}^P}_{\mu\nu}\cdot \partial^{{}^P}_{\rho} F^{{}^P}_{\nu\rho}
  +   \frac{e^2\cdot c_3}{4\cdot m^3}\cdot
F^{{}^P}_{\mu\nu}\cdot S^{{}^P}_{\nu\omega}\cdot
          F^{{}^P}_{\omega\rho}\cdot U^{{}^P}_{\rho}
  -  \frac{e^2\cdot c_4}{4\cdot m^3}\cdot
F^{{}^P}_{\mu\nu}\cdot U^{{}^P}_{\nu}\cdot
          S^{{}^P}_{\omega\rho}\cdot F^{{}^P}_{\rho\omega}
\nonumber\\&&\qquad
  +   \frac{e^2\cdot c_5}{4\cdot m^3}\cdot
S^{{}^P}_{\mu\nu}\cdot F^{{}^P}_{\nu\rho}\cdot
           F^{{}^P}_{\rho\omega}\cdot U^{{}^P}_{\omega}
                                       \; ,
 \qquad (\mu =1,...,4)
\label{5.2}
\end{eqnarray}
where $c_1,...,c_7$ are dimensionless real numbers. This can be
further simplified by using the vacuum Maxwell equations (3.11), so that
\begin{eqnarray}
&&   S^{{}^P}_{\nu\omega}\cdot \partial^{{}^P}_{\mu} F^{{}^P}_{\omega\nu} = -
2\cdot  S^{{}^P}_{\nu\omega}\cdot \partial^{{}^P}_{\nu}
               F^{{}^P}_{\mu\omega}\; .
 \qquad (\mu =1,...,4)
\label{5.3}
\end{eqnarray}
Hence by (3.11) the terms on the rhs of (5.2) which are proportional to
$c_6,c_7$ are not independent from the others so that the general ansatz
(5.2) finally simplifies to:
\begin{eqnarray}
&& Y^{{}^P}_{\mu} =
   -  \frac{e\cdot c_2}{4\cdot m^2}\cdot\biggl(
S^{{}^P}_{\nu\omega}\cdot \partial^{{}^P}_{\mu} F^{{}^P}_{\omega\nu}
+U^{{}^P}_{\mu}\cdot S^{{}^P}_{\nu\omega}\cdot U^{{}^P}_{\lambda}\cdot
                                          \partial^{{}^P}_{\lambda}
                            F^{{}^P}_{\omega\nu}  \biggr)
\nonumber\\&&\qquad
  +   \frac{e\cdot(c_2-c_1-2)}{2\cdot m^2}\cdot
S^{{}^P}_{\mu\nu}\cdot U^{{}^P}_{\omega}\cdot
   U^{{}^P}_{\lambda}\cdot\partial^{{}^P}_{\lambda}
                            F^{{}^P}_{\nu\omega}
  +   \frac{e^2\cdot c_3}{4\cdot m^3}\cdot
F^{{}^P}_{\mu\nu}\cdot S^{{}^P}_{\nu\omega}\cdot
            F^{{}^P}_{\omega\rho}\cdot U^{{}^P}_{\rho}
\nonumber\\&&\qquad
  -  \frac{e^2\cdot c_4}{4\cdot m^3}\cdot
F^{{}^P}_{\mu\nu}\cdot U^{{}^P}_{\nu}\cdot
           S^{{}^P}_{\omega\rho}\cdot F^{{}^P}_{\rho\omega}
  +   \frac{e^2\cdot c_5}{4\cdot m^3}\cdot
S^{{}^P}_{\mu\nu}\cdot F^{{}^P}_{\nu\rho}\cdot
            F^{{}^P}_{\rho\omega}\cdot U^{{}^P}_{\omega}  \; .
 \qquad (\mu =1,...,4) \qquad
\label{5.4}
\end{eqnarray}
Combining this with (5.1) I get the following general system of
equations:
\footnote{Note also that for every choice of the characteristic
                            parameters
$c_1,...,c_5$ the following equations of the previous sections remain valid:
(1.1-4),(1.5c),(1.6-7),(1.9),(2.1-9),(2.11a),(2.11c),(3.1-11),(3.13).}
\setcounter{INDEX}{1}
\begin{eqnarray}
 && \dot{X}^{{}^P}_{\mu} = U^{{}^P}_{\mu} \; ,
\\
\addtocounter{equation}{-1}
\addtocounter{INDEX}{1}
\label{5.5a}
&&\dot{U}^{{}^P}_{\mu}= \frac{e}{m}\cdot
 F^{{}^P}_{\mu\nu}\cdot U^{{}^P}_{\nu}
   -  \frac{e\cdot c_2}{4\cdot m^2}\cdot\biggl(
S^{{}^P}_{\nu\omega}\cdot \partial^{{}^P}_{\mu} F^{{}^P}_{\omega\nu}
+U^{{}^P}_{\mu}\cdot S^{{}^P}_{\nu\omega}\cdot U^{{}^P}_{\lambda}\cdot
                                          \partial^{{}^P}_{\lambda}
                            F^{{}^P}_{\omega\nu}  \biggr)
\nonumber\\&&\qquad
  +   \frac{e\cdot(c_2-c_1-2)}{2\cdot m^2}\cdot
S^{{}^P}_{\mu\nu}\cdot U^{{}^P}_{\omega}\cdot
   U^{{}^P}_{\lambda}\cdot\partial^{{}^P}_{\lambda}
                            F^{{}^P}_{\nu\omega}
  +   \frac{e^2\cdot c_3}{4\cdot m^3}\cdot
F^{{}^P}_{\mu\nu}\cdot S^{{}^P}_{\nu\omega}\cdot
            F^{{}^P}_{\omega\rho}\cdot U^{{}^P}_{\rho}
\nonumber\\&&\qquad
  -  \frac{e^2\cdot c_4}{4\cdot m^3}\cdot
F^{{}^P}_{\mu\nu}\cdot U^{{}^P}_{\nu}\cdot
           S^{{}^P}_{\omega\rho}\cdot F^{{}^P}_{\rho\omega}
  +   \frac{e^2\cdot c_5}{4\cdot m^3}\cdot
S^{{}^P}_{\mu\nu}\cdot F^{{}^P}_{\nu\rho}\cdot
            F^{{}^P}_{\rho\omega}\cdot U^{{}^P}_{\omega}
                                       \; ,
\\
\addtocounter{equation}{-1}
\addtocounter{INDEX}{1}
\label{5.5b}
 && \dot{S}^{{}^P}_{\mu\nu} =  \frac{e\cdot g}{2\cdot m}\cdot\biggl(
                   F^{{}^P}_{\mu\omega}\cdot S^{{}^P}_{\omega\nu}-
                   S^{{}^P}_{\mu\omega}\cdot F^{{}^P}_{\omega\nu} \biggr)
\nonumber\\&&
  -   \frac{e\cdot (g-2)}{2\cdot m}\cdot\biggl(
S^{{}^P}_{\mu\omega}\cdot F^{{}^P}_{\omega\lambda}\cdot U^{{}^P}_{\lambda}\cdot
                                            U^{{}^P}_{\nu}-
S^{{}^P}_{\nu\omega}\cdot F^{{}^P}_{\omega\lambda}\cdot U^{{}^P}_{\lambda}\cdot
                                            U^{{}^P}_{\mu}
                                           \biggr) \; .
 \qquad (\mu,\nu =1,...,4)
\label{5.5c}
\end{eqnarray}
\setcounter{INDEX}{0}
\par\noindent
The covariance of (5.5) under the $\rm{Poincar\acute{e}}$ group will be
demonstrated in subsection 5.3.
\par The Frenkel equations correspond to
\begin{eqnarray}
 && c_1 = 0 \; , \qquad  c_2 = g \; , \qquad
    c_3 = (g-2)\cdot g \; , \qquad
    c_4 = g \; , \qquad
    c_5 =-(g-2)^2  \; ,
\label{5.6}
\end{eqnarray}
and by neglecting the second order SG terms in the Frenkel equations one
gets the `reduced' Frenkel equations, which correspond to:
\begin{eqnarray}
 && c_1 = 0 \; , \qquad  c_2 = g \; , \qquad
     c_3 = c_4 =  c_5 = 0 \; .
\label{5.7}
\end{eqnarray}
The GNR equations, which are defined by \cite{Goo62,Nyb64,Raf70},
correspond to:
\begin{eqnarray}
 && c_1 = g-2 \; , \qquad  c_2 = g \; , \qquad
    c_3 = c_4 =  c_5 = 0 \; .
\label{5.8}
\end{eqnarray}
Another interesting choice is given by \cite{Cos94}:
\footnote{See equation [4.57] therein. Note that I neglect
second order spin terms.}
\begin{eqnarray}
&& c_1 = -2 \; , \qquad  c_2 = g \; , \qquad c_3 = g^2 \; , \qquad
   c_4 =  g \; , \qquad
   c_5 = -g^2+2\cdot g  \; .
 \label{5.9}
 \end{eqnarray}
This is a modification of a force given in \cite{CoM94}
by a `redshift term':
\footnote{See equation [15] in \cite{CoM94}. Note that I neglect second order
spin terms.}
\begin{eqnarray}
&& c_1 = -2 \; , \qquad  c_2 = g \; , \qquad c_3 = g^2 \; , \qquad
   c_4 =  0 \; , \qquad
   c_5 = -g^2  \; .
\label{5.10}
\end{eqnarray}
A very simple choice is
\begin{eqnarray}
&& c_1 = -2 \; , \qquad  c_2 = c_3 = c_4 =  c_5 = 0 \; ,
\label{5.11}
\end{eqnarray}
corresponding to a SG force, which vanishes for the `P' variables.
\par That the kinematic constraints
(3.8) allow the general form (5.4) of $Y^{{}^P}$ illustrates that
the five parameters $c_1,...,c_5$ are, on the classical level, just
phenomenological constants which can only be fixed by comparison with
experiments (e.g. in storage rings with polarized beams) $\lbrack o
\rbrack$. The physical implications of this plethora of possibilities
will be addressed later (see sections 8,9).
Note that in the above derivation of $Y^{{}^P}$
the assumption of first order dependence on spin was essential.
The reader who is interested in higher order spin terms is advised to
consult the large literature on classical relativistic spin. See, for
example, the book \cite{Cor68} for references.
A recent interesting treatment, nonlinear in spin, can be found in
\cite{Cos94}.
\subsection*{5.2}
It follows from (5.5b) and by using Appendix D that:
\begin{eqnarray*}
 && m\cdot (\gamma_{{}_P}\cdot\vec{v}_{{}_P})' =
    e\cdot (\vec{v}_{{}_P}\wtimes\vec{B}_{{}_P}) + e\cdot\vec{E}_{{}_P} +
\frac{e\cdot c_2}{2\cdot m\cdot\gamma_{{}_P}}\cdot \vec{\nabla}_{{}_P}
                                                               \biggl(
\vec{s}^{\,\dagger}\cdot\vec{B}_{{}_P}-\vec{E}_{{}_P}^\dagger
         \cdot(\vec{s}\wtimes\vec{v}_{{}_P}) \biggr)
\nonumber\\&& \qquad
+\frac{e\cdot \gamma_{{}_P}}{2\cdot m}\cdot \lbrack
 (c_1+2)\cdot\vec{s}^{\,\dagger}\cdot\vec{B}_{{}_P}'\cdot\vec{v}_{{}_P}
-c_2\cdot(\vec{s}\wtimes\vec{v}_{{}_P})^\dagger\cdot
  \vec{E}_{{}_P}'\cdot\vec{v}_{{}_P}
+(c_2-c_1-2)\cdot(\vec{E}_{{}_P}'\wtimes\vec{s})
\nonumber\\&&\qquad
+(c_2-c_1-2)\cdot\vec{v}_{{}_P}^\dagger\cdot\vec{s}\cdot\vec{B}_{{}_P}'
-(c_2-c_1-2)\cdot\vec{v}_{{}_P}^\dagger\cdot\vec{E}_{{}_P}'
     \cdot(\vec{v}_{{}_P}\wtimes\vec{s})\rbrack
\nonumber\\&&\qquad
+\frac{e^2}{4\cdot m^2}\cdot\lbrack
    c_5\cdot
   \vec{v}_{{}_P}^\dagger\cdot\vec{B}_{{}_P}\cdot
                                      (\vec{B}_{{}_P} \wtimes\vec{s})
   -c_5\cdot
   \vec{B}_{{}_P}^\dagger\cdot\vec{B}_{{}_P}\cdot
                                 (\vec{v}_{{}_P} \wtimes\vec{s})
   +c_5\cdot
   \vec{s}^{\,\dagger}\cdot\vec{E}_{{}_P}
\cdot\vec{B}_{{}_P}
\nonumber\\&&\qquad
+(-c_3\cdot \vec{v}_{{}_P}^\dagger\cdot\vec{v}_{{}_P}
-c_5+2\cdot c_4 -c_3   )\cdot
   \vec{s}^{\,\dagger}\cdot\vec{B}_{{}_P}\cdot\vec{E}_{{}_P}
   +c_5\cdot
   \vec{v}_{{}_P}^\dagger\cdot\vec{E}_{{}_P}\cdot
                                      (\vec{E}_{{}_P} \wtimes\vec{s})
\nonumber\\&&\qquad
   -c_5\cdot
       ( \vec{v}_{{}_P}\wtimes\vec{B}_{{}_P})^\dagger
 \cdot\vec{E}_{{}_P} \cdot(\vec{v}_{{}_P}\wtimes\vec{s})
   -c_5\cdot
       \vec{E}_{{}_P}^\dagger\cdot\vec{E}_{{}_P} \cdot
                                      (\vec{v}_{{}_P}\wtimes\vec{s})
\nonumber\\&&\qquad
   +(2\cdot c_4-c_3)\cdot
   \vec{s}^{\,\dagger}\cdot\vec{B}_{{}_P}\cdot
                               (\vec{v}_{{}_P} \wtimes\vec{B}_{{}_P})
   +c_3\cdot
   \vec{E}_{{}_P}^\dagger\cdot\vec{B}_{{}_P}\cdot \vec{s}
   -c_3\cdot
   \vec{v}_{{}_P}^\dagger\cdot\vec{E}_{{}_P}\cdot
          \vec{v}_{{}_P}^\dagger\cdot\vec{B}_{{}_P}\cdot\vec{s}
\nonumber\\&&\qquad
   +c_3\cdot
   \vec{v}_{{}_P}^\dagger\cdot\vec{E}_{{}_P}\cdot
           \vec{s}^{\,\dagger}\cdot\vec{B}_{{}_P}\cdot\vec{v}_{{}_P}
   +c_3\cdot
   \vec{v}_{{}_P}^\dagger\cdot\vec{s}\cdot
            \vec{v}_{{}_P}^\dagger\cdot\vec{B}_{{}_P}\cdot\vec{E}_{{}_P}
\nonumber\\&&\qquad
   +(2\cdot c_4-c_3)\cdot
  \vec{E}_{{}_P}^\dagger\cdot( \vec{v}_{{}_P}\wtimes
                                             \vec{s})\cdot\vec{E}_{{}_P}
   +2\cdot c_4\cdot
\vec{E}_{{}_P}^\dagger\cdot( \vec{v}_{{}_P}\wtimes
      \vec{s})\cdot(\vec{v}_{{}_P}\wtimes\vec{B}_{{}_P})
\rbrack   \; .
\end{eqnarray*}
Collecting this with (2.3a),(2.9) I get
\setcounter{INDEX}{1}
\begin{eqnarray}
 && \vec{r}^{\;{}_{'}}_{{}_P}
                              =  \vec{v}_{{}_P} \; ,
\\
\addtocounter{equation}{-1}
\addtocounter{INDEX}{1}
\label{5.12a}
 && m\cdot (\gamma_{{}_P}\cdot\vec{v}_{{}_P})' =
    e\cdot (\vec{v}_{{}_P}\wtimes\vec{B}_{{}_P}) + e\cdot\vec{E}_{{}_P} +
\frac{e\cdot c_2}{2\cdot m\cdot\gamma_{{}_P}}\cdot \vec{\nabla}_{{}_P}
                                                               \biggl(
\vec{s}^{\,\dagger}\cdot\vec{B}_{{}_P}-\vec{E}_{{}_P}^\dagger
         \cdot(\vec{s}\wtimes\vec{v}_{{}_P}) \biggr)
\nonumber\\&& \qquad
+\frac{e\cdot \gamma_{{}_P}}{2\cdot m}\cdot \lbrack
 (c_1+2)\cdot\vec{s}^{\,\dagger}\cdot\vec{B}_{{}_P}'\cdot\vec{v}_{{}_P}
-c_2\cdot(\vec{s}\wtimes\vec{v}_{{}_P})^\dagger\cdot
  \vec{E}_{{}_P}'\cdot\vec{v}_{{}_P}
+(c_2-c_1-2)\cdot(\vec{E}_{{}_P}'\wtimes\vec{s})
\nonumber\\&&\qquad
+(c_2-c_1-2)\cdot\vec{v}_{{}_P}^\dagger\cdot\vec{s}\cdot\vec{B}_{{}_P}'
-(c_2-c_1-2)\cdot\vec{v}_{{}_P}^\dagger\cdot\vec{E}_{{}_P}'
     \cdot(\vec{v}_{{}_P}\wtimes\vec{s})\rbrack
\nonumber\\&&\qquad
+\frac{e^2}{4\cdot m^2}\cdot\lbrack
    c_5\cdot
   \vec{v}_{{}_P}^\dagger\cdot\vec{B}_{{}_P}\cdot
                                      (\vec{B}_{{}_P} \wtimes\vec{s})
   -c_5\cdot
   \vec{B}_{{}_P}^\dagger\cdot\vec{B}_{{}_P}\cdot
                                 (\vec{v}_{{}_P} \wtimes\vec{s})
   +c_5\cdot
   \vec{s}^{\,\dagger}\cdot\vec{E}_{{}_P}
\cdot\vec{B}_{{}_P}
\nonumber\\&&\qquad
+(-c_3\cdot \vec{v}_{{}_P}^\dagger\cdot\vec{v}_{{}_P}
-c_5+2\cdot c_4 -c_3   )\cdot
   \vec{s}^{\,\dagger}\cdot\vec{B}_{{}_P}\cdot\vec{E}_{{}_P}
   +c_5\cdot
   \vec{v}_{{}_P}^\dagger\cdot\vec{E}_{{}_P}\cdot
                                      (\vec{E}_{{}_P} \wtimes\vec{s})
\nonumber\\&&\qquad
   -c_5\cdot
       ( \vec{v}_{{}_P}\wtimes\vec{B}_{{}_P})^\dagger
 \cdot\vec{E}_{{}_P} \cdot(\vec{v}_{{}_P}\wtimes\vec{s})
   -c_5\cdot
       \vec{E}_{{}_P}^\dagger\cdot\vec{E}_{{}_P} \cdot
                                      (\vec{v}_{{}_P}\wtimes\vec{s})
\nonumber\\&&\qquad
   +(2\cdot c_4-c_3)\cdot
   \vec{s}^{\,\dagger}\cdot\vec{B}_{{}_P}\cdot
                               (\vec{v}_{{}_P} \wtimes\vec{B}_{{}_P})
   +c_3\cdot
   \vec{E}_{{}_P}^\dagger\cdot\vec{B}_{{}_P}\cdot \vec{s}
   -c_3\cdot
   \vec{v}_{{}_P}^\dagger\cdot\vec{E}_{{}_P}\cdot
          \vec{v}_{{}_P}^\dagger\cdot\vec{B}_{{}_P}\cdot\vec{s}
\nonumber\\&&\qquad
   +c_3\cdot
   \vec{v}_{{}_P}^\dagger\cdot\vec{E}_{{}_P}\cdot
           \vec{s}^{\,\dagger}\cdot\vec{B}_{{}_P}\cdot\vec{v}_{{}_P}
   +c_3\cdot
   \vec{v}_{{}_P}^\dagger\cdot\vec{s}\cdot
            \vec{v}_{{}_P}^\dagger\cdot\vec{B}_{{}_P}\cdot\vec{E}_{{}_P}
\nonumber\\&&\qquad
   +(2\cdot c_4-c_3)\cdot
  \vec{E}_{{}_P}^\dagger\cdot( \vec{v}_{{}_P}\wtimes
                                             \vec{s})\cdot\vec{E}_{{}_P}
   +2\cdot c_4\cdot
\vec{E}_{{}_P}^\dagger\cdot( \vec{v}_{{}_P}\wtimes
      \vec{s})\cdot(\vec{v}_{{}_P}\wtimes\vec{B}_{{}_P})
\rbrack   \; ,
\\
\addtocounter{equation}{-1}
\addtocounter{INDEX}{1}
\label{5.12b}
 && \vec{s}\ ' =
      \frac{e}{m}\cdot \lbrack
 \frac{g-2}{2}\cdot\gamma_{{}_P}\cdot
 \vec{v}_{{}_P}^{\,\dagger}\cdot\vec{s}
\cdot
    (\vec{E}_{{}_P}+\vec{v}_{{}_P}\wtimes\vec{B}_{{}_P})
+\frac{1}{\gamma_{{}_P}}\cdot
 \vec{v}_{{}_P}^\dagger\cdot\vec{E}_{{}_P}\cdot\vec{s}
\nonumber\\&&
-\frac{g-2}{2}\cdot\gamma_{{}_P}\cdot
 \vec{v}_{{}_P}^{\,\dagger}\cdot\vec{s}
\cdot
        \vec{v}_{{}_P}^\dagger\cdot\vec{E}_{{}_P}\cdot \vec{v}_{{}_P}
+\frac{g}{2}\cdot\frac{1}{\gamma_{{}_P}}\cdot
            (\vec{s}\wtimes\vec{B}_{{}_P})
-\frac{g}{2}\cdot\frac{1}{\gamma_{{}_P}}\cdot
\vec{s}^{\,\dagger}\cdot\vec{E}_{{}_P}\cdot \vec{v}_{{}_P} \rbrack \; .
\label{5.12c}
\end{eqnarray}
\setcounter{INDEX}{0}
\par\noindent
It is shown in Appendix D that (5.12) is equivalent to (5.5).
For the special choice (5.6) this means that (2.11) is equivalent
to (3.12), i.e. that the DK equations are equivalent to
the Frenkel equations!
\subsection*{5.3}
In this subsection I show that equations (5.5) are
$\rm{Poincar\acute{e}}$ covariant. Note that $X^{{}^P}$ transforms
as a space-time position, $U^{{}^P}$ as a 4-vector, $S^{{}^P}$ as a tensor of
rank 2 and $F^{{}^P}$ as a
tensor field of rank 2. Thus (5.5) are covariant under the restricted
$\rm{Poincar\acute{e}}$ group $\lbrack p \rbrack$ because equations (5.5) are
constructed from covariant quantities. To show that (5.5) is covariant
under the whole $\rm{Poincar\acute{e}}$ group I first consider space
inversion (=parity transformation). This acts in the following way:
\begin{eqnarray}
 && \tau     \rightarrow    \tau                             \; ,
\nonumber\\
 && X^{{}^P}_{\mu}  \rightarrow
     (-\vec{r}_{{}_P}^{\,\dagger},X^{{}^P}_4)_{\mu}  \; ,
\nonumber\\
 && U^{{}^P}_{\mu}  \rightarrow
(-\gamma_{{}_P}\cdot\vec{v}_{{}_P}^\dagger,i\cdot
                                           \gamma_{{}_P})_{\mu} \; ,
\nonumber\\
&&  \partial^{{}^P}_{\mu} \rightarrow
    (-\frac{\partial}{\partial X^{{}^P}_1},
     -\frac{\partial}{\partial X^{{}^P}_2},
     -\frac{\partial}{\partial X^{{}^P}_3},
     \frac{\partial}{\partial X^{{}^P}_4})_{\mu} \; ,
\nonumber\\
 && \vec{B}_{{}_P}   \rightarrow     \vec{B}_{{}_P} \; ,
\nonumber\\
 && \vec{E}_{{}_P}   \rightarrow     -\vec{E}_{{}_P} \; ,
\nonumber\\
 && \vec{s}   \rightarrow     \vec{s} \; ,
\nonumber\\
 && \vec{q}   \rightarrow     -\vec{q} \; .
 \qquad (\mu =1,...,4)
\label{5.13}
\end{eqnarray}
Hence one sees that with the building blocks:
\begin{eqnarray*}
&& \frac{d}{d\tau} , \partial^{{}^P} , U^{{}^P} , F^{{}^P} ,
S
\end{eqnarray*}
no pseudotensors occur in (5.5), so that (5.5) is covariant under the
parity transformation.
\footnote{Note that the direct product or    the contraction of two
tensors is again a tensor, not a pseudotensor. For the distinction
between tensors and pseudotensors, see for example \cite{Moe72}.}
With (5.13) one sees that $\vec{r}_{{}_P},\vec{v}_{{}_P},\vec{E}_{{}_P},
\vec{q}$ are polar vectors whereas $\vec{B}_{{}_P},\vec{s}$ are
axial vectors.
\par
   Next I consider the time inversion (=time reversal transformation).
This  acts in the following way:
\begin{eqnarray}
 && \tau     \rightarrow      -         \tau        \; ,
\nonumber\\
 && X^{{}^P}_{\mu}  \rightarrow
     ( \vec{r}_{{}_P}^{\,\dagger},- X^{{}^P}_4)_{\mu}  \; ,
\nonumber\\
 && U^{{}^P}_{\mu}  \rightarrow
(-\gamma_{{}_P}\cdot\vec{v}_{{}_P}^\dagger,i\cdot
                                           \gamma_{{}_P})_{\mu} \; ,
\nonumber\\
&&  \partial^{{}^P}_{\mu} \rightarrow
    (\frac{\partial}{\partial X^{{}^P}_1},
     \frac{\partial}{\partial X^{{}^P}_2},
     \frac{\partial}{\partial X^{{}^P}_3},
    -\frac{\partial}{\partial X^{{}^P}_4})_{\mu} \; ,
\nonumber\\
 && \vec{B}_{{}_P}   \rightarrow    -\vec{B}_{{}_P} \; ,
\nonumber\\
 && \vec{E}_{{}_P}   \rightarrow      \vec{E}_{{}_P} \; ,
\nonumber\\
 && \vec{s}   \rightarrow    -\vec{s} \; ,
\nonumber\\
 && \vec{q}   \rightarrow      \vec{q} \; .
 \qquad (\mu =1,...,4)
\label{5.14}
\end{eqnarray}
Now I use the fact that (5.12) is equivalent to (5.5). In fact,
the application of (5.14) to (5.12b-c) shows first of all that
the spatial parts of (5.5a-c) are covariant under the time reversal
transformation. It then  follows by using (3.8) to get
\begin{eqnarray}
&&\dot{U}^{{}^P}_4 =-\frac{1}{U^{{}^P}_4}\cdot (U^{{}^P}_1\cdot
                                               \dot{U}^{{}^P}_1
                                     +U^{{}^P}_2\cdot\dot{U}^{{}^P}_2
                       +U^{{}^P}_3\cdot\dot{U}^{{}^P}_3) \; ,
\nonumber\\
 && \dot{\vec{q}}=  \dot{\vec{s}} \wtimes \vec{v}_{{}_P}
+\frac{e}{m}\cdot \vec{s} \wtimes
       ( \vec{v}_{{}_P}\wtimes \vec{B}_{{}_P} +   \vec{E}_{{}_P}  )
-\frac{e}{m}\cdot \vec{v}_{{}_P}^\dagger\cdot
 \vec{E}_{{}_P}\cdot\vec{q} \; ,
\label{5.15}
\end{eqnarray}
that the temporal parts of (5.5a-c) are covariant under the time
reversal transformation.
Hence one concludes that (5.5) is covariant under the time reversal
transformation. In summary:
\begin{itemize}
\item Equations (5.5) are covariant under the
$\rm{Poincar\acute{e}}$ group. Specifically, they are invariant under
$\rm{Poincar\acute{e}}$ transformations (except that the
electromagnetic field transforms as described in section 3).
\item As a special case the Frenkel equations (3.12) are
$\rm{Poincar\acute{e}}$ covariant from
which it follows that also (2.11) is $\rm{Poincar\acute{e}}$ covariant
$\lbrack q\rbrack$.
\end{itemize}
Because $F^{{}^P}$ transforms as a tensor field of rank 2 one also
finds   that (3.11) is invariant under the
$\rm{Poincar\acute{e}}$ group.
\subsection*{5.4}
So far I have assumed that $S^{{}^P}$ is a tensor. Now for completeness
and as announced in section 3
                   I demonstrate that the tensor property of
$S^{{}^P}$ is consistent with the transformation properties of
$\vec{v}_{{}_P}$ under the
$\rm{Poincar\acute{e}}$
                                   group,
i.e. the relation $\vec{q}=\vec{s} \wtimes \vec{v}_{{}_P}$ is conserved
under
$\rm{Poincar\acute{e}}$ transformations. In fact, by (5.13-14) one sees that it
is conserved under spatial rotations, space inversion and time reversal
and in the remaining part of this subsection I show that it is also
conserved under the proper Lorentz transformation.
\par The infinitesimal proper Lorentz transformation (=infinitesimal
Lorentz boost) is defined by
\begin{eqnarray}
&& L^{{}^{boost}}_{jk} \equiv  \delta_{jk} \; ,
\nonumber\\
&& L^{{}^{boost}}_{j4} \equiv i\cdot v_{{}_{boost},j} \; ,
\nonumber\\
&& L^{{}^{boost}}_{4j} \equiv -L^{{}^{boost}}_{j4} \; ,
\nonumber\\
&& L^{{}^{boost}}_{44} \equiv 1    \; ,
                                               \qquad (j,k=1,2,3)
\label{5.16}
\end{eqnarray}
where the infinitesimal vector $\vec{v}_{{}_{boost}}$ denotes the
relative velocity of the frames connected by $L^{{}^{boost}}$.
From (5.16) one obtains
\begin{eqnarray}
 && L^{{}^{boost}}_{\mu\nu} \cdot L^{{}^{boost}}_{\mu\rho} =
   \delta_{\nu\rho} \; ,
 \qquad (\nu,\rho =1,...,4)
\label{5.17}
\end{eqnarray}
which is consistent with the fact that
$L^{{}^{boost}}$ belongs to the Lorentz group.
The spin tensor transforms under $L^{{}^{boost}}$ via:
\begin{eqnarray}
 && S^{{}^P}_{\mu\nu} \rightarrow  S^{{}^{P,boost}}_{\mu\nu} \equiv
                            L^{{}^{boost}}_{\mu\rho} \cdot
               L^{{}^{boost}}_{\nu\omega}
                                                           \cdot
   S^{{}^P}_{\rho\omega}  \; .
 \qquad (\mu,\nu =1,...,4)
\label{5.18}
\end{eqnarray}
Abbreviating:
\begin{eqnarray*}
&& S^{{}^{P,boost}} \leftrightarrow
    (\vec{s}_{{}_{boost}},-i\cdot\vec{q}_{{}_{boost}})
                                                                  \; ,
\end{eqnarray*}
one sees by (5.18) that $\vec{s},\vec{q}$ transform under
$L^{{}^{boost}}$ via:
\begin{eqnarray}
&& \vec{s} \rightarrow \vec{s}_{{}_{boost}} =
   \vec{s} -  \vec{v}_{{}_{boost}}\wtimes \vec{q} \; ,
\nonumber\\
&& \vec{q} \rightarrow \vec{q}_{{}_{boost}} =
   \vec{q} + \vec{v}_{{}_{boost}}\wtimes \vec{s} \; .
\label{5.19}
\end{eqnarray}
Because $U^{{}^P}$ is a 4-vector, one observes that
$\vec{v}_{{}_P},\gamma_{{}_P}$ transform under $L^{{}^{boost}}$ via:
\begin{eqnarray}
&& \gamma_{{}_P} \rightarrow \gamma_{{}_{P,boost}} \equiv
   \gamma_{{}_P}\cdot  (1- \vec{v}_{{}_{boost}}^\dagger\cdot
     \vec{v}_{{}_P}) \; ,
\nonumber\\
&& \vec{v}_{{}_P}  \rightarrow \vec{v}_{{}_{P,boost}} \equiv
   \vec{v}_{{}_P} + \vec{v}_{{}_{boost}}^\dagger\cdot
\vec{v}_{{}_P}\cdot \vec{v}_{{}_P} - \vec{v}_{{}_{boost}} \; .
\label{5.20}
\end{eqnarray}
Collecting (5.19-20) one observes:
\begin{eqnarray*}
&& \vec{q}_{{}_{boost}} =  \vec{s}_{{}_{boost}}\wtimes
          \vec{v}_{{}_{P,boost}}\; ,
\end{eqnarray*}
so that the relation: $\vec{q}=\vec{s} \wtimes \vec{v}_{{}_P}$ is
conserved under the infinitesimal proper Lorentz transformation
$L^{{}^{boost}}$. This concludes the proof that this relation is conserved
under the whole $\rm{Poincar\acute{e}}$ group.
\subsection*{5.5}
Subsection 5.3 listed the transformation properties
of $\vec{r}_{{}_P},t,\vec{v}_{{}_P},\vec{s},\vec{B}_{{}_P},\vec{E}_{{}_P}$
under the $\rm{Poincar\acute{e}}$ group. Combining these with (2.1),(2.3a),
(2.4-5),(2.7)
one also obtains the transformation properties of
$\vec{r}_{{}_M},t,\vec{v}_{{}_M},\vec{\sigma},\vec{B}_{{}_M},\vec{E}_{{}_M}$.
One observes:
\begin{itemize}
\item The relations (2.1),(2.3a),(2.4-5),(2.7) are invariant under 
$\rm{Poincar\acute{e}}$ transformations.
\item By (2.7) the `P' fields depend on $\vec{r}_{{}_P},t$ in the same
way as the `M' fields depend on $\vec{r}_{{}_M},t$.
\end{itemize}
Because (5.12) is $\rm{Poincar\acute{e}}$ covariant, one concludes
that also the DK equations (1.5) are $\rm{Poincar\acute{e}}$ covariant.
Specifically,
they are invariant under $\rm{Poincar\acute{e}}$ transformations (except
that
the electromagnetic `M'
      fields transform  in the same way as the `P' fields)
$\lbrack r \rbrack$.
\par Note that $(\vec{r}_{{}_M}^{\,\dagger},i\cdot t)_{\mu}$ does not
transform as a space-time position and
$(\gamma_{{}_M}\cdot\vec{v}_{{}_M}^{\,\dagger},i\cdot\gamma_{{}_M})_{\mu}$
is not a 4-vector but both transform {\it nonlinearly} in a complicated
way. Analogously one obtains the well known property that
the rest frame spin vector $\vec{\sigma}$ is not the spatial part
of a 4-vector. Using (2.5) and the tensor property of $S^{{}^P}$ I now
discuss the transformation properties of $\vec{\sigma}$.
Firstly by (5.13),  $\vec{\sigma}$ is, like $\vec s$, an
axial vector and by (5.14) it transforms in the same way under time
reversal.
\par Secondly,  under the infinitesimal
proper Lorentz transformation (5.16) it transforms via:
\begin{eqnarray}
&&  \vec{\sigma}\rightarrow  \vec{\sigma}
      + \frac{\gamma_{{}_P}}{\gamma_{{}_P}+1}\cdot
   ( \vec{v}_{{}_{boost}} \wtimes \vec{v}_{{}_P}) \wtimes
     \vec{\sigma}  \; ,
\label{5.21}
\end{eqnarray}
which is simply a rotation associated with a change of orientation
of the reference frame due to the boost.
Thus   $\vec{\sigma}$ has in fact the
transformation properties of the rest frame spin vector
\cite{BMT59,Jac75,Tho27}.
\subsection*{5.6}
As already mentioned, $\vec{r}_{{}_M},t$ are the position and
time variables used in the everyday business of accelerator  physics.
In fact via (2.1) and by neglecting spin:
$\vec{r}_{{}_M}=\vec{r}_{{}_P}$, so that in this approximation
\begin{eqnarray}
 &&   (\vec{r}_{{}_M}^{\,\dagger},i\cdot t)_{\mu}
 \qquad (\mu =1,...,4)
\label{5.22}
\end{eqnarray}
transforms as a space-time position.
Also one observes in this approximation that:
\begin{eqnarray}
&& (\gamma_{{}_M}\cdot\vec{v}_{{}_M}^{\,\dagger},i\cdot\gamma_{{}_M})_{\mu}
 \qquad (\mu =1,...,4)
\label{5.23}
\end{eqnarray}
is a 4-vector. Even if one includes the spin one obtains these
transformation properties if the SG-force is neglected. In fact in that
approximation one gets:
\setcounter{INDEX}{1}
\begin{eqnarray}
 && \vec{r}^{\;{}_{'}}_{{}_M} =  \vec{v}_{{}_M},
\\
\addtocounter{equation}{-1}
\addtocounter{INDEX}{1}
\label{5.24a}
 && m\cdot (\gamma_{{}_M}\cdot\vec{v}_{{}_M})' =
e\cdot (\vec{v}_{{}_M}\wtimes\vec{B}_{{}_M}) + e\cdot\vec{E}_{{}_M} \; ,
\\
\addtocounter{equation}{-1}
\addtocounter{INDEX}{1}
\label{5.24b}
 && \vec{\sigma}\ ' =  \vec{\Omega}_{{}_M}\wtimes\vec{\sigma} \; .
\label{5.24c}
\end{eqnarray}
\setcounter{INDEX}{0}
\par\noindent
Note that (5.24) is studied in \cite{BMT59,Jac75,Tho27}.
\setcounter{equation}{0}
\section{Using the spin pseudo-4-vector}
\subsection*{6.1}
So far I have described the spin in terms of the spin tensor $S^{{}^P}$ but
because the particle only has     a intrinsic {\it magnetic} dipole moment
and  no intrinsic {\rm electric} dipole moment one can also describe the
spin just in terms of a   pseudo-4-vector $T^{{}^P}$ defined by:
\footnote{For more details on $T^{{}^P}$, see for example
\cite{BMT59,Cor68,FG61a,Nyb64}.}
\begin{eqnarray}
 && T^{{}^P}_{\mu} = -\frac{i}{2}\cdot
  \varepsilon_{\mu\nu\rho\omega}\cdot
   S^{{}^P}_{\nu\rho}\cdot U^{{}^P}_{\omega} \; ,
 \qquad (\mu =1,...,4)
\label{6.1}
\end{eqnarray}
where $\varepsilon_{\mu\nu\rho\omega}$ is the Levi-Civita symbol.
\footnote{Note that $\varepsilon$ is the totally
antisymmetric pseudotensor of rank 4 with $\varepsilon_{1234}=1.$}
From this it follows that
\begin{eqnarray}
 &&  S^{{}^P}_{\mu\nu} = - i\cdot
    \varepsilon_{\mu\nu\rho\omega}\cdot
   T^{{}^P}_{\rho}\cdot U^{{}^P}_{\omega} \; .
 \qquad (\mu,\nu =1,...,4)
\label{6.2}
\end{eqnarray}
The constraints (3.8) now read as:
\begin{eqnarray}
 && U^{{}^P}_{\mu}\cdot  U^{{}^P}_{\mu} = -1 \; ,
\nonumber\\
 && T^{{}^P}_{\mu}\cdot U^{{}^P}_{\mu} = 0 \; .
%\nonumber\\
% && \sqrt{T^{{}^P}_{\mu}\cdot T^{{}^P}_{\mu}} = \hbar/2 \; .
\label{6.3}
\end{eqnarray}
From (2.4-5),(3.2),(3.6) and (6.1) one has:
\begin{eqnarray}
 && T^{{}^P}_{\mu} = (\vec{T}_{{}_P}^\dagger , T^{{}^P}_4 )_{\mu} \; ,
 \qquad (\mu =1,...,4)
\nonumber\\
 && \vec{T}_{{}_P} =
  \frac{1}{\gamma_{{}_P}}\cdot \vec{s} + \gamma_{{}_P}\cdot
 \vec{v}_{{}_P}^{\,\dagger}\cdot\vec{s}
\cdot\vec{v}_{{}_P} =
   \vec{\sigma} + \frac{\gamma_{{}_P}^2}{\gamma_{{}_P}+1}\cdot
\vec{v}_{{}_P}^{\,\dagger}\cdot\vec{\sigma}
\cdot\vec{v}_{{}_P}\; ,
\nonumber\\
 &&    T^{{}^P}_4 = i\cdot  \vec{v}_{{}_P}^\dagger\cdot\vec{T}_{{}_P}
                = i\cdot \gamma_{{}_P}
     \cdot
 \vec{v}_{{}_P}^{\,\dagger}\cdot\vec{s}
  =  i\cdot \gamma_{{}_P}
     \cdot
\vec{v}_{{}_P}^{\,\dagger}\cdot\vec{\sigma}
\; .
\label{6.4}
\end{eqnarray}
Thus under the space inversion one gets by (5.13),(6.1):
\begin{eqnarray}
&& T^{{}^P}_{\mu}\rightarrow (\vec{T}_{{}_P}^\dagger , -T^{{}^P}_4 )_{\mu}\; ,
 \qquad (\mu =1,...,4)
\label{6.5}
\end{eqnarray}
so that in fact $T^{{}^P}$ is a pseudo-4-vector. Under the time reversal one
gets by (5.14),(6.1):
\begin{eqnarray}
 && T^{{}^P}_{\mu} \rightarrow (-\vec{T}_{{}_P}^\dagger , T^{{}^P}_4 )_{\mu}
\; .
 \qquad (\mu =1,...,4)
\label{6.6}
\end{eqnarray}
\subsection*{6.2}
Now I introduce the pseudotensor field $\tilde{F}^{{}^P}$ dual to $F^{{}^P}$
defined by
\begin{eqnarray}
 &&  \tilde{F}^{{}^P}_{\mu\nu} = -\frac{i}{2}\cdot
    \varepsilon_{\mu\nu\rho\omega}\cdot
   F^{{}^P}_{\rho\omega}  \; ,
\nonumber\\
 &&  F^{{}^P}_{\mu\nu} =    \frac{i}{2}\cdot
    \varepsilon_{\mu\nu\rho\omega}\cdot
   \tilde{F}^{{}^P}_{\rho\omega}  \; ,
 \qquad (\mu,\nu =1,...,4)
\label{6.7}
\end{eqnarray}
from which it follows that:
\begin{eqnarray*}
&& \tilde{F}^{{}^P}  \leftrightarrow
 (-\vec{E}_{{}_P},-i\cdot\vec{B}_{{}_P}) \; .
\end{eqnarray*}
In Appendix E it is shown that (5.5) is equivalent to:
\setcounter{INDEX}{1}
\begin{eqnarray}
 && \dot{X}^{{}^P}_{\mu} = U^{{}^P}_{\mu} \; ,
\\
\addtocounter{equation}{-1}
\addtocounter{INDEX}{1}
\label{6.8a}
 && \dot{U}^{{}^P}_{\mu} = \frac{e}{m}\cdot
  F^{{}^P}_{\mu\nu}\cdot U^{{}^P}_{\nu}
     - \frac{e\cdot c_2}{2\cdot m^2}\cdot
U^{{}^P}_{\alpha}\cdot T^{{}^P}_{\beta}\cdot\partial^{{}^P}_{\mu}
                                          \tilde{F}^{{}^P}_{\alpha\beta}
    - \frac{e\cdot (c_1+2)}{2\cdot m^2}\cdot
   U^{{}^P}_{\mu}\cdot U^{{}^P}_{\alpha}\cdot T^{{}^P}_{\beta}\cdot
 U^{{}^P}_{\lambda}\cdot\partial^{{}^P}_{\lambda}
          \tilde{F}^{{}^P}_{\alpha\beta}
\nonumber\\&&
   + \frac{e\cdot (c_2-c_1-2)}{2\cdot m^2}\cdot
T^{{}^P}_{\nu}\cdot U^{{}^P}_{\lambda}\cdot\partial^{{}^P}_{\lambda}
                                           \tilde{F}^{{}^P}_{\mu\nu}
   + \frac{e^2\cdot (c_3-2\cdot c_4+c_5)}{4\cdot m^3}\cdot
F^{{}^P}_{\mu\nu}\cdot U^{{}^P}_{\nu}\cdot
    U^{{}^P}_{\alpha}\cdot T^{{}^P}_{\beta}\cdot
  \tilde{F}^{{}^P}_{\alpha\beta}
\nonumber\\&&\qquad
   + \frac{e^2\cdot c_3}{4\cdot m^3}\cdot
  F^{{}^P}_{\mu\nu}\cdot \tilde{F}^{{}^P}_{\nu\alpha} \cdot T^{{}^P}_{\alpha}
   - \frac{e^2\cdot c_5}{4\cdot m^3}\cdot
\tilde{F}^{{}^P}_{\mu\nu}\cdot U^{{}^P}_{\nu}\cdot U^{{}^P}_{\alpha}
                                                   \cdot T^{{}^P}_{\beta}\cdot
  F^{{}^P}_{\alpha\beta} \; ,
\\
\addtocounter{equation}{-1}
\addtocounter{INDEX}{1}
\label{6.8b}
 && \dot{T}^{{}^P}_{\mu} =  \frac{e\cdot g}{2\cdot m}\cdot
                   F^{{}^P}_{\mu\nu}\cdot T^{{}^P}_{\nu}
  +   \frac{e\cdot (g-2)}{2\cdot m}\cdot
    U^{{}^P}_{\mu}\cdot
    T^{{}^P}_{\omega}\cdot F^{{}^P}_{\nu\omega}\cdot U^{{}^P}_{\nu} \; .
 \qquad (\mu =1,...,4)
\label{6.8c}
\end{eqnarray}
\setcounter{INDEX}{0}
\par\noindent
Note that (6.8) conserves the kinematic constraints (6.3) and (6.8c) is
the BMT equation \cite{BMT59}. Note also that (6.8) is covariant under the
$\rm{Poincar\acute{e}}$ group
\footnote{i.e. (6.8) is invariant under $\rm{Poincar\acute{e}}$
transformations (except that the electromagnetic field
transforms in
the prescribed way)}
because (5.5) is, too.
\par It follows from the normalization of the spin vector that:
\begin{eqnarray}
 && T^{{}^P}_{\mu}\cdot T^{{}^P}_{\mu} = \hbar^2/4 \; .
\label{6.9}
\end{eqnarray}
As in (1.9) this equation is of second order in spin so that it plays no
role in this paper. Note also that (6.9) is conserved under (6.8c).
\setcounter{equation}{0}
\section{The nonrelativistic limit. The rest frame}
\subsection*{7.1}
In this first subsection I consider the nonrelativistic limit (= zeroth
order in $1/c$)                                               and in
the remainder of this section I consider the particle rest frame.
I do this for the general case, i.e. for
arbitrary values of $c_1,...,c_5$.
In the nonrelativistic limit (5.12) leads to:
\footnote{The partial derivative $\partial/\partial t$ in
 (7.1), (7.2) and (7.5) acts on
functions depending on $\vec{r}_{{}_P},t$.}
\setcounter{INDEX}{1}
\begin{eqnarray}
 && \vec{r}^{\;{}_{'}}_{{}_P}  =  \vec{v}_{{}_P} \; ,
\\
\addtocounter{equation}{-1}
\addtocounter{INDEX}{1}
\label{7.1a}
  && m\cdot \vec{v}^{\;{}_{'}}_{{}_P}=
    e\cdot \vec{E}_{{}_P} + \frac{e\cdot c_2}{2\cdot m}\cdot
\vec{\nabla}_{{}_P} (\vec{s}^{\,\dagger}\cdot \vec{B}_{{}_P}) -
                     \frac{e\cdot (c_2-2-c_1)}{2\cdot m}\cdot
(\vec{s} \wtimes \frac{\partial \vec{E}_{{}_P}}{\partial t})
\nonumber\\&&
+   \frac{e^2\cdot c_3}{4\cdot m^2}\cdot
 \vec{E}_{{}_P}^\dagger\cdot\vec{B}_{{}_P}\cdot\vec{s}
+\frac{e^2}{4\cdot m^2}\cdot ( -c_3+2\cdot c_4-c_5 ) \cdot
 \vec{s}^{\,\dagger}\cdot\vec{B}_{{}_P}\cdot\vec{E}_{{}_P}
+\frac{e^2\cdot c_5}{4\cdot m^2}\cdot
 \vec{s}^{\,\dagger}\cdot\vec{E}_{{}_P}\cdot\vec{B}_{{}_P}  \; ,
\nonumber\\&&
\\
\addtocounter{equation}{-1}
\addtocounter{INDEX}{1}
\label{7.1b}
 && \vec{s}\ ' =
                     \frac{e\cdot g}{2\cdot m}\cdot
(\vec{s} \wtimes \vec{B}_{{}_P})  \; .
\label{7.1c}
\end{eqnarray}
\setcounter{INDEX}{0}
\par\noindent
One now sees that the second order SG terms survive even in the
nonrelativistic limit. This should come as no surprise: second order
terms are seen in the usual semi-relativistic Foldy-Wouthuysen
transformations $\lbrack b \rbrack$.
Note that the third term on the
rhs of (7.1b) is sometimes called the `magnetodynamic force'. In the
nonrelativistic limit the Frenkel equations lead via (5.6),(7.1) to:
%\footnote{The partial derivative $\partial/\partial t$ in (7.2) acts on
%functions depending on $\vec{r}_{{}_P},t$.}
%
\begin{eqnarray}
 && \vec{r}^{\;{}_{'}}_{{}_P}  =  \vec{v}_{{}_P} \; , \nonumber\\
 && m\cdot \vec{v}^{\;{}_{'}}_{{}_P}=
    e\cdot \vec{E}_{{}_P} + \frac{e\cdot g}{2\cdot m}\cdot
\vec{\nabla}_{{}_P} (\vec{s}^{\,\dagger}\cdot \vec{B}_{{}_P}) -
                     \frac{e\cdot (g-2)}{2\cdot m}\cdot
(\vec{s} \wtimes \frac{\partial \vec{E}_{{}_P}}{\partial t})
\nonumber\\&&\qquad
+   \frac{e^2\cdot (g-2)\cdot g}{4\cdot m^2}\cdot
 \vec{E}_{{}_P}^\dagger\cdot\vec{B}_{{}_P}\cdot\vec{s}
+\frac{e^2}{m^2}\cdot
 \vec{s}^{\,\dagger}\cdot\vec{B}_{{}_P}\cdot\vec{E}_{{}_P}
-\frac{e^2\cdot (g-2)^2}{4\cdot m^2}\cdot
 \vec{s}^{\,\dagger}\cdot\vec{E}_{{}_P}\cdot\vec{B}_{{}_P}  \; ,
\nonumber\\
 && \vec{s}\ ' =
                          \frac{e\cdot g}{2\cdot m}\cdot
(\vec{s} \wtimes \vec{B}_{{}_P})  \; ,
\label{7.2}
\end{eqnarray}
which is equivalent to:
\footnote{The partial derivative $\partial/\partial t$ in (7.3) acts on
functions depending on $\vec{r}_{{}_M},t$. Note also that in the
nonrelativistic limit one has: $\vec{s}=\vec{\sigma}$.}
\begin{eqnarray}
 && \vec{r}^{\;{}_{'}}_{{}_M}  =  \vec{v}_{{}_M} \; , \nonumber\\
&& m\cdot \vec{v}^{\;{}_{'}}_{{}_M}=
    e\cdot \vec{E}_{{}_M} + \frac{e\cdot g}{2\cdot m}\cdot
\vec{\nabla}_{{}_M} (\vec{\sigma}^{\,\dagger}\cdot \vec{B}_{{}_M}) -
                     \frac{e\cdot g}{2\cdot m}\cdot
(\vec{\sigma} \wtimes \frac{\partial \vec{E}_{{}_M}}{\partial t})
\nonumber\\&&  \qquad
+   \frac{e^2\cdot (g+2)\cdot g}{4\cdot m^2}\cdot
 \vec{E}_{{}_M}^\dagger\cdot\vec{B}_{{}_M}\cdot\vec{\sigma}
-\frac{e^2\cdot g^2}{4\cdot m^2}\cdot
 \vec{\sigma}^{\,\dagger}\cdot\vec{E}_{{}_M}\cdot\vec{B}_{{}_M}  \; ,
\nonumber\\
 && \vec{\sigma}\ ' =
                     \frac{e\cdot g}{2\cdot m}\cdot
(\vec{\sigma} \wtimes \vec{B}_{{}_M})  \; ,
\label{7.3}
\end{eqnarray}
In the nonrelativistic limit the GNR equations lead via
(5.8),(7.1) to:
\begin{eqnarray}
 && \vec{r}^{\;{}_{'}}_{{}_P}  =  \vec{v}_{{}_P} \; , \nonumber\\
 && m\cdot \vec{v}^{\;{}_{'}}_{{}_P}=
    e\cdot \vec{E}_{{}_P} + \frac{e\cdot g}{2\cdot m}\cdot
\vec{\nabla}_{{}_P} (\vec{s}^{\,\dagger}\cdot \vec{B}_{{}_P}) \; ,
\nonumber\\
 && \vec{s}\ ' =  \frac{e\cdot g}{2\cdot m}\cdot
(\vec{s} \wtimes \vec{B}_{{}_P})  \; .
\label{7.4}
\end{eqnarray}
The choice (5.9) leads in the nonrelativistic limit via (7.1) to:
%\footnote{The partial derivative $\partial/\partial t$ in (7.5) acts on
%functions depending on $\vec{r}_{{}_P},t$.}
%
\begin{eqnarray}
 && \vec{r}^{\;{}_{'}}_{{}_P}  =  \vec{v}_{{}_P} \; , \nonumber\\
 && m\cdot \vec{v}^{\;{}_{'}}_{{}_P}=
    e\cdot \vec{E}_{{}_P} + \frac{e\cdot g}{2\cdot m}\cdot
\vec{\nabla}_{{}_P} (\vec{s}^{\,\dagger}\cdot \vec{B}_{{}_P}) -
                     \frac{e\cdot g}{2\cdot m}\cdot
(\vec{s} \wtimes \frac{\partial \vec{E}_{{}_P}}{\partial t})
\nonumber\\&&\qquad
+   \frac{e^2\cdot g^2}{4\cdot m^2}\cdot
 \vec{E}_{{}_P}^\dagger\cdot\vec{B}_{{}_P}\cdot\vec{s}
-\frac{e^2\cdot (g^2-2\cdot g)}{4\cdot m^2}\cdot
 \vec{s}^{\,\dagger}\cdot\vec{E}_{{}_P}\cdot\vec{B}_{{}_P}  \; ,
\nonumber\\
 && \vec{s}\ ' =
                     \frac{e\cdot g}{2\cdot m}\cdot
(\vec{s} \wtimes \vec{B}_{{}_P})  \; ,
\label{7.5}
\end{eqnarray}
which agrees with [4.57] in \cite{Cos94}.
For the choice (5.11) one gets in the nonrelativistic limit via (7.1):
\begin{eqnarray}
 && \vec{r}^{\;{}_{'}}_{{}_P}  =  \vec{v}_{{}_P} \; , \nonumber\\
 && m\cdot \vec{v}^{\;{}_{'}}_{{}_P}=
    e\cdot \vec{E}_{{}_P}  \; ,
\nonumber\\
 && \vec{s}\ ' =
                      \frac{e\cdot g}{2\cdot m}\cdot
(\vec{s} \wtimes \vec{B}_{{}_P})  \; .
\label{7.6}
\end{eqnarray}
To summarize the above one observes that the
characteristic parameters  $c_3,-c_3+2\cdot
c_4-c_5,c_5$ only vanish if  $0=c_3=c_4=c_5$. Using
simple linear algebra it follows from (7.1b) that knowledge of the
nonrelativistic limit (7.1) uniquely determines the values of the
characteristic parameters
$c_1,...,c_5$. As an  immediate application I note that
the particle described in \cite{CPP95} (see eq. (1.1) thereof) has a
nonrelativistic limit identical to (7.4). If this particle belongs to the
class obeying (5.5), one can  conclude that the particle described in
\cite{CPP95} obeys the GNR equations, i.e. the characteristic parameters
$c_1,...,c_5$ for this particle assume the values given by (5.8).
Thus \cite{CPP95} provides a good example of how to identify
$c_1,...,c_5$. That in fact the particle described in \cite{CPP95} is
characterized by the choice (5.8) of the parameters
                                        $c_1,...,c_5$ is
supported by the results of section 8.
\subsection*{7.2}
Although the behaviour in the nonrelativistic limit is closely
related to that in the rest frame, which I define as the frame for
which the velocity $\vec{v}_{{}_P}$
vanishes, the two behaviours are nevertheless
distinct. In the nonrelativistic limit I am observing the motion in a
selected inertial frame but motion with respect to the rest frame is
actually motion with respect to an accelerated  frame.
The Lorentz transformation $L^{{}^R}$ to the particle rest frame
transforms $X^{{}^P}$ to the rest frame space-time position $X^{{}^R}$:
\begin{eqnarray}
 && X^{{}^R}_{\mu}= L^{{}^R}_{\mu\nu} \cdot X^{{}^P}_{\nu} \; ,
 \qquad (\mu =1,...,4)
\label{7.7}
\end{eqnarray}
where
\begin{eqnarray}
&& L^{{}^R}_{jk} = (\gamma_{{}_P}-1)\cdot
           v_{{}_{P,j}}\cdot v_{{}_{P,k}}\cdot
   \frac{1}{\vec{v}_{{}_P}^\dagger\cdot\vec{v}_{{}_P}}+ \delta_{jk} \; ,
\nonumber\\
&& L^{{}^R}_{j4} = i\cdot\gamma_{{}_P} \cdot v_{{}_{P,j}} \; ,
\nonumber\\
&& L^{{}^R}_{4j} = -L^{{}^R}_{j4} \; ,
\nonumber\\
&& L^{{}^R}_{44} = \gamma_{{}_P}  \; .  \qquad (j,k=1,2,3)
\label{7.8}
\end{eqnarray}
Since     the rest frame is an accelerated frame,
 $L^{{}^R}$ depends on the proper time.
Relative to the rest frame the particle motion vanishes, i.e.
\begin{eqnarray}
 && U^{{}^R}_{\mu} \equiv  L^{{}^R}_{\mu\nu} \cdot U^{{}^P}_{\nu}
   =  (0, i)_{\mu} \; .
 \qquad (\mu =1,...,4)
\label{7.9}
\end{eqnarray}
In the rest frame the quantities
$S^{{}^P},F^{{}^P}$ transform to:
\setcounter{INDEX}{1}
\begin{eqnarray}
 && S^{{}^R}_{\mu\nu} \equiv L^{{}^R}_{\mu\rho} \cdot L^{{}^R}_{\nu\omega}
                                                           \cdot
   S^{{}^P}_{\rho\omega} =   \left( \begin{array}{cccc}
      0 & \sigma_3 & -\sigma_2 & 0    \\
     -\sigma_3 & 0 & \sigma_1 & 0   \\
      \sigma_2 & -\sigma_1 & 0 & 0   \\
       0     &
       0     &
       0     &  0
                \end{array}
         \right)_{\mu\nu}  \; ,
\\
\addtocounter{equation}{-1}
\addtocounter{INDEX}{1}
\label{7.10a}
&& F^{{}^R}_{\mu\nu}(X^{{}^R}_1,...,X^{{}^R}_4) \equiv
 L^{{}^R}_{\mu\rho} \cdot L^{{}^R}_{\nu\omega} \cdot
   F^{{}^P}_{\rho\omega}(X^{{}^P}_1,...,X^{{}^P}_4) \; ,
 \qquad (\mu,\nu =1,...,4)
\label{7.10b}
\end{eqnarray}
\setcounter{INDEX}{0}
\par\noindent
where I also used (3.5). Note that on combining (7.8),(7.10a)
one obtains (3.6). Furthermore one concludes from (6.4),(7.8):
\begin{eqnarray}
 T^{{}^R}_{\mu} &=& L^{{}^R}_{\mu\nu} \cdot T^{{}^P}_{\nu}
   =  (\vec{\sigma}^{\,\dagger}, 0)_{\mu} \; ,
 \qquad (\mu =1,...,4)
\label{7.11}
\end{eqnarray}
where $T^{{}^R}$ denotes the spin pseudo-4-vector w.r.t. the rest frame.
\par If   $N^{{}^R}_{\mu}(\tau)$ transforms under the Lorentz group
as a 4-vector or as a space--time position
              and $N^{{}^R}_{\mu\nu}(\tau)$ as a tensor of rank 2, the
rest
frame proper time
derivative $(d/d\tau)_{{}_R}$ is defined by
\begin{eqnarray}
&& (\frac{d}{d\tau})_{{}_R} N^{{}^R}_{\mu} =
 L^{{}^R}_{\mu\nu} \cdot
     \frac{d}{d\tau}\; (L^{{}^R}_{\nu\rho}{}^{-1}\cdot
    N^{{}^R}_{\rho}) \;,
\nonumber\\
&& (\frac{d}{d\tau})_{{}_R} N^{{}^R}_{\mu\nu} =
 L^{{}^R}_{\mu\lambda} \cdot L^{{}^R}_{\nu\rho}\cdot
 \frac{d}{d\tau}\;(L^{{}^R}_{\lambda\alpha}{}^{-1}\cdot
                    L^{{}^R}_{\rho\beta}{}^{-1}\cdot
     N^{{}^R}_{\alpha\beta})  \; .
 \qquad (\mu,\nu =1,...,4)
\label{7.12}
\end{eqnarray}
The distinction between $(d/d\tau)_{{}_R}$ and $d/d\tau$ takes into
account the proper time dependence of the rest frame which occurs
because in general the 4-velocity $U^{{}^P}$ is not a constant of
motion. Applying (7.12) to the quantities
$X^{{}^R},U^{{}^R},S^{{}^R}$ yields:
\begin{eqnarray}
&& (\frac{d}{d\tau})_{{}_R} X^{{}^R}_{\mu} = L^{{}^R}_{\mu\nu}
   \cdot\dot{X}^{{}^P}_{\nu}  \; ,
\nonumber\\
&& (\frac{d}{d\tau})_{{}_R}
   U^{{}^R}_{\mu} =  L^{{}^R}_{\mu\nu} \cdot\dot{U}^{{}^P}_{\nu} \; ,
\nonumber\\
&& (\frac{d}{d\tau})_{{}_R}
    S^{{}^R}_{\mu\nu} =  L^{{}^R}_{\mu\rho} \cdot
                                         L^{{}^R}_{\nu\omega} \cdot
   \dot{S}^{{}^P}_{\rho\omega} \; .
 \qquad (\mu,\nu =1,...,4)
\label{7.13}
\end{eqnarray}
Combining (5.1) with (7.13) results in
\setcounter{INDEX}{1}
\begin{eqnarray}
&& (\frac{d}{d\tau})_{{}_R} X^{{}^R}_{\mu} = U^{{}^R}_{\mu} \; ,
\\
\addtocounter{equation}{-1}
\addtocounter{INDEX}{1}
\label{7.14a}
&& (\frac{d}{d\tau})_{{}_R}
 U^{{}^R}_{\mu} = \frac{e}{m}\cdot L^{{}^R}_{\mu\nu} \cdot
   F^{{}^P}_{\nu\rho}\cdot U^{{}^P}_{\rho}
  +  L^{{}^R}_{\mu\nu} \cdot Y^{{}^P}_{\nu} \; ,
\\
\addtocounter{equation}{-1}
\addtocounter{INDEX}{1}
\label{7.14b}
&& (\frac{d}{d\tau})_{{}_R}
    S^{{}^R}_{\mu\nu} =  L^{{}^R}_{\mu\alpha} \cdot
                                           L^{{}^R}_{\nu\beta} \cdot
\biggl(   \frac{e\cdot g}{2\cdot m}\cdot\lbrack
                   F^{{}^P}_{\alpha\omega}\cdot S^{{}^P}_{\omega\beta}-
                   S^{{}^P}_{\alpha\omega}\cdot F^{{}^P}_{\omega\beta} \rbrack
\nonumber\\&&
  -   \frac{e\cdot (g-2)}{2\cdot m}\cdot\lbrack
S^{{}^P}_{\alpha\omega}\cdot F^{{}^P}_{\omega\lambda}\cdot
                  U^{{}^P}_{\lambda}\cdot U^{{}^P}_{\beta}-
S^{{}^P}_{\beta\omega}\cdot F^{{}^P}_{\omega\lambda}\cdot
                  U^{{}^P}_{\lambda}\cdot U^{{}^P}_{\alpha}
                                           \rbrack \biggr) \; ,
 \;\; (\mu,\nu =1,...,4) \qquad
\label{7.14c}
\end{eqnarray}
\setcounter{INDEX}{0}
\par\noindent
where in (7.14a) I also used (7.9). To simplify (7.14) I introduce
the abbreviation
\begin{eqnarray}
 && \partial^{{}^R}_{\mu} \equiv
    (\frac{\partial}{\partial X^{{}^R}_1} \ , \
     \frac{\partial}{\partial X^{{}^R}_2} \ , \
     \frac{\partial}{\partial X^{{}^R}_3} \ , \
     \frac{\partial}{\partial X^{{}^R}_4})_{\mu} \; ,
 \qquad (\mu =1,...,4)
\label{7.15}
\end{eqnarray}
from which it follows by (7.10b) that:
\footnote{The partial derivatives $\partial^{{}^R}_{\mu}$
always act on functions depending on
$X^{{}^R}_1,X^{{}^R}_2,X^{{}^R}_3,X^{{}^R}_4$.}
\begin{eqnarray}
&&  \partial^{{}^P}_{\mu} F^{{}^P}_{\nu\rho} =
 L^{{}^R}_{\lambda\mu} \cdot L^{{}^R}_{\alpha\nu} \cdot
                                          L^{{}^R}_{\beta\rho} \cdot
    \partial^{{}^R}_{\lambda} F^{{}^R}_{\alpha\beta} \; .
 \qquad (\mu,\nu,\rho =1,...,4)
\label{7.16}
\end{eqnarray}
Combining (5.4),(7.10),(7.16) one gets
\begin{eqnarray}
 &&  L^{{}^R}_{\mu\nu} \cdot Y^{{}^P}_{\nu} =
   -  \frac{e\cdot c_2}{4\cdot m^2}\cdot\biggl(
S^{{}^R}_{\nu\omega}\cdot \partial^{{}^R}_{\mu} F^{{}^R}_{\omega\nu}
+U^{{}^R}_{\mu}\cdot S^{{}^R}_{\nu\omega}\cdot U^{{}^R}_{\lambda}\cdot
                                          \partial^{{}^R}_{\lambda}
                            F^{{}^R}_{\omega\nu}  \biggr)
\nonumber\\&&\qquad
  +   \frac{e\cdot(c_2-c_1-2)}{2\cdot m^2}\cdot
S^{{}^R}_{\mu\nu}\cdot U^{{}^R}_{\omega}\cdot
   U^{{}^R}_{\lambda}\cdot\partial^{{}^R}_{\lambda}
                            F^{{}^R}_{\nu\omega}
  +   \frac{e^2\cdot c_3}{4\cdot m^3}\cdot
F^{{}^R}_{\mu\nu}\cdot S^{{}^R}_{\nu\omega}\cdot
            F^{{}^R}_{\omega\rho}\cdot U^{{}^R}_{\rho}
\nonumber\\&&
  -  \frac{e^2\cdot c_4}{4\cdot m^3}\cdot
F^{{}^R}_{\mu\nu}\cdot U^{{}^R}_{\nu}\cdot
           S^{{}^R}_{\omega\rho}\cdot F^{{}^R}_{\rho\omega}
  +   \frac{e^2\cdot c_5}{4\cdot m^3}\cdot
S^{{}^R}_{\mu\nu}\cdot F^{{}^R}_{\nu\rho}\cdot
            F^{{}^R}_{\rho\omega}\cdot U^{{}^R}_{\omega}
                                       \; ,
 \qquad (\mu =1,...,4)
\label{7.17}
\end{eqnarray}
so that (7.14) specifies to:
\setcounter{INDEX}{1}
\begin{eqnarray}
&& (\frac{d}{d\tau})_{{}_R} X^{{}^R}_{\mu} = U^{{}^R}_{\mu} \; ,
\\
\addtocounter{equation}{-1}
\addtocounter{INDEX}{1}
\label{7.18a}
&& (\frac{d}{d\tau})_{{}_R}
 U^{{}^R}_{\mu} = \frac{e}{m}\cdot F^{{}^R}_{\mu\nu}\cdot
   U^{{}^R}_{\nu} -  \frac{e\cdot c_2}{4\cdot m^2}\cdot\biggl(
S^{{}^R}_{\nu\omega}\cdot \partial^{{}^R}_{\mu} F^{{}^R}_{\omega\nu}
+U^{{}^R}_{\mu}\cdot S^{{}^R}_{\nu\omega}\cdot U^{{}^R}_{\lambda}\cdot
                                          \partial^{{}^R}_{\lambda}
                            F^{{}^R}_{\omega\nu}  \biggr)
\nonumber\\&&\qquad
  +   \frac{e\cdot(c_2-c_1-2)}{2\cdot m^2}\cdot
S^{{}^R}_{\mu\nu}\cdot U^{{}^R}_{\omega}\cdot
   U^{{}^R}_{\lambda}\cdot\partial^{{}^R}_{\lambda}
                            F^{{}^R}_{\nu\omega}
  +   \frac{e^2\cdot c_3}{4\cdot m^3}\cdot
F^{{}^R}_{\mu\nu}\cdot S^{{}^R}_{\nu\omega}\cdot
            F^{{}^R}_{\omega\rho}\cdot U^{{}^R}_{\rho}
\nonumber\\&&\qquad
  -  \frac{e^2\cdot c_4}{4\cdot m^3}\cdot
F^{{}^R}_{\mu\nu}\cdot U^{{}^R}_{\nu}\cdot
           S^{{}^R}_{\omega\rho}\cdot F^{{}^R}_{\rho\omega}
  +   \frac{e^2\cdot c_5}{4\cdot m^3}\cdot
S^{{}^R}_{\mu\nu}\cdot F^{{}^R}_{\nu\rho}\cdot
            F^{{}^R}_{\rho\omega}\cdot U^{{}^R}_{\omega}
                                       \; ,
\\
\addtocounter{equation}{-1}
\addtocounter{INDEX}{1}
\label{7.18b}
&& (\frac{d}{d\tau})_{{}_R}
    S^{{}^R}_{\mu\nu} = \frac{e\cdot g}{2\cdot m}\cdot\biggl(
                   F^{{}^R}_{\mu\omega}\cdot S^{{}^R}_{\omega\nu}-
                   S^{{}^R}_{\mu\omega}\cdot F^{{}^R}_{\omega\nu} \biggr)
\nonumber\\&&
  -   \frac{e\cdot (g-2)}{2\cdot m}\cdot\biggl(
S^{{}^R}_{\mu\omega}\cdot F^{{}^R}_{\omega\lambda}\cdot
                                                 U^{{}^R}_{\lambda}\cdot
                                            U^{{}^R}_{\nu}-
S^{{}^R}_{\nu\omega}\cdot F^{{}^R}_{\omega\lambda}\cdot
                                                 U^{{}^R}_{\lambda}\cdot
                                            U^{{}^R}_{\mu}
                                           \biggr) \; . \;\;
 (\mu,\nu =1,...,4) \qquad
\label{7.18c}
\end{eqnarray}
\setcounter{INDEX}{0}
\par\noindent
Note that (7.18) has the same form as (5.5). Thus the use of
$(d/d\tau)_{{}_R}$ provides an economic formulation of the
rest frame equations of motion.
To compare the rest frame
equations (7.18) with the nonrelativistic behaviour studied in the
previous subsection I introduce the abbreviations
\begin{eqnarray}
 && X^{{}^R}_{\mu} =
   (\vec{r}_{{}_R}^{\,\dagger}, X^{{}^R}_4)_{\mu} \; , \qquad
    U^{{}^R}_{\mu} =
(\gamma_{{}_R}\cdot\vec{v}_{{}_R}^\dagger,i\cdot
                   \gamma_{{}_R})_{\mu} \; ,
\nonumber\\
&& F^{{}^R} \leftrightarrow
      (\vec{B}_{{}_R} ,-i\cdot\vec{E}_{{}_R})  \; .
 \qquad (\mu =1,...,4)
\label{7.19}
\end{eqnarray}
The simple structures of $U^{{}^R}$ and $S^{{}^R}$ allow
the spatial parts of equations (7.18a-c) to be easily obtained and one gets:
\footnote{The nabla operator $\vec{\nabla}_{{}_R}$ acts on functions
depending on $\vec{r}_{{}_R},-i\cdot X_4^{{}^R},\vec{\sigma}$ and it is
the gradient w.r.t. $\vec{r}_{{}_R}$.}
\setcounter{INDEX}{1}
\begin{eqnarray}
&& (\frac{d}{d\tau})_{{}_R} \vec{r}_{{}_R} = 0 \; ,
\\
\addtocounter{equation}{-1}
\addtocounter{INDEX}{1}
\label{7.20a}
 && m\cdot (\frac{d}{d\tau})_{{}_R} \vec{v}_{{}_R} =
  e\cdot \vec{E}_{{}_R} + \frac{e\cdot c_2}{2\cdot m}\cdot
\vec{\nabla}_{{}_R} (\vec{\sigma}^{\,\dagger}\cdot \vec{B}_{{}_R}) -
  \frac{i\cdot e\cdot (c_2-2-c_1)}{2\cdot m}\cdot
(\vec{\sigma} \wtimes \frac{\partial \vec{E}_{{}_R}}
                                              {\partial X^{{}^R}_4})
\nonumber\\&&
+   \frac{e^2\cdot c_3}{4\cdot m^2}\cdot
 \vec{E}_{{}_R}^\dagger\cdot\vec{B}_{{}_R}\cdot\vec{\sigma}
+\frac{e^2}{4\cdot m^2}\cdot ( -c_3+2\cdot c_4-c_5 ) \cdot
 \vec{\sigma}^{\,\dagger}\cdot\vec{B}_{{}_R}\cdot\vec{E}_{{}_R}
+\frac{e^2\cdot c_5}{4\cdot m^2}\cdot
 \vec{\sigma}^{\,\dagger}\cdot\vec{E}_{{}_R}\cdot\vec{B}_{{}_R}  \; ,
\nonumber\\&&
\\
\addtocounter{equation}{-1}
\addtocounter{INDEX}{1}
\label{7.20b}
&&  (\frac{d}{d\tau})_{{}_R} \vec{\sigma}  =
                                         \frac{e\cdot g}{2\cdot m}\cdot
(\vec{\sigma} \wtimes \vec{B}_{{}_R})  \; .
\label{7.20c}
\end{eqnarray}
\setcounter{INDEX}{0}
\par\noindent
By (7.20a) one sees that the rest frame is that frame where the
derivative $(d/d\tau)_{{}_R}$ of the position vanishes. Note
that (7.20) also displays the fact that the rest frame is an {\it accelerated}
frame. In fact, by the definition of $(d/d\tau)_{{}_R}$ one sees
that the rest frame proper time derivative of a vanishing quantity in
general is nonvanishing, which is exemplified by (7.20a-b):
\begin{eqnarray*}
&& (\frac{d}{d\tau})_{{}_R}  (\frac{d}{d\tau})_{{}_R} \vec{r}_{{}_R}
 = (\frac{d}{d\tau})_{{}_R}  \vec{v}_{{}_R}\neq 0 \; .
\end{eqnarray*}
\subsection*{7.3}
In the special case of the Frenkel equations it is interesting to consider
the rest frame behaviour of $\Pi^{{}^P}$
instead of $U^{{}^P}$. Abbreviating:
\begin{eqnarray}
 && \Pi^{{}^R}_{\mu} \equiv  L^{{}^R}_{\mu\nu} \cdot \Pi^{{}^P}_{\nu}
\equiv (\vec{\pi}_{{}_R}^\dagger, \Pi^{{}^R}_4)_{\mu} \; ,
 \qquad (\mu =1,...,4)
\label{7.21}
\end{eqnarray}
one obtains by (4.9),(7.12):
\begin{eqnarray}
&& (\frac{d}{d\tau})_{{}_R}
 \Pi^{{}^R}_{\mu} = \frac{e}{m}\cdot F^{{}^R}_{\mu\nu}\cdot
   U^{{}^R}_{\nu} -  \frac{e\cdot g}{4\cdot m^2}\cdot
S^{{}^R}_{\nu\omega}\cdot \partial^{{}^R}_{\mu} F^{{}^R}_{\omega\nu} \; .
 \qquad (\mu =1,...,4)
\label{7.22}
\end{eqnarray}
Together with (7.20a),(7.20c) one then gets:
\setcounter{INDEX}{1}
\begin{eqnarray}
&& (\frac{d}{d\tau})_{{}_R} \vec{r}_{{}_R} = 0 \; ,
\\
\addtocounter{equation}{-1}
\addtocounter{INDEX}{1}
\label{7.23a}
 &&  (\frac{d}{d\tau})_{{}_R} \vec{\pi}_{{}_R} =
   e\cdot \vec{E}_{{}_R} + \frac{e\cdot g}{2\cdot m}\cdot
\vec{\nabla}_{{}_R} (\vec{\sigma}^{\,\dagger}\cdot \vec{B}_{{}_R}) \; ,
\\
\addtocounter{equation}{-1}
\addtocounter{INDEX}{1}
\label{7.23b}
&&  (\frac{d}{d\tau})_{{}_R} \vec{\sigma}  =
                                         \frac{e\cdot g}{2\cdot m}\cdot
(\vec{\sigma} \wtimes \vec{B}_{{}_R})  \; .
\label{7.23c}
\end{eqnarray}
\setcounter{INDEX}{0}
\par\noindent
Comparing with (7.20) one sees that in the rest frame the canonical
momentum vector for the Frenkel equations fulfills the same equation of
motion as the kinetic momentum vector of the GNR equations.
Note that (7.23) was obtained also in \cite{Pla66a}.
\setcounter{equation}{0}
\section{Estimating the strength of the SG force in magnets}
\subsection*{8.1}
This paper is partly motivated by the suggestion \cite{CPP95,NR87} that the
SG force can be used to separate spin states, either in real space or
`energy space', in (anti-)proton storage rings. Now that I have general
forms for the relativistic SG force I am in a position to carry this
study further. In this section I will apply my equations of motion to
the HERA proton ring (HERA-p) \cite{Bri95}.
I do this for the general case, i.e. for
arbitrary values of $c_1,...,c_5$. Note that for (anti-)protons one has:
$(g-2)/2\approx 1.79$.
\subsection*{8.2}
In this subsection I study the equations of motion of the Pryce
coordinates. I will only treat the case of static (i.e. time
independent) magnetic fields and vanishing electric fields, i.e.:
\footnote{The partial derivative $\partial/\partial t$ in (8.1) acts on
functions depending on $\vec{r}_{{}_P},t$.}
\begin{eqnarray}
 &&\frac{\partial\vec{B}_{{}_P}}{\partial t} = 0 \; , \qquad
 \vec{E}_{{}_P} = 0 \; .
\label{8.1}
\end{eqnarray}
I leave it to the reader to investigate other cases. To facilitate the
numerical evaluations I will use Gaussian units
\footnote{See for example \cite{Jac75}.}
so that I drop the convention: $c=1$. In the case of a static magnetic
field (5.12b) reduces to:
\begin{eqnarray}
 && m\cdot (\gamma_{{}_P}\cdot\vec{v}_{{}_P})' =
\frac{e}{c}\cdot (\vec{v}_{{}_P}\wtimes\vec{B}_{{}_P})
+\frac{e\cdot c_2}{2\cdot m\cdot c\cdot\gamma_{{}_P}}\cdot
 \vec{\nabla}_{{}_P}
(\vec{s}^{\,\dagger}\cdot\vec{B}_{{}_P})
\nonumber\\&&\qquad
+\frac{e\cdot \gamma_{{}_P}}{2\cdot m\cdot c^3}\cdot \lbrack
 (c_1+2)\cdot\vec{s}^{\,\dagger}\cdot\vec{B}_{{}_P}'\cdot\vec{v}_{{}_P}
+(c_2-c_1-2)\cdot\vec{v}_{{}_P}^\dagger\cdot\vec{s}\cdot
    \vec{B}_{{}_P}' \rbrack
\nonumber\\&&\qquad
+\frac{e^2}{4\cdot m^2\cdot c^4}\cdot\lbrack
    c_5\cdot
   \vec{v}_{{}_P}^\dagger\cdot\vec{B}_{{}_P}\cdot
                                      (\vec{B}_{{}_P} \wtimes\vec{s})
   -c_5\cdot
   \vec{B}_{{}_P}^\dagger\cdot\vec{B}_{{}_P}\cdot
                                 (\vec{v}_{{}_P} \wtimes\vec{s})
\nonumber\\&&\qquad
   +(2\cdot c_4-c_3)\cdot
   \vec{s}^{\,\dagger}\cdot\vec{B}_{{}_P}\cdot
                               (\vec{v}_{{}_P} \wtimes\vec{B}_{{}_P})
\rbrack   \; .
\label{8.2}
\end{eqnarray}
To simplify the comparison with \cite{CPP95}, I express $\vec{s}$ in terms of
$\vec{\sigma}$ in (8.2). This leads by (2.4) to:
\footnote{The nabla operator $\hat{\vec{\nabla}}_{{}_P}$ in this paper
always acts on functions depending on
$\vec{r}_{{}_P},t,\vec{v}_{{}_P},\vec{\sigma}$ and it is the gradient
w.r.t. $\vec{r}_{{}_P}$.}
\begin{eqnarray}
 && m\cdot (\gamma_{{}_P}\cdot\vec{v}_{{}_P})' =
   \frac{e}{c}\cdot (\vec{v}_{{}_P}\wtimes\vec{B}_{{}_P})
\nonumber\\&&\qquad
+\frac{e\cdot c_2}{2\cdot m\cdot c\cdot\gamma_{{}_P}}\cdot
   \hat{\vec{\nabla}}_{{}_P} \biggl(
   \vec{B}_{{}_P}^\dagger\cdot \lbrack
     \gamma_{{}_P} \cdot \vec{\sigma} -
      \frac{\gamma_{{}_P}^2}{c^2\cdot(\gamma_{{}_P}+1)}
\cdot\vec{\sigma}^{\,\dagger}\cdot\vec{v}_{{}_P}\cdot
                                                \vec{v}_{{}_P}\rbrack
\biggr)
\nonumber\\&&\qquad
+\frac{e\cdot \gamma_{{}_P}}{2\cdot m\cdot c^3}\cdot \lbrack
 (c_1+2)\cdot
   ( \gamma_{{}_P} \cdot \vec{\sigma} -
    \frac{\gamma_{{}_P}^2}{c^2\cdot(\gamma_{{}_P}+1)}
\cdot \vec{\sigma}^{\,\dagger}\cdot
      \vec{v}_{{}_P}\cdot\vec{v}_{{}_P})^{\,\dagger}\cdot
                                  \vec{B}_{{}_P}'\cdot\vec{v}_{{}_P}
\nonumber\\&&\qquad
+(c_2-c_1-2)\cdot\vec{v}_{{}_P}^\dagger\cdot\vec{\sigma}\cdot
       \vec{B}_{{}_P}' \rbrack
+\frac{e^2}{4\cdot m^2\cdot c^4}\cdot\lbrack
    c_5\cdot
   \vec{v}_{{}_P}^\dagger\cdot\vec{B}_{{}_P}\cdot
   \vec{B}_{{}_P} \wtimes
   ( \gamma_{{}_P} \cdot \vec{\sigma}
\nonumber\\&&\qquad
   -            \frac{\gamma_{{}_P}^2}{c^2\cdot(\gamma_{{}_P}+1)}
  \cdot \vec{\sigma}^{\,\dagger}\cdot\vec{v}_{{}_P}\cdot\vec{v}_{{}_P})
   -c_5\cdot
     \gamma_{{}_P} \cdot
   \vec{B}_{{}_P}^\dagger\cdot\vec{B}_{{}_P}\cdot
                                 (\vec{v}_{{}_P} \wtimes\vec{\sigma})
\nonumber\\&&\qquad
   +(2\cdot c_4-c_3)\cdot
   \vec{B}_{{}_P}^\dagger\cdot (
     \gamma_{{}_P} \cdot \vec{\sigma} -
      \frac{\gamma_{{}_P}^2}{c^2\cdot(\gamma_{{}_P}+1)}
\cdot\vec{\sigma}^{\,\dagger}\cdot\vec{v}_{{}_P}\cdot\vec{v}_{{}_P} )
      \cdot  (\vec{v}_{{}_P} \wtimes\vec{B}_{{}_P})
\rbrack
\nonumber\\
&=& \frac{e}{c}\cdot (\vec{v}_{{}_P}\wtimes\vec{B}_{{}_P})
+\frac{e\cdot c_2}{2\cdot m\cdot c\cdot\gamma_{{}_P}}\cdot
   \hat{\vec{\nabla}}_{{}_P}\biggl(
     \gamma_{{}_P} \cdot
   \vec{B}_{{}_P}^\dagger\cdot\vec{\sigma} -
      \frac{\gamma_{{}_P}^2}{c^2\cdot(\gamma_{{}_P}+1)}
\cdot\vec{\sigma}^{\,\dagger}\cdot\vec{v}_{{}_P}\cdot
   \vec{B}_{{}_P}^\dagger\cdot\vec{v}_{{}_P}\biggr)
\nonumber\\&&\qquad
+\frac{e\cdot \gamma_{{}_P}}{2\cdot m\cdot c^3}\cdot \lbrack
 (c_1+2)\cdot
   ( \gamma_{{}_P} \cdot \vec{\sigma} -
    \frac{\gamma_{{}_P}^2}{c^2\cdot(\gamma_{{}_P}+1)}
\cdot \vec{\sigma}^{\,\dagger}\cdot
      \vec{v}_{{}_P}\cdot\vec{v}_{{}_P})^{\,\dagger}\cdot
      \vec{B}_{{}_P}'\cdot\vec{v}_{{}_P}
\nonumber\\&&\qquad
+(c_2-c_1-2)\cdot\vec{v}_{{}_P}^\dagger\cdot
                       \vec{\sigma}\cdot\vec{B}_{{}_P}'\rbrack
+\frac{e^2}{4\cdot m^2\cdot c^4}\cdot\lbrack
    c_5\cdot \gamma_{{}_P} \cdot
   \vec{v}_{{}_P}^\dagger\cdot\vec{B}_{{}_P}\cdot
  (\vec{B}_{{}_P} \wtimes \vec{\sigma})
\nonumber\\&&\qquad
   + \frac{\gamma_{{}_P}^2}{c^2\cdot(\gamma_{{}_P}+1)}\cdot
    (-c_5+2\cdot c_4-c_3)\cdot
        \vec{\sigma}^{\,\dagger}\cdot\vec{v}_{{}_P}\cdot
   \vec{v}_{{}_P}^\dagger\cdot\vec{B}_{{}_P}\cdot
  (\vec{B}_{{}_P} \wtimes \vec{v}_{{}_P})
\nonumber\\&&\qquad
   -c_5\cdot
     \gamma_{{}_P} \cdot
   \vec{B}_{{}_P}^\dagger\cdot\vec{B}_{{}_P}\cdot
                                 (\vec{v}_{{}_P} \wtimes\vec{\sigma})
   +(2\cdot c_4-c_3)\cdot
     \gamma_{{}_P} \cdot
   \vec{B}_{{}_P}^\dagger\cdot  \vec{\sigma}
      \cdot  (\vec{v}_{{}_P} \wtimes\vec{B}_{{}_P})
\rbrack
\nonumber\\
&=& \frac{e}{c}\cdot (\vec{v}_{{}_P}\wtimes\vec{B}_{{}_P})
+\frac{e\cdot c_2}{2\cdot m\cdot c}\cdot
   \hat{\vec{\nabla}}_{{}_P}
  (\vec{\sigma}^{\,\dagger}\cdot\vec{B}_{{}_P})
+\frac{e}{2\cdot m\cdot c^3}\cdot \lbrack
\gamma_{{}_P}^2\cdot(c_1+2)\cdot\vec{\sigma}^{\,\dagger}\cdot
  \vec{B}_{{}_P}'\cdot\vec{v}_{{}_P}
\nonumber\\&&\qquad
-(c_1+2)\cdot\frac{\gamma_{{}_P}^3}{c^2\cdot(\gamma_{{}_P}+1)}
\cdot
\vec{v}_{{}_P}^{\,\dagger}\cdot\vec{\sigma}
\cdot\vec{v}_{{}_P}^{\,\dagger}\cdot
      \vec{B}_{{}_P}'\cdot\vec{v}_{{}_P}
\nonumber\\&&\qquad
+ \gamma_{{}_P}\cdot
(c_2\cdot\frac{\gamma_{{}_P}}{\gamma_{{}_P}+1}
     -c_1-2)\cdot\vec{v}_{{}_P}^\dagger\cdot
                       \vec{\sigma}\cdot\vec{B}_{{}_P}'\rbrack
+\frac{e^2}{4\cdot m^2\cdot c^4}\cdot\lbrack
    c_5\cdot \gamma_{{}_P} \cdot
   \vec{v}_{{}_P}^\dagger\cdot\vec{B}_{{}_P}\cdot
  (\vec{B}_{{}_P} \wtimes \vec{\sigma})
\nonumber\\&&\qquad
   + \frac{\gamma_{{}_P}^2}{c^2\cdot(\gamma_{{}_P}+1)}\cdot
    (-c_5+2\cdot c_4-c_3)\cdot
 \vec{v}_{{}_P}^{\,\dagger}\cdot\vec{\sigma}
 \cdot
   \vec{v}_{{}_P}^\dagger\cdot\vec{B}_{{}_P}\cdot
  (\vec{B}_{{}_P} \wtimes \vec{v}_{{}_P})
\nonumber\\&&\qquad
   -c_5\cdot
     \gamma_{{}_P} \cdot
   \vec{B}_{{}_P}^\dagger\cdot\vec{B}_{{}_P}\cdot
                                 (\vec{v}_{{}_P} \wtimes\vec{\sigma})
   +(2\cdot c_4-c_3)\cdot
     \gamma_{{}_P} \cdot
 \vec{\sigma}^{\,\dagger}\cdot\vec{B}_{{}_P}
  \cdot(\vec{v}_{{}_P} \wtimes \vec{B}_{{}_P})\rbrack   \; .
\label{8.3}
\end{eqnarray}
I say that the magnetic field is `transverse' if
\footnote{This corresponds, for example, to the case of a particle
travelling instantaneously parallel to the axis in a pure quadrupole
magnetic field or to a particle travelling instantaneously perpendicular
to a pure dipole field.}
\begin{eqnarray}
&& \vec{v}_{{}_P}^\dagger\cdot\vec{B}_{{}_P} = 0 \; ,
\nonumber\\
&& \vec{B}_{{}_P}' = 0 \; .
\label{8.4}
\end{eqnarray}
Hence for a static, transverse magnetic field (8.3) simplifies to:
\begin{eqnarray}
 && m\cdot (\gamma_{{}_P}\cdot\vec{v}_{{}_P})' =
          \frac{e}{c}\cdot (\vec{v}_{{}_P}\wtimes\vec{B}_{{}_P})
+\frac{e\cdot c_2}{2\cdot m\cdot c}\cdot
   \hat{\vec{\nabla}}_{{}_P}
 (\vec{\sigma}^{\,\dagger}\cdot\vec{B}_{{}_P})
\nonumber\\&&
+\frac{e^2}{4\cdot m^2\cdot c^4}\cdot\lbrack
   -c_5\cdot
     \gamma_{{}_P} \cdot
   \vec{B}_{{}_P}^\dagger\cdot\vec{B}_{{}_P}\cdot
                                 (\vec{v}_{{}_P} \wtimes\vec{\sigma})
   +(2\cdot c_4-c_3)\cdot
     \gamma_{{}_P} \cdot
  \vec{\sigma}^{\,\dagger}\cdot\vec{B}_{{}_P}
      \cdot  (\vec{v}_{{}_P} \wtimes\vec{B}_{{}_P})
\rbrack   \; . \; \; \;
%\nonumber\\&&
\label{8.5}
\end{eqnarray}
If the magnetic field is transverse and the spin is parallel to the
magnetic field then (8.5) simplifies to:
\begin{eqnarray}
 && m\cdot (\gamma_{{}_P}\cdot\vec{v}_{{}_P})' =
          \frac{e}{c}\cdot (\vec{v}_{{}_P}\wtimes\vec{B}_{{}_P})
+\frac{e\cdot c_2}{2\cdot m\cdot c}\cdot
   \hat{\vec{\nabla}}_{{}_P}
  (\vec{\sigma}^{\,\dagger}\cdot\vec{B}_{{}_P})
\nonumber\\&&\qquad
+\frac{e^2}{4\cdot m^2\cdot c^4}\cdot \gamma_{{}_P} \cdot
  (-c_5+ 2\cdot c_4-c_3)\cdot
  \vec{\sigma}^{\,\dagger}\cdot\vec{B}_{{}_P}
      \cdot  (\vec{v}_{{}_P} \wtimes\vec{B}_{{}_P})  \; .
\label{8.6}
\end{eqnarray}
Thus with $c_2=g$ and when using Pryce coordinates the first order SG
force equals the SG force given in \cite{CPP95}.
Note that the SG force in (8.6) does not depend on $c_1$.
\par For protons one has
\begin{eqnarray}
   \frac{e\cdot \hbar}{m\cdot c} &=&
                    10^{-23}\cdot ergs\cdot Gauss^{-1} =
           6.3\cdot 10^{-18}\cdot MeV\cdot Gauss^{-1} \;  .
\label{8.7}
\end{eqnarray}
In the HERA proton ring at about 800 GeV the magnetic field $B_{HD}$ in
the bending magnets is about 45 kGauss and the quadrupole magnetic field
gradient $G_{HQ}$ is about 9 kGauss/cm. Thus one has
\begin{eqnarray}
&& \frac{e^2\cdot B_{HD}^2\cdot\hbar}{m^2\cdot c^3}
   = 6.5\cdot 10^{-21}\cdot dyne =
     4.1\cdot 10^{-15}\cdot MeV\cdot cm^{-1} \;  ,
\nonumber\\
&& \frac{e\cdot G_{HQ}\cdot\hbar}{m\cdot c}
   =  9.1\cdot 10^{-20}\cdot dyne
   =  5.7\cdot 10^{-14}\cdot MeV\cdot cm^{-1} \;  .
\label{8.8}
\end{eqnarray}
For the static, transverse magnetic field with spin vector
$\vec{\sigma}$ parallel to the magnetic field the numerical values (8.8)
lead via (8.6) to the following maximal values of the SG force:
\begin{eqnarray}
&&  \frac{e\cdot c_2}{2\cdot m\cdot c}\cdot
   \hat{\vec{\nabla}}_{{}_P}
  (\vec{\sigma}^{\,\dagger}\cdot\vec{B}_{{}_P})\;
     \rightarrow\;
    c_2\cdot 2.3\cdot 10^{-20}\cdot dyne =
    c_2\cdot 1.4\cdot 10^{-14}\cdot MeV\cdot cm^{-1} \;  ,
\nonumber\\&&
 \frac{e^2}{4\cdot m^2\cdot c^4}\cdot
    (-c_5+2\cdot c_4-c_3)\cdot
     \gamma_{{}_P} \cdot
   \vec{\sigma}^{\,\dagger}\cdot\vec{B}_{{}_P}
      \cdot  (\vec{v}_{{}_P} \wtimes\vec{B}_{{}_P})\;     \rightarrow
\nonumber\\&&
      (-c_5+2\cdot c_4-c_3)\cdot  \gamma_{{}_P} \cdot
              8.2\cdot 10^{-22}\cdot dyne =
      (-c_5+2\cdot c_4-c_3)\cdot  \gamma_{{}_P} \cdot
              5.1\cdot 10^{-16}\cdot MeV\cdot cm^{-1}
\nonumber\\ &&
     = (-c_5+2\cdot c_4-c_3)\cdot
              7.3\cdot 10^{-19}\cdot dyne
     = (-c_5+2\cdot c_4-c_3)\cdot
              4.6\cdot 10^{-13}\cdot MeV\cdot cm^{-1} \;  ,
\nonumber\\ &&
\label{8.9}
\end{eqnarray}
where I assumed $\gamma_{{}_P}=900$.
\par Thus at HERA-p energies and for the Frenkel equations (where one
has: $-c_5+2\cdot c_4-c_3=4)$ (see (5.6)), the second order SG force in
the dipoles completely outweighs the first order SG force in the
quadrupoles. It is also simple to show that the second order force for a
particle travelling about 1 mm off axis through a quadrupole is
negligible compared to both of the above. In the case of the GNR force
(see (5.8)) the second order force is zero. Clearly, if the SG force
is to be utilized in high energy proton storage rings, one must first
decide which equation of motion appertains.
\par I say that the magnetic field is `longitudinal' if
\begin{eqnarray}
&& \vec{v}_{{}_P} \wtimes \vec{B}_{{}_P} = 0 \; ,
\nonumber\\
&& \vec{B}_{{}_P}' =
  \frac{1}{\vec{v}_{{}_P}^\dagger\cdot\vec{v}_{{}_P}}\cdot
   \vec{v}_{{}_P}^\dagger\cdot\vec{B}_{{}_P}'\cdot
   \vec{v}_{{}_P} \; ,
\label{8.10}
\end{eqnarray}
so that for a static, longitudinal magnetic field one gets:
\begin{eqnarray*}
&& \vec{B}_{{}_P} =
  \frac{1}{\vec{v}_{{}_P}^\dagger\cdot\vec{v}_{{}_P}}\cdot
   \vec{v}_{{}_P}^\dagger\cdot\vec{B}_{{}_P}\cdot
   \vec{v}_{{}_P} \; .
\end{eqnarray*}
Hence for a static, longitudinal magnetic field (8.3) simplifies to:
\begin{eqnarray}
 && m\cdot (\gamma_{{}_P}\cdot\vec{v}_{{}_P})' =
 \frac{e\cdot c_2}{2\cdot m\cdot c}\cdot
  \frac{1}{\vec{v}_{{}_P}^\dagger\cdot\vec{v}_{{}_P}}\cdot
  \gamma_{{}_P}\cdot
\vec{v}_{{}_P}^{\,\dagger}\cdot\vec{\sigma}
\cdot
                                   \vec{B}_{{}_P}' \; .
\label{8.11}
\end{eqnarray}
Note that the second order SG terms have disappeared and that the force
only depends on $c_2$. If the spin is parallel to the field this also
agrees with the form given in \cite{CPP95} if $c_2 = g$.
However, because $c_1$ does not appear in (8.6) and (8.11), one sees
that for these special magnetic fields, the GNR and reduced
Frenkel equations are identical.
\par I consider a gradient $G_{long}$ of the longitudinal field along the
longitudinal direction of about 9 kGauss/cm\cite{CPP95}, so that:
\begin{eqnarray}
&& \frac{e\cdot G_{long}\cdot\hbar}{m\cdot c}
   =  9.1\cdot 10^{-20}\cdot dyne
   =  5.7\cdot 10^{-14}\cdot MeV\cdot cm^{-1} \;  .
\label{8.12}
\end{eqnarray}
Thus for the static, longitudinal magnetic field with spin vector
$\vec{\sigma}$ parallel to
the magnetic field the numerical value (8.12) leads via (8.11) to the
following maximal value of the SG force:
\begin{eqnarray}
&& \frac{e\cdot c_2}{2\cdot m\cdot c}\cdot
  \frac{1}{\vec{v}_{{}_P}^\dagger\cdot\vec{v}_{{}_P}}\cdot
  \gamma_{{}_P}\cdot
\vec{v}_{{}_P}^{\,\dagger}\cdot\vec{\sigma}
\cdot
                                   \vec{B}_{{}_P}' \;
     \rightarrow\;
  c_2\cdot\gamma_{{}_P}\cdot
                     2.3\cdot 10^{-20}\cdot dyne
\nonumber\\ &&
 = c_2\cdot\gamma_{{}_P}\cdot
                     1.4\cdot 10^{-14}\cdot MeV\cdot cm^{-1}
= c_2\cdot 2\cdot 10^{-17}\cdot dyne
= c_2\cdot 1.3\cdot 10^{-11}\cdot MeV\cdot cm^{-1} \; .
\nonumber\\ &&
\label{8.13}
\end{eqnarray}
This agrees with \cite{CPP95} where it is pointed out that the longitudinal
SG force is much larger than the first order transverse force. It is also
much larger than the second order transverse force.
However, before this force could be used to separate the spin ensemble
into two parts \cite{CPP95}, a way must be found to overcome the severe
mixing \cite{Hof96} caused by incoherent synchrotron oscillations.
For a survey of other problems see \cite{Der95,Der90b}.
\par Because for the GNR equations  $c_2=g$,
the identity of (8.6),(8.11) with the corresponding results in \cite{CPP95}
is consistent with the supposition in  section 7 that the particle described in
\cite{CPP95} obeys the GNR equations, i.e. belongs to the choice
(5.8) of the characteristic parameters $c_1,...,c_5$.
\subsection*{8.3}
This paper opened with a description of the DK
Hamiltonian. This is based on the `M' variables and so far I have only
used these to provide the missing link connecting the
DK equations to special relativity. However, since
many other investigations \cite{BHR94a,BHR94b,Der90a,Der90b} have been based
on this Hamiltonian it is natural that one inspects the equations of
motion for the `M' variables in more detail.
As in the previous subsection I will only treat the case of static (i.e. time
independent) magnetic fields and vanishing electric fields, i.e.:
\footnote{The partial derivative $\partial/\partial t$ in (8.14) acts on
functions depending on $\vec{r}_{{}_M},t$.}
\begin{eqnarray}
 &&\frac{\partial\vec{B}_{{}_M}}{\partial t} = 0 \; , \qquad
 \vec{E}_{{}_M} = 0 \; .
\label{8.14}
\end{eqnarray}
Note that by (2.7) the conditions (8.1),(8.14) are equivalent.
\par I begin by using Appendix C to rewrite (8.3) as:
\begin{eqnarray}
 && m\cdot (\gamma_{{}_M}\cdot\vec{v}_{{}_M})'
 =   \frac{e}{c}\cdot (\vec{v}_{{}_M}\wtimes\vec{B}_{{}_M})
 + \frac{e}{2\cdot m\cdot c}\cdot
  (c_2-2+\frac{2}{\gamma_{{}_M}})\cdot
  \vec{\nabla}_{{}_M} (\vec{\sigma}^{\,\dagger}\cdot \vec{B}_{{}_M})
\nonumber\\&&\qquad
 + \frac{e}{2\cdot m\cdot c^3}\cdot\biggl(
 \lbrack(c_2-g)\cdot\frac{\gamma_{{}_M}^2}{\gamma_{{}_M}+1}-c_1\cdot
\gamma_{{}_M}\rbrack
\cdot    \vec{\sigma}^{\,\dagger}\cdot \vec{v}_{{}_M}\cdot\vec{B}_{{}_M}'
\nonumber\\&&\qquad
+ (c_1\cdot\gamma_{{}_M}^2+2\cdot\gamma_{{}_M})\cdot
  \vec{\sigma}^{\,\dagger}\cdot \vec{B}_{{}_M}'\cdot\vec{v}_{{}_M}
\nonumber\\&&\qquad
+\frac{1}{c^2}\cdot \frac{\gamma_{{}_M}^2}{\gamma_{{}_M}+1}\cdot\lbrack
\frac{\gamma_{{}_M}}{\gamma_{{}_M}+1}\cdot(g-2)
- c_1\cdot\gamma_{{}_M}\rbrack\cdot
      \vec{\sigma}^{\,\dagger}\cdot
      \vec{v}_{{}_M}\cdot\vec{v}_{{}_M}^{\,\dagger}\cdot
      \vec{B}_{{}_M}'\cdot\vec{v}_{{}_M}
\nonumber\\&&\qquad
   + (g-2)\cdot\frac{\gamma_{{}_M}}{\gamma_{{}_M}+1}\cdot
   \vec{v}_{{}_M}^{\,\dagger}\cdot\vec{B}_{{}_M}'\cdot\vec{\sigma}\biggr)
\nonumber\\&&\qquad
+\frac{e^2}{4\cdot m^2\cdot c^4}\cdot\biggl(
\lbrack 4-4\cdot\gamma_{{}_M}+\gamma_{{}_M}\cdot
   (2\cdot c_4-c_3)\rbrack\cdot
    \vec{\sigma}^{\,\dagger}\cdot\vec{B}_{{}_M}
      \cdot  (\vec{v}_{{}_M} \wtimes\vec{B}_{{}_M})
\nonumber\\&&
 + \frac{1}{c^2}\cdot\frac{\gamma_{{}_M}^2}{\gamma_{{}_M}+1}\cdot\lbrack
   -2\cdot c_4+c_3+c_5+4+(g^2-2\cdot g)\cdot
 \frac{1}{\gamma_{{}_M}+1}\rbrack\cdot
 \vec{v}_{{}_M}^{\,\dagger}\cdot\vec{\sigma}\cdot
 \vec{v}_{{}_M}^{\,\dagger}\cdot\vec{B}_{{}_M}
       \cdot(\vec{v}_{{}_M}\wtimes \vec{B}_{{}_M})
\nonumber\\&&\qquad
+\lbrack(g-2)\cdot g\cdot\frac{1}{\gamma_{{}_M}+1}-\gamma_{{}_M}\cdot c_5
\rbrack\cdot \vec{v}_{{}_M}^{\,\dagger}\cdot\vec{B}_{{}_M}\cdot
  (\vec{\sigma}\wtimes\vec{B}_{{}_M})
   -c_5\cdot\gamma_{{}_M}\cdot
   \vec{B}_{{}_M}^\dagger\cdot\vec{B}_{{}_M}\cdot
                                 (\vec{v}_{{}_M} \wtimes\vec{\sigma})
\nonumber\\&&\qquad
+(g-2)^2\cdot\frac{\gamma_{{}_M}^2}{\gamma_{{}_M}+1}\cdot
(\vec{\sigma}\wtimes\vec{v}_{{}_M})^\dagger\cdot\vec{B}_{{}_M}\cdot
\vec{B}_{{}_M}
\nonumber\\&&\qquad
-\frac{1}{c^2}\cdot(g^2-2\cdot g)\cdot
   \frac{\gamma_{{}_M}^2}{\gamma_{{}_M}+1}\cdot
 (\vec{\sigma}\wtimes\vec{v}_{{}_M})^\dagger\cdot\vec{B}_{{}_M}
   \cdot\vec{v}_{{}_M}^{\,\dagger}\cdot\vec{B}_{{}_M}
                                      \cdot\vec{v}_{{}_M}\biggr) \; .
\label{8.15}
\end{eqnarray}
If the magnetic field is transverse and the spin
is parallel to the magnetic field then (8.15) simplifies to:
\begin{eqnarray}
 && m\cdot (\gamma_{{}_M}\cdot\vec{v}_{{}_M})'
 =
   \frac{e}{c}\cdot (\vec{v}_{{}_M}\wtimes\vec{B}_{{}_M})
 + \frac{e}{2\cdot m\cdot c}\cdot
  (c_2-2+\frac{2}{\gamma_{{}_M}})\cdot
  \vec{\nabla}_{{}_M} (\vec{\sigma}^{\,\dagger}\cdot \vec{B}_{{}_M})
\nonumber\\&&\qquad
+\frac{e^2}{4\cdot m^2\cdot c^4}\cdot
\lbrack 4-4\cdot\gamma_{{}_M}+\gamma_{{}_M}\cdot
   (2\cdot c_4-c_3-c_5)\rbrack\cdot
   \vec{\sigma}^{\,\dagger}\cdot\vec{B}_{{}_M}
      \cdot  (\vec{v}_{{}_M} \wtimes\vec{B}_{{}_M}) \; ,
\label{8.16}
\end{eqnarray}
which for the Frenkel equations (see (5.6)) becomes:
\begin{eqnarray}
 && m\cdot (\gamma_{{}_M}\cdot\vec{v}_{{}_M})'
 =
   \frac{e}{c}\cdot (\vec{v}_{{}_M}\wtimes\vec{B}_{{}_M})
 + \frac{e}{2\cdot m\cdot c}\cdot
  (g -2+\frac{2}{\gamma_{{}_M}})\cdot
  \vec{\nabla}_{{}_M} (\vec{\sigma}^{\,\dagger}\cdot \vec{B}_{{}_M})
\nonumber\\&&\qquad
+\frac{e^2}{m^2\cdot c^4}\cdot
  \vec{\sigma}^{\,\dagger}\cdot\vec{B}_{{}_M}
       \cdot  (\vec{v}_{{}_M} \wtimes\vec{B}_{{}_M})     \; .
\label{8.17}
\end{eqnarray}
For the GNR case (see (5.8)) equation (8.16) becomes:
\begin{eqnarray}
 && m\cdot (\gamma_{{}_M}\cdot\vec{v}_{{}_M})'
 =  \frac{e}{c}\cdot (\vec{v}_{{}_M}\wtimes\vec{B}_{{}_M})
 + \frac{e}{2\cdot m\cdot c}\cdot
  (g -2+\frac{2}{\gamma_{{}_M}})\cdot
  \vec{\nabla}_{{}_M} (\vec{\sigma}^{\,\dagger}\cdot \vec{B}_{{}_M})
\nonumber\\&&\qquad
-\frac{e^2}{m^2\cdot c^4}\cdot (\gamma_{{}_M}-1)\cdot
   \vec{\sigma}^{\,\dagger}\cdot\vec{B}_{{}_M}
      \cdot  (\vec{v}_{{}_M} \wtimes\vec{B}_{{}_M})     \; .
\label{8.18}
\end{eqnarray}
The first order parts of (8.17) and (8.18) are identical.
\par Although I have obtained the equations for the kinetic momentum,
Hamiltonians lead more naturally to equations of motion for canonical
momenta. If the magnetic field is transverse and the spin is parallel to
the magnetic field then by (B.4),(B.6),(B.17),(B.25-26) and by taking
just the first order SG terms the Frenkel case gives:
\begin{eqnarray}
  \vec{\pi}_{{}_M}'
 &=&  \frac{e}{c}\cdot (\vec{v}_{{}_M}\wtimes\vec{B}_{{}_M})
 + \frac{e}{2\cdot m\cdot c}\cdot
  (g -2+\frac{2}{\gamma_{{}_M}})\cdot
  \vec{\nabla}_{{}_M} (\vec{\sigma}^{\,\dagger}\cdot \vec{B}_{{}_M})
\nonumber\\&& \qquad
- \frac{e\cdot(g-2)}{2\cdot m\cdot c^3}\cdot
\frac{\gamma_{{}_M}}{\gamma_{{}_M}+1}
\cdot
\vec{v}_{{}_M}^{\,\dagger}\cdot\vec{\sigma}
\cdot\vec{B}_{{}_M}' \; ,
\label{8.19}
\end{eqnarray}
and
\begin{eqnarray}
 \vec{p}^{\;{}_{'}}_{{}_M} &=&
   \frac{e}{c}\cdot (\vec{v}_{{}_M}\wtimes\vec{B}_{{}_M})
   -e \cdot\vec{\nabla}_{{}_M}\phi_{{}_M} +
    \frac{e}{c}\cdot
 (\vec{v}_{{}_M}^\dagger\cdot\vec{\nabla}_{{}_M}) \vec{A}_{{}_M} \nonumber\\
 &+&  \frac{e}{2\cdot m\cdot c}\cdot
  (g -2+\frac{2}{\gamma_{{}_M}})\cdot
  \vec{\nabla}_{{}_M} (\vec{\sigma}^{\,\dagger}\cdot \vec{B}_{{}_M})
 - \frac{e\cdot(g-2)}{2\cdot m\cdot c^3}\cdot
\frac{\gamma_{{}_M}}{\gamma_{{}_M}+1}
\cdot\vec{v}_{{}_M}^{\,\dagger}\cdot\vec{\sigma}
\cdot\vec{B}_{{}_M}' \; .\qquad
\label{8.20}
\end{eqnarray}
In equations (8.15-20) the first order SG piece contains the factor
$(g/2 -1 + 1/\gamma_{{}_M})$ which differs from the
corresponding term for `P' variables in (8.6)
by the term $(-1 + 1/\gamma_{{}_M})$
\cite{Hof95}. This term can be directly traced back to a similar term in
 $\vec{\Omega}_{{}_M}$ in (1.6c) associated with Thomas precession
\cite{Jac76}.
\footnote{Note that in the nonrelativistic limit there is no difference
(see (7.2-3) or put ${\gamma_{{}_M}} = 1$) .}
\par Further differences appear for the second order terms both for the
Frenkel and GNR equations. Thus naive use of the DK
Hamiltonian to obtain estimates of the relativistic SG force could lead
one to the conclusion that it gives a first order SG force very
different from that predicted by \cite{CPP95}.
In particular, for electrons one has: $(g-2)/2\approx 0.00116$ so that
the Thomas term causes a massive relative change.
\par However, it is clear that the difference is only an artifact of the
choice of position coordinates \cite{DH95}:
the rest frame implied by the `M'
variables is different from that of the `P' variables and the
corresponding Thomas precession terms are different.
Furthermore, although the forces on
$\vec{r}_{{}_M}$ and $\vec{r}_{{}_P}$ can be rather different, the
position variable $\vec{r}_{{}_M}$ is always closer to $\vec{r}_{{}_P}$
than the particle Compton wave length (see (2.1)). Of course, the
corresponding equations of motion conserve this property in time.
\par The equation of motion (2.36) given in \cite{CJKP96} also contains the
$(-1 + 1/\gamma_{{}_M})$ term and thereby appears to differ
from \cite{CPP95}. This should now come as no surprise
since the former works with Newton-Wigner coordinates.
\setcounter{equation}{0}
\section{Estimating the strength of the SG force in electromagnetic
traps}
\subsection*{9.1}
Most accounts of the SG force emphasize the first order
                                              component. But now
that one has seen that the second order force can be important at high
energy in large fields it is interesting to estimate their effect under
other circumstances where the SG forces play a central role. An obvious
case is that of the nonrelativistic SG force in electromagnetic traps
\cite{DSV86}. As in the previous section I use Gaussian units
and I consider the general case, i.e. arbitrary values of $c_1,...,c_5$.
By (7.1b) one has:
\footnote{The partial derivative $\partial/\partial t$ in
(9.1) and (9.2) acts on
functions depending on $\vec{r}_{{}_P},t$.}
\begin{eqnarray}
 && m\cdot \vec{v}^{\;{}_{'}}_{{}_P}=
    e\cdot \vec{E}_{{}_P} + \frac{e\cdot c_2}{2\cdot m\cdot c}\cdot
\vec{\nabla}_{{}_P} (\vec{\sigma}^{\,\dagger}\cdot \vec{B}_{{}_P}) -
                     \frac{e\cdot (c_2-2-c_1)}{2\cdot m\cdot c^2}\cdot
(\vec{\sigma} \wtimes \frac{\partial \vec{E}_{{}_P}}{\partial t})
\nonumber\\&&
+   \frac{e^2\cdot c_3}{4\cdot m^2\cdot c^3}\cdot
 \vec{E}_{{}_P}^\dagger\cdot\vec{B}_{{}_P}\cdot\vec{\sigma}
+\frac{e^2}{4\cdot m^2\cdot c^3}\cdot ( -c_3+2\cdot c_4-c_5 ) \cdot
 \vec{\sigma}^{\,\dagger}\cdot\vec{B}_{{}_P}\cdot\vec{E}_{{}_P}
\nonumber\\&&
+\frac{e^2\cdot c_5}{4\cdot m^2\cdot c^3}\cdot
\vec{\sigma}^{\,\dagger}\cdot\vec{E}_{{}_P}\cdot\vec{B}_{{}_P}\; .
\label{9.1}
\end{eqnarray}
If the magnetic field and electric field are parallel to the spin then
(9.1) simplifies to:
%\footnote{The partial derivative $\partial/\partial t$ in (9.2) acts on
%functions depending on $\vec{r}_{{}_P},t$.}
%
\begin{eqnarray}
 && m\cdot \vec{v}^{\;{}_{'}}_{{}_P}=
    e\cdot \vec{E}_{{}_P} + \frac{e\cdot c_2}{2\cdot m\cdot c}\cdot
\vec{\nabla}_{{}_P} (\vec{\sigma}^{\,\dagger}\cdot \vec{B}_{{}_P}) -
                     \frac{e\cdot (c_2-2-c_1)}{2\cdot m\cdot c^2}\cdot
(\vec{\sigma} \wtimes \frac{\partial \vec{E}_{{}_P}}{\partial t})
\nonumber\\&&\qquad
+   \frac{e^2\cdot c_4}{2\cdot m^2\cdot c^3}\cdot
 \vec{E}_{{}_P}^\dagger\cdot\vec{B}_{{}_P}\cdot\vec{\sigma} \; .
\label{9.2}
\end{eqnarray}
\subsection*{9.2}
To apply my equations to the case of the electron I calculate
\begin{eqnarray}
   \frac{e\cdot \hbar}{m\cdot c} &=&
           1.9\cdot 10^{-20}\cdot ergs\cdot Gauss^{-1} =
           1.16\cdot 10^{-14}\cdot MeV\cdot Gauss^{-1} \;  .
\label{9.3}
\end{eqnarray}
In one of the traps used in \cite{DSV86} the magnetic field $B_{trap}$
is about 20 kGauss and the electric field $E_{trap}$ is about
0.000033 Statvolt/cm=1000 V/meter $\widehat{=}$ 61 Gauss.   Then I get
\begin{eqnarray}
&& \frac{e^2\cdot B_{trap}\cdot E_{trap}\cdot
\hbar}{m^2\cdot c^3}=7.3\cdot 10^{-21}\cdot dyne =
                     4.5\cdot 10^{-15}\cdot MeV\cdot cm^{-1} \;  .
\label{9.4}
\end{eqnarray}
If the electric and magnetic fields are static and parallel to the spin
then the numerical value (9.4) leads via (9.2) to the following maximal
value of the second order SG force
\begin{eqnarray}
&& \frac{e^2\cdot c_4}{2\cdot m^2\cdot c^3}\cdot
 \vec{E}_{{}_P}^\dagger\cdot\vec{B}_{{}_P}\cdot\vec{\sigma}
     \rightarrow\;
  c_4\cdot 1.8\cdot 10^{-21}\cdot dyne =
  c_4\cdot 1.1\cdot 10^{-15}\cdot MeV\cdot cm^{-1} \; .\qquad
\label{9.5}
\end{eqnarray}
The magnetic field has a gradient $G_{trap}$ of about 2.4 Gauss/cm so
that
\begin{eqnarray}
&& \frac{e\cdot G_{trap}\cdot\hbar}{m\cdot c}
       =   4.5\cdot 10^{-20}\cdot dyne
       =   2.8\cdot 10^{-14}\cdot MeV\cdot cm^{-1} \;  .
\label{9.6}
\end{eqnarray}
For the static, transverse magnetic field when the spin vector
$\vec{\sigma}$ is parallel to the magnetic field the numerical value
(9.6) leads via (9.1) to the following maximal values of the first order
SG force:
\begin{eqnarray}
&&  \frac{e\cdot c_2}{2\cdot m\cdot c}\cdot
   \hat{\vec{\nabla}}_{{}_P}
  (\vec{\sigma}^{\,\dagger}\cdot\vec{B}_{{}_P})\;
     \rightarrow\;
    c_2\cdot 1.1 \cdot 10^{-20}\cdot dyne =
    c_2\cdot 6.9 \cdot 10^{-15}\cdot MeV\cdot cm^{-1} \;  . \qquad
\label{9.7}
\end{eqnarray}
Thus for the Frenkel case ((5.6) with $g \approx 2)$ the second order SG
force is comparable to, but still smaller than, the first order force.
For the GNR case ((5.8) with $g \approx 2)$ the first order
force is unchanged but the second order force vanishes.
\par For protons the values corresponding to (9.5),(9.7) are smaller and the
second order SG force is much smaller than the first order force.
\par My main purpose in presenting these numbers is to compare the
different terms of the SG force. I do not claim that they give a good
representation of the real SG force in a trap because my formalism only
applies to the semiclassical regime and this may not always be directly
applicable to traps.
\setcounter{equation}{0}
\section{A transformation of the GNR equations}
\subsection*{10.1}
In this section I deal with transformations  which in particular allow
the GNR equations to be transformed into the reduced Frenkel equations.
This would, for example,   make  it possible to solve the GNR
equations by symplectic methods.
\subsection*{10.2}
The general equations (5.5) in the approximation that the second order
SG terms are neglected read as:
\setcounter{INDEX}{1}
\begin{eqnarray}
 && \dot{X}^{{}^P}_{\mu} = U^{{}^P}_{\mu} \; ,
\\
\addtocounter{equation}{-1}
\addtocounter{INDEX}{1}
\label{10.1a}
&&\dot{U}^{{}^P}_{\mu} =
  \frac{e}{m}\cdot F^{{}^P}_{\mu\nu}\cdot U^{{}^P}_{\nu}
   -  \frac{e\cdot c_2}{4\cdot m^2}\cdot\biggl(
              S^{{}^P}_{\nu\omega}\cdot
          \partial^{{}^P}_{\mu} F^{{}^P}_{\omega\nu}
+U^{{}^P}_{\mu}\cdot S^{{}^P}_{\nu\omega}\cdot U^{{}^P}_{\lambda}\cdot
                                            \partial^{{}^P}_{\lambda}
                            F^{{}^P}_{\omega\nu}  \biggr)
\nonumber\\&&\qquad
  +   \frac{e\cdot(c_2-c_1-2)}{2\cdot m^2}\cdot
S^{{}^P}_{\mu\nu}\cdot U^{{}^P}_{\omega}\cdot
      U^{{}^P}_{\lambda}\cdot\partial^{{}^P}_{\lambda}
                            U^{{}^P}_{\nu\omega}  \; ,
\\
\addtocounter{equation}{-1}
\addtocounter{INDEX}{1}
\label{10.1b}
 && \dot{S}^{{}^P}_{\mu\nu} =  \frac{e\cdot g}{2\cdot m}\cdot\biggl(
                   F^{{}^P}_{\mu\omega}\cdot S^{{}^P}_{\omega\nu}-
                   S^{{}^P}_{\mu\omega}\cdot F^{{}^P}_{\omega\nu} \biggr)
\nonumber\\&&\qquad
  -   \frac{e\cdot (g-2)}{2\cdot m}\cdot\biggl(
S^{{}^P}_{\mu\omega}\cdot F^{{}^P}_{\omega\lambda}\cdot U^{{}^P}_{\lambda}
                                             \cdot U^{{}^P}_{\nu}-
S^{{}^P}_{\nu\omega}\cdot F^{{}^P}_{\omega\lambda}\cdot
  U^{{}^P}_{\lambda}\cdot U^{{}^P}_{\mu}
                                           \biggr) \; .
 \qquad (\mu,\nu =1,...,4)
\nonumber\\&&
\label{10.1c}
\end{eqnarray}
\setcounter{INDEX}{0}
\par\noindent
Equations (10.1) are obtained by setting $c_3=c_4=c_5=0$ in (5.5).
In the case where $c_2=g$ equations (10.1) simplify to:
\setcounter{INDEX}{1}
\begin{eqnarray}
 && \dot{X}^{{}^P}_{\mu} = U^{{}^P}_{\mu} \; ,
\\
\addtocounter{equation}{-1}
\addtocounter{INDEX}{1}
\label{10.2a}
 && \dot{U}^{{}^P}_{\mu} =
  \frac{e}{m}\cdot F^{{}^P}_{\mu\nu}\cdot U^{{}^P}_{\nu}
  -   \frac{e\cdot g}{4\cdot m^2}\cdot\biggl(
              S^{{}^P}_{\nu\omega}\cdot
           \partial^{{}^P}_{\mu} F^{{}^P}_{\omega\nu}
+U^{{}^P}_{\mu}\cdot S^{{}^P}_{\nu\omega}\cdot U^{{}^P}_{\lambda}\cdot
                                           \partial^{{}^P}_{\lambda}
                            F^{{}^P}_{\omega\nu}  \biggr)
\nonumber\\&&\qquad
  +   \frac{e\cdot(g-2-c_1)}{2\cdot m^2}\cdot
S^{{}^P}_{\mu\nu}\cdot U^{{}^P}_{\omega}\cdot U^{{}^P}_{\lambda}\cdot
                    \partial^{{}^P}_{\lambda}  F^{{}^P}_{\nu\omega} \; ,
\\
\addtocounter{equation}{-1}
\addtocounter{INDEX}{1}
\label{10.2b}
 && \dot{S}^{{}^P}_{\mu\nu} =  \frac{e\cdot g}{2\cdot m}\cdot\biggl(
                   F^{{}^P}_{\mu\omega}\cdot S^{{}^P}_{\omega\nu}-
                   S^{{}^P}_{\mu\omega}\cdot F^{{}^P}_{\omega\nu} \biggr)
\nonumber\\&&
  -   \frac{e\cdot (g-2)}{2\cdot m}\cdot\biggl(
S^{{}^P}_{\mu\omega}\cdot F^{{}^P}_{\omega\lambda}\cdot U^{{}^P}_{\lambda}
                                             \cdot U^{{}^P}_{\nu} -
S^{{}^P}_{\nu\omega}\cdot F^{{}^P}_{\omega\lambda}\cdot
 U^{{}^P}_{\lambda}\cdot U^{{}^P}_{\mu}
                                           \biggr) \; .
 \;\; (\mu,\nu =1,...,4) \qquad
\label{10.2c}
\end{eqnarray}
\setcounter{INDEX}{0}
\par\noindent
For $c_1=g-2$, equations (10.2) are the GNR equations
and for $c_1=0$ they are the reduced Frenkel equations.
\subsection*{10.3}
I now demonstrate how to relate equations (10.2) corresponding to
different values of $c_1$. I do this by defining a transformation
of $X^{{}^P}$ and $S^{{}^P}$ via the rule:
\setcounter{INDEX}{1}
\begin{eqnarray}
&& X^{{}^P}_{\mu} \longrightarrow   X_{\mu} \; , \qquad
S^{{}^P}_{\mu\nu} \longrightarrow   S_{\mu\nu}=S^{{}^P}_{\mu\nu} \; ,
\qquad (\mu,\nu =1,...,4)
\label{10.3a}
\end{eqnarray}
\addtocounter{equation}{-1}
\addtocounter{INDEX}{1}
\par\noindent
with
\begin{eqnarray}
&& U_{\mu} \equiv \dot{X}_{\mu} \equiv
     U^{{}^P}_{\mu} +\frac{e\cdot \Delta c_1}{2\cdot m^2}\cdot
S^{{}^P}_{\mu\nu}\cdot
            F^{{}^P}_{\nu\omega}\cdot U^{{}^P}_{\omega}  \; ,
 \qquad (\mu =1,...,4)
\label{10.3b}
\end{eqnarray}
\setcounter{INDEX}{0}
\par\noindent
where $\Delta c_1$ is a real number. One sees that the spin tensor does
not change under (10.3), so that this transformation only has an effect on
the SG force, as seen below.
The transformation (10.3) is a straightforward generalization
of a transformation given in \cite{Pla66a}.
The constraints (3.8) are equivalent  to:
\setcounter{INDEX}{1}
\begin{eqnarray}
 && U_{\mu}\cdot  U_{\mu} = -1 \; ,
\\
\addtocounter{equation}{-1}
\addtocounter{INDEX}{1}
\label{10.4a}
 && S_{\mu\nu}\cdot U_{\nu} = 0 \; .
 \qquad (\mu =1,...,4)
\label{10.4b}
\end{eqnarray}
\setcounter{INDEX}{0}
\par\noindent
On introducing the abbreviations
\footnote{The partial derivatives $\partial_{\mu}$ always act on
functions depending on $X_1,X_2,X_3,X_4$.}
\begin{eqnarray}
 && X_{\mu} =  (\vec{r}^{\,\dagger}, X_4)_{\mu} \; ,
\nonumber\\
 && \partial_{\mu} \equiv
    (\frac{\partial}{\partial X_1},
     \frac{\partial}{\partial X_2},
     \frac{\partial}{\partial X_3},
     \frac{\partial}{\partial X_4})_{\mu} \; ,
 \qquad (\mu =1,...,4)
\label{10.5}
\end{eqnarray}
I define:
\footnote{My notation is chosen so as to indicate that the functional
dependence of $\vec{E}$ on $\vec{r},-i\cdot X_4$ is the same as the
dependence of $\vec{E}_{{}_M}$ on $\vec{r}_{{}_M},t$ and likewise for
$\vec{B}_{{}_M}$. Note that $F$ depends on $X_1,X_2,X_3,X_4$ in
the same way as $F^{{}^P}$ depends on
$X^{{}^P}_1,X^{{}^P}_2,X^{{}^P}_3,X^{{}^P}_4$.}
\begin{eqnarray}
&&\vec{E} \equiv  \vec{E}_{{}_M}(\vec{r},-i\cdot X_4) \; , \qquad
  \vec{B} \equiv  \vec{B}_{{}_M}(\vec{r},-i\cdot X_4) \; ,
\nonumber\\
 && F \leftrightarrow (\vec{B},-i\cdot\vec{E}) \; .
\label{10.6}
\end{eqnarray}
Now I come to the main conclusion of this section: if
$X^{{}^P},U^{{}^P},S^{{}^P}$ obey (10.2), then $X,U,S$ obey
\setcounter{INDEX}{1}
\begin{eqnarray}
 && \dot{X}_{\mu} = U_{\mu} \; ,
\\
\addtocounter{equation}{-1}
\addtocounter{INDEX}{1}
\label{10.7a}
 && \dot{U}_{\mu} =
  \frac{e}{m}\cdot F_{\mu\nu}\cdot U_{\nu}
  -   \frac{e\cdot g}{4\cdot m^2}\cdot\biggl(
              S_{\nu\omega}\cdot
           \partial_{\mu} F_{\omega\nu}
+U_{\mu}\cdot S_{\nu\omega}\cdot U_{\lambda}\cdot
                                           \partial_{\lambda}
                            F_{\omega\nu}  \biggr)
\nonumber\\&&\qquad
  +   \frac{e\cdot(g-2-c_1+\Delta c_1)}{2\cdot m^2}\cdot
S_{\mu\nu}\cdot U_{\omega}\cdot U_{\lambda}\cdot
                    \partial_{\lambda}  F_{\nu\omega} \; ,
\\
\addtocounter{equation}{-1}
\addtocounter{INDEX}{1}
\label{10.7b}
 && \dot{S}_{\mu\nu} =  \frac{e\cdot g}{2\cdot m}\cdot\biggl(
                   F_{\mu\omega}\cdot S_{\omega\nu}-
                   S_{\mu\omega}\cdot F_{\omega\nu} \biggr)
\nonumber\\&&
  -   \frac{e\cdot (g-2)}{2\cdot m}\cdot\biggl(
S_{\mu\omega}\cdot F_{\omega\lambda}\cdot U_{\lambda}
                                             \cdot U_{\nu} -
S_{\nu\omega}\cdot F_{\omega\lambda}\cdot
 U_{\lambda}\cdot U_{\mu}
                                           \biggr) \; ,
 \;\; (\mu,\nu =1,...,4) \qquad
\label{10.7c}
\end{eqnarray}
\setcounter{INDEX}{0}
\par\noindent
where all second order SG terms are neglected in (10.7b).
In the special case  where $c_1=g-2$ equations (10.7) read as:
\setcounter{INDEX}{1}
\begin{eqnarray}
 && \dot{X}_{\mu} = U_{\mu} \; ,
\\
\addtocounter{equation}{-1}
\addtocounter{INDEX}{1}
\label{10.8a}
 && \dot{U}_{\mu} =
  \frac{e}{m}\cdot F_{\mu\nu}\cdot U_{\nu}
  -   \frac{e\cdot g}{4\cdot m^2}\cdot\biggl(
              S_{\nu\omega}\cdot
           \partial_{\mu} F_{\omega\nu}
+U_{\mu}\cdot S_{\nu\omega}\cdot U_{\lambda}\cdot
                                           \partial_{\lambda}
                            F_{\omega\nu}  \biggr)
\nonumber\\&&\qquad
  +   \frac{e\cdot\Delta c_1}{2\cdot m^2}\cdot
S_{\mu\nu}\cdot U_{\omega}\cdot U_{\lambda}\cdot
                    \partial_{\lambda}  F_{\nu\omega} \; ,
\\
\addtocounter{equation}{-1}
\addtocounter{INDEX}{1}
\label{10.8b}
 && \dot{S}_{\mu\nu} =  \frac{e\cdot g}{2\cdot m}\cdot\biggl(
                   F_{\mu\omega}\cdot S_{\omega\nu}-
                   S_{\mu\omega}\cdot F_{\omega\nu} \biggr)
\nonumber\\&&
  -   \frac{e\cdot (g-2)}{2\cdot m}\cdot\biggl(
S_{\mu\omega}\cdot F_{\omega\lambda}\cdot U_{\lambda}
                                             \cdot U_{\nu} -
S_{\nu\omega}\cdot F_{\omega\lambda}\cdot
 U_{\lambda}\cdot U_{\mu}
                                           \biggr) \; ,
 \;\; (\mu,\nu =1,...,4) \qquad
\label{10.8c}
\end{eqnarray}
\setcounter{INDEX}{0}
\par\noindent
where all second order SG terms are neglected in (10.8b).
Therefore the GNR equations are transformed under (10.3) into
equations (10.8). In particular, with  the choice: $\Delta c_1=g-2$, one has
transformed (10.8)
into   the reduced Frenkel equations. Thus one has transformed the GNR
equations into the reduced Frenkel equations.
Therefore  the GNR equations can be solved by solving the DK equations
and inverting (10.3b) so that one can use symplectic methods
\cite{BHR}. For practical applications in accelerator physics it is helpful
that (10.1-8) contain the charge $e$ only up to first order.
\par It follows from the normalization of the spin vector that:
\begin{eqnarray}
 && S_{\mu\nu}\cdot S_{\mu\nu} =
  \hbar^2/2 \; .
\label{10.9}
\end{eqnarray}
As in (1.9) this equation is of second order in spin so that it plays no
role
in this paper. Note also that (10.9) is conserved under (10.7c).
\section*{Summary}
\addcontentsline{toc}{section}{Summary}
I have studied classical spin-orbit systems at first order in spin and,
by applying dimensional analysis and imposing $\rm{Poincar\acute{e}}$
covariance, I have found that these spin-orbit systems are
characterized by five dimensionless parameters
$c_1,...,c_5$. My axiomatic approach is supported by the observation
that the most prominent spin-orbit systems, namely those of Frenkel and
GNR, are special cases of my scheme.
\par In this approach, i.e. at first order in spin,
                      the five parameters are purely phenomenological
and are to be determined by experiment.
For example the Frenkel and GNR equations give very different SG forces
at high energy in proton storage rings. There are also differences for
high fields in traps.                  Theory is of little help. For
example, as I pointed out in $\lbrack o \rbrack$ even the Dirac
equation cannot deliver unambiguous answers. In the three cases mentioned
the parameters all depend on $g$. However, protons, for
example, have substructure and one should not assume a priori that the
dependence of the $c's$ on the $g's$ is the same for all particles.
\par In this paper I have concentrated on spin 1/2 particles. Nevertheless
my results are formulated classically so that they can be applied to
particles of arbitrary spin.
\par In addition I have devoted special attention to the
DK equations and have found a transformation of the
particle coordinates which relates these equations to the Frenkel
equations. The DK equations are therefore
(nonmanifestly) $\rm{Poincar\acute{e}}$ covariant. The new coordinates
differ from the original coordinates by less than the Compton wave
length and the corresponding time variables are the same. Thus one
concludes that the particles described by both equations are effectively
indistinguishable.
\par As I have just pointed out, different values of the $c's$,
correspond to different systems of spin-orbit equations and can lead
to very different SG forces. Thus, before proposing techniques which
rely on SG forces at high energy, one must decide which equations are
applicable. Alternatively one can take the view that a measurement of
the forces is in itself a way of discovering which equations to use.
\renewcommand{\thesection}{\Alph{section}}
\setcounter{section}{1}
\setcounter{subsection}{0}
\setcounter{equation}{0}
\section*{Appendix A}
\addcontentsline{toc}{section}{Appendix A}
\subsection*{A.1}
In this Appendix I derive (2.9) from section 1 and subsection 2.1.
I introduce the abbreviation:
\begin{eqnarray}
 && \vec{\Omega}_{{}_P}\equiv
                        -\frac{e}{m}\cdot\biggl(
 \lbrack \frac{1}{\gamma_{{}_P}}+
                            \frac{g-2}{2} \rbrack \cdot\vec{B}_{{}_P}
 -\frac{g-2}{2}\cdot\frac{\gamma_{{}_P}}{\gamma_{{}_P}+1}\cdot
  \vec{v}_{{}_P}^\dagger\cdot\vec{B}_{{}_P}\cdot\vec{v}_{{}_P}
\nonumber\\ &&\qquad
-\lbrack \frac{g}{2} -\frac{\gamma_{{}_P}}{\gamma_{{}_P}+1}\rbrack\cdot
 (\vec{v}_{{}_P}\wtimes\vec{E}_{{}_P}) \biggr)    \; ,
 \label{A.1}
\end{eqnarray}
and conclude from (2.4),(A.1):
\begin{eqnarray}
&&\vec{s}\ '=\biggl(\gamma_{{}_M}\cdot\vec{\sigma}-
            \frac{\gamma_{{}_M}^2}{\gamma_{{}_M}+1}
  \cdot \vec{\sigma}^{\,\dagger}\cdot\vec{v}_{{}_M}\cdot
                                               \vec{v}_{{}_M} \biggr)'
 =   \gamma_{{}_M}'\cdot \vec{\sigma}  +
                                     \gamma_{{}_M}\cdot \vec{\sigma}\ '
   - \biggl( \frac{\gamma_{{}_M}^2}{\gamma_{{}_M}+1}
  \cdot \vec{\sigma}^{\,\dagger}\cdot\vec{v}_{{}_M}\cdot
                                               \vec{v}_{{}_M} \biggr)'
 \nonumber\\
&=&\frac{e}{m}\cdot\vec{E}_{{}_M}^\dagger\cdot
                                        \vec{v}_{{}_M}\cdot\vec{\sigma}
+ \gamma_{{}_M}\cdot (\vec{\Omega}_{{}_M}\wtimes\vec{\sigma})
+ \frac{e}{m}\cdot
  \frac{\gamma_{{}_M}^2}{(\gamma_{{}_M}+1)^2}
\cdot \vec{E}_{{}_M}^\dagger\cdot\vec{v}_{{}_M}
 \cdot\vec{\sigma}^{\,\dagger}\cdot\vec{v}_{{}_M}\cdot \vec{v}_{{}_M}
\nonumber\\&&\qquad
   -  \frac{\gamma_{{}_M}}{\gamma_{{}_M}+1}\cdot \biggl(
     \gamma_{{}_M}\cdot \vec{v}_{{}_M}^{\,\dagger}\cdot
  (\vec{\Omega}_{{}_M}\wtimes \vec{\sigma})\cdot\vec{v}_{{}_M}
 + \vec{\sigma}^{\,\dagger}\cdot\vec{v}_{{}_M}\cdot
    (\gamma_{{}_M}\cdot \vec{v}_{{}_M})'
 + \vec{\sigma}^{\,\dagger}\cdot
  (\gamma_{{}_M}\cdot \vec{v}_{{}_M})' \cdot \vec{v}_{{}_M} \biggr)
 \nonumber\\
&=&  \frac{e}{m}\cdot \vec{E}_{{}_M}^\dagger\cdot\vec{v}_{{}_M}\cdot
\lbrack \frac{1}{\gamma_{{}_M}}\cdot \vec{s} +
           \frac{\gamma_{{}_M}}{\gamma_{{}_M}+1}
\cdot \vec{s}^{\,\dagger}\cdot\vec{v}_{{}_M}\cdot\vec{v}_{{}_M} \rbrack
+ \gamma_{{}_M}\cdot \vec{\Omega}_{{}_M}\wtimes
\lbrack \frac{1}{\gamma_{{}_M}}\cdot \vec{s} +
           \frac{\gamma_{{}_M}}{\gamma_{{}_M}+1}
\cdot \vec{s}^{\,\dagger}\cdot\vec{v}_{{}_M}\cdot\vec{v}_{{}_M} \rbrack
\nonumber\\&&
+ \frac{e}{m}\cdot
  \frac{\gamma_{{}_M}^2}{(\gamma_{{}_M}+1)^2}
\cdot \vec{E}_{{}_M}^\dagger\cdot\vec{v}_{{}_M}
 \cdot\vec{s}^{\,\dagger}\cdot\vec{v}_{{}_M}\cdot \vec{v}_{{}_M}
\nonumber\\&&\qquad
 -\frac{\gamma_{{}_M}^2}{\gamma_{{}_M}+1}\cdot
      (\vec{v}_{{}_M}\wtimes\vec{\Omega}_{{}_M})^\dagger
\cdot\lbrack\frac{1}{\gamma_{{}_M}}\cdot\vec{s}+
          \frac{\gamma_{{}_M}}{\gamma_{{}_M}+1}
\cdot \vec{s}^{\,\dagger}\cdot\vec{v}_{{}_M}\cdot\vec{v}_{{}_M} \rbrack
             \cdot \vec{v}_{{}_M}
\nonumber\\&&\qquad
-\frac{e}{m}\cdot \frac{\gamma_{{}_M}}{\gamma_{{}_M}+1}\cdot
   \vec{s}^{\,\dagger}\cdot\vec{v}_{{}_M}\cdot
\lbrack    \vec{v}_{{}_M}\wtimes\vec{B}_{{}_M} + \vec{E}_{{}_M}\rbrack
-\frac{e}{m}\cdot \frac{\gamma_{{}_M}}{\gamma_{{}_M}+1}\cdot
                   \vec{\sigma}^{\,\dagger}\cdot
\lbrack    \vec{v}_{{}_M}\wtimes\vec{B}_{{}_M} +
    \vec{E}_{{}_M}\rbrack\cdot\vec{v}_{{}_M}
 \nonumber\\
&=&  \frac{e}{m}\cdot \vec{E}_{{}_P}^\dagger\cdot\vec{v}_{{}_P}\cdot
\lbrack \frac{1}{\gamma_{{}_P}}\cdot
   \vec{s} + \frac{\gamma_{{}_P}}{\gamma_{{}_P}+1}
\cdot \vec{s}^{\,\dagger}\cdot\vec{v}_{{}_P}\cdot\vec{v}_{{}_P} \rbrack
+ \gamma_{{}_P}\cdot \vec{\Omega}_{{}_P}\wtimes
\lbrack \frac{1}{\gamma_{{}_P}}\cdot
   \vec{s} + \frac{\gamma_{{}_P}}{\gamma_{{}_P}+1}
\cdot \vec{s}^{\,\dagger}\cdot\vec{v}_{{}_P}\cdot\vec{v}_{{}_P} \rbrack
\nonumber\\&&\qquad
       +  \frac{e}{m}\cdot\frac{\gamma_{{}_P}^2}{(\gamma_{{}_P}+1)^2}
\cdot\vec{E}_{{}_P}^\dagger\cdot\vec{v}_{{}_P}\cdot
   \vec{s}^{\,\dagger}\cdot\vec{v}_{{}_P}\cdot\vec{v}_{{}_P}
\nonumber\\&&\qquad
+\frac{\gamma_{{}_P}}{\gamma_{{}_P}+1}\cdot
                    (\vec{v}_{{}_P}\wtimes\vec{s})^\dagger\cdot
   \vec{\Omega}_{{}_P} \cdot \vec{v}_{{}_P}
-\frac{e}{m}\cdot\frac{\gamma_{{}_P}}{\gamma_{{}_P}+1}\cdot
                                 \vec{s}^{\,\dagger}
   \cdot\vec{v}_{{}_P}\cdot
\lbrack    \vec{v}_{{}_P}\wtimes\vec{B}_{{}_P} + \vec{E}_{{}_P}\rbrack
 \nonumber\\ &&\qquad
-\frac{e}{m}\cdot\frac{1}{\gamma_{{}_P}+1}\cdot\vec{s}^{\,\dagger}\cdot
(\vec{v}_{{}_P}\wtimes\vec{B}_{{}_P})\cdot\vec{v}_{{}_P}
-\frac{e}{m}\cdot\frac{\gamma_{{}_P}}{\gamma_{{}_P}+1}
                             \cdot\vec{E}_{{}_P}^\dagger\cdot
\lbrack \frac{1}{\gamma_{{}_P}}\cdot \vec{s}
 \nonumber\\ &&\qquad
+ \frac{\gamma_{{}_P}}{\gamma_{{}_P}+1}
   \cdot \vec{s}^{\,\dagger}\cdot\vec{v}_{{}_P}
   \cdot\vec{v}_{{}_P} \rbrack \cdot\vec{v}_{{}_P}
 \nonumber\\
&=&  \frac{e}{m}\cdot\frac{1}{\gamma_{{}_P}}
         \cdot\vec{E}_{{}_P}^\dagger\cdot\vec{v}_{{}_P}\cdot\vec{s}
+\frac{e}{m}\cdot\frac{2\cdot\gamma_{{}_P}^2+
   \gamma_{{}_P}}{(\gamma_{{}_P}+1)^2}
\cdot\vec{E}_{{}_P}^\dagger\cdot\vec{v}_{{}_P}
    \cdot\vec{s}^{\,\dagger}\cdot\vec{v}_{{}_P}\cdot\vec{v}_{{}_P}
\nonumber\\&&\qquad
+\frac{e}{m}\cdot \lbrack \vec{s}
 +  \frac{\gamma_{{}_P}^2}{\gamma_{{}_P}+1}
 \cdot \vec{s}^{\,\dagger}\cdot\vec{v}_{{}_P}\cdot\vec{v}_{{}_P} \rbrack
\wtimes  \biggl(
 \lbrack \frac{1}{\gamma_{{}_P}}+
                            \frac{g-2}{2} \rbrack \cdot\vec{B}_{{}_P}
 -\frac{g-2}{2}\cdot\frac{\gamma_{{}_P}}{\gamma_{{}_P}+1}\cdot
  \vec{v}_{{}_P}^\dagger\cdot\vec{B}_{{}_P}\cdot\vec{v}_{{}_P}
 \nonumber\\ &&\qquad
-\lbrack \frac{g}{2} -\frac{\gamma_{{}_P}}{\gamma_{{}_P}+1}\rbrack\cdot
 (\vec{v}_{{}_P}\wtimes\vec{E}_{{}_P}) \biggr)
 \nonumber\\ &&\qquad
+\frac{e}{m}\cdot \frac{\gamma_{{}_P}}{\gamma_{{}_P}+1}\cdot
(\vec{s}\wtimes\vec{v}_{{}_P})^\dagger\cdot
        \biggl(
 \lbrack \frac{1}{\gamma_{{}_P}}+
                            \frac{g-2}{2} \rbrack \cdot\vec{B}_{{}_P}
 -\frac{g-2}{2}\cdot\frac{\gamma_{{}_P}}{\gamma_{{}_P}+1}\cdot
  \vec{v}_{{}_P}^\dagger\cdot\vec{B}_{{}_P}\cdot\vec{v}_{{}_P}
\nonumber\\&&\qquad
-\lbrack \frac{g}{2} -\frac{\gamma_{{}_P}}{\gamma_{{}_P}+1}\rbrack\cdot
 (\vec{v}_{{}_P}\wtimes\vec{E}_{{}_P}) \biggr)\cdot\vec{v}_{{}_P}
-\frac{e}{m}\cdot\frac{\gamma_{{}_P}}{\gamma_{{}_P}+1}\cdot
                                    \vec{s}^{\,\dagger}
   \cdot\vec{v}_{{}_P}\cdot
\lbrack      \vec{v}_{{}_P}\wtimes\vec{B}_{{}_P} + \vec{E}_{{}_P}\rbrack
 \nonumber\\ &&\qquad
-\frac{e}{m}\cdot\frac{1}{\gamma_{{}_P}+1}\cdot\vec{s}^{\,\dagger}\cdot
(\vec{v}_{{}_P}\wtimes\vec{B}_{{}_P})\cdot\vec{v}_{{}_P}
-\frac{e}{m}\cdot
  \frac{\gamma_{{}_P}}{\gamma_{{}_P}+1}\cdot\vec{E}_{{}_P}^\dagger\cdot
\lbrack \frac{1}{\gamma_{{}_P}}\cdot \vec{s}
 \nonumber\\ &&\qquad
+ \frac{\gamma_{{}_P}}{\gamma_{{}_P}+1}
 \cdot \vec{s}^{\,\dagger}\cdot\vec{v}_{{}_P}
      \cdot\vec{v}_{{}_P} \rbrack \cdot\vec{v}_{{}_P}
 \nonumber\\
&=&  \frac{e}{m}\cdot\frac{1}{\gamma_{{}_P}}
       \cdot\vec{E}_{{}_P}^\dagger\cdot\vec{v}_{{}_P}\cdot\vec{s}
+\frac{e}{m}\cdot\frac{2\cdot
    \gamma_{{}_P}^2+\gamma_{{}_P}}{(\gamma_{{}_P}+1)^2}
\cdot\vec{E}_{{}_P}^\dagger\cdot\vec{v}_{{}_P}
   \cdot\vec{s}^{\,\dagger}\cdot\vec{v}_{{}_P}\cdot\vec{v}_{{}_P}
\nonumber\\&&\qquad
+\frac{e}{m}\cdot \lbrack \frac{1}{\gamma_{{}_P}}+
                                             \frac{g-2}{2} \rbrack\cdot
 (\vec{s}\wtimes \vec{B}_{{}_P})
+\frac{e}{m}\cdot \frac{\gamma_{{}_P}^2}{\gamma_{{}_P}+1}\cdot\lbrack
 \frac{g-2}{2} + \frac{1}{\gamma_{{}_P}}\rbrack
             \cdot \vec{s}^{\,\dagger}\cdot\vec{v}_{{}_P}\cdot
 (\vec{v}_{{}_P}\wtimes \vec{B}_{{}_P})
\nonumber\\&&
+\frac{e\cdot(g-2)}{2\cdot m}\cdot \frac{\gamma_{{}_P}}{\gamma_{{}_P}+1}
             \cdot \vec{v}_{{}_P}^\dagger\cdot\vec{B}_{{}_P}\cdot
 (\vec{v}_{{}_P}\wtimes \vec{s})
-\frac{e}{m}\cdot
 \lbrack \frac{g}{2} -
    \frac{\gamma_{{}_P}}{\gamma_{{}_P}+1}\rbrack\cdot\lbrack
   \vec{s}^{\,\dagger}\cdot\vec{E}_{{}_P}\cdot\vec{v}_{{}_P} -
   \vec{s}^{\,\dagger}\cdot\vec{v}_{{}_P}\cdot\vec{E}_{{}_P} \rbrack
\nonumber\\&&\qquad
-\frac{e}{m}\cdot \frac{\gamma_{{}_P}^2}{\gamma_{{}_P}+1}\cdot
 \lbrack \frac{g}{2} -
    \frac{\gamma_{{}_P}}{\gamma_{{}_P}+1}\rbrack\cdot\lbrack
   \vec{v}_{{}_P}^\dagger\cdot\vec{E}_{{}_P}\cdot\vec{v}_{{}_P} -
   \vec{v}_{{}_P}^\dagger\cdot\vec{v}_{{}_P}\cdot\vec{E}_{{}_P} \rbrack
              \cdot\vec{s}^{\,\dagger}\cdot\vec{v}_{{}_P}
\nonumber\\&&\qquad
+\frac{e}{m}\cdot \frac{\gamma_{{}_P}}{\gamma_{{}_P}+1}\cdot
   \lbrack \frac{1}{\gamma_{{}_P}}+\frac{g-2}{2} \rbrack\cdot
 \vec{B}_{{}_P}^\dagger\cdot(\vec{s}\wtimes
                                     \vec{v}_{{}_P}) \cdot\vec{v}_{{}_P}
\nonumber\\&&\qquad
-\frac{e}{m}\cdot \frac{\gamma_{{}_P}}{\gamma_{{}_P}+1}\cdot
 \lbrack \frac{g}{2} -
   \frac{\gamma_{{}_P}}{\gamma_{{}_P}+1}\rbrack\cdot\lbrack
   \vec{v}_{{}_P}^\dagger\cdot\vec{s}\cdot
                      \vec{v}_{{}_P}^\dagger\cdot\vec{E}_{{}_P}
  -\vec{v}_{{}_P}^\dagger\cdot\vec{v}_{{}_P}
        \cdot\vec{s}^{\,\dagger}\cdot\vec{E}_{{}_P} \rbrack
                          \cdot\vec{v}_{{}_P}
\nonumber\\&&\qquad
-\frac{e}{m}\cdot
     \frac{\gamma_{{}_P}}{\gamma_{{}_P}+1}\cdot\vec{s}^{\,\dagger}
   \cdot\vec{v}_{{}_P}\cdot
  (\vec{v}_{{}_P}\wtimes\vec{B}_{{}_P} +\vec{E}_{{}_P} )
\nonumber\\&&\qquad
-\frac{e}{m}\cdot\frac{1}{\gamma_{{}_P}+1}\cdot\vec{s}^{\,\dagger}\cdot
(\vec{v}_{{}_P}\wtimes\vec{B}_{{}_P})\cdot\vec{v}_{{}_P}
-\frac{e}{m}\cdot\frac{\gamma_{{}_P}}{\gamma_{{}_P}+1}\cdot
                \vec{E}_{{}_P}^\dagger\cdot
\lbrack \frac{1}{\gamma_{{}_P}}\cdot \vec{s}
\nonumber\\&&\qquad
   + \frac{\gamma_{{}_P}}{\gamma_{{}_P}+1}
  \cdot \vec{s}^{\,\dagger}\cdot\vec{v}_{{}_P}
  \cdot\vec{v}_{{}_P} \rbrack \cdot\vec{v}_{{}_P}
 \nonumber\\
&=&  \frac{e}{m}\cdot\frac{1}{\gamma_{{}_P}}
    \cdot\vec{E}_{{}_P}^\dagger\cdot\vec{v}_{{}_P}\cdot\vec{s}
-\frac{e\cdot(g-2)}{2\cdot m}\cdot \gamma_{{}_P}
\cdot\vec{E}_{{}_P}^\dagger\cdot\vec{v}_{{}_P}
  \cdot\vec{s}^{\,\dagger}\cdot\vec{v}_{{}_P}\cdot\vec{v}_{{}_P}
\nonumber\\&&\qquad
+\frac{e}{m}\cdot \lbrack \frac{1}{\gamma_{{}_P}}+
                                             \frac{g-2}{2} \rbrack\cdot
 (\vec{s}\wtimes \vec{B}_{{}_P})
+\frac{e\cdot(g-2)}{2\cdot m}\cdot
    \frac{\gamma_{{}_P}^2}{\gamma_{{}_P}+1}
             \cdot \vec{s}^{\,\dagger}\cdot\vec{v}_{{}_P}\cdot
 (\vec{v}_{{}_P}\wtimes \vec{B}_{{}_P})
\nonumber\\&&
+\frac{e\cdot(g-2)}{2\cdot m}\cdot \frac{\gamma_{{}_P}}{\gamma_{{}_P}+1}
             \cdot \vec{v}_{{}_P}^\dagger\cdot\vec{B}_{{}_P}\cdot
 (\vec{v}_{{}_P}\wtimes \vec{s})
-\frac{e\cdot g}{2\cdot m}\cdot\frac{1}{\gamma_{{}_P}}\cdot
    \vec{s}^{\,\dagger}\cdot\vec{E}_{{}_P}\cdot\vec{v}_{{}_P}
\nonumber\\&&\qquad
+\frac{e\cdot (g-2)}{2\cdot m}\cdot\gamma_{{}_P}\cdot
        \vec{s}^{\,\dagger}\cdot\vec{v}_{{}_P}\cdot\vec{E}_{{}_P}
+\frac{e\cdot (g-2)}{2\cdot m}\cdot
     \frac{\gamma_{{}_P}}{\gamma_{{}_P}+1}\cdot
 \vec{B}_{{}_P}^\dagger\cdot(\vec{s}\wtimes
                                \vec{v}_{{}_P}) \cdot\vec{v}_{{}_P} \; .
\label{A.2}
\end{eqnarray}
Introducing the abbreviations
\footnote{Note that all $\Delta's$ are first order in spin.}
\begin{eqnarray}
&&\vec{\Delta}_0  \equiv
     \frac{e}{m}\cdot\frac{1}{\gamma_{{}_P}}
     \cdot\vec{E}_{{}_P}^\dagger\cdot\vec{v}_{{}_P}\cdot\vec{s}
-\frac{e\cdot(g-2)}{2\cdot m}\cdot \gamma_{{}_P}
\cdot\vec{E}_{{}_P}^\dagger\cdot\vec{v}_{{}_P}
   \cdot\vec{s}^{\,\dagger}\cdot\vec{v}_{{}_P}\cdot\vec{v}_{{}_P}
\nonumber\\&&\qquad
+\frac{e}{m}\cdot \lbrack \frac{1}{\gamma_{{}_P}}+
                                             \frac{g-2}{2} \rbrack\cdot
 (\vec{s}\wtimes \vec{B}_{{}_P})
+\frac{e\cdot(g-2)}{2\cdot m}\cdot
    \frac{\gamma_{{}_P}^2}{\gamma_{{}_P}+1}
             \cdot \vec{s}^{\,\dagger}\cdot\vec{v}_{{}_P}\cdot
 (\vec{v}_{{}_P}\wtimes \vec{B}_{{}_P})
\nonumber\\&&
+\frac{e\cdot(g-2)}{2\cdot m}\cdot \frac{\gamma_{{}_P}}{\gamma_{{}_P}+1}
             \cdot \vec{v}_{{}_P}^\dagger\cdot\vec{B}_{{}_P}\cdot
 (\vec{v}_{{}_P}\wtimes \vec{s})
-\frac{e\cdot g}{2\cdot m}\cdot\frac{1}{\gamma_{{}_P}}\cdot
    \vec{s}^{\,\dagger}\cdot\vec{E}_{{}_P}\cdot\vec{v}_{{}_P}
\nonumber\\&&\qquad
+\frac{e\cdot (g-2)}{2\cdot m}\cdot\gamma_{{}_P}\cdot
        \vec{s}^{\,\dagger}\cdot\vec{v}_{{}_P}\cdot\vec{E}_{{}_P}
+\frac{e\cdot (g-2)}{2\cdot m}\cdot
       \frac{\gamma_{{}_P}}{\gamma_{{}_P}+1}\cdot
 \vec{B}_{{}_P}^\dagger\cdot(\vec{s}\wtimes
                                \vec{v}_{{}_P}) \cdot\vec{v}_{{}_P} \; ,
\nonumber\\
&&\vec{\Delta}_1  \equiv
     \frac{e}{m}\cdot \lbrack
 \frac{g-2}{2}\cdot\gamma_{{}_P}\cdot
     \vec{s}^{\,\dagger}\cdot\vec{v}_{{}_P}\cdot
    (\vec{E}_{{}_P}+\vec{v}_{{}_P}\wtimes\vec{B}_{{}_P})
+\frac{1}{\gamma_{{}_P}}\cdot \vec{v}_{{}_P}^\dagger\cdot
 \vec{E}_{{}_P}\cdot\vec{s}
\nonumber\\&&
-\frac{g-2}{2}\cdot\gamma_{{}_P}\cdot
     \vec{s}^{\,\dagger}\cdot\vec{v}_{{}_P}\cdot
        \vec{v}_{{}_P}^\dagger\cdot\vec{E}_{{}_P}\cdot \vec{v}_{{}_P}
+\frac{g}{2}\cdot\frac{1}{\gamma_{{}_P}}\cdot
                      (\vec{s}\wtimes\vec{B}_{{}_P})
-\frac{g}{2}\cdot\frac{1}{\gamma_{{}_P}}\cdot
\vec{s}^{\,\dagger}\cdot\vec{E}_{{}_P}\cdot \vec{v}_{{}_P} \rbrack \; ,
\nonumber\\
&&\vec{\Delta}_2 \equiv  \vec{\Delta}_1 - \vec{\Delta}_0
=  \frac{e}{2\cdot m}\cdot\biggl(
   (g-2)\cdot\frac{\gamma_{{}_P}}{\gamma_{{}_P}+1}\cdot
\vec{s}^{\,\dagger}\cdot\vec{v}_{{}_P}\cdot
                                  (\vec{v}_{{}_P}\wtimes\vec{B}_{{}_P})
\nonumber\\&&\qquad
  +(g-2)\cdot\frac{1-\gamma_{{}_P}}{\gamma_{{}_P}}\cdot
 (\vec{s}\wtimes \vec{B}_{{}_P})
-(g-2)\cdot\frac{\gamma_{{}_P}}{\gamma_{{}_P}+1}
             \cdot \vec{v}_{{}_P}^\dagger\cdot\vec{B}_{{}_P}\cdot
 (\vec{v}_{{}_P}\wtimes \vec{s})
\nonumber\\&&
- (g-2)\cdot\frac{\gamma_{{}_P}}{\gamma_{{}_P}+1}\cdot
 \vec{B}_{{}_P}^\dagger\cdot(\vec{s}\wtimes
    \vec{v}_{{}_P}) \cdot\vec{v}_{{}_P} \biggr) \; ,
\label{A.3}
\end{eqnarray}
one sees by (A.2) that (2.9) is valid if
$\vec{\Delta}_2$ vanishes. Hence the remaining task of Appendix A is to
show that $\vec{\Delta}_2$ vanishes.
\subsection*{A.2}
For the case  where
$\vec{v}_{{}_P},\vec{B}_{{}_P}$ are linearly independent (e.g.
nonparallel), one has the following 3 linearly independent vectors:
\begin{eqnarray*}
&& \vec{v}_{{}_P},\vec{B}_{{}_P},\vec{v}_{{}_P}\wtimes\vec{B}_{{}_P} \; .
\end{eqnarray*}
One concludes from (A.3):
\begin{eqnarray}
&& \vec{v}_{{}_P}^\dagger\cdot \vec{\Delta}_2 =
 \frac{e\cdot (g-2)}{2\cdot m}\cdot
             \vec{v}_{{}_P}^\dagger\cdot(\vec{s}\wtimes\vec{B}_{{}_P})
\cdot \lbrack \frac{1-\gamma_{{}_P}}{\gamma_{{}_P}}
    + \frac{\gamma_{{}_P}}{\gamma_{{}_P}+1}\cdot
              \vec{v}_{{}_P}^\dagger\cdot
                                    \vec{v}_{{}_P} \rbrack = 0 \; ,
\nonumber\\
&& \vec{B}_{{}_P}^\dagger\cdot \vec{\Delta}_2 =  0 \; ,
\nonumber\\
&& (\vec{v}_{{}_P}\wtimes\vec{B}_{{}_P})^\dagger\cdot
\vec{\Delta}_2 =
   \frac{e}{2\cdot m}\cdot(\vec{v}_{{}_P}\wtimes\vec{B}_{{}_P})^\dagger
  \cdot \biggl(
   (g-2)\cdot\frac{\gamma_{{}_P}}{\gamma_{{}_P}+1}\cdot
\vec{s}^{\,\dagger}\cdot\vec{v}_{{}_P}\cdot
                                  (\vec{v}_{{}_P}\wtimes\vec{B}_{{}_P})
\nonumber\\&&\qquad
  +(g-2)\cdot\frac{1-\gamma_{{}_P}}{\gamma_{{}_P}}\cdot
 (\vec{s}\wtimes \vec{B}_{{}_P})
-(g-2)\cdot\frac{\gamma_{{}_P}}{\gamma_{{}_P}+1}
             \cdot \vec{v}_{{}_P}^\dagger\cdot\vec{B}_{{}_P}\cdot
 (\vec{v}_{{}_P}\wtimes \vec{s})
\nonumber\\&&
- (g-2)\cdot\frac{\gamma_{{}_P}}{\gamma_{{}_P}+1}\cdot
 \vec{B}_{{}_P}^\dagger\cdot(\vec{s}\wtimes
     \vec{v}_{{}_P}) \cdot\vec{v}_{{}_P} \biggr)
\nonumber\\
&=& \frac{e\cdot(g-2)}{2\cdot m}\cdot\biggl(
\frac{\gamma_{{}_P}}{\gamma_{{}_P}+1}\cdot
            \vec{s}^{\,\dagger}\cdot\vec{v}_{{}_P}\cdot
  \lbrack \vec{v}_{{}_P}^\dagger\cdot\vec{v}_{{}_P}
     \cdot \vec{B}_{{}_P}^\dagger\cdot\vec{B}_{{}_P}
-\vec{B}_{{}_P}^\dagger\cdot\vec{v}_{{}_P}\cdot
                \vec{v}_{{}_P}^\dagger\cdot\vec{B}_{{}_P}\rbrack
\nonumber\\&& \qquad
  +\frac{1-\gamma_{{}_P}}{\gamma_{{}_P}}\cdot
  \lbrack \vec{v}_{{}_P}^\dagger\cdot
 \vec{s}\cdot \vec{B}_{{}_P}^\dagger\cdot\vec{B}_{{}_P}
-\vec{s}^{\,\dagger}\cdot\vec{B}_{{}_P}\cdot
  \vec{v}_{{}_P}^\dagger\cdot\vec{B}_{{}_P}\rbrack
\nonumber\\&& \qquad
-\frac{\gamma_{{}_P}}{\gamma_{{}_P}+1}\cdot
                \vec{v}_{{}_P}^\dagger\cdot\vec{B}_{{}_P}\cdot
  \lbrack \vec{v}_{{}_P}^\dagger\cdot
 \vec{v}_{{}_P}\cdot \vec{B}_{{}_P}^\dagger\cdot\vec{s}
-\vec{v}_{{}_P}^\dagger\cdot\vec{s}\cdot
    \vec{v}_{{}_P}^\dagger\cdot\vec{B}_{{}_P}\rbrack\biggr)
 = 0 \; .
\label{A.4}
\end{eqnarray}
Because $\vec{v}_{{}_P},\vec{B}_{{}_P},\vec{v}_{{}_P}\wtimes\vec{B}_{{}_P}$
constitute a basis of vectors, one concludes by (A.4)
that $\vec{\Delta}_2$ vanishes for the case  where
$\vec{v}_{{}_P},\vec{B}_{{}_P}$ are linearly independent.
\par To discuss the case  where $\vec{v}_{{}_P},\vec{B}_{{}_P}$ are
linearly dependent, one first observes by (A.3) that $\vec{\Delta}_2$
vanishes, if $\vec{v}_{{}_P}=0$ or $\vec{B}_{{}_P}=0$. It remains to consider
the subcase  with: $\vec{B}_{{}_P}=\lambda\cdot\vec{v}_{{}_P}$, where $\lambda$
is a constant which balances the dimensions. Then from (A.3) follows
\begin{eqnarray}
&& \vec{\Delta}_2
=  \frac{e\cdot\lambda\cdot(g-2)}{2\cdot m}\cdot\biggl(
  \frac{1-\gamma_{{}_P}}{\gamma_{{}_P}}\cdot
 (\vec{s}\wtimes \vec{v}_{{}_P})
-\frac{\gamma_{{}_P}}{\gamma_{{}_P}+1}
             \cdot \vec{v}_{{}_P}^\dagger\cdot\vec{v}_{{}_P}\cdot
 (\vec{v}_{{}_P}\wtimes \vec{s}) \biggr) =0 \; .
\label{A.5}
\end{eqnarray}
Hence I have shown that $\vec{\Delta}_2$ vanishes. This completes
the proof of (2.9).
\setcounter{section}{2}
\setcounter{subsection}{0}
\setcounter{equation}{0}
\section*{Appendix B}
\addcontentsline{toc}{section}{Appendix B}
\subsection*{B.1}
In this Appendix I derive (2.10) from subsections 1.1-2 and subsections
2.1-2. This task is tedious but straightforward. In fact I only have to
consider $m\cdot(\gamma_{{}_P}\cdot\vec{v}_{{}_P})'$ as determined by the
Hamiltonian $H_{{}_M}$ through the relation:
\footnote{Here the partial derivative $\partial/\partial t$ acts on
functions depending on $\vec{r}_{{}_M},t,\vec{p}_{{}_M}, \vec{\sigma}$.}
\begin{eqnarray*}
&& m\cdot (\gamma_{{}_P}\cdot\vec{v}_{{}_P})' =
 \lbrace m\cdot\gamma_{{}_P}\cdot
    \vec{v}_{{}_P},H_{{}_M}\rbrace_{{}_M}  +
\frac{\partial}{\partial t}(m\cdot \gamma_{{}_P}\cdot\vec{v}_{{}_P})\; .
\end{eqnarray*}
Hence $m\cdot (\gamma_{{}_P}\cdot\vec{v}_{{}_P})'$ is a well defined
function of $\vec{r}_{{}_M},t,\vec{p}_{{}_M},\vec{\sigma}$ and the main
task is to reexpress it as a function of
$\vec{r}_{{}_P},t,\vec{v}_{{}_P},\vec{s}$.
\subsection*{B.2}
First of all I express $\vec{v}^{\;{}_{'}}_{{}_M}$ in terms of
$\vec{r}_{{}_M},t,\vec{v}_{{}_M},\vec{\sigma}$ and I abbreviate:
\footnote{The nabla operator $\check{\vec{\nabla}}_{{}_M}$ always acts
on functions depending on $\vec{r}_{{}_M},t,\vec{p}_{{}_M},
\vec{\sigma}$ and it is the gradient w.r.t. $\vec{p}_{{}_M}$.}
\begin{eqnarray}
&& K_{{}_M} \equiv m\cdot\gamma_{{}_M}  \; ,
\nonumber\\
&&\vec{\Delta}_3\equiv\lbrace\vec{r}_{{}_M},
                         \vec{\sigma}^{\,\dagger}\cdot
           \vec{W}_{{}_M}\rbrace_{{}_M}
=\check{\vec{\nabla}}_{{}_M} (\vec{\sigma}^{\,\dagger}\cdot
                                          \vec{W}_{{}_M}) \; .
 \label{B.1}
\end{eqnarray}
To simplify (B.1) I calculate for an arbitrary function
$f(\vec{\pi}_{{}_M})$:
\begin{eqnarray}
&& \check{\vec{\nabla}}_{{}_M} (f\cdot\vec{\sigma}^{\,\dagger}\cdot
                                                  \vec{B}_{{}_M}) =
\vec{\sigma}^{\,\dagger}\cdot \vec{B}_{{}_M}\cdot
                                             \check{\vec{\nabla}}_{{}_M}
                                                               f   \; ,
\nonumber\\
&& \check{\vec{\nabla}}_{{}_M} (f\cdot\vec{\sigma}^{\,\dagger}\cdot
                                                  \vec{\pi}_{{}_M}) =
\vec{\sigma}^{\,\dagger}\cdot\vec{\pi}_{{}_M}\cdot
                                             \check{\vec{\nabla}}_{{}_M}
                                                                 f +
         f\cdot\vec{\sigma} \; ,
\nonumber\\
&& \check{\vec{\nabla}}_{{}_M} (f\cdot\vec{\sigma}^{\,\dagger}\cdot
               (\vec{\pi}_{{}_M}\wtimes\vec{E}_{{}_M}))
   =\check{\vec{\nabla}}_{{}_M} (f\cdot\vec{\pi}_{{}_M}^\dagger\cdot
               (\vec{E}_{{}_M}\wtimes\vec{\sigma}))
   = \vec{\sigma}^{\,\dagger}\cdot(\vec{\pi}_{{}_M}\wtimes
             \vec{E}_{{}_M})\cdot \check{\vec{\nabla}}_{{}_M} f
    + f\cdot(\vec{E}_{{}_M}\wtimes\vec{\sigma}) \; ,
\nonumber\\
&& \check{\vec{\nabla}}_{{}_M} J_{{}_M} =
   \frac{\vec{\pi}_{{}_M}}{J_{{}_M}} \; , \qquad
 \check{\vec{\nabla}}_{{}_M} \frac{1}{J_{{}_M}} =
     -\frac{\vec{\pi}_{{}_M}}{J_{{}_M}^3} \; ,
 \label{B.2}
\end{eqnarray}
from which follows by (B.1):
\begin{eqnarray}
&&\vec{\Delta}_3 = \check{\vec{\nabla}}_{{}_M} (\vec{\sigma}^{\,\dagger}
                                                  \cdot \vec{W}_{{}_M})
= -\frac{e}{m}\cdot\biggl( \vec{\sigma}^{\,\dagger}\cdot
             \vec{B}_{{}_M}\cdot\check{\vec{\nabla}}_{{}_M}
 \lbrack \frac{m}{J_{{}_M}}+\frac{g-2}{2} \rbrack
\nonumber\\ &&\qquad
-\frac{g-2}{2}\cdot\lbrack\vec{\sigma}^{\,\dagger}\cdot
      \vec{\pi}_{{}_M}\cdot\check{\vec{\nabla}}_{{}_M}
 (\frac{1}{J_{{}_M}\cdot(J_{{}_M}+m)}\cdot
     \vec{\pi}_{{}_M}^\dagger\cdot\vec{B}_{{}_M} )
+\frac{1}{J_{{}_M}\cdot(J_{{}_M}+m)}\cdot
     \vec{\pi}_{{}_M}^\dagger\cdot\vec{B}_{{}_M}
                    \cdot\vec{\sigma}\rbrack
\nonumber\\ &&\qquad
   - \vec{\sigma}^{\,\dagger}\cdot(\vec{\pi}_{{}_M}\wtimes
                   \vec{E}_{{}_M})\cdot \check{\vec{\nabla}}_{{}_M}
 \lbrack \frac{g}{2\cdot J_{{}_M}} -\frac{1}{J_{{}_M}+m}\rbrack
-\lbrack \frac{g}{2\cdot J_{{}_M}} -\frac{1}{J_{{}_M}+m}\rbrack\cdot
 (\vec{E}_{{}_M}\wtimes\vec{\sigma}) \biggr)
\nonumber\\
&=& -\frac{e}{m}\cdot
   \biggl(-\frac{m}{J_{{}_M}^3}\cdot\vec{\sigma}^{\,\dagger}\cdot
                  \vec{B}_{{}_M}\cdot\vec{\pi}_{{}_M}
-\frac{g-2}{2}\cdot \vec{\sigma}^{\,\dagger}\cdot
                                            \vec{\pi}_{{}_M}\cdot\lbrack
  \frac{1}{J_{{}_M}\cdot(J_{{}_M}+m)}\cdot \vec{B}_{{}_M}
\nonumber\\ &&\qquad
-\frac{m+2\cdot J_{{}_M}}{J_{{}_M}^3\cdot(J_{{}_M}+m)^2}\cdot
 \vec{\pi}_{{}_M}^\dagger\cdot
                        \vec{B}_{{}_M}\cdot\vec{\pi}_{{}_M} \rbrack
-\frac{g-2}{2}\cdot \frac{1}{J_{{}_M}\cdot(J_{{}_M}+m)}\cdot
      \vec{\pi}_{{}_M}^\dagger\cdot\vec{B}_{{}_M}\cdot\vec{\sigma}
\nonumber\\ &&
 + \lbrack
  \frac{g}{2\cdot J_{{}_M}^3}-
                         \frac{1}{J_{{}_M}\cdot(J_{{}_M}+m)^2}\rbrack
 \cdot\vec{\sigma}^{\,\dagger}\cdot(\vec{\pi}_{{}_M}\wtimes
     \vec{E}_{{}_M})   \cdot\vec{\pi}_{{}_M}
-\lbrack \frac{g}{2\cdot J_{{}_M}} -\frac{1}{J_{{}_M}+m}\rbrack\cdot
 (\vec{E}_{{}_M}\wtimes\vec{\sigma}) \biggr)
\nonumber\\
&=& \frac{e}{J_{{}_M}^3} \cdot
                         \vec{\sigma}^{\,\dagger}\cdot
     \vec{B}_{{}_M}\cdot\vec{\pi}_{{}_M}
+\frac{e}{m\cdot J_{{}_M}\cdot(J_{{}_M}+m)^2}\cdot
                    \vec{\sigma}^{\,\dagger}
                                                                 \cdot
    (\vec{\pi}_{{}_M}\wtimes\vec{E}_{{}_M})\cdot\vec{\pi}_{{}_M}
\nonumber\\ &&
- \frac{e}{m\cdot(J_{{}_M}+m)}\cdot(\vec{E}_{{}_M}\wtimes\vec{\sigma})
+ \frac{e\cdot g}{2\cdot m\cdot J_{{}_M}}\cdot
       (\vec{E}_{{}_M}\wtimes\vec{\sigma})
-\frac{e\cdot g}{2\cdot m\cdot J_{{}_M}^3}\cdot
                   \vec{\sigma}^{\,\dagger}\cdot
    (\vec{\pi}_{{}_M}\wtimes\vec{E}_{{}_M})\cdot\vec{\pi}_{{}_M}
\nonumber\\ &&\qquad
+\frac{g-2}{2}\cdot \biggl(
 \frac{e}{m}\cdot\frac{1}{J_{{}_M}\cdot(J_{{}_M}+m)}\cdot
  \vec{\sigma}^{\,\dagger}\cdot\vec{\pi}_{{}_M}\cdot\vec{B}_{{}_M}
\nonumber\\ &&
-\frac{e}{m}\cdot
 \frac{m+2\cdot J_{{}_M}}{J_{{}_M}^3\cdot(J_{{}_M}+m)^2}\cdot
\vec{\sigma}^{\,\dagger}\cdot\vec{\pi}_{{}_M}\cdot
               \vec{\pi}_{{}_M}^\dagger\cdot\vec{B}_{{}_M}
                               \cdot\vec{\pi}_{{}_M}
+\frac{e}{m}\cdot\frac{1}{J_{{}_M}\cdot(J_{{}_M}+m)}\cdot
\vec{\pi}_{{}_M}^\dagger\cdot\vec{B}_{{}_M}\cdot\vec{\sigma} \biggr)
\nonumber\\
&=& \frac{e}{K_{{}_M}^2} \cdot\vec{\sigma}^{\,\dagger}\cdot
                 \vec{B}_{{}_M}\cdot\vec{v}_{{}_M}
+\frac{e\cdot K_{{}_M}}{m\cdot(K_{{}_M}+m)^2}\cdot
                                        \vec{\sigma}^{\,\dagger}\cdot
    (\vec{v}_{{}_M}\wtimes\vec{E}_{{}_M})\cdot\vec{v}_{{}_M}
- \frac{e}{m\cdot(K_{{}_M}+m)}\cdot(\vec{E}_{{}_M}\wtimes\vec{\sigma})
\nonumber\\ &&\qquad
+ \frac{e\cdot g}{2\cdot m\cdot K_{{}_M}}\cdot
         (\vec{E}_{{}_M}\wtimes\vec{\sigma})
-\frac{e\cdot g}{2\cdot m\cdot K_{{}_M}}\cdot
                    \vec{\sigma}^{\,\dagger}\cdot
    (\vec{v}_{{}_M}\wtimes\vec{E}_{{}_M})\cdot\vec{v}_{{}_M}
\nonumber\\ &&\qquad
+\frac{g-2}{2}\cdot \biggl(
 \frac{e}{m}\cdot\frac{1}{K_{{}_M}+m}\cdot
  \vec{\sigma}^{\,\dagger}\cdot\vec{v}_{{}_M}\cdot\vec{B}_{{}_M}
-\frac{e}{m}\cdot\frac{m+2\cdot K_{{}_M}}{(K_{{}_M}+m)^2}\cdot
  \vec{\sigma}^{\,\dagger}\cdot\vec{v}_{{}_M}\cdot
                                             \vec{v}_{{}_M}^{\,\dagger}\cdot
      \vec{B}_{{}_M}\cdot\vec{v}_{{}_M}
\nonumber\\ &&\qquad
+\frac{e}{m}\cdot\frac{1}{K_{{}_M}+m}\cdot
      \vec{v}_{{}_M}^{\,\dagger}\cdot\vec{B}_{{}_M}\cdot
                      \vec{\sigma} \biggr) \; .
 \label{B.3}
\end{eqnarray}
Also one gets from (1.4),(B.1-2):
\begin{eqnarray}
&& \frac{\vec{\pi}_{{}_M}}{J_{{}_M}} =
   \check{\vec{\nabla}}_{{}_M} J_{{}_M} =
     \lbrace\vec{r}_{{}_M}, J_{{}_M}  \rbrace_{{}_M}
=\lbrace\vec{r}_{{}_M},H_{{}_M}\rbrace_{{}_M} -\lbrace\vec{r}_{{}_M},
                    \vec{\sigma}^{\,\dagger}\cdot
                \vec{W}_{{}_M}\rbrace_{{}_M}
= \vec{v}_{{}_M} - \vec{\Delta}_3 \; .
 \label{B.4}
\end{eqnarray}
Next I abbreviate
\begin{eqnarray}
&&\vec{\Delta}_4\equiv\lbrace\vec{\pi}_{{}_M},
                    \vec{\sigma}^{\,\dagger}\cdot
                 \vec{W}_{{}_M}\rbrace_{{}_M} \;,
\nonumber\\
&&\vec{\Delta}_5\equiv \vec{\Delta}_3' \; ,
\nonumber\\
&&\Delta_6\equiv m\cdot\gamma_{{}_M}^3\cdot
      \vec{v}_{{}_M}^{\,\dagger}\cdot\vec{\Delta}_3\; ,
 \label{B.5}
\end{eqnarray}
from which follows by (1.3-4):
\footnote{The partial derivative $\partial/\partial t$ in
(B.6)
                                 acts
on functions depending on $\vec{r}_{{}_M},t,\vec{p}_{{}_M}$.}
\begin{eqnarray}
 && \vec{\pi}_{{}_M}' =   \lbrace
                             \vec{\pi}_{{}_M} , H_{{}_M} \rbrace_{{}_M}
   + \frac{\partial\vec{\pi}_{{}_M}}{\partial t} =
\frac{e}{J_{{}_M}}\cdot(\vec{\pi}_{{}_M}\wtimes\vec{B}_{{}_M}) +
                                                 e\cdot\vec{E}_{{}_M}
 + \lbrace\vec{\pi}_{{}_M},\vec{\sigma}^{\,\dagger}\cdot
                                          \vec{W}_{{}_M} \rbrace_{{}_M}
\nonumber\\&&\qquad
 =  \frac{e}{J_{{}_M}}\cdot(\vec{\pi}_{{}_M}\wtimes
               \vec{B}_{{}_M}) + e\cdot\vec{E}_{{}_M}
 + \vec{\Delta}_4 \; ,
\nonumber\\
 && J_{{}_M}\cdot J_{{}_M}' = \vec{\pi}_{{}_M}^\dagger\cdot
                            \vec{\pi}_{{}_M}' =
e\cdot\vec{E}_{{}_M}^\dagger\cdot\vec{\pi}_{{}_M}+
      \vec{\pi}_{{}_M}^\dagger\cdot\vec{\Delta}_4 \;,
\nonumber\\
 &&  J_{{}_M}' = \frac{e}{J_{{}_M}}\cdot
                               \vec{E}_{{}_M}^\dagger\cdot
                                                    \vec{\pi}_{{}_M} +
 \frac{1}{J_{{}_M}}\cdot \vec{\pi}_{{}_M}^\dagger\cdot\vec{\Delta}_4
 =  \frac{e}{J_{{}_M}}\cdot \vec{E}_{{}_M}^\dagger\cdot
                                                 \vec{\pi}_{{}_M} +
         \vec{v}_{{}_M}^{\,\dagger}\cdot\vec{\Delta}_4 \; ,
 \label{B.6}
\end{eqnarray}
so that one gets from (B.1),(B.4):
\begin{eqnarray}
 && \vec{v}^{\;{}_{'}}_{{}_M}
                    =   \frac{\vec{\pi}_{{}_M}'}{J_{{}_M}}
 - \frac{J_{{}_M}'}{J_{{}_M}^2}\cdot\vec{\pi}_{{}_M} +  \vec{\Delta}_3'
 =  \frac{e}{J_{{}_M}^2}\cdot(\vec{\pi}_{{}_M}\wtimes\vec{B}_{{}_M}) +
    \frac{e}{J_{{}_M}}\cdot\vec{E}_{{}_M} +
  \frac{1}{J_{{}_M}}\cdot \vec{\Delta}_4
 - \frac{e}{J_{{}_M}^3}\cdot \vec{E}_{{}_M}^\dagger\cdot
       \vec{\pi}_{{}_M}\cdot\vec{\pi}_{{}_M}
\nonumber\\&&\qquad
 - \frac{1}{J_{{}_M}^2}\cdot \vec{v}_{{}_M}^{\,\dagger}\cdot
       \vec{\Delta}_4\cdot\vec{\pi}_{{}_M}
                                               +  \vec{\Delta}_5
 =  \frac{e}{J_{{}_M}^2}\cdot(\vec{\pi}_{{}_M}\wtimes\vec{B}_{{}_M}) +
\frac{e}{J_{{}_M}}\cdot\vec{E}_{{}_M} +\frac{1}{K_{{}_M}}\cdot
                                                          \vec{\Delta}_4
\nonumber\\&&\qquad
 - \frac{e}{J_{{}_M}^3}\cdot \vec{E}_{{}_M}^\dagger\cdot
       \vec{\pi}_{{}_M}\cdot\vec{\pi}_{{}_M}
 - \frac{1}{K_{{}_M}}\cdot \vec{v}_{{}_M}^{\,\dagger}\cdot
                                      \vec{\Delta}_4\cdot\vec{v}_{{}_M}
                      + \vec{\Delta}_5 \; .
 \label{B.7}
\end{eqnarray}
To eliminate $\vec{\pi}_{{}_M}$ from the rhs of (B.7) I use (B.4)
to calculate:
\begin{eqnarray}
 && \vec{v}_{{}_M}^{\,\dagger} \cdot \vec{v}_{{}_M}  =
                                              \frac{1}{J_{{}_M}^2}\cdot
    \vec{\pi}_{{}_M}^\dagger \cdot \vec{\pi}_{{}_M} +
                                                \frac{2}{J_{{}_M}}\cdot
\vec{\pi}_{{}_M}^\dagger\cdot\vec{\Delta}_3
 =    \frac{1}{J_{{}_M}^2}\cdot
     \vec{\pi}_{{}_M}^\dagger \cdot \vec{\pi}_{{}_M}
 +  2\cdot \vec{v}_{{}_M}^{\,\dagger}\cdot\vec{\Delta}_3
\nonumber\\ &&\qquad
 =   1 -  \frac{m^2}{J_{{}_M}^2}
 +  2\cdot \vec{v}_{{}_M}^{\,\dagger}\cdot\vec{\Delta}_3  \; ,
 \label{B.8}
\end{eqnarray}
so that
\begin{eqnarray}
 && J_{{}_M}^2 = m^2\cdot
                    \biggl( 1 - \vec{v}_{{}_M}^{\,\dagger} \cdot
                                                       \vec{v}_{{}_M}
 +  2\cdot \vec{v}_{{}_M}^{\,\dagger}\cdot\vec{\Delta}_3 \biggr)^{-1}
    = m^2\cdot(1- \vec{v}_{{}_M}^{\,\dagger} \cdot \vec{v}_{{}_M})^{-1}\cdot
\biggl(1 + \frac{2\cdot \vec{v}_{{}_M}^{\,\dagger}\cdot\vec{\Delta}_3}
              {1- \vec{v}_{{}_M}^{\,\dagger} \cdot \vec{v}_{{}_M}}
                                                      \biggr)^{-1}
\nonumber\\ &&\qquad
 =    m^2\cdot\gamma_{{}_M}^2\cdot \biggl(1 -
                                          2\cdot\gamma_{{}_M}^2\cdot
\vec{v}_{{}_M}^{\,\dagger}\cdot\vec{\Delta}_3 \biggr)  \; ,
\nonumber\\
&& J_{{}_M}= m\cdot\gamma_{{}_M}\cdot
\biggl(1 - \gamma_{{}_M}^2\cdot\vec{v}_{{}_M}^{\,\dagger}\cdot
                                          \vec{\Delta}_3\biggr)
    = K_{{}_M} - \Delta_6  \; ,
\nonumber\\
&& \frac{1}{J_{{}_M}}= (K_{{}_M}-\Delta_6)^{-1}
 = \frac{1}{K_{{}_M}}\cdot (1-\Delta_6/K_{{}_M})^{-1}
 = \frac{1}{K_{{}_M}}+ \frac{\Delta_6}{K_{{}_M}^2} \; ,
\nonumber\\
&& \frac{1}{J_{{}_M}^2}= \frac{1}{K_{{}_M}^2}+
                \frac{2\cdot\Delta_6}{K_{{}_M}^3} \; ,
\nonumber\\
&& \frac{\vec{\pi}_{{}_M}}{J_{{}_M}^2} =
                                     \frac{\vec{v}_{{}_M}}{J_{{}_M}}
      - \frac{\vec{\Delta}_3}{J_{{}_M}} = \vec{v}_{{}_M}\cdot
  \lbrack \frac{1}{K_{{}_M}}+ \frac{\Delta_6}{K_{{}_M}^2}\rbrack
      - \frac{\vec{\Delta}_3}{K_{{}_M}}  \; .
 \label{B.9}
\end{eqnarray}
Inserting (B.4),(B.9) into (B.7) yields
\begin{eqnarray}
 && \vec{v}^{\;{}_{'}}_{{}_M}
                    =  (\frac{e}{K_{{}_M}}+
             \frac{e\cdot\Delta_6}{K_{{}_M}^2} )\cdot
                      (\vec{v}_{{}_M}\wtimes\vec{B}_{{}_M})
      - \frac{e}{K_{{}_M}}\cdot (\vec{\Delta}_3 \wtimes \vec{B}_{{}_M})
 +(\frac{e}{K_{{}_M}}+ \frac{e\cdot\Delta_6}{K_{{}_M}^2} )\cdot
                                                      \vec{E}_{{}_M}
                        +\frac{1}{K_{{}_M}}\cdot \vec{\Delta}_4
\nonumber\\ &&\qquad
 - e\cdot \vec{E}_{{}_M}^\dagger\cdot\lbrack\vec{v}_{{}_M} -
                                                  \vec{\Delta}_3 \rbrack
   \cdot\lbrack  \vec{v}_{{}_M}\cdot(\frac{1}{K_{{}_M}}+
  \frac{\Delta_6}{K_{{}_M}^2} )
      - \frac{\vec{\Delta}_3}{K_{{}_M}} \rbrack
 - \frac{1}{K_{{}_M}}\cdot \vec{v}_{{}_M}^{\,\dagger}\cdot
                                                 \vec{\Delta}_4\cdot
                            \vec{v}_{{}_M}
                      + \vec{\Delta}_5
\nonumber\\&&\qquad
 =  \frac{e}{K_{{}_M}}\cdot(\vec{v}_{{}_M}\wtimes\vec{B}_{{}_M})
 +\frac{e}{K_{{}_M}}\cdot \vec{E}_{{}_M}
      - \frac{e}{K_{{}_M}}\cdot \vec{E}_{{}_M}^\dagger
               \cdot\vec{v}_{{}_M}\cdot \vec{v}_{{}_M}
  + \frac{e}{K_{{}_M}^2}\cdot\Delta_6 \cdot
                  (\vec{v}_{{}_M}\wtimes\vec{B}_{{}_M})
\nonumber\\ &&\qquad
      - \frac{e}{K_{{}_M}}\cdot (\vec{\Delta}_3 \wtimes \vec{B}_{{}_M})
      + \frac{e}{K_{{}_M}^2}\cdot \Delta_6\cdot \vec{E}_{{}_M}
 + \frac{1}{K_{{}_M}}\cdot \vec{\Delta}_4
  + \frac{e}{K_{{}_M}}\cdot \vec{E}_{{}_M}^\dagger\cdot
                                       \vec{\Delta}_3\cdot\vec{v}_{{}_M}
\nonumber\\ &&\qquad
-\frac{e}{K_{{}_M}^2}\cdot\Delta_6\cdot
          \vec{E}_{{}_M}^\dagger\cdot\vec{v}_{{}_M}\cdot
   \vec{v}_{{}_M}
+ \frac{e}{K_{{}_M}}\cdot \vec{E}_{{}_M}^\dagger\cdot
                                     \vec{v}_{{}_M}\cdot\vec{\Delta}_3
 - \frac{1}{K_{{}_M}}\cdot \vec{v}_{{}_M}^{\,\dagger}
                                 \cdot\vec{\Delta}_4\cdot\vec{v}_{{}_M}
                      + \vec{\Delta}_5\; .
 \label{B.10}
\end{eqnarray}
\subsection*{B.3}
Now I begin to evaluate $m\cdot (\gamma_{{}_P}\cdot\vec{v}_{{}_P})'$ and
it is clear by (2.3b) that:
\begin{eqnarray}
 && m\cdot (\gamma_{{}_P}\cdot\vec{v}_{{}_P})' =
  m\cdot \gamma_{{}_P}\cdot
    \vec{v}^{\;{}_{'}}_{{}_P}
                                           +
m\cdot \gamma_{{}_P}^3\cdot\vec{v}_{{}_P}^\dagger
   \cdot \vec{v}^{\;{}_{'}}_{{}_P}
                       \cdot\vec{v}_{{}_P} \; .
 \label{B.11}
\end{eqnarray}
The remaining task in this Appendix is to reexpress the rhs of
(B.11) in terms of $\vec{r}_{{}_P},t,\vec{v}_{{}_P},\vec{s}$ in order to
demonstrate that it is identical with the rhs of (2.10).
\par Introducing the abbreviations:
\begin{eqnarray}
&&\vec{\Delta}_7\equiv \vec{v}_{{}_P}-\vec{v}_{{}_M} \; ,
\nonumber\\
&&\vec{\Delta}_8\equiv \vec{\Delta}_7' \; ,
 \label{B.12}
\end{eqnarray}
one gets by (1.3),(1.6),(2.3a),(2.5),(A.1):
\begin{eqnarray}
&&  \vec{\Omega}_{{}_P}^\dagger\cdot\vec{v}_{{}_P}  =
       -  \frac{e\cdot g}{2\cdot m}\cdot\frac{1}{\gamma_{{}_P}}
\cdot \vec{B}_{{}_P}^\dagger\cdot\vec{v}_{{}_P}  \; ,
\nonumber\\
&&  \vec{\Omega}_{{}_M}^\dagger\cdot\vec{v}_{{}_M}  =
       -  \frac{e\cdot g}{2\cdot m}\cdot\frac{1}{\gamma_{{}_M}}
\cdot \vec{B}_{{}_M}^\dagger\cdot\vec{v}_{{}_M}  \; ,
\nonumber\\
&&\vec{\Delta}_7 =
    \biggl(   \frac{1}{m}\cdot\frac{\gamma_{{}_M}}{\gamma_{{}_M}+1}
\cdot (\vec{\sigma}\wtimes\vec{v}_{{}_M}) \biggr)'
 =  \biggl(   \frac{1}{m^2}\cdot\frac{1}{\gamma_{{}_M}+1}
\cdot (\vec{\sigma}\wtimes\vec{\pi}_{{}_M}) \biggr)'
\nonumber\\&&\qquad
 =  -\frac{1}{m^2}\cdot \gamma_{{}_M}'\cdot
                                       \frac{1}{(\gamma_{{}_M}+1)^2}
\cdot (\vec{\sigma}\wtimes\vec{\pi}_{{}_M})
       +  \frac{1}{m^2}\cdot\frac{1}{\gamma_{{}_M}+1}
\cdot (\vec{\sigma}\wtimes\vec{\pi}_{{}_M})'
\nonumber\\&&\qquad
 =  -\frac{e}{m^3}\cdot \vec{E}_{{}_M}^\dagger\cdot\vec{v}_{{}_M}
  \cdot \frac{1}{(\gamma_{{}_M}+1)^2} \cdot
     (\vec{\sigma}\wtimes\vec{\pi}_{{}_M})
       +  \frac{1}{m^2}\cdot\frac{1}{\gamma_{{}_M}+1}
\cdot (\vec{\sigma}\wtimes\vec{\pi}_{{}_M})'
\nonumber\\ &&\qquad
 =  -\frac{e}{m^2}\cdot \vec{E}_{{}_M}^\dagger\cdot\vec{v}_{{}_M}
\cdot \frac{\gamma_{{}_M}}{(\gamma_{{}_M}+1)^2}
              \cdot (\vec{\sigma}\wtimes\vec{v}_{{}_M})
       +  \frac{1}{m^2}\cdot\frac{1}{\gamma_{{}_M}+1}
\cdot (\vec{\sigma}\wtimes\vec{\pi}_{{}_M})'
\nonumber\\
&=& -\frac{e}{m^2}\cdot \vec{E}_{{}_M}^\dagger\cdot\vec{v}_{{}_M}
\cdot \frac{\gamma_{{}_M}}{(\gamma_{{}_M}+1)^2}
               \cdot (\vec{\sigma}\wtimes\vec{v}_{{}_M})
       +  \frac{1}{m^2}\cdot\frac{1}{\gamma_{{}_M}+1}
\cdot \biggl( (\vec{\Omega}_{{}_M}\wtimes \vec{\sigma})
                                                 \wtimes\vec{\pi}_{{}_M}
 + \vec{\sigma}\wtimes\vec{\pi}_{{}_M}' \biggr)
\nonumber\\ &&\qquad
 =  -\frac{e}{m^2}\cdot \vec{E}_{{}_M}^\dagger\cdot\vec{v}_{{}_M}
  \cdot \frac{1}{(\gamma_{{}_M}+1)^2} \cdot
                                       (\vec{s}\wtimes\vec{v}_{{}_M})
       +  \frac{1}{m^2}\cdot\frac{1}{\gamma_{{}_M}+1}
\cdot \biggl(\vec{\Omega}_{{}_M}^\dagger\cdot\vec{\pi}_{{}_M}\cdot
 \lbrack  \frac{1}{\gamma_{{}_M}}\cdot \vec{s}
\nonumber\\ &&\qquad
  + \frac{\gamma_{{}_M}}{\gamma_{{}_M}+1}
\cdot \vec{s}^{\,\dagger}\cdot\vec{v}_{{}_M}\cdot
                                       \vec{v}_{{}_M} \rbrack
   -  \vec{s}^{\,\dagger}\cdot\vec{\pi}_{{}_M}\cdot\vec{\Omega}_{{}_M}
+\lbrack  \frac{1}{\gamma_{{}_M}}\cdot
     \vec{s} + \frac{\gamma_{{}_M}}{\gamma_{{}_M}+1}
       \cdot \vec{s}^{\,\dagger}\cdot\vec{v}_{{}_M}
                                    \cdot\vec{v}_{{}_M} \rbrack\wtimes
                \vec{\pi}_{{}_M}' \biggr)
\nonumber\\&&\qquad
 =  -\frac{e}{m^2}\cdot \vec{E}_{{}_M}^\dagger\cdot\vec{v}_{{}_M}
  \cdot \frac{1}{(\gamma_{{}_M}+1)^2} \cdot
                                       (\vec{s}\wtimes\vec{v}_{{}_M})
\nonumber\\ &&
       -  \frac{e\cdot g}{2\cdot m^2}\cdot\frac{1}{\gamma_{{}_M}+1}
\cdot \vec{B}_{{}_M}^\dagger\cdot\vec{v}_{{}_M} \cdot
 \lbrack  \frac{1}{\gamma_{{}_M}}\cdot \vec{s}
  + \frac{\gamma_{{}_M}}{\gamma_{{}_M}+1}
       \cdot \vec{s}^{\,\dagger}\cdot\vec{v}_{{}_M}\cdot
                                              \vec{v}_{{}_M} \rbrack
   -  \frac{1}{m}\cdot \frac{\gamma_{{}_M}}{\gamma_{{}_M}+1}\cdot
      \vec{s}^{\,\dagger}\cdot\vec{v}_{{}_M}\cdot\vec{\Omega}_{{}_M}
\nonumber\\ &&\qquad
    +  \frac{1}{m^2}\cdot \frac{1}{\gamma_{{}_M}+1}\cdot
\lbrack\frac{1}{\gamma_{{}_M}}\cdot \vec{s} +
       \frac{\gamma_{{}_M}}{\gamma_{{}_M}+1}
\cdot \vec{s}^{\,\dagger}\cdot\vec{v}_{{}_M}\cdot
                                       \vec{v}_{{}_M} \rbrack\wtimes
\lbrack  e\cdot(\vec{v}_{{}_M}\wtimes\vec{B}_{{}_M})
 +e\cdot \vec{E}_{{}_M} \rbrack
\nonumber\\&&\qquad
 =  -\frac{e}{m^2}\cdot \vec{E}_{{}_M}^\dagger\cdot\vec{v}_{{}_P}
  \cdot \frac{1}{(\gamma_{{}_P}+1)^2} \cdot
                                       (\vec{s}\wtimes\vec{v}_{{}_P})
\nonumber\\ &&
       -  \frac{e\cdot g}{2\cdot m^2}\cdot\frac{1}{\gamma_{{}_P}+1}
\cdot \vec{B}_{{}_M}^\dagger\cdot\vec{v}_{{}_P} \cdot
 \lbrack  \frac{1}{\gamma_{{}_P}}\cdot \vec{s}
  + \frac{\gamma_{{}_P}}{\gamma_{{}_P}+1}
       \cdot \vec{s}^{\,\dagger}\cdot\vec{v}_{{}_P}\cdot
                                              \vec{v}_{{}_P} \rbrack
   -  \frac{1}{m}\cdot \frac{\gamma_{{}_P}}{\gamma_{{}_P}+1}\cdot
      \vec{s}^{\,\dagger}\cdot\vec{v}_{{}_P}\cdot\vec{\Omega}_{{}_P}
\nonumber\\ &&\qquad
    +  \frac{1}{m^2}\cdot \frac{1}{\gamma_{{}_P}+1}\cdot
 \lbrack  \frac{1}{\gamma_{{}_P}}\cdot
    \vec{s}+\frac{\gamma_{{}_P}}{\gamma_{{}_P}+1}
\cdot \vec{s}^{\,\dagger}\cdot\vec{v}_{{}_P}\cdot
                                       \vec{v}_{{}_P} \rbrack\wtimes
\lbrack  e\cdot(\vec{v}_{{}_P}\wtimes\vec{B}_{{}_M})
 +e\cdot \vec{E}_{{}_M} \rbrack
\nonumber\\&&\qquad
 =  -\frac{e}{m^2}\cdot \vec{E}_{{}_M}^\dagger\cdot\vec{v}_{{}_P}
  \cdot \frac{1}{(\gamma_{{}_P}+1)^2} \cdot
                                       (\vec{s}\wtimes\vec{v}_{{}_P})
\nonumber\\ &&\qquad
       -  \frac{e\cdot g}{2\cdot m^2}\cdot\frac{1}{\gamma_{{}_P}+1}
\cdot \vec{B}_{{}_M}^\dagger\cdot\vec{v}_{{}_P} \cdot
 \lbrack  \frac{1}{\gamma_{{}_P}}\cdot \vec{s}
  + \frac{\gamma_{{}_P}}{\gamma_{{}_P}+1}
       \cdot \vec{s}^{\,\dagger}\cdot\vec{v}_{{}_P}\cdot
                                              \vec{v}_{{}_P} \rbrack
\nonumber\\ && \qquad
+\frac{e}{m^2}\cdot\frac{\gamma_{{}_P}}{\gamma_{{}_P}+1}\cdot
      \vec{s}^{\,\dagger}\cdot\vec{v}_{{}_P}\cdot \biggl(
 \lbrack \frac{1}{\gamma_{{}_P}}+
                            \frac{g-2}{2} \rbrack \cdot\vec{B}_{{}_M}
 -\frac{g-2}{2}\cdot\frac{\gamma_{{}_P}}{\gamma_{{}_P}+1}\cdot
  \vec{v}_{{}_P}^\dagger\cdot\vec{B}_{{}_M}\cdot\vec{v}_{{}_P}
\nonumber\\ &&\qquad
-\lbrack \frac{g}{2} -\frac{\gamma_{{}_P}}{\gamma_{{}_P}+1}\rbrack\cdot
 (\vec{v}_{{}_P}\wtimes\vec{E}_{{}_M}) \biggr)
    +  \frac{e}{m^2}\cdot
    \frac{1}{\gamma_{{}_P}\cdot(\gamma_{{}_P}+1)}\cdot
             \vec{s}^{\,\dagger}\cdot\vec{B}_{{}_M}\cdot\vec{v}_{{}_P}
\nonumber\\ &&\qquad
    -  \frac{e}{m^2}\cdot
    \frac{1}{\gamma_{{}_P}\cdot(\gamma_{{}_P}+1)}\cdot
             \vec{s}^{\,\dagger}\cdot\vec{v}_{{}_P}\cdot\vec{B}_{{}_M}
    +  \frac{e}{m^2}\cdot \frac{\gamma_{{}_P}}{(\gamma_{{}_P}+1)^2}\cdot
             \vec{v}_{{}_P}^\dagger\cdot\vec{B}_{{}_M}\cdot
             \vec{s}^{\,\dagger}\cdot\vec{v}_{{}_P} \cdot\vec{v}_{{}_P}
\nonumber\\ &&\qquad
    -  \frac{e}{m^2}\cdot \frac{\gamma_{{}_P}}{(\gamma_{{}_P}+1)^2}\cdot
             \vec{v}_{{}_P}^\dagger\cdot\vec{v}_{{}_P}\cdot
             \vec{s}^{\,\dagger}\cdot\vec{v}_{{}_P}\cdot  \vec{B}_{{}_M}
    +  \frac{e}{m^2}\cdot
    \frac{1}{\gamma_{{}_P}\cdot(\gamma_{{}_P}+1)}\cdot
            (\vec{s}\wtimes \vec{E}_{{}_M})
\nonumber\\ &&\qquad
    +  \frac{e}{m^2}\cdot \frac{\gamma_{{}_P}}{(\gamma_{{}_P}+1)^2}\cdot
\vec{s}^{\,\dagger}\cdot\vec{v}_{{}_P}\cdot
                                  (\vec{v}_{{}_P}\wtimes \vec{E}_{{}_M})
\nonumber\\&&\qquad
 =  -\frac{e}{m^2}\cdot \vec{E}_{{}_M}^\dagger\cdot\vec{v}_{{}_P}
  \cdot \frac{1}{(\gamma_{{}_P}+1)^2} \cdot
                                       (\vec{s}\wtimes\vec{v}_{{}_P})
- \frac{e\cdot g}{2\cdot m^2}\cdot
   \frac{1}{\gamma_{{}_P}\cdot(\gamma_{{}_P}+1)}
\cdot \vec{B}_{{}_M}^\dagger\cdot\vec{v}_{{}_P} \cdot \vec{s}
\nonumber\\ &&\qquad
- \frac{e\cdot(g-2)}{2\cdot m^2}\cdot
    \frac{\gamma_{{}_P}}{\gamma_{{}_P}+1}
\cdot \vec{B}_{{}_M}^\dagger\cdot\vec{v}_{{}_P} \cdot
       \vec{s}^{\,\dagger}\cdot\vec{v}_{{}_P}\cdot\vec{v}_{{}_P}
   +\frac{e\cdot(g-2)}{2\cdot m^2}\cdot
\frac{\gamma_{{}_P}}{\gamma_{{}_P}+1}\cdot
    \vec{s}^{\,\dagger}\cdot\vec{v}_{{}_P}\cdot\vec{B}_{{}_M}
\nonumber\\ &&\qquad
+  \frac{e}{m^2}\cdot \frac{1}{\gamma_{{}_P}\cdot(\gamma_{{}_P}+1)}\cdot
             \vec{s}^{\,\dagger}\cdot\vec{B}_{{}_M}\cdot\vec{v}_{{}_P}
-\frac{e\cdot(g-2)}{2\cdot m^2}\cdot
   \frac{\gamma_{{}_P}}{\gamma_{{}_P}+1}\cdot
\vec{s}^{\,\dagger}\cdot\vec{v}_{{}_P}\cdot
                                  (\vec{v}_{{}_P}\wtimes \vec{E}_{{}_M})
\nonumber\\ &&\qquad
+  \frac{e}{m^2}\cdot \frac{1}{\gamma_{{}_P}\cdot(\gamma_{{}_P}+1)}\cdot
            (\vec{s}\wtimes \vec{E}_{{}_M})  \; .
 \label{B.13}
\end{eqnarray}
Next one concludes from (B.10),(B.12):
\begin{eqnarray}
 &&  \vec{v}^{\;{}_{'}}_{{}_P}
  =  \vec{v}^{\;{}_{'}}_{{}_M}
                      +  \vec{\Delta}_8
\nonumber\\
&=& \frac{e}{K_{{}_M}}\cdot(\vec{v}_{{}_M}\wtimes\vec{B}_{{}_M})
 +\frac{e}{K_{{}_M}}\cdot \vec{E}_{{}_M}
 - \frac{e}{K_{{}_M}}\cdot \vec{E}_{{}_M}^\dagger\cdot
                                      \vec{v}_{{}_M}\cdot \vec{v}_{{}_M}
  + \frac{e}{K_{{}_M}^2}\cdot\Delta_6 \cdot
                (\vec{v}_{{}_M}\wtimes\vec{B}_{{}_M})
\nonumber\\ &&\qquad
      - \frac{e}{K_{{}_M}}\cdot (\vec{\Delta}_3 \wtimes \vec{B}_{{}_M})
      + \frac{e}{K_{{}_M}^2}\cdot \Delta_6\cdot \vec{E}_{{}_M}
 + \frac{1}{K_{{}_M}}\cdot \vec{\Delta}_4
 + \frac{e}{K_{{}_M}}\cdot \vec{E}_{{}_M}^\dagger\cdot
                                      \vec{\Delta}_3\cdot\vec{v}_{{}_M}
\nonumber\\&&
-\frac{e}{K_{{}_M}^2}\cdot\Delta_6\cdot
   \vec{E}_{{}_M}^\dagger\cdot\vec{v}_{{}_M}\cdot\vec{v}_{{}_M}
    + \frac{e}{K_{{}_M}}\cdot \vec{E}_{{}_M}^\dagger\cdot
                                     \vec{v}_{{}_M}\cdot\vec{\Delta}_3
-\frac{1}{K_{{}_M}}\cdot\vec{v}_{{}_M}^{\,\dagger}
                   \cdot\vec{\Delta}_4\cdot\vec{v}_{{}_M}
                      + \vec{\Delta}_5 + \vec{\Delta}_8 \; .
\nonumber\\&&
 \label{B.14}
\end{eqnarray}
\subsection*{B.4}
If one inserts (B.14) into (B.11) then the rhs of (B.11) depends
explicitly on
$\vec{v}_{{}_M}$. In this subsection I show how the variable
$\vec{v}_{{}_M}$ on the rhs of (B.11),(B.14) can be replaced by
$\vec{v}_{{}_P}$. I abbreviate:
\begin{eqnarray}
&& K_{{}_P} \equiv m\cdot\gamma_{{}_P}   \; ,
\nonumber\\
&& \Delta_9\equiv -K_{{}_P}\cdot\gamma_{{}_P}^2
          \cdot\vec{v}_{{}_P}^\dagger\cdot\vec{\Delta}_7  \; .
 \label{B.15}
\end{eqnarray}
From this follows by (B.1),(B.12):
\begin{eqnarray}
&& \frac{1}{K_{{}_M}^2}= \frac{1}{m^2}\cdot
               (1-\vec{v}_{{}_M}^{\,\dagger}\cdot\vec{v}_{{}_M}) =
 \frac{1}{m^2}\cdot \lbrack 1 -
    (\vec{v}_{{}_P}-\vec{\Delta}_7)^\dagger\cdot
                                (\vec{v}_{{}_P}-\vec{\Delta}_7)\rbrack
\nonumber\\
&=& \frac{1}{m^2}\cdot \lbrack 1 -
    \vec{v}_{{}_P}^\dagger\cdot\vec{v}_{{}_P}
      +2\cdot\vec{v}_{{}_P}^\dagger\cdot\vec{\Delta}_7\rbrack
 = \frac{1}{m^2}\cdot \lbrack 1-
  \vec{v}_{{}_P}^\dagger\cdot\vec{v}_{{}_P}\rbrack\cdot
\lbrack 1+2\cdot\gamma_{{}_P}^2\cdot
 \vec{v}_{{}_P}^\dagger\cdot\vec{\Delta}_7\rbrack
\nonumber\\
&=& \frac{1}{K_{{}_P}^2}\cdot
\lbrack 1+2\cdot\gamma_{{}_P}^2\cdot
  \vec{v}_{{}_P}^\dagger\cdot\vec{\Delta}_7\rbrack
 =  \frac{1}{K_{{}_P}^2}\cdot
\lbrack 1 -  \frac{2\cdot\Delta_9}{K_{{}_P}} \rbrack
 = \frac{1}{K_{{}_P}^2}- \frac{2\cdot\Delta_9}{K_{{}_P}^3} \; ,
\nonumber\\
&& \frac{1}{K_{{}_M}}= \frac{1}{K_{{}_P}}\cdot
\lbrack 1 -  \frac{\Delta_9}{K_{{}_P}} \rbrack=
                   \frac{1}{K_{{}_P}} - \frac{\Delta_9}{K_{{}_P}^2} \; ,
\nonumber\\
&& K_{{}_M} = K_{{}_P}  + \Delta_9  \; ,
\nonumber\\
&&\frac{\vec{v}_{{}_M}}{K_{{}_M}}=
     \lbrack\frac{1}{K_{{}_P}}-\frac{\Delta_9}{K_{{}_P}^2}\rbrack\cdot
  \lbrack \vec{v}_{{}_P}-\vec{\Delta}_7\rbrack
   = \frac{1}{K_{{}_P}}\cdot \vec{v}_{{}_P}
    - \frac{\Delta_9}{K_{{}_P}^2}\cdot\vec{v}_{{}_P}
   - \frac{1}{K_{{}_P}}\cdot\vec{\Delta}_7
 \; .
 \label{B.16}
\end{eqnarray}
Also one has by (2.5),(B.1),(B.3),(B.12-13),(B.15):
\begin{eqnarray}
&&\vec{\Delta}_3 =
    \frac{e}{K_{{}_P}^2} \cdot\vec{v}_{{}_P}\cdot
                             \vec{B}_{{}_M}^\dagger\cdot
 \lbrack  \frac{1}{\gamma_{{}_P}}\cdot
     \vec{s} + \frac{\gamma_{{}_P}}{\gamma_{{}_P}+1}
 \cdot \vec{s}^{\,\dagger}\cdot\vec{v}_{{}_P}\cdot\vec{v}_{{}_P} \rbrack
+\frac{e}{(K_{{}_P}+m)^2}\cdot\vec{s}^{\,\dagger}\cdot
    (\vec{v}_{{}_P}\wtimes\vec{E}_{{}_M})\cdot\vec{v}_{{}_P}
\nonumber\\ &&\qquad
- \frac{e}{m\cdot(K_{{}_P}+m)}\cdot \vec{E}_{{}_M}\wtimes
 \lbrack  \frac{1}{\gamma_{{}_P}}\cdot
      \vec{s} + \frac{\gamma_{{}_P}}{\gamma_{{}_P}+1}
 \cdot \vec{s}^{\,\dagger}\cdot\vec{v}_{{}_P}\cdot\vec{v}_{{}_P} \rbrack
\nonumber\\ &&\qquad
+ \frac{e\cdot g}{2\cdot m\cdot K_{{}_P}}\cdot  \vec{E}_{{}_M}\wtimes
 \lbrack  \frac{1}{\gamma_{{}_P}}\cdot
     \vec{s} + \frac{\gamma_{{}_P}}{\gamma_{{}_P}+1}
 \cdot \vec{s}^{\,\dagger}\cdot\vec{v}_{{}_P}\cdot\vec{v}_{{}_P} \rbrack
-\frac{e\cdot g}{2\cdot K_{{}_P}^2}\cdot\vec{s}^{\,\dagger}\cdot
    (\vec{v}_{{}_P}\wtimes\vec{E}_{{}_M})\cdot\vec{v}_{{}_P}
\nonumber\\ &&\qquad
+\frac{g-2}{2}\cdot \biggl(
 \frac{e}{m}\cdot\frac{1}{K_{{}_P}+m}\cdot
  \vec{s}^{\,\dagger}\cdot\vec{v}_{{}_P}\cdot\vec{B}_{{}_M}
-\frac{e}{m}\cdot\frac{m+2\cdot K_{{}_P}}{(K_{{}_P}+m)^2}\cdot
  \vec{s}^{\,\dagger}\cdot\vec{v}_{{}_P}\cdot
  \vec{v}_{{}_P}^\dagger\cdot\vec{B}_{{}_M}\cdot\vec{v}_{{}_P}
\nonumber\\ &&\qquad
+\frac{e}{m}\cdot\frac{1}{K_{{}_P}+m}\cdot
      \vec{v}_{{}_P}^\dagger\cdot\vec{B}_{{}_M}\cdot
 \lbrack  \frac{1}{\gamma_{{}_P}}\cdot
     \vec{s} + \frac{\gamma_{{}_P}}{\gamma_{{}_P}+1}
  \cdot \vec{s}^{\,\dagger}\cdot\vec{v}_{{}_P}\cdot
                                               \vec{v}_{{}_P} \rbrack
                                          \biggr)
\nonumber\\
&=& \frac{e}{K_{{}_P}^2} \cdot\vec{v}_{{}_P}\cdot
                           \vec{B}_{{}_M}^\dagger\cdot
 \lbrack  \frac{1}{\gamma_{{}_P}}\cdot
     \vec{s} + \frac{\gamma_{{}_P}}{\gamma_{{}_P}+1}
 \cdot \vec{s}^{\,\dagger}\cdot\vec{v}_{{}_P}\cdot\vec{v}_{{}_P} \rbrack
+\frac{e}{(K_{{}_P}+m)^2}\cdot\vec{s}^{\,\dagger}\cdot
    (\vec{v}_{{}_P}\wtimes\vec{E}_{{}_M})\cdot\vec{v}_{{}_P}
\nonumber\\ &&\qquad
- \frac{e}{m\cdot(K_{{}_P}+m)}\cdot \vec{E}_{{}_M}\wtimes
 \lbrack  \frac{1}{\gamma_{{}_P}}\cdot
     \vec{s} + \frac{\gamma_{{}_P}}{\gamma_{{}_P}+1}
 \cdot \vec{s}^{\,\dagger}\cdot\vec{v}_{{}_P}\cdot\vec{v}_{{}_P} \rbrack
\nonumber\\ &&\qquad
+ \frac{e\cdot g}{2\cdot m\cdot K_{{}_P}}\cdot  \vec{E}_{{}_M}\wtimes
 \lbrack  \frac{1}{\gamma_{{}_P}}\cdot
    \vec{s} + \frac{\gamma_{{}_P}}{\gamma_{{}_P}+1}
  \cdot \vec{s}^{\,\dagger}\cdot\vec{v}_{{}_P}\cdot
                                               \vec{v}_{{}_P} \rbrack
-\frac{e\cdot g}{2\cdot K_{{}_P}^2}\cdot\vec{s}^{\,\dagger}\cdot
    (\vec{v}_{{}_P}\wtimes\vec{E}_{{}_M})\cdot\vec{v}_{{}_P}
\nonumber\\ &&\qquad
+\frac{g-2}{2}\cdot \biggl(
 \frac{e}{m}\cdot\frac{1}{K_{{}_P}+m}\cdot
  \vec{s}^{\,\dagger}\cdot\vec{v}_{{}_P}\cdot\vec{B}_{{}_M}
-\frac{e}{m^2}\cdot\frac{1}{\gamma_{{}_P} + 1}\cdot
  \vec{s}^{\,\dagger}\cdot\vec{v}_{{}_P}\cdot
  \vec{v}_{{}_P}^\dagger\cdot\vec{B}_{{}_M}\cdot\vec{v}_{{}_P}
 \nonumber\\ &&\qquad
+\frac{e}{m^2}\cdot\frac{1}{\gamma_{{}_P}\cdot(\gamma_{{}_P}+1)}\cdot
      \vec{v}_{{}_P}^\dagger\cdot\vec{B}_{{}_M}\cdot \vec{s} \biggr)
                                                  \; ,
\nonumber\\
&&\vec{\Delta}_3 + \vec{\Delta}_7 =
    \frac{e}{m^2} \cdot
\frac{\gamma_{{}_P}^2+\gamma_{{}_P}+1}{\gamma_{{}_P}^3\cdot
                                            (\gamma_{{}_P} + 1)}
         \cdot\vec{B}_{{}_M}^\dagger\cdot\vec{s}\cdot\vec{v}_{{}_P}
 \nonumber\\ &&\qquad
+\frac{e}{m^2}\cdot \lbrack \frac{1}{\gamma_{{}_P}\cdot(\gamma_{{}_P}
                   +1)}-\frac{g-2}{2} \rbrack
\cdot\vec{s}^{\,\dagger}\cdot\vec{v}_{{}_P}
 \cdot\vec{v}_{{}_P}^\dagger\cdot\vec{B}_{{}_M}\cdot\vec{v}_{{}_P}
 \nonumber\\ &&\qquad
+\frac{e}{m^2}\cdot \lbrack \frac{1}{(\gamma_{{}_P}+1)^2}
     -\frac{g}{2\cdot\gamma_{{}_P}^2}\rbrack
\cdot\vec{s}^{\,\dagger}\cdot (\vec{v}_{{}_P}\wtimes
       \vec{E}_{{}_M})\cdot\vec{v}_{{}_P}
 \nonumber\\ &&\qquad
+ \frac{e}{m^2}\cdot \lbrack -
     \frac{2}{\gamma_{{}_P}\cdot(\gamma_{{}_P}+1)}
     +\frac{g}{2\cdot\gamma_{{}_P}^2}\rbrack
              \cdot (\vec{E}_{{}_M}\wtimes \vec{s})
 \nonumber\\ &&\qquad
+\frac{e}{m^2}\cdot \lbrack
  -\frac{\gamma_{{}_P}^2+2\cdot
     \gamma_{{}_P}}{(\gamma_{{}_P} +1)^2}+ \frac{g}{2}\rbrack
  \cdot \vec{s}^{\,\dagger}\cdot\vec{v}_{{}_P}\cdot
                    (\vec{E}_{{}_M}\wtimes \vec{v}_{{}_P})
+\frac{e\cdot(g-2)}{2\cdot m^2}\cdot \vec{s}^{\,\dagger}\cdot
                                                         \vec{v}_{{}_P}
                                               \cdot\vec{B}_{{}_M}
\nonumber\\ &&\qquad
    -\frac{e}{m^2}\cdot \vec{E}_{{}_M}^\dagger\cdot\vec{v}_{{}_P}
  \cdot \frac{1}{(\gamma_{{}_P}+1)^2} \cdot
                                       (\vec{s}\wtimes\vec{v}_{{}_P})
-  \frac{e}{m^2}\cdot \frac{1}{\gamma_{{}_P}\cdot(\gamma_{{}_P}+1)}\cdot
             \vec{B}_{{}_M}^\dagger\cdot\vec{v}_{{}_P}\cdot\vec{s} \; ,
\nonumber\\
&& \Delta_6 = m\cdot\gamma_{{}_P}^3
    \cdot \vec{v}_{{}_P}^\dagger\cdot\vec{\Delta}_3 \; .
 \label{B.17}
\end{eqnarray}
Combining (B.16-17) I can replace on the rhs of (B.14) the variable
$\vec{v}_{{}_M}$ by $\vec{v}_{{}_P}$:
\begin{eqnarray}
&&   \vec{v}^{\;{}_{'}}_{{}_P}
                    =  \lbrack  \frac{e}{K_{{}_P}}\cdot \vec{v}_{{}_P}
    - \frac{e\cdot\Delta_9}{K_{{}_P}^2}\cdot\vec{v}_{{}_P}
   - \frac{e}{K_{{}_P}}\cdot\vec{\Delta}_7\rbrack  \wtimes\vec{B}_{{}_M}
+\lbrack\frac{e}{K_{{}_P}} -\frac{e\cdot
          \Delta_9}{K_{{}_P}^2}\rbrack\cdot\vec{E}_{{}_M}
\nonumber\\ &&\qquad
 - e\cdot \vec{E}_{{}_M}^\dagger\cdot
      \lbrack \frac{1}{K_{{}_P}}\cdot \vec{v}_{{}_P}
    - \frac{\Delta_9}{K_{{}_P}^2}\cdot\vec{v}_{{}_P}
   - \frac{1}{K_{{}_P}}\cdot\vec{\Delta}_7\rbrack \cdot
  \lbrack \vec{v}_{{}_P}-\vec{\Delta}_7\rbrack
   + \frac{e}{K_{{}_P}^2}\cdot\Delta_6 \cdot
                  (\vec{v}_{{}_P}\wtimes\vec{B}_{{}_M})
\nonumber\\ &&\qquad
      - \frac{e}{K_{{}_P}}\cdot (\vec{\Delta}_3 \wtimes \vec{B}_{{}_M})
      + \frac{e}{K_{{}_P}^2}\cdot \Delta_6\cdot \vec{E}_{{}_M}
 + \frac{1}{K_{{}_P}}\cdot \vec{\Delta}_4
+ \frac{e}{K_{{}_P}}\cdot \vec{E}_{{}_M}^\dagger\cdot
                                     \vec{\Delta}_3\cdot\vec{v}_{{}_P}
\nonumber\\ &&\qquad
-\frac{e}{K_{{}_P}^2}\cdot\Delta_6\cdot
               \vec{E}_{{}_M}^\dagger\cdot\vec{v}_{{}_P}
                                 \cdot\vec{v}_{{}_P}
 + \frac{e}{K_{{}_P}}\cdot \vec{E}_{{}_M}^\dagger\cdot
                                      \vec{v}_{{}_P}\cdot\vec{\Delta}_3
 - \frac{1}{K_{{}_P}}\cdot \vec{v}_{{}_P}^\dagger\cdot
                                      \vec{\Delta}_4\cdot\vec{v}_{{}_P}
                      + \vec{\Delta}_5 + \vec{\Delta}_8
\nonumber\\
&=& \frac{e}{K_{{}_P}}\cdot(\vec{v}_{{}_P}\wtimes\vec{B}_{{}_M})
 +\frac{e}{K_{{}_P}}\cdot \vec{E}_{{}_M}
  - \frac{e}{K_{{}_P}}\cdot \vec{E}_{{}_M}^\dagger\cdot
  \vec{v}_{{}_P}\cdot \vec{v}_{{}_P}
    - \frac{e\cdot\Delta_9}{K_{{}_P}^2}\cdot
                   (\vec{v}_{{}_P}\wtimes\vec{B}_{{}_M})
\nonumber\\ &&\qquad
   - \frac{e}{K_{{}_P}}\cdot(\vec{\Delta}_7 \wtimes\vec{B}_{{}_M})
  -\frac{e\cdot\Delta_9}{K_{{}_P}^2}\cdot\vec{E}_{{}_M}
 + \frac{e}{K_{{}_P}}\cdot  \vec{E}_{{}_M}^\dagger\cdot
                                       \vec{v}_{{}_P}\cdot\vec{\Delta}_7
\nonumber\\ &&\qquad
+\frac{e}{K_{{}_P}^2}\cdot\Delta_9\cdot
         \vec{E}_{{}_M}^\dagger\cdot\vec{v}_{{}_P}
   \cdot\vec{v}_{{}_P}
 + \frac{e}{K_{{}_P}}\cdot \vec{E}_{{}_M}^\dagger\cdot
                                      \vec{\Delta}_7 \cdot\vec{v}_{{}_P}
   + \frac{e}{K_{{}_P}^2}\cdot\Delta_6 \cdot
               (\vec{v}_{{}_P}\wtimes\vec{B}_{{}_M})
\nonumber\\ &&\qquad
      - \frac{e}{K_{{}_P}}\cdot (\vec{\Delta}_3 \wtimes \vec{B}_{{}_M})
      + \frac{e}{K_{{}_P}^2}\cdot \Delta_6\cdot \vec{E}_{{}_M}
 + \frac{1}{K_{{}_P}}\cdot \vec{\Delta}_4
+ \frac{e}{K_{{}_P}}\cdot \vec{E}_{{}_M}^\dagger\cdot
                                     \vec{\Delta}_3\cdot\vec{v}_{{}_P}
\nonumber\\ &&\qquad
-\frac{e}{K_{{}_P}^2}\cdot\Delta_6\cdot\vec{E}_{{}_M}^\dagger\cdot
     \vec{v}_{{}_P}\cdot\vec{v}_{{}_P}
 + \frac{e}{K_{{}_P}}\cdot \vec{E}_{{}_M}^\dagger\cdot
                                      \vec{v}_{{}_P}\cdot\vec{\Delta}_3
 - \frac{1}{K_{{}_P}}\cdot \vec{v}_{{}_P}^\dagger\cdot
                                      \vec{\Delta}_4\cdot\vec{v}_{{}_P}
                      + \vec{\Delta}_5 + \vec{\Delta}_8 \; ,
\nonumber\\
 \label{B.18}
\end{eqnarray}
so that (B.11) reads as:
\begin{eqnarray}
 && m\cdot (\gamma_{{}_P}\cdot\vec{v}_{{}_P})' =
  m\cdot \gamma_{{}_P}\cdot
    \vec{v}^{\;{}_{'}}_{{}_P}
                                           +
m\cdot \gamma_{{}_P}^3\cdot\vec{v}_{{}_P}^\dagger
   \cdot \vec{v}^{\;{}_{'}}_{{}_P}
                       \cdot\vec{v}_{{}_P}
\nonumber\\
&=& e\cdot(\vec{v}_{{}_P}\wtimes\vec{B}_{{}_M})
 + e\cdot \vec{E}_{{}_M}
\nonumber\\ &&\qquad
    + \frac{e}{K_{{}_P}}\cdot \lbrack \Delta_6 - \Delta_9 \rbrack \cdot
   \lbrack \vec{v}_{{}_P}\wtimes\vec{B}_{{}_M} + \vec{E}_{{}_M} \rbrack
- e\cdot \lbrack  1 + \gamma_{{}_P}^2\cdot
    \vec{v}_{{}_P}\cdot\vec{v}_{{}_P}^\dagger\rbrack
\cdot\lbrack (\vec{\Delta}_3  +
              \vec{\Delta}_7)\wtimes\vec{B}_{{}_M}\rbrack
\nonumber\\ &&\qquad
+ e\cdot\lbrack \vec{E}_{{}_M}^\dagger\cdot\vec{v}_{{}_P}
+\gamma_{{}_P}^2\cdot\vec{v}_{{}_P}\cdot\vec{E}_{{}_M}^\dagger\cdot
  \vec{v}_{{}_P}\cdot\vec{v}_{{}_P}^\dagger
+\gamma_{{}_P}^2\cdot\vec{v}_{{}_P}\cdot
                \vec{E}_{{}_M}^\dagger \rbrack \cdot
    \lbrack \vec{\Delta}_3  + \vec{\Delta}_7 \rbrack
\nonumber\\ &&\qquad
 +                    \vec{\Delta}_4
 + K_{{}_P}\cdot\lbrack \vec{\Delta}_5 +  \vec{\Delta}_8 \rbrack
+K_{{}_P}\cdot\gamma_{{}_P}^2\cdot
       \vec{v}_{{}_P}^\dagger\cdot \lbrack \vec{\Delta}_5 +
   \vec{\Delta}_8 \rbrack \cdot \vec{v}_{{}_P}  \; .
 \label{B.19}
\end{eqnarray}
\subsection*{B.5}
In this subsection I show how the variable $\vec{r}_{{}_M}$ on the rhs
of (B.19) can be replaced by $\vec{r}_{{}_P}$. The dependence
on $\vec{r}_{{}_M}$ comes in only via the field vectors
$\vec{E}_{{}_M},\vec{B}_{{}_M}$ and their first derivatives. First of
all by using (2.1),(2.5) I abbreviate
\begin{eqnarray}
\vec{\Delta}_{10} &\equiv& \vec{r}_{{}_M}-\vec{r}_{{}_P}
   = -        \frac{1}{m}\cdot\frac{\gamma_{{}_M}}{\gamma_{{}_M}+1}
\cdot (\vec{\sigma}\wtimes\vec{v}_{{}_M})
   = -        \frac{1}{m}\cdot\frac{1}{\gamma_{{}_M}+1}
\cdot (\vec{s}\wtimes\vec{v}_{{}_M})
\nonumber\\
&=&  -        \frac{1}{m}\cdot\frac{1}{\gamma_{{}_P}+1}
\cdot (\vec{s}\wtimes\vec{v}_{{}_P}) \; ,
\nonumber\\
\vec{\Delta}_{11} &\equiv&
  (\vec{\Delta}_{10}^\dagger\cdot\vec{\nabla}_{{}_M})\vec{E}_{{}_M}\;,
\nonumber\\
\vec{\Delta}_{12} &\equiv&
  (\vec{\Delta}_{10}^\dagger\cdot\vec{\nabla}_{{}_M})\vec{B}_{{}_M} \; ,
 \label{B.20}
\end{eqnarray}
from which follows:
\begin{eqnarray}
&&\vec{E}_{{}_M} =
  \vec{E}_{{}_M}(\vec{r}_{{}_M},t) = \vec{E}_{{}_M}(\vec{r}_{{}_P}+
        \vec{\Delta}_{10},t) =
 \vec{E}_{{}_M}(\vec{r}_{{}_P},t) + \vec{\Delta}_{11}
=\vec{E}_{{}_P}  + \vec{\Delta}_{11} \; ,
\nonumber\\
&&\vec{B}_{{}_M} =
  \vec{B}_{{}_M}(\vec{r}_{{}_M},t) = \vec{B}_{{}_M}(\vec{r}_{{}_P}+
                           \vec{\Delta}_{10},t) =
 \vec{B}_{{}_M}(\vec{r}_{{}_P},t) + \vec{\Delta}_{12}
=\vec{B}_{{}_P}  + \vec{\Delta}_{12} \; .
 \label{B.21}
\end{eqnarray}
Inserting this into (B.19) yields:
\begin{eqnarray}
 && m\cdot (\gamma_{{}_P}\cdot\vec{v}_{{}_P})' =
    \underbrace{
    e\cdot(\vec{v}_{{}_P}\wtimes\vec{B}_{{}_P})
 + e\cdot \vec{E}_{{}_P}
                       }_{{\rm Lorentz\; force}} \; +
    \underbrace{
     e\cdot(\vec{v}_{{}_P} \wtimes\vec{\Delta}_{12})
 + e\cdot\vec{\Delta}_{11}
         }_{{\rm first\; order\;SG\;terms}}
\nonumber\\ &&\qquad
    \underbrace{
 +                    \vec{\Delta}_4
 + K_{{}_P}\cdot\lbrack \vec{\Delta}_5 +  \vec{\Delta}_8 \rbrack
+K_{{}_P}\cdot\gamma_{{}_P}^2\cdot
       \vec{v}_{{}_P}^\dagger\cdot \lbrack \vec{\Delta}_5 +
   \vec{\Delta}_8 \rbrack \cdot \vec{v}_{{}_P}
         }_{{\rm first\; order\;and\;second\; order\;SG\;terms}}
\nonumber\\ &&\qquad
    \underbrace{
    + \frac{e}{K_{{}_P}}\cdot \lbrack \Delta_6 - \Delta_9 \rbrack \cdot
   \lbrack \vec{v}_{{}_P}\wtimes\vec{B}_{{}_P} + \vec{E}_{{}_P} \rbrack
- e\cdot \lbrack  1 + \gamma_{{}_P}^2\cdot
  \vec{v}_{{}_P}\cdot\vec{v}_{{}_P}^\dagger\rbrack
\cdot\lbrack (\vec{\Delta}_3  +
           \vec{\Delta}_7) \wtimes\vec{B}_{{}_P}\rbrack
         }_{{\rm second\; order\;SG\;terms}}
\nonumber\\ &&\qquad
    \underbrace{
+ e\cdot\lbrack \vec{E}_{{}_P}^\dagger\cdot\vec{v}_{{}_P}
+\gamma_{{}_P}^2\cdot\vec{v}_{{}_P}\cdot\vec{E}_{{}_P}^\dagger
  \cdot\vec{v}_{{}_P}\cdot\vec{v}_{{}_P}^\dagger
+\gamma_{{}_P}^2\cdot\vec{v}_{{}_P}\cdot\vec{E}_{{}_P}^\dagger
                                                         \rbrack \cdot
    \lbrack \vec{\Delta}_3  + \vec{\Delta}_7 \rbrack
         }_{{\rm second\; order\;SG\;terms}} \; .
 \label{B.22}
\end{eqnarray}
Introducing the abbreviation
\begin{eqnarray}
&&\vec{\Delta}_{13}\equiv
     e\cdot(\vec{v}_{{}_P} \wtimes\vec{\Delta}_{12})
 + e\cdot\vec{\Delta}_{11}
 +                    \vec{\Delta}_4
 + K_{{}_P}\cdot\lbrack \vec{\Delta}_5 +  \vec{\Delta}_8 \rbrack
+K_{{}_P}\cdot\gamma_{{}_P}^2\cdot
       \vec{v}_{{}_P}^\dagger\cdot \lbrack \vec{\Delta}_5 +
   \vec{\Delta}_8 \rbrack \cdot \vec{v}_{{}_P}
\nonumber\\ &&\qquad
    + \frac{e}{K_{{}_P}}\cdot \lbrack \Delta_6 - \Delta_9 \rbrack \cdot
   \lbrack \vec{v}_{{}_P}\wtimes\vec{B}_{{}_P} + \vec{E}_{{}_P} \rbrack
- e\cdot \lbrack  1 + \gamma_{{}_P}^2\cdot
   \vec{v}_{{}_P}\cdot\vec{v}_{{}_P}^\dagger\rbrack
\cdot\lbrack (\vec{\Delta}_3  +
            \vec{\Delta}_7) \wtimes\vec{B}_{{}_P}\rbrack
\nonumber\\ &&\qquad
+ e\cdot\lbrack \vec{E}_{{}_P}^\dagger\cdot\vec{v}_{{}_P}
+\gamma_{{}_P}^2\cdot\vec{v}_{{}_P}\cdot\vec{E}_{{}_P}^\dagger
   \cdot\vec{v}_{{}_P}\cdot\vec{v}_{{}_P}^\dagger
+\gamma_{{}_P}^2\cdot\vec{v}_{{}_P}\cdot\vec{E}_{{}_P}^\dagger
                                                         \rbrack \cdot
    \lbrack \vec{\Delta}_3  + \vec{\Delta}_7 \rbrack\; ,
 \label{B.23}
\end{eqnarray}
one then gets
\begin{eqnarray}
 && m\cdot (\gamma_{{}_P}\cdot\vec{v}_{{}_P})' =
    e\cdot(\vec{v}_{{}_P}\wtimes\vec{B}_{{}_P})
 + e\cdot \vec{E}_{{}_P}  +
  \vec{\Delta}_{13} \; .
 \label{B.24}
\end{eqnarray}
I now have to show that (B.24) is identical with
(2.10). Therefore the remaining task of this Appendix is to simplify
$\vec{\Delta}_{13}$.
The first two terms on the rhs of (B.24) constitute the Lorentz
force whereas the remaining part constitutes the SG force. To
disentangle $\vec{\Delta}_{13}$ it is important to notice that the SG
force occurs in two different forms. The `first order part' contains the
field vectors $\vec{E}_{{}_P},\vec{B}_{{}_P}$ only linearly;
more specifically it is linear in the first derivatives of the
field vectors. The `second order part' contains the  field
vectors $\vec{E}_{{}_P},\vec{B}_{{}_P}$ quadratically. Note that in the
second order part no derivatives of the field vectors occur.
In equation (B.22) I have  indicated which of the two forms of
the SG force occurs in a term.
\par Accordingly one can split the SG force $\vec{\Delta}_{13}$ into a
first order part plus a second order part. First I abbreviate by using (2.8):
\footnote{The partial derivative $\partial/\partial t$ in
(B.25), (B.30) and (B.32)  acts
on a function depending on $\vec{r}_{{}_P},t$.}
\begin{eqnarray}
&&\vec{\Delta}_{14}\equiv
-\vec{\nabla}_{{}_P}   (\vec{\sigma}^{\,\dagger}\cdot
                                                 \vec{\Omega}_{{}_P})
=-\vec{\nabla}_{{}_P}
  (\vec{\Omega}_{{}_P}^\dagger\cdot
\lbrack \frac{1}{\gamma_{{}_P}}\cdot
   \vec{s} + \frac{\gamma_{{}_P}}{\gamma_{{}_P}+1}
    \cdot \vec{s}^{\,\dagger}\cdot\vec{v}_{{}_P}\cdot
                         \vec{v}_{{}_P} \rbrack )
\nonumber\\
&=&-\frac{1}{\gamma_{{}_P}}\cdot
\vec{\nabla}_{{}_P}(\vec{\Omega}_{{}_P}^\dagger\cdot\vec{s})
       +  \frac{e\cdot g}{2\cdot m}\cdot\frac{1}{\gamma_{{}_P}+1}\cdot
\vec{s}^{\,\dagger}\cdot\vec{v}_{{}_P} \cdot
\vec{\nabla}_{{}_P}(\vec{B}_{{}_P}^\dagger\cdot\vec{v}_{{}_P})
\nonumber\\
&=& \frac{e}{m}\cdot
   \frac{1}{\gamma_{{}_P}}\cdot
             \vec{\nabla}_{{}_P}\biggl(\vec{s}^{\,\dagger}\cdot
 \biggl\lbrack
 \lbrack \frac{1}{\gamma_{{}_P}}+
                            \frac{g-2}{2} \rbrack \cdot\vec{B}_{{}_P}
 +\frac{\gamma_{{}_P}}{\gamma_{{}_P}+1}\cdot
  \vec{v}_{{}_P}^\dagger\cdot\vec{B}_{{}_P}\cdot\vec{v}_{{}_P}
-\lbrack \frac{g}{2} -\frac{\gamma_{{}_P}}{\gamma_{{}_P}+1}\rbrack\cdot
 (\vec{v}_{{}_P}\wtimes\vec{E}_{{}_P}) \biggr\rbrack\biggr)
\nonumber\\
&=& \frac{e}{m}\cdot
    \frac{1}{\gamma_{{}_P}}\cdot
          \vec{\nabla}_{{}_P}(\vec{s}^{\,\dagger}\cdot
 \biggl(
 \lbrack \frac{1}{\gamma_{{}_P}}+
                            \frac{g-2}{2} \rbrack \cdot\vec{B}_{{}_P}
-\lbrack \frac{g}{2} -\frac{\gamma_{{}_P}}{\gamma_{{}_P}+1}\rbrack\cdot
 (\vec{v}_{{}_P}\wtimes\vec{E}_{{}_P}) \biggr))
\nonumber\\ &&\qquad
+\frac{e}{m}\cdot\frac{1}{\gamma_{{}_P}+1}\cdot
                 \vec{v}_{{}_P}^\dagger\cdot\vec{s}
\cdot\biggl(  (\vec{v}_{{}_P}^\dagger\cdot
                                   \vec{\nabla}_{{}_P})\vec{B}_{{}_P}
+ \vec{v}_{{}_P}\wtimes \frac{\partial
   \vec{E}_{{}_P}}{\partial t} \biggr)
                                     \; ,
\nonumber\\
&&\vec{\Delta}_{15} \equiv
    \frac{e}{K_{{}_P}^2} \cdot\vec{v}_{{}_P}\cdot
 \lbrack  \frac{1}{\gamma_{{}_P}}\cdot
   \vec{s} + \frac{\gamma_{{}_P}}{\gamma_{{}_P}+1}
 \cdot \vec{s}^{\,\dagger}\cdot\vec{v}_{{}_P}\cdot
                                        \vec{v}_{{}_P} \rbrack^\dagger
               \cdot \vec{B}_{{}_P}'
+\frac{e}{(K_{{}_P}+m)^2}\cdot\vec{s}^{\,\dagger}\cdot
    (\vec{v}_{{}_P}\wtimes\vec{E}_{{}_P}')\cdot\vec{v}_{{}_P}
\nonumber\\ &&\qquad
- \frac{e}{m\cdot(K_{{}_P}+m)}\cdot \vec{E}_{{}_P}'\wtimes
 \lbrack  \frac{1}{\gamma_{{}_P}}\cdot
   \vec{s} + \frac{\gamma_{{}_P}}{\gamma_{{}_P}+1}
 \cdot \vec{s}^{\,\dagger}\cdot\vec{v}_{{}_P}\cdot\vec{v}_{{}_P} \rbrack
\nonumber\\ &&\qquad
+ \frac{e\cdot g}{2\cdot m\cdot K_{{}_P}}\cdot  \vec{E}_{{}_P}'\wtimes
 \lbrack  \frac{1}{\gamma_{{}_P}}\cdot
    \vec{s} + \frac{\gamma_{{}_P}}{\gamma_{{}_P}+1}
  \cdot \vec{s}^{\,\dagger}\cdot
                            \vec{v}_{{}_P}\cdot\vec{v}_{{}_P} \rbrack
-\frac{e\cdot g}{2\cdot K_{{}_P}^2}\cdot\vec{s}^{\,\dagger}\cdot
    (\vec{v}_{{}_P}\wtimes\vec{E}_{{}_P}')\cdot\vec{v}_{{}_P}
\nonumber\\ &&\qquad
+\frac{g-2}{2}\cdot \biggl(
 \frac{e}{m}\cdot\frac{1}{K_{{}_P}+m}\cdot
  \vec{s}^{\,\dagger}\cdot\vec{v}_{{}_P}\cdot\vec{B}_{{}_P}'
-\frac{e}{m^2}\cdot\frac{1}{\gamma_{{}_P} + 1}\cdot
  \vec{s}^{\,\dagger}\cdot\vec{v}_{{}_P}\cdot
  \vec{v}_{{}_P}^\dagger\cdot\vec{B}_{{}_P}'\cdot\vec{v}_{{}_P}
 \nonumber\\ &&\qquad
+\frac{e}{m^2}\cdot\frac{1}{\gamma_{{}_P}\cdot(\gamma_{{}_P}+1)}\cdot
 \vec{v}_{{}_P}^\dagger\cdot\vec{B}_{{}_P}'\cdot \vec{s} \biggr) \; ,
\nonumber\\
&&\vec{\Delta}_{16}\equiv
    -\frac{e}{m^2}\cdot \vec{v}_{{}_P}^\dagger\cdot\vec{E}_{{}_P}'
  \cdot \frac{1}{(\gamma_{{}_P}+1)^2} \cdot
                                       (\vec{s}\wtimes\vec{v}_{{}_P})
- \frac{e\cdot g}{2\cdot m^2}\cdot
    \frac{1}{\gamma_{{}_P}\cdot(\gamma_{{}_P}+1)}
\cdot \vec{v}_{{}_P}^\dagger\cdot\vec{B}_{{}_P}'\cdot \vec{s}
\nonumber\\ &&
- \frac{e\cdot(g-2)}{2\cdot m^2}\cdot
    \frac{\gamma_{{}_P}}{\gamma_{{}_P}+1}
\cdot \vec{v}_{{}_P}^\dagger\cdot\vec{B}_{{}_P}'\cdot
       \vec{s}^{\,\dagger}\cdot\vec{v}_{{}_P}\cdot\vec{v}_{{}_P}
   +\frac{e\cdot(g-2)}{2\cdot m^2}\cdot
\frac{\gamma_{{}_P}}{\gamma_{{}_P}+1}\cdot
    \vec{s}^{\,\dagger}\cdot\vec{v}_{{}_P}\cdot\vec{B}_{{}_P}'
\nonumber\\ &&
+  \frac{e}{m^2}\cdot \frac{1}{\gamma_{{}_P}\cdot(\gamma_{{}_P}+1)}\cdot
             \vec{s}^{\,\dagger}\cdot\vec{B}_{{}_P}'\cdot\vec{v}_{{}_P}
-\frac{e\cdot(g-2)}{2\cdot m^2}\cdot
   \frac{\gamma_{{}_P}}{\gamma_{{}_P}+1}\cdot
             \vec{s}^{\,\dagger}\cdot\vec{v}_{{}_P}\cdot
  (\vec{v}_{{}_P}\wtimes \vec{E}_{{}_P}')
\nonumber\\ &&
    +  \frac{e}{m^2}\cdot
   \frac{1}{\gamma_{{}_P}\cdot(\gamma_{{}_P}+1)}\cdot
            (\vec{s}\wtimes \vec{E}_{{}_P}')  \; ,
\nonumber\\
&&\vec{\Delta}_{17}\equiv  \vec{\Delta}_4 - \vec{\Delta}_{14} \; ,
\nonumber\\
&&\vec{\Delta}_{18}\equiv  \vec{\Delta}_5 - \vec{\Delta}_{15} \; ,
\nonumber\\
&&\vec{\Delta}_{19}\equiv  \vec{\Delta}_8 - \vec{\Delta}_{16} \; ,
 \label{B.25}
\end{eqnarray}
from which follows
\begin{eqnarray}
&&\vec{\Delta}_4 = \vec{\Delta}_{14} +  \vec{\Delta}_{17} \; ,
\nonumber\\
&&\vec{\Delta}_5 = \vec{\Delta}_{15} +  \vec{\Delta}_{18} \; ,
\nonumber\\
&&\vec{\Delta}_8 = \vec{\Delta}_{16} +  \vec{\Delta}_{19} \; .
 \label{B.26}
\end{eqnarray}
Note that $\vec{\Delta}_{14},\vec{\Delta}_{15},\vec{\Delta}_{16}$ are
linear in the electromagnetic field vectors, whereas
$\vec{\Delta}_{17},\vec{\Delta}_{18},$\\
$\vec{\Delta}_{19}$ are quadratic.
With (B.25-26) I can now abbreviate
\setcounter{INDEX}{1}
\begin{eqnarray}
&&\vec{\Delta}_{20} \equiv   e\cdot
 (\vec{v}_{{}_P} \wtimes\vec{\Delta}_{12})
 + e\cdot\vec{\Delta}_{11}
\nonumber\\ &&\qquad
 +                    \vec{\Delta}_{14}
 + K_{{}_P}\cdot\lbrack \vec{\Delta}_{15} +  \vec{\Delta}_{16} \rbrack
+K_{{}_P}\cdot\gamma_{{}_P}^2\cdot\vec{v}_{{}_P}^\dagger\cdot
                                         \lbrack \vec{\Delta}_{15} +
   \vec{\Delta}_{16} \rbrack \cdot \vec{v}_{{}_P}  \; ,
 \label{B.27a}
\end{eqnarray}
\addtocounter{equation}{-1}
\addtocounter{INDEX}{1}
which denotes the first order part of the SG force and
\begin{eqnarray}
&&\vec{\Delta}_{21} \equiv
                      \vec{\Delta}_{17}
 + K_{{}_P}\cdot\lbrack \vec{\Delta}_{18} +  \vec{\Delta}_{19} \rbrack
+K_{{}_P}\cdot\gamma_{{}_P}^2\cdot\vec{v}_{{}_P}^\dagger\cdot
                                         \lbrack \vec{\Delta}_{18} +
   \vec{\Delta}_{19} \rbrack \cdot \vec{v}_{{}_P}
\nonumber\\ &&\qquad
    + \frac{e}{K_{{}_P}}\cdot \lbrack \Delta_6 - \Delta_9 \rbrack \cdot
   \lbrack \vec{v}_{{}_P}\wtimes\vec{B}_{{}_P} + \vec{E}_{{}_P} \rbrack
- e\cdot \lbrack  1 + \gamma_{{}_P}^2\cdot
 \vec{v}_{{}_P}\cdot\vec{v}_{{}_P}^\dagger\rbrack
\cdot\lbrack (\vec{\Delta}_3  +
        \vec{\Delta}_7) \wtimes\vec{B}_{{}_P}\rbrack
\nonumber\\ &&\qquad
+ e\cdot\lbrack \vec{E}_{{}_P}^\dagger\cdot\vec{v}_{{}_P}
+\gamma_{{}_P}^2\cdot\vec{v}_{{}_P}\cdot
  \vec{E}_{{}_P}^\dagger\cdot\vec{v}_{{}_P}\cdot\vec{v}_{{}_P}^\dagger
+\gamma_{{}_P}^2\cdot\vec{v}_{{}_P}\cdot\vec{E}_{{}_P}^\dagger
                                                         \rbrack \cdot
    \lbrack \vec{\Delta}_3  + \vec{\Delta}_7 \rbrack\; ,
 \label{B.27b}
\end{eqnarray}
\setcounter{INDEX}{0}
which denotes the second order part of the SG force.
Then
\begin{eqnarray}
&&\vec{\Delta}_{13}  =
  \vec{\Delta}_{20}
+ \vec{\Delta}_{21} \; .
 \label{B.28}
\end{eqnarray}
Thus (B.24) reads as:
\begin{eqnarray}
 && m\cdot (\gamma_{{}_P}\cdot\vec{v}_{{}_P})' =
    e\cdot(\vec{v}_{{}_P}\wtimes\vec{B}_{{}_P}) + e\cdot \vec{E}_{{}_P}
 +  \; \underbrace{ \vec{\Delta}_{20} }_{{\rm first\;order\;SG\;terms}}
\; + \; \underbrace{ \vec{\Delta}_{21} }_{{\rm second\;order\;SG\;terms}}
 \;  . \qquad
 \label{B.29}
\end{eqnarray}
\subsection*{B.6}
In this subsection I simplify the first order part
$\vec{\Delta}_{20}$ of the SG force. First of all I calculate by using
(2.8),(B.20):
%\footnote{The partial derivative $\partial/\partial t$ in (B.30)
%acts on functions depending on $\vec{r}_{{}_P},t$.}
%
\begin{eqnarray}
&& \vec{\nabla}_{{}_P}\wtimes (\vec{v}_{{}_P} \wtimes \vec{B}_{{}_P}) =
- (\vec{v}_{{}_P}^\dagger\cdot \vec{\nabla}_{{}_P}) \vec{B}_{{}_P} \; ,
\nonumber\\
&& \vec{\nabla}_{{}_P}(\vec{v}_{{}_P}^\dagger\cdot \vec{B}_{{}_P})=
 (\vec{v}_{{}_P}^\dagger\cdot \vec{\nabla}_{{}_P})  \vec{B}_{{}_P} +
   \vec{v}_{{}_P} \wtimes (\vec{\nabla}_{{}_P}\wtimes \vec{B}_{{}_P})
=(\vec{v}_{{}_P}^\dagger\cdot \vec{\nabla}_{{}_P})  \vec{B}_{{}_P} +
 \vec{v}_{{}_P} \wtimes \frac{\partial \vec{E}_{{}_P}}{\partial t} \; ,
\nonumber\\
&&   e\cdot(\vec{v}_{{}_P} \wtimes\vec{\Delta}_{12}) =
e\cdot  \vec{v}_{{}_P} \wtimes\lbrack
       (\vec{\Delta}_{10}^\dagger\cdot\vec{\nabla}_{{}_P})
  \vec{B}_{{}_P}\rbrack
= e\cdot(\vec{\Delta}_{10}^\dagger\cdot
  \vec{\nabla}_{{}_P})(\vec{v}_{{}_P}\wtimes\vec{B}_{{}_P})
\nonumber\\
&=& e\cdot\vec{\nabla}_{{}_P}\lbrack
\vec{\Delta}_{10}^\dagger\cdot(\vec{v}_{{}_P}\wtimes\vec{B}_{{}_P})\rbrack
- e\cdot \vec{\Delta}_{10} \wtimes \lbrack \vec{\nabla}_{{}_P} \wtimes
  (\vec{v}_{{}_P} \wtimes \vec{B}_{{}_P})\rbrack
\nonumber\\
&=&  - \frac{e}{m}\cdot\frac{1}{\gamma_{{}_P}+1}\cdot
                           \vec{\nabla}_{{}_P}
((\vec{s}\wtimes\vec{v}_{{}_P})^\dagger\cdot
            (\vec{v}_{{}_P}\wtimes\vec{B}_{{}_P}))
+e\cdot\vec{\Delta}_{10}^\dagger \wtimes\lbrack
  (\vec{v}_{{}_P}^\dagger\cdot \vec{\nabla}_{{}_P}) \vec{B}_{{}_P}\rbrack
\nonumber\\
&=&  -        \frac{e}{m}\cdot\frac{1}{\gamma_{{}_P}+1}
\cdot\vec{\nabla}_{{}_P}(\vec{s}^{\,\dagger}\cdot
 \vec{v}_{{}_P}\cdot\vec{v}_{{}_P}^\dagger\cdot\vec{B}_{{}_P})
     +        \frac{e}{m}\cdot\frac{1}{\gamma_{{}_P}+1}
\cdot \vec{\nabla}_{{}_P}
     (\vec{v}_{{}_P}^\dagger\cdot\vec{v}_{{}_P}\cdot
                    \vec{s}^{\,\dagger}\cdot\vec{B}_{{}_P})
\nonumber\\&&\qquad
+e\cdot(\vec{v}_{{}_P}^\dagger\cdot \vec{\nabla}_{{}_P})
      (\vec{\Delta}_{10}  \wtimes  \vec{B}_{{}_P})
\nonumber\\
&=& \frac{e}{m}\cdot\frac{1}{\gamma_{{}_P}+1}\biggl(
 - \vec{\nabla}_{{}_P}( \vec{s}^{\,\dagger}\cdot
 \vec{v}_{{}_P}\cdot\vec{v}_{{}_P}^\dagger\cdot\vec{B}_{{}_P})
+ \vec{\nabla}_{{}_P}(\vec{v}_{{}_P}^\dagger\cdot
  \vec{v}_{{}_P}\cdot\vec{s}^{\,\dagger}\cdot\vec{B}_{{}_P})
 + (\vec{v}_{{}_P}^\dagger\cdot \vec{\nabla}_{{}_P})
              (\vec{v}_{{}_P}^\dagger\cdot\vec{B}_{{}_P}\cdot\vec{s})
\nonumber\\&&\qquad
 - (\vec{v}_{{}_P}^\dagger\cdot \vec{\nabla}_{{}_P})
(\vec{s}^{\,\dagger}\cdot\vec{B}_{{}_P}\cdot\vec{v}_{{}_P}) \biggr)
\nonumber\\
&=& \frac{e}{m}\cdot\frac{1}{\gamma_{{}_P}+1}\biggl(
- \vec{s}^{\,\dagger}\cdot\vec{v}_{{}_P}\cdot
  (\vec{v}_{{}_P}^\dagger\cdot\vec{\nabla}_{{}_P}) \vec{B}_{{}_P}
- \vec{s}^{\,\dagger}\cdot\vec{v}_{{}_P}\cdot\vec{v}_{{}_P}\wtimes
  \frac{\partial \vec{E}_{{}_P}}{\partial t}
+ \vec{v}_{{}_P}^\dagger\cdot\vec{v}_{{}_P}\cdot
          \vec{\nabla}_{{}_P}(\vec{s}^{\,\dagger}\cdot\vec{B}_{{}_P})
\nonumber\\ &&\qquad
 + (\vec{v}_{{}_P}^\dagger\cdot \vec{\nabla}_{{}_P})
              (\vec{v}_{{}_P}^\dagger\cdot\vec{B}_{{}_P}\cdot\vec{s})
 - (\vec{v}_{{}_P}^\dagger\cdot \vec{\nabla}_{{}_P})
 (\vec{s}^{\,\dagger}\cdot\vec{B}_{{}_P}\cdot\vec{v}_{{}_P})\biggr) \; ,
\nonumber\\
&& e\cdot \vec{\Delta}_{11} =
     -        \frac{e}{m}\cdot\frac{1}{\gamma_{{}_P}+1}
\cdot\lbrack
 (\vec{s}\wtimes\vec{v}_{{}_P})^\dagger\cdot \vec{\nabla}_{{}_P}\rbrack
                                                          \vec{E}_{{}_P}
   = \frac{e}{m}\cdot\frac{1}{\gamma_{{}_P}+1}\cdot\biggl(
  -  \vec{\nabla}_{{}_P}(\vec{E}_{{}_P}^\dagger\cdot
                           (\vec{s}\wtimes\vec{v}_{{}_P}))
\nonumber\\ &&\qquad
  +  (\vec{s}\wtimes\vec{v}_{{}_P}) \wtimes
   (\vec{\nabla}_{{}_P} \wtimes \vec{E}_{{}_P} )
\biggr)
\nonumber\\
&=& \frac{e}{m}\cdot\frac{1}{\gamma_{{}_P}+1}\biggl(
  -  \vec{\nabla}_{{}_P}\lbrack\vec{E}_{{}_P}^\dagger\cdot
                            (\vec{s}\wtimes\vec{v}_{{}_P})\rbrack
-(\vec{s}\wtimes\vec{v}_{{}_P})\wtimes
    \frac{\partial\vec{B}_{{}_P}}{\partial t}
\biggr)
\nonumber\\
&=& \frac{e}{m}\cdot\frac{1}{\gamma_{{}_P}+1}\biggl(
  -  \vec{\nabla}_{{}_P}\lbrack\vec{E}_{{}_P}^\dagger\cdot
                                  (\vec{s}\wtimes\vec{v}_{{}_P})\rbrack
+\vec{v}_{{}_P}^\dagger\cdot
  \frac{\partial\vec{B}_{{}_P}}{\partial t}\cdot \vec{s}
-\vec{s}^{\,\dagger}\cdot
  \frac{\partial\vec{B}_{{}_P}}{\partial t}\cdot\vec{v}_{{}_P}
\biggr) \; .
 \label{B.30}
\end{eqnarray}
Here I used the fact that the spatial derivatives of
$\vec{E}_{{}_P}$ and $\vec{B}_{{}_P}$ only   appears in the SG terms,
i.e. in leading order spin. Therefore one can always approximate:
\begin{eqnarray*}
&&\frac{\partial  E_{{}_{P,k}}}{\partial  r_{{}_{P,j}}} =
\frac{\partial  E_{{}_{M,k}}}{\partial  r_{{}_{M,j}}} \; ,\qquad
\frac{\partial  B_{{}_{P,k}}}{\partial  r_{{}_{P,j}}} =
\frac{\partial  B_{{}_{M,k}}}{\partial  r_{{}_{M,j}}} \; . \qquad (j,k=1,2,3)
\end{eqnarray*}
Secondly I conclude from (B.25)
\begin{eqnarray}
&&\vec{\Delta}_{15} + \vec{\Delta}_{16} =
    \frac{e}{m^2} \cdot
\frac{\gamma^2_{{}_P}+\gamma_{{}_P}+1}{\gamma_{{}_P}^3\cdot
                                            (\gamma_{{}_P} + 1)}
         \cdot\vec{s}^{\,\dagger}\cdot\vec{B}_{{}_P}'\cdot\vec{v}_{{}_P}
 \nonumber\\ &&\qquad
+\frac{e}{m^2}\cdot \lbrack \frac{1}{\gamma_{{}_P}\cdot(\gamma_{{}_P}
                   +1)}-\frac{g-2}{2} \rbrack
\cdot\vec{s}^{\,\dagger}\cdot\vec{v}_{{}_P}\cdot
        \vec{v}_{{}_P}^\dagger\cdot\vec{B}_{{}_P}'
 \cdot\vec{v}_{{}_P}
 \nonumber\\ &&\qquad
+\frac{e}{m^2}\cdot \lbrack \frac{1}{(\gamma_{{}_P}+1)^2}
     -\frac{g}{2\cdot\gamma_{{}_P}^2}\rbrack
\cdot\vec{s}^{\,\dagger}\cdot (\vec{v}_{{}_P}\wtimes
   \vec{E}_{{}_P}')\cdot\vec{v}_{{}_P}
 \nonumber\\ &&\qquad
+ \frac{e}{m^2}\cdot \lbrack -
   \frac{2}{\gamma_{{}_P}\cdot(\gamma_{{}_P}+1)}
     +\frac{g}{2\cdot\gamma_{{}_P}^2}\rbrack
              \cdot (\vec{E}_{{}_P}'\wtimes \vec{s})
 \nonumber\\ &&\qquad
+\frac{e}{m^2}\cdot \lbrack
-\frac{\gamma_{{}_P}^2+2\cdot
   \gamma_{{}_P}}{(\gamma_{{}_P} +1)^2}+ \frac{g}{2}\rbrack
  \cdot \vec{s}^{\,\dagger}\cdot\vec{v}_{{}_P}\cdot
                    (\vec{E}_{{}_P}'\wtimes \vec{v}_{{}_P})
+\frac{e\cdot(g-2)}{2\cdot m^2}\cdot\vec{s}^{\,\dagger}\cdot
                                                  \vec{v}_{{}_P}\cdot
   \vec{B}_{{}_P}'
 \cdot\vec{v}_{{}_P}
\nonumber\\ &&\qquad
    -\frac{e}{m^2}\cdot \vec{v}_{{}_P}^\dagger\cdot\vec{E}_{{}_P}'
\cdot \frac{1}{(\gamma_{{}_P}+1)^2} \cdot
                                     (\vec{s}\wtimes\vec{v}_{{}_P})
-  \frac{e}{m^2}\cdot \frac{1}{\gamma_{{}_P}\cdot(\gamma_{{}_P}+1)}\cdot
             \vec{v}_{{}_P}^\dagger\cdot\vec{B}_{{}_P}'\cdot\vec{s} \; ,
\nonumber\\
&&  K_{{}_P}\cdot\lbrack \vec{\Delta}_{15} +  \vec{\Delta}_{16} \rbrack
+K_{{}_P}\cdot\gamma_{{}_P}^2\cdot\vec{v}_{{}_P}^\dagger\cdot
                                         \lbrack \vec{\Delta}_{15} +
   \vec{\Delta}_{16} \rbrack \cdot \vec{v}_{{}_P}
=   \frac{e}{m} \cdot
\frac{\gamma^2_{{}_P}+\gamma_{{}_P}+1}{\gamma_{{}_P} + 1}
         \cdot\vec{s}^{\,\dagger}\cdot\vec{B}_{{}_P}'\cdot\vec{v}_{{}_P}
 \nonumber\\ &&\qquad
-\frac{e}{m}\cdot
  \frac{\gamma_{{}_P}^3+2\cdot\gamma_{{}_P}^2}{(\gamma+1)^2}
\cdot\vec{s}^{\,\dagger}\cdot (\vec{v}_{{}_P}\wtimes
   \vec{E}_{{}_P}')\cdot\vec{v}_{{}_P}
 \nonumber\\ &&\qquad
+ \frac{e}{m}\cdot \lbrack -\frac{2}{\gamma_{{}_P}+1}
     +\frac{g}{2\cdot\gamma_{{}_P}}\rbrack
              \cdot (\vec{E}_{{}_P}'\wtimes \vec{s})
 \nonumber\\ &&\qquad
+\frac{e}{m}\cdot \lbrack
-\frac{\gamma_{{}_P}^3+2\cdot\gamma_{{}_P}^2}{(\gamma_{{}_P} +1)^2}
 + \frac{g}{2}\cdot\gamma_{{}_P} \rbrack
  \cdot \vec{s}^{\,\dagger}\cdot\vec{v}_{{}_P}\cdot
                    (\vec{E}_{{}_P}'\wtimes \vec{v}_{{}_P})
+\frac{e\cdot(g-2)}{2\cdot m}\cdot\gamma_{{}_P} \cdot
          \vec{s}^{\,\dagger}\cdot\vec{v}_{{}_P}\cdot\vec{B}_{{}_P}'
\nonumber\\ &&\qquad
    -\frac{e}{m}\cdot \vec{v}_{{}_P}^\dagger\cdot\vec{E}_{{}_P}'
  \cdot \frac{\gamma_{{}_P}}{(\gamma_{{}_P}+1)^2} \cdot
 (\vec{s}\wtimes\vec{v}_{{}_P})
-  \frac{e}{m}\cdot \frac{1}{\gamma_{{}_P}+1}\cdot
             \vec{v}_{{}_P}^\dagger\cdot\vec{B}_{{}_P}'\cdot\vec{s} \; .
 \label{B.31}
\end{eqnarray}
Combining (B.25),(B.27),(B.30-31) one gets
%\footnote{The partial derivative $\partial/\partial t$ in (B.32) acts
%on functions depending on $\vec{r}_{{}_P},t$.}
%
\begin{eqnarray}
&&\vec{\Delta}_{20}  =
    \frac{e}{m}\cdot\frac{1}{\gamma_{{}_P}+1}\biggl(
- \vec{s}^{\,\dagger}\cdot\vec{v}_{{}_P}\cdot(\vec{v}_{{}_P}^\dagger\cdot
          \vec{\nabla}_{{}_P}) \vec{B}_{{}_P}
- \vec{s}^{\,\dagger}\cdot\vec{v}_{{}_P}\cdot (\vec{v}_{{}_P}\wtimes
  \frac{\partial \vec{E}_{{}_P}}{\partial t})
+ \vec{v}_{{}_P}^\dagger\cdot\vec{v}_{{}_P}\cdot
         \vec{\nabla}_{{}_P}(\vec{s}^{\,\dagger}\cdot\vec{B}_{{}_P})
\nonumber\\  &&\qquad
 + (\vec{v}_{{}_P}^\dagger\cdot \vec{\nabla}_{{}_P})
              (\vec{v}_{{}_P}^\dagger\cdot\vec{B}_{{}_P}\cdot\vec{s})
 - \vec{v}_{{}_P}^\dagger\cdot \vec{\nabla}_{{}_P}
              (\vec{s}^{\,\dagger}\cdot\vec{B}_{{}_P}\cdot
                                        \vec{v}_{{}_P})\biggr)
\nonumber\\  &&\qquad
 +  \frac{e}{m}\cdot\frac{1}{\gamma_{{}_P}+1}\biggl(
  -  \vec{\nabla}_{{}_P}\lbrack\vec{E}_{{}_P}^\dagger\cdot
                             (\vec{s}\wtimes\vec{v}_{{}_P})\rbrack
+\vec{v}_{{}_P}^\dagger\cdot
  \frac{\partial\vec{B}_{{}_P}}{\partial t}\cdot \vec{s}
-\vec{s}^{\,\dagger}\cdot
  \frac{\partial\vec{B}_{{}_P}}{\partial t}\cdot\vec{v}_{{}_P}
\biggr)
\nonumber\\  &&\qquad
+\frac{e}{m}\cdot\frac{1}{\gamma_{{}_P}}\cdot
       \vec{\nabla}_{{}_P}(\vec{s}^{\,\dagger}\cdot
 \biggl(
 \lbrack \frac{1}{\gamma_{{}_P}}+
                            \frac{g-2}{2} \rbrack \cdot\vec{B}_{{}_P}
-\lbrack \frac{g}{2} -\frac{\gamma_{{}_P}}{\gamma_{{}_P}+1}\rbrack\cdot
 (\vec{v}_{{}_P}\wtimes\vec{E}_{{}_P}) \biggr))
\nonumber\\  &&\qquad
+\frac{e}{m}\cdot\frac{1}{\gamma_{{}_P}+1}\cdot
             \vec{v}_{{}_P}^\dagger\cdot\vec{s}
\cdot\biggl(  (\vec{v}_{{}_P}^\dagger\cdot
                                   \vec{\nabla}_{{}_P})\vec{B}_{{}_P}
+ \vec{v}_{{}_P}\wtimes \frac{\partial
  \vec{E}_{{}_P}}{\partial t} \biggr)
\nonumber\\  &&\qquad
 +  \frac{e}{m} \cdot
\frac{\gamma^2_{{}_P}+\gamma_{{}_P}+1}{\gamma_{{}_P} + 1}
         \cdot\vec{s}^{\,\dagger}\cdot\vec{B}_{{}_P}'\cdot\vec{v}_{{}_P}
 \nonumber\\  &&\qquad
-\frac{e}{m}\cdot \frac{\gamma_{{}_P}^3+2\cdot
                                         \gamma_{{}_P}^2}{(\gamma+1)^2}
\cdot\vec{s}^{\,\dagger}\cdot (\vec{v}_{{}_P}\wtimes
  \vec{E}_{{}_P}')\cdot\vec{v}_{{}_P}
 \nonumber\\  &&\qquad
+ \frac{e}{m}\cdot \lbrack -\frac{2}{\gamma_{{}_P}+1}
     +\frac{g}{2\cdot\gamma_{{}_P}}\rbrack
              \cdot (\vec{E}_{{}_P}'\wtimes \vec{s})
 \nonumber\\  &&\qquad
+\frac{e}{m}\cdot \lbrack
-\frac{\gamma_{{}_P}^3+2\cdot\gamma_{{}_P}^2}{(\gamma_{{}_P} +1)^2}
 + \frac{g}{2}\cdot\gamma_{{}_P} \rbrack
  \cdot \vec{s}^{\,\dagger}\cdot\vec{v}_{{}_P}\cdot
                    (\vec{E}_{{}_P}'\wtimes \vec{v}_{{}_P})
+\frac{e\cdot(g-2)}{2\cdot m}\cdot\gamma_{{}_P} \cdot
       \vec{s}^{\,\dagger}\cdot\vec{v}_{{}_P}\cdot\vec{B}_{{}_P}'
\nonumber\\  &&\qquad
    -\frac{e}{m}\cdot \vec{v}_{{}_P}^\dagger\cdot\vec{E}_{{}_P}'
  \cdot \frac{\gamma_{{}_P}}{(\gamma_{{}_P}+1)^2} \cdot
  (\vec{s}\wtimes\vec{v}_{{}_P})
-  \frac{e}{m}\cdot \frac{1}{\gamma_{{}_P}+1}\cdot
             \vec{v}_{{}_P}^\dagger\cdot\vec{B}_{{}_P}'\cdot\vec{s}
\nonumber\\
&=& \frac{e\cdot g}{2\cdot m}\cdot\frac{1}{\gamma_{{}_P}}\cdot
\vec{\nabla}_{{}_P}(\vec{s}^{\,\dagger}\cdot\vec{B}_{{}_P})
+   \frac{e}{m}\cdot\frac{1}{\gamma_{{}_P}+1}\biggl(
   \vec{v}_{{}_P}^\dagger\cdot \vec{B}_{{}_P}'\cdot\vec{s}
-\vec{v}_{{}_P}^\dagger\cdot
  \frac{\partial\vec{B}_{{}_P}}{\partial t}\cdot \vec{s}
 - \vec{s}^{\,\dagger}\cdot \vec{B}_{{}_P}'\cdot\vec{v}_{{}_P}
\nonumber\\  &&\qquad
+\vec{s}^{\,\dagger}\cdot
   \frac{\partial\vec{B}_{{}_P}}{\partial t}\cdot \vec{v}_{{}_P}
                                         \biggr)
 +  \frac{e}{m}\cdot\frac{1}{\gamma_{{}_P}+1}\biggl(
  -  \vec{\nabla}_{{}_P}\lbrack\vec{E}_{{}_P}^\dagger\cdot
                                   (\vec{s}\wtimes\vec{v}_{{}_P})\rbrack
+\vec{v}_{{}_P}^\dagger\cdot
  \frac{\partial\vec{B}_{{}_P}}{\partial t}\cdot \vec{s}
-\vec{s}^{\,\dagger}\cdot
  \frac{\partial\vec{B}_{{}_P}}{\partial t}\cdot\vec{v}_{{}_P}
\biggr)
\nonumber\\  &&\qquad
+\frac{e}{m}\cdot\lbrack\frac{1}{\gamma_{{}_P}+1}-
  \frac{g}{2\cdot\gamma_{{}_P}}
\rbrack \cdot\vec{\nabla}_{{}_P}\lbrack\vec{s}^{\,\dagger}\cdot
  (\vec{v}_{{}_P}\wtimes\vec{E}_{{}_P})\rbrack
\nonumber\\  &&\qquad
 +  \frac{e}{m} \cdot
\frac{\gamma^2_{{}_P}+\gamma_{{}_P}+1}{\gamma_{{}_P} + 1}
         \cdot\vec{s}^{\,\dagger}\cdot\vec{B}_{{}_P}'\cdot\vec{v}_{{}_P}
 \nonumber\\  &&\qquad
-\frac{e}{m}\cdot \frac{\gamma_{{}_P}^3+2\cdot
                                         \gamma_{{}_P}^2}{(\gamma+1)^2}
\cdot\vec{s}^{\,\dagger}\cdot (\vec{v}_{{}_P}\wtimes
  \vec{E}_{{}_P}')\cdot\vec{v}_{{}_P}
 \nonumber\\  &&\qquad
+ \frac{e}{m}\cdot \lbrack -\frac{2}{\gamma_{{}_P}+1}
     +\frac{g}{2\cdot\gamma_{{}_P}}\rbrack
              \cdot (\vec{E}_{{}_P}'\wtimes \vec{s})
 \nonumber\\  &&\qquad
+\frac{e}{m}\cdot \lbrack
-\frac{\gamma_{{}_P}^3+2\cdot\gamma_{{}_P}^2}{(\gamma_{{}_P} +1)^2}
 + \frac{g}{2}\cdot\gamma_{{}_P} \rbrack
  \cdot \vec{s}^{\,\dagger}\cdot\vec{v}_{{}_P}\cdot
                    (\vec{E}_{{}_P}'\wtimes \vec{v}_{{}_P})
+\frac{e\cdot(g-2)}{2\cdot m}\cdot\gamma_{{}_P} \cdot
       \vec{s}^{\,\dagger}\cdot\vec{v}_{{}_P}\cdot\vec{B}_{{}_P}'
\nonumber\\  &&\qquad
    -\frac{e}{m}\cdot \vec{v}_{{}_P}^\dagger\cdot\vec{E}_{{}_P}'
  \cdot \frac{\gamma_{{}_P}}{(\gamma_{{}_P}+1)^2} \cdot
   (\vec{s}\wtimes\vec{v}_{{}_P})
-  \frac{e}{m}\cdot \frac{1}{\gamma_{{}_P}+1}\cdot
       \vec{v}_{{}_P}^\dagger\cdot\vec{B}_{{}_P}'\cdot\vec{s}
\nonumber\\
&=& \frac{e\cdot g}{2\cdot m}\cdot\frac{1}{\gamma_{{}_P}}\cdot
\vec{\nabla}_{{}_P}(\vec{s}^{\,\dagger}\cdot\vec{B}_{{}_P})
+\frac{e\cdot(g-2)}{2\cdot m}\cdot\gamma_{{}_P} \cdot
            \vec{s}^{\,\dagger}\cdot\vec{v}_{{}_P}\cdot\vec{B}_{{}_P}'
+\frac{e}{m}\cdot\gamma_{{}_P} \cdot \vec{s}^{\,\dagger}\cdot
 \vec{B}_{{}_P}'\cdot\vec{v}_{{}_P}
\nonumber\\  &&\qquad
-\frac{e\cdot g}{2\cdot m}\cdot\frac{1}{\gamma_{{}_P}}
  \cdot\vec{\nabla}_{{}_P}\lbrack\vec{s}^{\,\dagger}\cdot
                              (\vec{v}_{{}_P}\wtimes\vec{E}_{{}_P})\rbrack
-\frac{e}{m}\cdot \frac{\gamma_{{}_P}^3+
                                   2\cdot\gamma_{{}_P}^2}{(\gamma+1)^2}
\cdot\vec{s}^{\,\dagger}\cdot (\vec{v}_{{}_P}\wtimes
  \vec{E}_{{}_P}')\cdot\vec{v}_{{}_P}
 \nonumber\\  &&\qquad
+ \frac{e}{m}\cdot \lbrack -\frac{2}{\gamma_{{}_P}+1}
     +\frac{g}{2\cdot\gamma_{{}_P}}\rbrack
              \cdot (\vec{E}_{{}_P}'\wtimes \vec{s})
 \nonumber\\  &&\qquad
    -\frac{e}{m}\cdot \vec{v}_{{}_P}^\dagger\cdot\vec{E}_{{}_P}'
\cdot \frac{\gamma_{{}_P}}{(\gamma_{{}_P}+1)^2} \cdot
 (\vec{s}\wtimes\vec{v}_{{}_P})
\nonumber\\  &&\qquad
+\frac{e}{m}\cdot \lbrack
-\frac{\gamma_{{}_P}^3+2\cdot\gamma_{{}_P}^2}{(\gamma_{{}_P} +1)^2}
 + \frac{g}{2}\cdot\gamma_{{}_P} \rbrack
  \cdot \vec{s}^{\,\dagger}\cdot\vec{v}_{{}_P}\cdot
                    (\vec{E}_{{}_P}'\wtimes \vec{v}_{{}_P}) \; .
 \label{B.32}
\end{eqnarray}
This can be further simplified by calculating
\begin{eqnarray}
&& \vec{v}_{{}_P}^\dagger\cdot\vec{E}_{{}_P}'\cdot
                                        (\vec{s}\wtimes\vec{v}_{{}_P})
 - \vec{v}_{{}_P}^\dagger\cdot\vec{v}_{{}_P}\cdot
       (\vec{s}\wtimes\vec{E}_{{}_P}')
= \vec{s}\wtimes (\vec{v}_{{}_P} \wtimes
      (\vec{v}_{{}_P}\wtimes\vec{E}_{{}_P}'))
\nonumber\\  &&\qquad
 = \vec{v}_{{}_P}\cdot\vec{s}^{\,\dagger}\cdot
     (\vec{v}_{{}_P}\wtimes\vec{E}_{{}_P}')
 - \vec{s}^{\,\dagger}\cdot\vec{v}_{{}_P}\cdot
   (\vec{v}_{{}_P}\wtimes\vec{E}_{{}_P}')\; ,
 \label{B.33}
\end{eqnarray}
from which follows
\begin{eqnarray}
&&\vec{\Delta}_{20}  =
    \frac{e\cdot g}{2\cdot m}\cdot\frac{1}{\gamma_{{}_P}}\cdot
\vec{\nabla}_{{}_P}(\vec{s}^{\,\dagger}\cdot\vec{B}_{{}_P})
+\frac{e\cdot(g-2)}{2\cdot m}\cdot\gamma_{{}_P} \cdot
          \vec{s}^{\,\dagger}\cdot\vec{v}_{{}_P}\cdot\vec{B}_{{}_P}'
+\frac{e}{m}\cdot\gamma_{{}_P} \cdot
  \vec{s}^{\,\dagger}\cdot\vec{B}_{{}_P}'\cdot\vec{v}_{{}_P}
\nonumber\\  &&\qquad
-\frac{e\cdot g}{2\cdot m}\cdot\frac{1}{\gamma_{{}_P}}
          \cdot\vec{\nabla}_{{}_P}\lbrack\vec{s}^{\,\dagger}\cdot
  (\vec{v}_{{}_P}\wtimes\vec{E}_{{}_P})\rbrack
\nonumber\\  &&\qquad
- \frac{e\cdot g}{2\cdot m}\cdot
               \gamma_{{}_P} \cdot\vec{s}^{\,\dagger}\cdot
           (\vec{v}_{{}_P}\wtimes\vec{E}_{{}_P}')\cdot\vec{v}_{{}_P}
 \nonumber\\  &&\qquad
+\frac{e}{m}\cdot\lbrack-
   \frac{2}{\gamma_{{}_P}+1}+\frac{g}{2\cdot\gamma_{{}_P}}
 - \vec{v}_{{}_P}^\dagger\cdot\vec{v}_{{}_P}\cdot
 \frac{\gamma_{{}_P}^3+2\cdot\gamma_{{}_P}^2}{(\gamma_{{}_P} +1)^2}
+ \frac{g}{2}\cdot\gamma_{{}_P} \cdot
 \vec{v}_{{}_P}^\dagger\cdot\vec{v}_{{}_P} \rbrack \cdot
                    (\vec{E}_{{}_P}'\wtimes \vec{s})
 \nonumber\\  &&\qquad
    +\frac{e}{m}\cdot  \lbrack
 -  \frac{\gamma_{{}_P}}{(\gamma_{{}_P}+1)^2}
-\frac{\gamma_{{}_P}^3+2\cdot\gamma_{{}_P}^2}{(\gamma_{{}_P} +1)^2}
+ \frac{g}{2}\cdot\gamma_{{}_P} \rbrack
 \cdot   \vec{v}_{{}_P}^\dagger\cdot\vec{E}_{{}_P}' \cdot
                                          (\vec{s}\wtimes\vec{v}_{{}_P})
 \nonumber\\
&=&
\frac{e\cdot g}{2\cdot m\cdot\gamma_{{}_P}}\cdot \vec{\nabla}_{{}_P}
                                  \biggl(
\vec{s}^{\,\dagger}\cdot\vec{B}_{{}_P}
 -\vec{E}_{{}_P}^\dagger\cdot(\vec{s}\wtimes\vec{v}_{{}_P}) \biggr)
+\frac{e\cdot \gamma_{{}_P}}{2\cdot m}\cdot \biggl(
 g\cdot \lbrack
  \vec{s}^{\,\dagger}\cdot\vec{B}_{{}_P}'-
    (\vec{s}\wtimes\vec{v}_{{}_P})^\dagger\cdot
      \vec{E}_{{}_P}' \rbrack \cdot\vec{v}_{{}_P}
\nonumber\\&&
   +(g-2)\cdot \lbrack
\vec{E}_{{}_P}'+\vec{v}_{{}_P}\wtimes\vec{B}_{{}_P}'
   -\vec{v}_{{}_P}^\dagger\cdot\vec{E}_{{}_P}'\cdot\vec{v}_{{}_P}
\rbrack \wtimes\vec{s} \biggr)  \; .
 \label{B.34}
\end{eqnarray}
Inserting (B.34) into (B.29) one has thus obtained:
\begin{eqnarray}
 && m\cdot (\gamma_{{}_P}\cdot\vec{v}_{{}_P})' =
    e\cdot (\vec{v}_{{}_P}\wtimes\vec{B}_{{}_P}) + e\cdot\vec{E}_{{}_P} +
\frac{e\cdot g}{2\cdot m\cdot\gamma_{{}_P}}\cdot \vec{\nabla}_{{}_P}
                                                               \biggl(
\vec{s}^{\,\dagger}\cdot\vec{B}_{{}_P}-\vec{E}_{{}_P}^\dagger\cdot
(\vec{s}\wtimes\vec{v}_{{}_P}) \biggr)
\nonumber\\&&
+\frac{e\cdot \gamma_{{}_P}}{2\cdot m}\cdot \biggl(
 g\cdot\lbrack \vec{s}^{\,\dagger}\cdot\vec{B}_{{}_P}'-
     (\vec{s}\wtimes\vec{v}_{{}_P})^\dagger\cdot
      \vec{E}_{{}_P}' \rbrack  \cdot\vec{v}_{{}_P}
   +(g-2)\cdot \lbrack
\vec{E}_{{}_P}'+\vec{v}_{{}_P}\wtimes\vec{B}_{{}_P}'
  -\vec{v}_{{}_P}^\dagger\cdot\vec{E}_{{}_P}'\cdot\vec{v}_{{}_P}
\rbrack \wtimes\vec{s} \biggr)
 \nonumber\\&&\qquad
 +\vec{\Delta}_{21} \; .
\label{B.35}
\end{eqnarray}
With (B.35) I have simplified the Lorentz and the first order
SG terms and have derived the first order terms of  (2.10).
\footnote{One sees by (B.35) that by neglecting second order SG terms the
charge $e$ appears only up to first order. Thus one could have derived
the first order SG terms in an alternative way by making first order
perturbation theory w.r.t. the charge.
This approach is chosen in \cite{DS70}, so that from
this point of view the first 6 subsections of Appendix B are just a check
of \cite{DS70}.}
In the above derivation it was essential that the
electromagnetic
field obeys (2.8), i.e. is a solution of the
vacuum Maxwell equations.
The second order SG terms are simplified below.
\subsection*{B.7}
In the remaining subsections of this Appendix I complete the derivation of
(2.10) by disentangling $\vec{\Delta}_{21}$, i.e. I have to deal with
the second order SG terms.
\footnote{To the knowledge of the author this is the first treatment which
takes the second order SG force into account.}
\par First of all I simplify the rhs of (B.23) by collecting its terms
in a convenient way and to do this I calculate by using
(2.12),(B.5),(B.12):
\begin{eqnarray}
 &&  \biggl( K_{{}_P}\cdot\lbrack  \vec{\Delta}_3  +
                                                 \gamma_{{}_P}^2\cdot
    \vec{v}_{{}_P}^\dagger\cdot\vec{\Delta}_3\cdot
                                       \vec{v}_{{}_P} \rbrack \biggr)'
=  \frac{1}{m}\cdot
 \biggl( K_{{}_P}\cdot\lbrack  m\cdot\vec{\Delta}_3  +
                                          m\cdot \gamma_{{}_P}^2\cdot
    \vec{v}_{{}_P}^\dagger\cdot\vec{\Delta}_3\cdot
                                       \vec{v}_{{}_P} \rbrack \biggr)'
\nonumber\\&&\qquad
 = \frac{1}{m}\cdot
   K_{{}_P}'\cdot
\lbrack  m\cdot\vec{\Delta}_3  +   m\cdot \gamma_{{}_P}^2\cdot
    \vec{v}_{{}_P}^\dagger\cdot\vec{\Delta}_3\cdot
                                       \vec{v}_{{}_P} \rbrack
 + \frac{1}{m}\cdot K_{{}_P}\cdot \lbrack m\cdot\vec{\Delta}_5
\nonumber\\&&
+  m\cdot \gamma_{{}_P}^2\cdot
    \vec{v}_{{}_P}^\dagger\cdot\vec{\Delta}_5\cdot
                                       \vec{v}_{{}_P}
  +m\cdot \gamma_{{}_P}\cdot
   \vec{\Delta}_3^\dagger\cdot(\gamma_{{}_P}\cdot
    \vec{v}_{{}_P})'\cdot\vec{v}_{{}_P}
 + m\cdot \gamma_{{}_P}\cdot
    \vec{v}_{{}_P}^\dagger\cdot\vec{\Delta}_3\cdot
  (\gamma_{{}_P}\cdot
    \vec{v}_{{}_P})' \rbrack
 \nonumber\\
&=& e\cdot\vec{E}_{{}_P}^\dagger\cdot
                               \vec{v}_{{}_P}\cdot\vec{\Delta}_3  +
    e\cdot \gamma_{{}_P}^2\cdot
        \vec{E}_{{}_P}^\dagger\cdot\vec{v}_{{}_P}\cdot
          \vec{v}_{{}_P}^\dagger\cdot\vec{\Delta}_3\cdot\vec{v}_{{}_P}
 + K_{{}_P}\cdot\vec{\Delta}_5  +
K_{{}_P}\cdot\gamma_{{}_P}^2\cdot\vec{v}_{{}_P}^\dagger\cdot
                                       \vec{\Delta}_5\cdot\vec{v}_{{}_P}
\nonumber\\&&\qquad
+e\cdot \gamma_{{}_P}^2\cdot \vec{\Delta}_3^\dagger\cdot \lbrack
   \vec{v}_{{}_P}\wtimes\vec{B}_{{}_P} +
  \vec{E}_{{}_P} \rbrack\cdot \vec{v}_{{}_P}
+e\cdot \gamma_{{}_P}^2\cdot \vec{v}_{{}_P}^\dagger\cdot
    \vec{\Delta}_3\cdot \lbrack
   \vec{v}_{{}_P}\wtimes\vec{B}_{{}_P} + \vec{E}_{{}_P} \rbrack  \; ,
\nonumber\\
 &&  \biggl( K_{{}_P}\cdot\lbrack  \vec{\Delta}_7  +
                                                 \gamma_{{}_P}^2\cdot
    \vec{v}_{{}_P}^\dagger\cdot\vec{\Delta}_7\cdot
                                       \vec{v}_{{}_P} \rbrack \biggr)'
=  \frac{1}{m}\cdot
 \biggl( K_{{}_P}\cdot\lbrack  m\cdot\vec{\Delta}_7  +
                                          m\cdot \gamma_{{}_P}^2\cdot
    \vec{v}_{{}_P}^\dagger\cdot\vec{\Delta}_7\cdot
                                       \vec{v}_{{}_P} \rbrack \biggr)'
\nonumber\\&&\qquad
 = \frac{1}{m}\cdot
   K_{{}_P}'\cdot
\lbrack  m\cdot\vec{\Delta}_7  +   m\cdot \gamma_{{}_P}^2\cdot
    \vec{v}_{{}_P}^\dagger\cdot\vec{\Delta}_7\cdot
                                       \vec{v}_{{}_P} \rbrack
 + \frac{1}{m}\cdot K_{{}_P}\cdot \lbrack m\cdot\vec{\Delta}_8
\nonumber\\&&
+  m\cdot \gamma_{{}_P}^2\cdot
    \vec{v}_{{}_P}^\dagger\cdot\vec{\Delta}_8\cdot
                                       \vec{v}_{{}_P}
  +m\cdot \gamma_{{}_P}\cdot
   \vec{\Delta}_7^\dagger\cdot(\gamma_{{}_P}\cdot
    \vec{v}_{{}_P})'\cdot\vec{v}_{{}_P}
 + m\cdot \gamma_{{}_P}\cdot
    \vec{v}_{{}_P}^\dagger\cdot\vec{\Delta}_7\cdot
  (\gamma_{{}_P}\cdot
    \vec{v}_{{}_P})' \rbrack
 \nonumber\\
&=& e\cdot\vec{E}_{{}_P}^\dagger\cdot
                               \vec{v}_{{}_P}\cdot\vec{\Delta}_7  +
    e\cdot \gamma_{{}_P}^2\cdot
        \vec{E}_{{}_P}^\dagger\cdot\vec{v}_{{}_P}\cdot
          \vec{v}_{{}_P}^\dagger\cdot\vec{\Delta}_7\cdot\vec{v}_{{}_P}
 + K_{{}_P}\cdot\vec{\Delta}_8  +
K_{{}_P}\cdot\gamma_{{}_P}^2\cdot\vec{v}_{{}_P}^\dagger\cdot
                                       \vec{\Delta}_8\cdot\vec{v}_{{}_P}
\nonumber\\&&\qquad
+e\cdot \gamma_{{}_P}^2\cdot \vec{\Delta}_7^\dagger\cdot \lbrack
   \vec{v}_{{}_P}\wtimes\vec{B}_{{}_P} +
  \vec{E}_{{}_P} \rbrack\cdot \vec{v}_{{}_P}
+e\cdot \gamma_{{}_P}^2\cdot \vec{v}_{{}_P}^\dagger\cdot
    \vec{\Delta}_7\cdot \lbrack
   \vec{v}_{{}_P}\wtimes\vec{B}_{{}_P} + \vec{E}_{{}_P} \rbrack  \; .
\label{B.36}
\end{eqnarray}
Thus (B.23) can be rewritten as:
\begin{eqnarray}
&&\vec{\Delta}_{13} =
    \underbrace{
     e\cdot(\vec{v}_{{}_P} \wtimes\vec{\Delta}_{12})
 + e\cdot\vec{\Delta}_{11}
         }_{{\rm first\; order\;SG\;terms}}
+   \underbrace{
                      \vec{\Delta}_4
+ \biggl( K_{{}_P}\cdot
 \lbrack 1 + \gamma_{{}_P}^2\cdot
    \vec{v}_{{}_P}\cdot\vec{v}_{{}_P}^\dagger \rbrack
\cdot \lbrack \vec{\Delta}_3 +  \vec{\Delta}_7 \rbrack \biggr)'
         }_{{\rm first\; order\;and\;second\; order\;SG\;terms}}
\nonumber\\&&\qquad
  - \underbrace{
   e\cdot\lbrack\vec{\Delta}_3 +
   \vec{\Delta}_7\rbrack\wtimes\vec{B}_{{}_P}
         }_{{\rm second\; order\;SG\;terms}} \; .
 \label{B.37}
\end{eqnarray}
To obtain $\vec{\Delta}_{21}$, i.e. to identify the second order SG terms on
the rhs of (B.37), one first observes by (B.25) that $\vec{\Delta}_4$ contains
the second order term $\vec{\Delta}_{17}$. Also one has by (B.5),(B.12):
\begin{eqnarray*}
&& \biggl( K_{{}_P}\cdot
 \lbrack 1 + \gamma_{{}_P}^2\cdot
    \vec{v}_{{}_P}\cdot\vec{v}_{{}_P}^\dagger \rbrack
\cdot \lbrack \vec{\Delta}_3 +  \vec{\Delta}_7 \rbrack \biggr)' =
 K_{{}_P}\cdot
 \lbrack 1 + \gamma_{{}_P}^2\cdot
    \vec{v}_{{}_P}\cdot\vec{v}_{{}_P}^\dagger \rbrack
\cdot \lbrack \vec{\Delta}_5 +  \vec{\Delta}_8 \rbrack \nonumber\\
&&\qquad + \biggl( K_{{}_P}\cdot
 \lbrack 1 + \gamma_{{}_P}^2\cdot
    \vec{v}_{{}_P}\cdot\vec{v}_{{}_P}^\dagger \rbrack \biggr)'
\cdot \lbrack \vec{\Delta}_3 +  \vec{\Delta}_7 \rbrack  \; .
\end{eqnarray*}
On the rhs of this equation only the first part contains first order SG terms
and by (B.25) one observes that the first order terms are given by
$K_{{}_P}\cdot \lbrack 1 + \gamma_{{}_P}^2\cdot
    \vec{v}_{{}_P}\cdot\vec{v}_{{}_P}^\dagger \rbrack
\cdot \lbrack \vec{\Delta}_{15}+\vec{\Delta}_{16} \rbrack$.
Introducing the abbreviations:
\begin{eqnarray}
&&\vec{\Delta}_{22} \equiv
  \biggl( K_{{}_P}\cdot
 \lbrack 1 + \gamma_{{}_P}^2\cdot
      \vec{v}_{{}_P}\cdot\vec{v}_{{}_P}^\dagger \rbrack
\cdot \lbrack \vec{\Delta}_3 +  \vec{\Delta}_7 \rbrack \biggr)'
-  K_{{}_P}\cdot \lbrack 1 + \gamma_{{}_P}^2\cdot
    \vec{v}_{{}_P}\cdot\vec{v}_{{}_P}^\dagger \rbrack
\cdot \lbrack \vec{\Delta}_{15}+\vec{\Delta}_{16} \rbrack \; ,
\nonumber\\
&&\vec{\Delta}_{23} \equiv
- e\cdot\lbrack\vec{\Delta}_3 +
   \vec{\Delta}_7\rbrack\wtimes\vec{B}_{{}_P} \; ,
 \label{B.38}
\end{eqnarray}
one thus can simplify the second order SG terms as follows:
\begin{eqnarray}
&&\vec{\Delta}_{21} = \vec{\Delta}_{17}
                    + \vec{\Delta}_{22}
                    + \vec{\Delta}_{23} \; .
 \label{B.39}
\end{eqnarray}
\subsection*{B.8}
In this subsection I simplify $\vec{\Delta}_{22}$ and I first
of all calculate
\begin{eqnarray}
&&\vec{\Delta}_3 + \vec{\Delta}_7 =
    \frac{e}{m^2} \cdot
\frac{\gamma_{{}_P}^2+\gamma_{{}_P}+1}{\gamma_{{}_P}^3\cdot
                                            (\gamma_{{}_P} + 1)}
         \cdot\vec{B}_{{}_P}^\dagger\cdot\vec{s}\cdot\vec{v}_{{}_P}
 \nonumber\\ &&\qquad
+\frac{e}{m^2}\cdot \lbrack \frac{1}{\gamma_{{}_P}\cdot(\gamma_{{}_P}
                   +1)}-\frac{g-2}{2} \rbrack
\cdot\vec{s}^{\,\dagger}\cdot\vec{v}_{{}_P}
 \cdot\vec{v}_{{}_P}^\dagger\cdot\vec{B}_{{}_P}\cdot\vec{v}_{{}_P}
 \nonumber\\ &&\qquad
+\frac{e}{m^2}\cdot \lbrack \frac{1}{(\gamma_{{}_P}+1)^2}
     -\frac{g}{2\cdot\gamma_{{}_P}^2}\rbrack
\cdot\vec{s}^{\,\dagger}\cdot (\vec{v}_{{}_P}\wtimes
       \vec{E}_{{}_P})\cdot\vec{v}_{{}_P}
 \nonumber\\ &&\qquad
+ \frac{e}{m^2}\cdot \lbrack -
     \frac{2}{\gamma_{{}_P}\cdot(\gamma_{{}_P}+1)}
     +\frac{g}{2\cdot\gamma_{{}_P}^2}\rbrack
              \cdot (\vec{E}_{{}_P}\wtimes \vec{s})
 \nonumber\\ &&\qquad
+\frac{e}{m^2}\cdot \lbrack
  -\frac{\gamma_{{}_P}^2+2\cdot
     \gamma_{{}_P}}{(\gamma_{{}_P} +1)^2}+ \frac{g}{2}\rbrack
  \cdot \vec{s}^{\,\dagger}\cdot\vec{v}_{{}_P}\cdot
                    (\vec{E}_{{}_P}\wtimes \vec{v}_{{}_P})
+\frac{e\cdot(g-2)}{2\cdot m^2}\cdot \vec{s}^{\,\dagger}\cdot
                                                         \vec{v}_{{}_P}
                                               \cdot\vec{B}_{{}_P}
\nonumber\\ &&\qquad
    -\frac{e}{m^2}\cdot \vec{E}_{{}_P}^\dagger\cdot\vec{v}_{{}_P}
  \cdot \frac{1}{(\gamma_{{}_P}+1)^2} \cdot
                                       (\vec{s}\wtimes\vec{v}_{{}_P})
-  \frac{e}{m^2}\cdot \frac{1}{\gamma_{{}_P}\cdot(\gamma_{{}_P}+1)}\cdot
             \vec{B}_{{}_M}^\dagger\cdot\vec{v}_{{}_P}\cdot\vec{s} \; ,
\nonumber\\
&& K_{{}_P}\cdot
 \lbrack 1 + \gamma_{{}_P}^2\cdot
      \vec{v}_{{}_P}\cdot\vec{v}_{{}_P}^\dagger \rbrack
\cdot \lbrack \vec{\Delta}_3 +  \vec{\Delta}_7 \rbrack =
    \frac{e}{m} \cdot
\frac{\gamma^2_{{}_P}+\gamma_{{}_P}+1}{\gamma_{{}_P} + 1}
         \cdot\vec{s}^{\,\dagger}\cdot\vec{B}_{{}_P}\cdot\vec{v}_{{}_P}
 \nonumber\\ &&\qquad
-\frac{e}{m}\cdot \frac{\gamma_{{}_P}^3
                                  +2\cdot\gamma_{{}_P}^2}{(\gamma+1)^2}
\cdot\vec{s}^{\,\dagger}\cdot (\vec{v}_{{}_P}\wtimes
   \vec{E}_{{}_P})\cdot\vec{v}_{{}_P}
 \nonumber\\ &&\qquad
+ \frac{e}{m}\cdot \lbrack -\frac{2}{\gamma_{{}_P}+1}
     +\frac{g}{2\cdot\gamma_{{}_P}}\rbrack
              \cdot (\vec{E}_{{}_P}\wtimes \vec{s})
 \nonumber\\ &&\qquad
+\frac{e}{m}\cdot \lbrack
-\frac{\gamma_{{}_P}^3+2\cdot\gamma_{{}_P}^2}{(\gamma_{{}_P} +1)^2}
 + \frac{g}{2}\cdot\gamma_{{}_P} \rbrack
  \cdot \vec{s}^{\,\dagger}\cdot\vec{v}_{{}_P}\cdot
                    (\vec{E}_{{}_P}\wtimes \vec{v}_{{}_P})
+\frac{e\cdot(g-2)}{2\cdot m}\cdot\gamma_{{}_P} \cdot
       \vec{s}^{\,\dagger}\cdot\vec{v}_{{}_P}\cdot\vec{B}_{{}_P}
\nonumber\\ &&\qquad
    -\frac{e}{m}\cdot \vec{v}_{{}_P}^\dagger\cdot\vec{E}_{{}_P}
  \cdot \frac{\gamma_{{}_P}}{(\gamma_{{}_P}+1)^2} \cdot
    (\vec{s}\wtimes\vec{v}_{{}_P})
-  \frac{e}{m}\cdot \frac{1}{\gamma_{{}_P}+1}\cdot
             \vec{v}_{{}_P}^\dagger\cdot\vec{B}_{{}_P}\cdot\vec{s} \; ,
 \label{B.40}
\end{eqnarray}
from which follows by (B.38):
\begin{eqnarray}
&& \vec{\Delta}_{22} =
    \frac{e}{m} \cdot  \biggl(
\frac{\gamma^2_{{}_P}+\gamma_{{}_P}+1}{\gamma_{{}_P} + 1}
 \cdot   \vec{v}_{{}_P}\cdot\vec{s}^{\,\dagger} \biggr)'\cdot
                                                         \vec{B}_{{}_P}
-\frac{e}{m}\cdot \biggl(
   \frac{\gamma_{{}_P}^3+2\cdot\gamma_{{}_P}^2}{(\gamma_{{}_P}+1)^2}
\cdot\vec{v}_{{}_P}\cdot (\vec{s}\wtimes
 \vec{v}_{{}_P})^\dagger\biggr)' \cdot\vec{E}_{{}_P}
 \nonumber\\ &&\qquad
- \frac{e}{m}\cdot \biggl(
                   \lbrack -\frac{2}{\gamma_{{}_P}+1}
     +\frac{g}{2\cdot\gamma_{{}_P}}\rbrack
              \cdot \vec{s} \biggr)'\wtimes \vec{E}_{{}_P}
 \nonumber\\ &&\qquad
-\frac{e}{m}\cdot \biggl(
                  \lbrack
-\frac{\gamma_{{}_P}^3+2\cdot\gamma_{{}_P}^2}{(\gamma_{{}_P} +1)^2}
 + \frac{g}{2}\cdot\gamma_{{}_P} \rbrack
  \cdot \vec{s}^{\,\dagger}\cdot\vec{v}_{{}_P}\cdot
   \vec{v}_{{}_P} \biggr)'\wtimes\vec{E}_{{}_P}
+\frac{e\cdot(g-2)}{2\cdot m}\cdot\biggl(
        \gamma_{{}_P} \cdot  \vec{s}^{\,\dagger}\cdot
   \vec{v}_{{}_P} \biggr)' \cdot\vec{B}_{{}_P}
\nonumber\\ &&\qquad
    -\frac{e}{m}\cdot \biggl(
        \frac{\gamma_{{}_P}}{(\gamma_{{}_P}+1)^2} \cdot
  (\vec{s}\wtimes\vec{v}_{{}_P})
               \cdot\vec{v}_{{}_P}^\dagger \biggr)' \cdot\vec{E}_{{}_P}
-  \frac{e}{m}\cdot \biggl(
                    \frac{1}{\gamma_{{}_P}+1}\cdot
  \vec{s}\cdot \vec{v}_{{}_P}^\dagger \biggr)' \cdot\vec{B}_{{}_P} \; .
 \label{B.41}
\end{eqnarray}
To simplify this, I calculate by (2.3a),(2.5),(2.9),(B.12-13),(B.18):
\begin{eqnarray}
&& \vec{B}_{{}_P}^\dagger\cdot\vec{s}\ ' =
   \vec{B}_{{}_P}^\dagger\cdot\biggl\lbrack
                  \frac{e\cdot g}{2\cdot m\cdot\gamma_{{}_P}}\cdot
\biggl( \vec{s}\wtimes\vec{B}_{{}_P} +(\vec{v}_{{}_P}\wtimes
     \vec{s})\wtimes\vec{E}_{{}_P}
\biggr)
\nonumber\\&&
  + \frac{e\cdot (g-2)}{2\cdot m\cdot\gamma_{{}_P}}\cdot\biggl(
    \gamma_{{}_P}^2 \cdot
\vec{s}^{\,\dagger}\cdot\vec{v}_{{}_P}\cdot \lbrack \vec{E}_{{}_P}+
  \vec{v}_{{}_P}\wtimes\vec{B}_{{}_P}
\rbrack - \vec{v}_{{}_P}^\dagger\cdot\vec{E}_{{}_P}\cdot\vec{s}
  - \gamma_{{}_P}^2 \cdot
\vec{s}^{\,\dagger}\cdot\vec{v}_{{}_P}\cdot \vec{v}_{{}_P}^\dagger\cdot
    \vec{E}_{{}_P}\cdot \vec{v}_{{}_P}
\biggr) \biggr\rbrack
\nonumber\\
 &=&  \frac{e\cdot g}{2\cdot m\cdot\gamma_{{}_P}}\cdot
\biggl( \vec{E}_{{}_P}^\dagger\cdot
    \vec{v}_{{}_P}\cdot\vec{B}_{{}_P}^\dagger\cdot\vec{s}
       -\vec{E}_{{}_P}^\dagger\cdot
   \vec{s}\cdot\vec{v}_{{}_P}^\dagger\cdot\vec{B}_{{}_P}
\biggr)
  + \frac{e\cdot (g-2)}{2\cdot m\cdot\gamma_{{}_P}}\cdot\biggl(
\gamma_{{}_P}^2\cdot\vec{s}^{\,\dagger}\cdot
   \vec{v}_{{}_P}\cdot \vec{E}_{{}_P}^\dagger\cdot\vec{B}_{{}_P}
\nonumber\\&&
- \vec{E}_{{}_P}^\dagger\cdot\vec{v}_{{}_P}\cdot
               \vec{s}^{\,\dagger}\cdot\vec{B}_{{}_P}
-\gamma_{{}_P}^2\cdot\vec{s}^{\,\dagger}\cdot
  \vec{v}_{{}_P}\cdot \vec{E}_{{}_P}^\dagger\cdot\vec{v}_{{}_P}
             \cdot\vec{v}_{{}_P}^\dagger\cdot \vec{B}_{{}_P}
\biggr)
\nonumber\\
 &=& -\frac{e\cdot g}{2\cdot m\cdot\gamma_{{}_P}}\cdot
        \vec{E}_{{}_P}^\dagger\cdot
    \vec{s}\cdot\vec{v}_{{}_P}^\dagger\cdot\vec{B}_{{}_P}
  + \frac{e\cdot (g-2)}{2\cdot m}\cdot\gamma_{{}_P}\cdot\biggl(
   \vec{s}^{\,\dagger}\cdot\vec{v}_{{}_P}\cdot
                                           \vec{E}_{{}_P}^\dagger\cdot
                                                          \vec{B}_{{}_P}
\nonumber\\&&\qquad
-   \vec{s}^{\,\dagger}\cdot\vec{v}_{{}_P}\cdot
   \vec{E}_{{}_P}^\dagger\cdot\vec{v}_{{}_P}
            \cdot\vec{v}_{{}_P}^\dagger\cdot \vec{B}_{{}_P}
\biggr)
  +  \frac{e}{m\cdot\gamma_{{}_P}}\cdot
        \vec{E}_{{}_P}^\dagger\cdot
    \vec{v}_{{}_P}\cdot\vec{s}^{\,\dagger}\cdot\vec{B}_{{}_P}
           \; ,
\nonumber\\
&& ( \vec{s}\wtimes\vec{v}_{{}_P})' =
  \biggl(\gamma_{{}_P}\cdot ( \vec{\sigma}\wtimes
                                           \vec{v}_{{}_P})\biggr)' =
  \biggl(\gamma_{{}_M}\cdot ( \vec{\sigma}\wtimes
                                           \vec{v}_{{}_M})\biggr)' =
  m\cdot (\gamma_{{}_M} +1)\cdot \vec{\Delta}_7
\nonumber\\ &&
- \gamma_{{}_M}\cdot(\gamma_{{}_M} +1)\cdot
               ( \vec{\sigma}\wtimes\vec{v}_{{}_M})\cdot
  \biggl(\frac{1}{\gamma_{{}_M}+1} \biggr)'
\nonumber\\ &&
 =  m\cdot (\gamma_{{}_M} +1)\cdot \vec{\Delta}_7
   + \frac{e}{m}\cdot\frac{\gamma_{{}_M}}{\gamma_{{}_M}+1}
\cdot \vec{v}_{{}_M}^{\,\dagger}\cdot\vec{E}_{{}_P}\cdot
        ( \vec{\sigma}\wtimes\vec{v}_{{}_M})
\nonumber\\
&=& m\cdot (\gamma_{{}_P} +1)\cdot \vec{\Delta}_7
   + \frac{e}{m}\cdot\frac{1}{\gamma_{{}_P}+1}
\cdot \vec{v}_{{}_P}^\dagger\cdot\vec{E}_{{}_P}\cdot
                                         ( \vec{s}\wtimes\vec{v}_{{}_P})
\nonumber\\
&=&
- \frac{e\cdot g}{2\cdot m}\cdot\frac{1}{\gamma_{{}_P}}
\cdot \vec{B}_{{}_P}^\dagger\cdot\vec{v}_{{}_P} \cdot \vec{s}
\nonumber\\ &&
- \frac{e\cdot(g-2)}{2\cdot m}\cdot\gamma_{{}_P}
\cdot \vec{B}_{{}_P}^\dagger\cdot\vec{v}_{{}_P} \cdot
       \vec{s}^{\,\dagger}\cdot\vec{v}_{{}_P}\cdot\vec{v}_{{}_P}
   +\frac{e\cdot(g-2)}{2\cdot m}\cdot\gamma_{{}_P}\cdot
    \vec{s}^{\,\dagger}\cdot\vec{v}_{{}_P}\cdot\vec{B}_{{}_P}
\nonumber\\ &&
+  \frac{e}{m}\cdot \frac{1}{\gamma_{{}_P}}\cdot
             \vec{s}^{\,\dagger}\cdot\vec{B}_{{}_P}\cdot\vec{v}_{{}_P}
-\frac{e\cdot(g-2)}{2\cdot m}\cdot\gamma_{{}_P}\cdot
             \vec{s}^{\,\dagger}\cdot\vec{v}_{{}_P}\cdot
  (\vec{v}_{{}_P}\wtimes \vec{E}_{{}_P})
\nonumber\\ &&
    +  \frac{e}{m}\cdot \frac{1}{\gamma_{{}_P}}\cdot
            (\vec{s}\wtimes \vec{E}_{{}_P})  \; ,
\nonumber\\
&&  \vec{E}_{{}_P}^\dagger\cdot( \vec{s}\wtimes\vec{v}_{{}_P})' =
- \frac{e\cdot g}{2\cdot m}\cdot\frac{1}{\gamma_{{}_P}}
\cdot \vec{B}_{{}_P}^\dagger\cdot\vec{v}_{{}_P} \cdot
                   \vec{s}^{\,\dagger}\cdot\vec{E}_{{}_P}
\nonumber\\ &&
- \frac{e\cdot(g-2)}{2\cdot m}\cdot\gamma_{{}_P}
\cdot \vec{B}_{{}_P}^\dagger\cdot\vec{v}_{{}_P} \cdot
       \vec{s}^{\,\dagger}\cdot\vec{v}_{{}_P}\cdot
  \vec{v}_{{}_P}^\dagger\cdot\vec{E}_{{}_P}
   +\frac{e\cdot(g-2)}{2\cdot m}\cdot\gamma_{{}_P}\cdot
    \vec{s}^{\,\dagger}\cdot\vec{v}_{{}_P}\cdot
            \vec{B}_{{}_P}^\dagger\cdot\vec{E}_{{}_P}
\nonumber\\ &&
+  \frac{e}{m}\cdot \frac{1}{\gamma_{{}_P}}\cdot
    \vec{s}^{\,\dagger}\cdot\vec{B}_{{}_P}\cdot
          \vec{v}_{{}_P}^\dagger\cdot\vec{E}_{{}_P} \; ,
\nonumber\\
&&  \vec{E}_{{}_P}\wtimes\vec{s}\ ' =
   \vec{E}_{{}_P}\wtimes \biggl\lbrack
                  \frac{e\cdot g}{2\cdot m\cdot\gamma_{{}_P}}\cdot
\biggl( \vec{s}\wtimes\vec{B}_{{}_P} +
  (\vec{v}_{{}_P}\wtimes\vec{s})\wtimes\vec{E}_{{}_P}
\biggr)
\nonumber\\&&
  + \frac{e\cdot (g-2)}{2\cdot m\cdot\gamma_{{}_P}}\cdot\biggl(
    \gamma_{{}_P}^2 \cdot
\vec{s}^{\,\dagger}\cdot\vec{v}_{{}_P}\cdot
  \lbrack \vec{E}_{{}_P}+\vec{v}_{{}_P}\wtimes\vec{B}_{{}_P}
\rbrack - \vec{v}^\dagger\cdot\vec{E}_{{}_P}\cdot\vec{s}
  - \gamma_{{}_P}^2 \cdot
\vec{s}^{\,\dagger}\cdot\vec{v}_{{}_P}\cdot
  \vec{v}_{{}_P}^\dagger\cdot\vec{E}_{{}_P}\cdot \vec{v}_{{}_P}
\biggr) \biggr\rbrack
\nonumber\\
&=&               \frac{e\cdot g}{2\cdot m\cdot\gamma_{{}_P}}\cdot
\biggl( \vec{E}_{{}_P}^\dagger\cdot\vec{B}_{{}_P}\cdot\vec{s}
      - \vec{s}^{\,\dagger}\cdot\vec{E}_{{}_P}\cdot\vec{B}_{{}_P}
 + \vec{E}_{{}_P}^\dagger\cdot\vec{E}_{{}_P}\cdot
                                           (\vec{v}_{{}_P}\wtimes\vec{s})
 - \vec{E}_{{}_P}^\dagger\cdot(\vec{v}_{{}_P}\wtimes
  \vec{s})\cdot\vec{E}_{{}_P} \biggr)
\nonumber\\&&\qquad
  + \frac{e\cdot (g-2)}{2\cdot m\cdot\gamma_{{}_P}}\cdot\biggl(
    \gamma_{{}_P}^2 \cdot
\vec{s}^{\,\dagger}\cdot\vec{v}_{{}_P}\cdot
  \vec{E}_{{}_P}^\dagger\cdot\vec{B}_{{}_P}\cdot\vec{v}_{{}_P}
   -\gamma_{{}_P}^2 \cdot
\vec{s}^{\,\dagger}\cdot\vec{v}_{{}_P}\cdot
   \vec{E}_{{}_P}^\dagger\cdot\vec{v}_{{}_P}\cdot\vec{B}_{{}_P}
\nonumber\\&&\qquad
- \vec{v}_{{}_P}^\dagger\cdot \vec{E}_{{}_P}\cdot
                                           (\vec{E}_{{}_P}\wtimes\vec{s})
   -\gamma_{{}_P}^2 \cdot
\vec{s}^{\,\dagger}\cdot\vec{v}_{{}_P}\cdot
   \vec{E}_{{}_P}^\dagger\cdot\vec{v}_{{}_P}\cdot
                               (\vec{E}_{{}_P}\wtimes\vec{v}_{{}_P})
\biggr) \; ,
\nonumber\\
&& (\vec{s}^{\,\dagger}\cdot\vec{v}_{{}_P})' =
   (\vec{\sigma}^{\,\dagger}\cdot\vec{v}_{{}_P})' =
\vec{v}_{{}_P}^\dagger\cdot(\vec{\Omega}_{{}_M}\wtimes
   \vec{\sigma}) +\vec{\sigma}^{\,\dagger}\cdot\vec{v}_{{}_P} '
\nonumber\\
&=& \frac{1}{\gamma_{{}_P}}\cdot (\vec{s}\wtimes
                       \vec{v}_{{}_P})^\dagger\cdot
 \vec{\Omega}_{{}_P} + \frac{e}{K_{{}_P}}\cdot
                   \vec{\sigma}^{\,\dagger}\cdot
  \lbrack \vec{v}_{{}_P}\wtimes\vec{B}_{{}_P} + \vec{E}_{{}_P}
     - \vec{E}_{{}_P}^\dagger\cdot
                            \vec{v}_{{}_P}\cdot\vec{v}_{{}_P}\rbrack
\nonumber\\
&=&-\frac{e}{m}\cdot\frac{1}{\gamma_{{}_P}}\cdot
 (\vec{s}\wtimes\vec{v}_{{}_P})^\dagger
  \cdot \biggl(
 \lbrack \frac{1}{\gamma_{{}_P}}+
                            \frac{g-2}{2} \rbrack \cdot\vec{B}_{{}_P}
 -\frac{g-2}{2}\cdot\frac{\gamma_{{}_P}}{\gamma_{{}_P}+1}\cdot
  \vec{v}_{{}_P}^\dagger\cdot\vec{B}_{{}_P}\cdot\vec{v}_{{}_P}
 \nonumber\\ &&\qquad
-\lbrack \frac{g}{2} -\frac{\gamma_{{}_P}}{\gamma_{{}_P}+1}\rbrack\cdot
 (\vec{v}_{{}_P}\wtimes\vec{E}_{{}_P}) \biggr)
 \nonumber\\ &&\qquad
+\frac{e}{m}\cdot\frac{1}{\gamma_{{}_P}^2}\cdot\vec{s}^{\,\dagger}\cdot
         (\vec{v}_{{}_P}\wtimes\vec{B}_{{}_P})
  + \frac{e}{K_{{}_P}}\cdot\vec{E}_{{}_P}^\dagger\cdot
  \lbrack  \frac{1}{\gamma_{{}_P}}\cdot
   \vec{s} + \frac{\gamma_{{}_P}}{\gamma_{{}_P}+1}
\cdot \vec{s}^{\,\dagger}\cdot\vec{v}_{{}_P}\cdot\vec{v}_{{}_P} \rbrack
\nonumber\\&&\qquad
-\frac{e}{m}\cdot\frac{1}{\gamma_{{}_P}}\cdot
              \vec{s}^{\,\dagger}\cdot\vec{v}_{{}_P}
 \cdot\vec{E}_{{}_P}^\dagger\cdot\vec{v}_{{}_P}
\nonumber\\
&=&-\frac{e\cdot(g-2)}{2\cdot m}\cdot \frac{1}{\gamma_{{}_P}}\cdot
 (\vec{s}\wtimes\vec{v}_{{}_P})^\dagger\cdot\vec{B}_{{}_P}
  +\frac{e\cdot(g-2)}{2\cdot m}\cdot \frac{1}{\gamma_{{}_P}}\cdot
  \vec{s}^{\,\dagger}\cdot\vec{v}_{{}_P}\cdot
  \vec{E}_{{}_P}^\dagger\cdot\vec{v}_{{}_P}
 \nonumber\\ &&\qquad
  +\frac{e}{m}\cdot\lbrack \frac{1}{\gamma_{{}_P}}
 -\frac{g}{2}\cdot
     \frac{\gamma_{{}_P}^2-1}{\gamma_{{}_P}^3} \rbrack \cdot
          \vec{E}_{{}_P}^\dagger\cdot\vec{s} \; ,
\nonumber\\
&& s_j\cdot \vec{B}_{{}_P}^\dagger\cdot
                                 \vec{v}_{{}_P} ' = s_j \cdot \biggl(
\frac{e}{m}\cdot \frac{1}{\gamma_{{}_P}}\cdot
    \vec{B}_{{}_P}^\dagger\cdot\vec{E}_{{}_P}
-\frac{e}{m}\cdot \frac{1}{\gamma_{{}_P}}\cdot
 \vec{v}_{{}_P}^\dagger\cdot\vec{E}_{{}_P}
    \cdot\vec{v}_{{}_P}^\dagger\cdot\vec{B}_{{}_P}\biggr) \; , \;\;
(j=1,2,3) \qquad
 \label{B.42}
\end{eqnarray}
from which follows by (B.41),(2.9):
\begin{eqnarray}
&& \vec{\Delta}_{22} =
    \frac{e^2}{m^3} \cdot
\frac{\gamma_{{}_P}\cdot(\gamma_{{}_P}+1)\cdot(2\cdot\gamma_{{}_P}+1)
   - (\gamma_{{}_P}^2+\gamma_{{}_P}+1)\cdot(2\cdot\gamma_{{}_P}+1)}
     {\gamma_{{}_P}^2\cdot(\gamma_{{}_P}+1)^2}
 \cdot\vec{v}_{{}_P}^\dagger\cdot\vec{E}_{{}_P}\cdot
                \vec{s}^{\,\dagger}\cdot\vec{B}_{{}_P}
                           \cdot\vec{\pi}_{{}_M}
\nonumber\\&&
+   \frac{e}{m^2} \cdot
\frac{\gamma_{{}_P}^2+\gamma_{{}_P}+1}{\gamma_{{}_P}\cdot
                                          (\gamma_{{}_P}+1)}
 \cdot\vec{s}^{\,\dagger}\cdot\vec{B}_{{}_P}\cdot\vec{\pi}_{{}_M}'
+   \frac{e}{m} \cdot
\frac{\gamma_{{}_P}^2+\gamma_{{}_P}+1}{\gamma_{{}_P}+1}
 \cdot\vec{v}_{{}_P}\cdot\vec{B}_{{}_P}^\dagger\cdot \vec{s}\ '
\nonumber\\&&
  - \frac{e^2}{m^3} \cdot
\frac{(2\cdot\gamma_{{}_P}+2)\cdot(\gamma_{{}_P}+1)^2
-2\cdot(\gamma_{{}_P}+1)\cdot(\gamma_{{}_P}^2+2\cdot\gamma_{{}_P})}
     {(\gamma_{{}_P}+1)^4}
 \cdot\vec{v}_{{}_P}^\dagger\cdot\vec{E}_{{}_P}\cdot
           (\vec{s}\wtimes\vec{v}_{{}_P})^\dagger\cdot
                 \vec{E}_{{}_P}\cdot\vec{\pi}_{{}_M}
\nonumber\\&&
-   \frac{e}{m^2} \cdot
\frac{\gamma_{{}_P}^2+2\cdot\gamma_{{}_P}}{(\gamma_{{}_P}+1)^2}
 \cdot(\vec{s}\wtimes\vec{v}_{{}_P})^\dagger\cdot
    \vec{E}_{{}_P}\cdot\vec{\pi}_{{}_M}'
-   \frac{e}{m} \cdot
\frac{\gamma_{{}_P}^3+2\cdot\gamma_{{}_P}^2}{(\gamma_{{}_P}+1)^2}
 \cdot\vec{v}_{{}_P}\cdot\vec{E}_{{}_P}^\dagger\cdot
                                         (\vec{s}\wtimes\vec{v}_{{}_P})'
 \nonumber\\ &&
- \frac{e^2}{m^2}\cdot
                   \lbrack \frac{2}{(\gamma_{{}_P}+1)^2}
     -\frac{g}{2\cdot\gamma_{{}_P}^2}\rbrack
   \cdot \vec{v}_{{}_P}^\dagger\cdot\vec{E}_{{}_P}\cdot
    (\vec{s}\wtimes \vec{E}_{{}_P})
+ \frac{e}{m}\cdot \lbrack -\frac{2}{\gamma_{{}_P}+1}
     +\frac{g}{2\cdot\gamma_{{}_P}}\rbrack
              \cdot \vec{E}_{{}_P} \wtimes \vec{s}\ '
\nonumber\\&&
  + \frac{e^2}{m^3} \cdot
\frac{(\gamma_{{}_P}+1)^2\cdot(2\cdot\gamma_{{}_P}+2)
  -2\cdot(\gamma_{{}_P}+1)\cdot(\gamma_{{}_P}^2+2\cdot\gamma_{{}_P})}
     {(\gamma_{{}_P}+1)^4}
 \cdot\vec{v}_{{}_P}^\dagger\cdot\vec{E}_{{}_P}\cdot
              \vec{s}^{\,\dagger}\cdot\vec{v}_{{}_P}\cdot
                                  (\vec{\pi}_{{}_M}\wtimes\vec{E}_{{}_P})
\nonumber\\&&
-\frac{e}{m^2}\cdot \lbrack
-\frac{\gamma_{{}_P}^2+2\cdot\gamma_{{}_P}}{(\gamma_{{}_P} +1)^2}
                                                   + \frac{g}{2} \rbrack
  \cdot \vec{s}^{\,\dagger}\cdot\vec{v}_{{}_P}\cdot
     \vec{\pi}_{{}_M}'\wtimes\vec{E}_{{}_P}
\nonumber\\&&
-\frac{e}{m}\cdot \lbrack
-\frac{\gamma_{{}_P}^3+2\cdot\gamma_{{}_P}^2}{(\gamma_{{}_P} +1)^2}
 + \frac{g}{2}\cdot\gamma_{{}_P} \rbrack
  \cdot (\vec{s}^{\,\dagger}\cdot\vec{v}_{{}_P})'\cdot
     (\vec{v}_{{}_P} \wtimes\vec{E}_{{}_P})
\nonumber\\&&
+\frac{e^2\cdot(g-2)}{2\cdot m^2}\cdot
 \vec{v}_{{}_P}^\dagger\cdot\vec{E}_{{}_P} \cdot\vec{s}^{\,\dagger}\cdot
                \vec{v}_{{}_P}\cdot\vec{B}_{{}_P}
+\frac{e\cdot(g-2)}{2\cdot m}\cdot
 \gamma_{{}_P} \cdot (\vec{s}^{\,\dagger}\cdot
                                    \vec{v}_{{}_P})' \cdot\vec{B}_{{}_P}
\nonumber\\ &&
    +\frac{2\cdot e^2}{m^3}\cdot\frac{1}{(\gamma_{{}_P}+1)^3} \cdot
       \vec{v}_{{}_P}^\dagger\cdot\vec{E}_{{}_P}\cdot
     \vec{\pi}_{{}_M}^\dagger \cdot\vec{E}_{{}_P}
                              \cdot (\vec{s}\wtimes\vec{v}_{{}_P})
    -\frac{e}{m^2}\cdot\frac{1}{(\gamma_{{}_P}+1)^2} \cdot
    \vec{E}_{{}_P}^\dagger\cdot\vec{\pi}_{{}_M}'\cdot
                                           (\vec{s}\wtimes\vec{v}_{{}_P})
\nonumber\\ &&
    -\frac{e}{m}\cdot
        \frac{\gamma_{{}_P}}{(\gamma_{{}_P}+1)^2} \cdot
             \vec{v}_{{}_P}^\dagger  \cdot\vec{E}_{{}_P}
                    \cdot (\vec{s}\wtimes\vec{v}_{{}_P})'
    +\frac{e^2}{m^2}\cdot
 \frac{1}{(\gamma_{{}_P}+1)^2} \cdot
    \vec{v}_{{}_P}^\dagger  \cdot\vec{E}_{{}_P}
                    \cdot \vec{v}_{{}_P}^\dagger\cdot
                  \vec{B}_{{}_P}\cdot\vec{s}
\nonumber\\ &&
-  \frac{e}{m}\cdot \frac{1}{\gamma_{{}_P}+1}\cdot
             \vec{s}\cdot \vec{B}_{{}_P}^\dagger\cdot\vec{v}_{{}_P}'
-  \frac{e}{m}\cdot \frac{1}{\gamma_{{}_P}+1}\cdot
             \vec{v}_{{}_P}^\dagger\cdot \vec{B}_{{}_P}\cdot\vec{s}\ '
\nonumber\\
&=&-\frac{e^2}{m^2} \cdot
\frac{2\cdot\gamma_{{}_P}+1}{\gamma_{{}_P}\cdot(\gamma_{{}_P}+1)^2}
 \cdot\vec{v}_{{}_P}^\dagger\cdot\vec{E}_{{}_P}\cdot
               \vec{s}^{\,\dagger}\cdot\vec{B}_{{}_P}
                    \cdot\vec{v}_{{}_P}
+   \frac{e^2}{m^2} \cdot
\frac{\gamma_{{}_P}^2+\gamma_{{}_P}+1}{\gamma_{{}_P}\cdot
                                          (\gamma_{{}_P}+1)}
 \cdot\vec{s}^{\,\dagger}\cdot\vec{B}_{{}_P}\cdot\lbrack
               \vec{v}_{{}_P}\wtimes\vec{B}_{{}_P}
 + \vec{E}_{{}_P} \rbrack
\nonumber\\&&
+   \frac{e}{m} \cdot
\frac{\gamma_{{}_P}^2+\gamma_{{}_P}+1}{\gamma_{{}_P}+1}
 \cdot\vec{v}_{{}_P}\cdot  \biggl\lbrack
     -\frac{e\cdot g}{2\cdot m\cdot\gamma_{{}_P}}\cdot
        \vec{E}_{{}_P}^\dagger\cdot\vec{s}\cdot
                \vec{v}_{{}_P}^\dagger\cdot\vec{B}_{{}_P}
\nonumber\\&&
  + \frac{e\cdot (g-2)}{2\cdot m}\cdot\gamma_{{}_P}\cdot\biggl(
   \vec{s}^{\,\dagger}\cdot\vec{v}_{{}_P}\cdot
         \vec{E}_{{}_P}^\dagger\cdot\vec{B}_{{}_P}
-   \vec{s}^{\,\dagger}\cdot\vec{v}_{{}_P}\cdot
   \vec{E}_{{}_P}^\dagger\cdot\vec{v}_{{}_P}
        \cdot\vec{v}_{{}_P}^\dagger\cdot \vec{B}_{{}_P}
\biggr)
\nonumber\\&&
  +  \frac{e}{m\cdot\gamma_{{}_P}}\cdot
   \vec{E}_{{}_P}^\dagger\cdot\vec{v}_{{}_P}\cdot
        \vec{s}^{\,\dagger}\cdot\vec{B}_{{}_P} \biggr\rbrack
  - \frac{e^2}{m^2} \cdot
\frac{2\cdot\gamma_{{}_P}}{(\gamma_{{}_P}+1)^3}
\cdot\vec{v}_{{}_P}^\dagger\cdot\vec{E}_{{}_P}\cdot
          (\vec{s}\wtimes\vec{v}_{{}_P})^\dagger\cdot
                 \vec{E}_{{}_P}\cdot\vec{v}_{{}_P}
\nonumber\\&&
-   \frac{e^2}{m^2} \cdot
\frac{\gamma_{{}_P}^2+2\cdot\gamma_{{}_P}}{(\gamma_{{}_P}+1)^2}
 \cdot(\vec{s}\wtimes\vec{v}_{{}_P})^\dagger\cdot\vec{E}_{{}_P}\cdot
   \lbrack \vec{v}_{{}_P}\wtimes\vec{B}_{{}_P} + \vec{E}_{{}_P} \rbrack
\nonumber\\&&
-   \frac{e}{m} \cdot
\frac{\gamma_{{}_P}^3+2\cdot\gamma_{{}_P}^2}{(\gamma_{{}_P}+1)^2}
 \cdot\vec{v}_{{}_P}\cdot\biggl(
- \frac{e\cdot g}{2\cdot m}\cdot\frac{1}{\gamma_{{}_P}}
\cdot \vec{B}_{{}_P}^\dagger\cdot\vec{v}_{{}_P} \cdot
                  \vec{s}^{\,\dagger}\cdot\vec{E}_{{}_P}
\nonumber\\ &&
- \frac{e\cdot(g-2)}{2\cdot m}\cdot\gamma_{{}_P}
\cdot \vec{B}_{{}_P}^\dagger\cdot\vec{v}_{{}_P} \cdot
       \vec{s}^{\,\dagger}\cdot\vec{v}_{{}_P}\cdot
  \vec{v}_{{}_P}^\dagger\cdot\vec{E}_{{}_P}
   +\frac{e\cdot(g-2)}{2\cdot m}\cdot\gamma_{{}_P}\cdot
    \vec{s}^{\,\dagger}\cdot\vec{v}_{{}_P}\cdot
    \vec{B}_{{}_P}^\dagger\cdot\vec{E}_{{}_P}
\nonumber\\ &&
+  \frac{e}{m}\cdot \frac{1}{\gamma_{{}_P}}\cdot
     \vec{s}^{\,\dagger}\cdot\vec{B}_{{}_P}\cdot
  \vec{v}_{{}_P}^\dagger\cdot\vec{E}_{{}_P} \biggr)
- \frac{e^2}{m^2}\cdot
                   \lbrack \frac{2}{(\gamma_{{}_P}+1)^2}
     -\frac{g}{2\cdot\gamma_{{}_P}^2}\rbrack
   \cdot \vec{v}_{{}_P}^\dagger\cdot\vec{E}_{{}_P}\cdot
   (\vec{s}\wtimes \vec{E}_{{}_P})
 \nonumber\\ &&
+ \frac{e}{m}\cdot \lbrack -\frac{2}{\gamma_{{}_P}+1}
     +\frac{g}{2\cdot\gamma_{{}_P}}\rbrack
              \cdot \biggl\lbrack
                  \frac{e\cdot g}{2\cdot m\cdot\gamma_{{}_P}}\cdot
\biggl( \vec{E}_{{}_P}^\dagger\cdot\vec{B}_{{}_P}\cdot\vec{s}
      - \vec{s}^{\,\dagger}\cdot\vec{E}_{{}_P}\cdot\vec{B}_{{}_P}
\nonumber\\&&
\vec{E}_{{}_P}^\dagger\cdot\vec{E}_{{}_P}\cdot
                                        (\vec{v}_{{}_P}\wtimes\vec{s})
 - \vec{E}_{{}_P}^\dagger\cdot(\vec{v}_{{}_P}\wtimes
  \vec{s})\cdot\vec{E}_{{}_P} \biggr)
  + \frac{e\cdot (g-2)}{2\cdot m\cdot\gamma_{{}_P}}\cdot\biggl(
    \gamma_{{}_P}^2 \cdot
\vec{s}^{\,\dagger}\cdot\vec{v}_{{}_P}\cdot \vec{E}_{{}_P}^\dagger\cdot
          \vec{B}_{{}_P}\cdot\vec{v}_{{}_P}
\nonumber\\&&
   -\gamma_{{}_P}^2 \cdot
\vec{s}^{\,\dagger}\cdot\vec{v}_{{}_P}\cdot \vec{E}_{{}_P}^\dagger\cdot
        \vec{v}_{{}_P}\cdot\vec{B}_{{}_P}
- \vec{v}_{{}_P}^\dagger\cdot \vec{E}_{{}_P}\cdot
                                           (\vec{E}_{{}_P}\wtimes\vec{s})
   -\gamma_{{}_P}^2 \cdot
\vec{s}^{\,\dagger}\cdot\vec{v}_{{}_P}\cdot
 \vec{E}_{{}_P}^\dagger\cdot\vec{v}_{{}_P}\cdot
                               (\vec{E}_{{}_P}\wtimes\vec{v}_{{}_P})
\biggr) \biggr\rbrack
\nonumber\\&&
  + \frac{e^2}{m^2} \cdot\frac{2\cdot\gamma_{{}_P}}{(\gamma_{{}_P}+1)^3}
\cdot\vec{v}_{{}_P}^\dagger\cdot
                   \vec{E}_{{}_P}\cdot\vec{s}^{\,\dagger}\cdot
                                                  \vec{v}_{{}_P}\cdot
                                  (\vec{v}_{{}_P}\wtimes\vec{E}_{{}_P})
\nonumber\\&&
-\frac{e^2}{m^2}\cdot \lbrack
-\frac{\gamma_{{}_P}^2+2\cdot\gamma_{{}_P}}{(\gamma_{{}_P} +1)^2} +
                                                     \frac{g}{2} \rbrack
  \cdot \vec{s}^{\,\dagger}\cdot\vec{v}_{{}_P}\cdot\lbrack
        \vec{v}_{{}_P}^\dagger\cdot\vec{E}_{{}_P}\cdot\vec{B}_{{}_P}
       -\vec{B}_{{}_P}^\dagger\cdot\vec{E}_{{}_P}\cdot
                                                \vec{v}_{{}_P} \rbrack
\nonumber\\&&
-\frac{e}{m}\cdot \lbrack
-\frac{\gamma_{{}_P}^3+2\cdot\gamma_{{}_P}^2}{(\gamma_{{}_P} +1)^2}
 + \frac{g}{2}\cdot\gamma_{{}_P} \rbrack
  \cdot (\vec{v}_{{}_P} \wtimes\vec{E}_{{}_P})\cdot\biggl(
   -\frac{e\cdot(g-2)}{2\cdot m}\cdot \frac{1}{\gamma_{{}_P}}\cdot
 (\vec{s}\wtimes\vec{v}_{{}_P})^\dagger\cdot\vec{B}_{{}_P}
\nonumber\\&&
  +\frac{e\cdot(g-2)}{2\cdot m}\cdot \frac{1}{\gamma_{{}_P}}\cdot
  \vec{s}^{\,\dagger}\cdot\vec{v}_{{}_P}\cdot\vec{E}_{{}_P}^\dagger\cdot
                                                        \vec{v}_{{}_P}
  +\frac{e}{m}\cdot\lbrack \frac{1}{\gamma_{{}_P}}
 -\frac{g}{2}\cdot
\frac{\gamma_{{}_P}^2-1}{\gamma_{{}_P}^3} \rbrack \cdot
          \vec{E}_{{}_P}^\dagger\cdot\vec{s} \biggr)
\nonumber\\&&
+\frac{e^2\cdot(g-2)}{2\cdot m^2}\cdot
 \vec{v}_{{}_P}^\dagger\cdot\vec{E}_{{}_P} \cdot\vec{s}^{\,\dagger}\cdot
          \vec{v}_{{}_P}\cdot\vec{B}_{{}_P}
\nonumber\\&&
+\frac{e\cdot(g-2)}{2\cdot m}\cdot\gamma_{{}_P}\cdot
                                  \vec{B}_{{}_P}\cdot\biggl(
   -\frac{e\cdot(g-2)}{2\cdot m}\cdot \frac{1}{\gamma_{{}_P}}\cdot
 (\vec{s}\wtimes\vec{v}_{{}_P})^\dagger\cdot\vec{B}_{{}_P}
\nonumber\\&&
  +\frac{e\cdot(g-2)}{2\cdot m}\cdot \frac{1}{\gamma_{{}_P}}\cdot
  \vec{s}^{\,\dagger}\cdot\vec{v}_{{}_P}\cdot\vec{E}_{{}_P}^\dagger\cdot
                                                        \vec{v}_{{}_P}
  +\frac{e}{m}\cdot\lbrack \frac{1}{\gamma_{{}_P}}
 -\frac{g}{2}\cdot
   \frac{\gamma_{{}_P}^2-1}{\gamma_{{}_P}^3} \rbrack \cdot
          \vec{E}_{{}_P}^\dagger\cdot\vec{s} \biggr)
\nonumber\\ &&
    +\frac{2\cdot e^2}{m^2}\cdot
  \frac{\gamma_{{}_P}}{(\gamma_{{}_P}+1)^3} \cdot
       \vec{v}_{{}_P}^\dagger\cdot\vec{E}_{{}_P}\cdot
 \vec{v}_{{}_P}^\dagger \cdot\vec{E}_{{}_P}
                              \cdot (\vec{s}\wtimes\vec{v}_{{}_P})
\nonumber\\ &&
-\frac{e^2}{m^2}\cdot\frac{1}{(\gamma_{{}_P}+1)^2}\cdot
 (\vec{s}\wtimes\vec{v}_{{}_P})\cdot\vec{E}_{{}_P}^\dagger\cdot\lbrack
   \vec{v}_{{}_P}\wtimes\vec{B}_{{}_P} + \vec{E}_{{}_P} \rbrack
\nonumber\\ &&
    -\frac{e}{m}\cdot
        \frac{\gamma_{{}_P}}{(\gamma_{{}_P}+1)^2} \cdot
               \vec{v}_{{}_P}^\dagger  \cdot\vec{E}_{{}_P}
                    \cdot \biggl(
- \frac{e\cdot g}{2\cdot m}\cdot\frac{1}{\gamma_{{}_P}}
\cdot \vec{B}_{{}_P}^\dagger\cdot\vec{v}_{{}_P} \cdot \vec{s}
\nonumber\\ &&
- \frac{e\cdot(g-2)}{2\cdot m}\cdot\gamma_{{}_P}
\cdot \vec{B}_{{}_P}^\dagger\cdot\vec{v}_{{}_P} \cdot
       \vec{s}^{\,\dagger}\cdot\vec{v}_{{}_P}\cdot\vec{v}_{{}_P}
   +\frac{e\cdot(g-2)}{2\cdot m}\cdot\gamma_{{}_P}\cdot
    \vec{s}^{\,\dagger}\cdot\vec{v}_{{}_P}\cdot\vec{B}_{{}_P}
\nonumber\\ &&
+  \frac{e}{m}\cdot \frac{1}{\gamma_{{}_P}}\cdot
             \vec{s}^{\,\dagger}\cdot\vec{B}_{{}_P}\cdot\vec{v}_{{}_P}
-\frac{e\cdot(g-2)}{2\cdot m}\cdot\gamma_{{}_P}\cdot
             \vec{s}^{\,\dagger}\cdot\vec{v}_{{}_P}\cdot
                (\vec{v}_{{}_P}\wtimes \vec{E}_{{}_P})
    +  \frac{e}{m}\cdot \frac{1}{\gamma_{{}_P}}\cdot
            (\vec{s}\wtimes \vec{E}_{{}_P})  \biggr)
\nonumber\\ &&
    +\frac{e^2}{m^2}\cdot
 \frac{1}{(\gamma_{{}_P}+1)^2} \cdot
    \vec{v}_{{}_P}^\dagger  \cdot\vec{E}_{{}_P}
           \cdot \vec{v}_{{}_P}^\dagger\cdot\vec{B}_{{}_P}\cdot\vec{s}
\nonumber\\ &&
-  \frac{e}{m}\cdot \frac{1}{\gamma_{{}_P}+1}\cdot
             \vec{s}\cdot \biggl(
\frac{e}{m}\cdot \frac{1}{\gamma_{{}_P}}\cdot
      \vec{B}_{{}_P}^\dagger\cdot\vec{E}_{{}_P}
-\frac{e}{m}\cdot \frac{1}{\gamma_{{}_P}}\cdot
   \vec{v}_{{}_P}^\dagger\cdot\vec{E}_{{}_P}
           \cdot\vec{v}_{{}_P}^\dagger\cdot\vec{B}_{{}_P} \biggr)
\nonumber\\ &&
-  \frac{e}{m}\cdot \frac{1}{\gamma_{{}_P}+1}\cdot
                \vec{v}_{{}_P}^\dagger\cdot\vec{B}_{{}_P}
\cdot\biggl\lbrack
                  \frac{e\cdot g}{2\cdot m\cdot\gamma_{{}_P}}\cdot
\biggl( \vec{s}\wtimes\vec{B}_{{}_P} +(\vec{v}_{{}_P}\wtimes
   \vec{s})\wtimes\vec{E}_{{}_P}
\biggr)
\nonumber\\&&
  + \frac{e\cdot (g-2)}{2\cdot m\cdot\gamma_{{}_P}}\cdot\biggl(
    \gamma_{{}_P}^2 \cdot
\vec{s}^{\,\dagger}\cdot\vec{v}_{{}_P}\cdot \lbrack \vec{E}_{{}_P}+
                \vec{v}_{{}_P}\wtimes\vec{B}_{{}_P}
\rbrack - \vec{v}_{{}_P}^\dagger\cdot\vec{E}_{{}_P}\cdot\vec{s}
  - \gamma_{{}_P}^2 \cdot
\vec{s}^{\,\dagger}\cdot\vec{v}_{{}_P}\cdot \vec{v}_{{}_P}^\dagger\cdot
                  \vec{E}_{{}_P}\cdot \vec{v}_{{}_P}
\biggr)  \biggr\rbrack
\nonumber\\
&=&
%A
   -\frac{e^2}{m^2} \cdot \frac{1}{(\gamma_{{}_P}+1)^2}
 \cdot\vec{v}_{{}_P}^\dagger\cdot\vec{E}_{{}_P}\cdot
                 \vec{s}^{\,\dagger}\cdot\vec{B}_{{}_P}
                    \cdot\vec{v}_{{}_P}
%B
+   \frac{e^2}{m^2} \cdot
\frac{\gamma_{{}_P}^2+\gamma_{{}_P}+1}{\gamma_{{}_P}\cdot
                                          (\gamma_{{}_P}+1)}
 \cdot\vec{s}^{\,\dagger}\cdot\vec{B}_{{}_P}\cdot
           \lbrack \vec{v}_{{}_P}\wtimes\vec{B}_{{}_P}
 + \vec{E}_{{}_P} \rbrack
\nonumber\\&&\qquad
%C
-   \frac{e^2\cdot g}{2\cdot m^2} \cdot
      \frac{1}{(\gamma_{{}_P}+1)^2}\cdot
  \vec{E}_{{}_P}^\dagger\cdot\vec{s}\cdot\vec{v}_{{}_P}^\dagger\cdot
                \vec{B}_{{}_P}\cdot\vec{v}_{{}_P}
\nonumber\\&&\qquad
%D
+   \frac{e^2}{m^2} \cdot\lbrack \frac{g^2}{4}
-\frac{g}{2}\cdot \frac{\gamma_{{}_P}}{(\gamma_{{}_P}+1)^2}
 - \frac{\gamma_{{}_P}}{\gamma_{{}_P}+1}\rbrack\cdot
 \vec{s}^{\,\dagger}\cdot\vec{v}_{{}_P}\cdot\vec{E}_{{}_P}^\dagger\cdot
                 \vec{B}_{{}_P}\cdot\vec{v}_{{}_P}
\nonumber\\&&\qquad
%F
  - \frac{e^2}{m^2} \cdot
\frac{2\cdot\gamma_{{}_P}}{(\gamma_{{}_P}+1)^3}
 \cdot\vec{v}_{{}_P}^\dagger\cdot\vec{E}_{{}_P}\cdot
    (\vec{s}\wtimes\vec{v}_{{}_P})^\dagger\cdot
                 \vec{E}_{{}_P}\cdot\vec{v}_{{}_P}
\nonumber\\&&\qquad
%G
-   \frac{e^2}{m^2} \cdot
\frac{\gamma_{{}_P}^2+2\cdot\gamma_{{}_P}}{(\gamma_{{}_P}+1)^2}
 \cdot(\vec{s}\wtimes\vec{v}_{{}_P})^\dagger\cdot\vec{E}_{{}_P}\cdot
 (\vec{v}_{{}_P}\wtimes\vec{B}_{{}_P})
\nonumber\\&&\qquad
%H
+   \frac{e^2}{m^2} \cdot \lbrack
 \frac{g^2}{4}\cdot \frac{1}{\gamma_{{}_P}^2}
- g\cdot \frac{1}{\gamma_{{}_P}\cdot(\gamma_{{}_P}+1)}
-\frac{\gamma_{{}_P}^2+2\cdot\gamma_{{}_P}}{(\gamma_{{}_P}+1)^2}\rbrack
\cdot \vec{E}_{{}_P}^\dagger\cdot(\vec{s}\wtimes
                                    \vec{v}_{{}_P})\cdot\vec{E}_{{}_P}
\nonumber\\&&\qquad
%I
+   \frac{e^2}{m^2} \cdot \lbrack
 \frac{g^2}{4}\cdot \frac{1}{\gamma_{{}_P}^2}
   - g\cdot\frac{1}{\gamma_{{}_P}\cdot(\gamma_{{}_P}+1)}
   -\frac{\gamma_{{}_P}-2}{\gamma_{{}_P}\cdot(\gamma_{{}_P}+1)^2}\rbrack
\cdot  \vec{v}_{{}_P}^\dagger\cdot\vec{E}_{{}_P}\cdot
   (\vec{s}\wtimes \vec{E}_{{}_P})
\nonumber\\&&\qquad
%J
+   \frac{e^2}{m^2} \cdot \lbrack
 \frac{g^2}{4}\cdot \frac{1}{\gamma_{{}_P}^2}
- g \cdot \frac{1}{\gamma_{{}_P}\cdot(\gamma_{{}_P}+1)}
  - \frac{1}{\gamma_{{}_P}\cdot(\gamma_{{}_P}+1)} \rbrack
\cdot  \vec{E}_{{}_P}^\dagger\cdot\vec{B}_{{}_P}\cdot\vec{s}
\nonumber\\&&\qquad
%K
+   \frac{e^2}{m^2} \cdot \lbrack
-\frac{g^2}{4}
+\frac{g}{2} \cdot \frac{2\cdot
   \gamma_{{}_P}^3+2\cdot\gamma_{{}_P}^2+\gamma_{{}_P}-1}
 {\gamma_{{}_P}^2\cdot(\gamma_{{}_P}+1)}-1\rbrack
\cdot  \vec{s}^{\,\dagger}\cdot\vec{E}_{{}_P}\cdot\vec{B}_{{}_P}
\nonumber\\&&\qquad
%L
+   \frac{e^2}{m^2} \cdot \lbrack
 \frac{g^2}{4}\cdot \frac{1}{\gamma_{{}_P}^2}
- g\cdot \frac{1}{\gamma_{{}_P}\cdot(\gamma_{{}_P}+1)}
+\frac{1}{(\gamma_{{}_P}+1)^2}\rbrack
\cdot  \vec{E}_{{}_P}^\dagger\cdot\vec{E}_{{}_P}\cdot
                                     (\vec{v}_{{}_P}\wtimes\vec{s})
\nonumber\\&&\qquad
%M
-   \frac{e^2\cdot g}{2\cdot m^2} \cdot \frac{1}{(1+\gamma_{{}_P})^2}
\cdot\vec{s}^{\,\dagger}\cdot\vec{v}_{{}_P}\cdot
    \vec{E}_{{}_P}^\dagger\cdot\vec{v}_{{}_P}\cdot
  \vec{B}_{{}_P}
\nonumber\\&&\qquad
%N
-   \frac{2\cdot e^2}{m^2} \cdot
      \frac{\gamma_{{}_P}}{(\gamma_{{}_P}+1)^3}
\cdot \vec{s}^{\,\dagger}\cdot\vec{v}_{{}_P}\cdot
    \vec{E}_{{}_P}^\dagger\cdot\vec{v}_{{}_P}\cdot
    (\vec{E}_{{}_P}\wtimes \vec{v}_{{}_P})
\nonumber\\&&\qquad
%O
+   \frac{e^2}{m^2} \cdot \lbrack
 \frac{g^2}{4}
 - \frac{g}{2}\cdot
 \frac{2\cdot\gamma_{{}_P}^2+4\cdot\gamma_{{}_P}+1}{(\gamma_{{}_P}+1)^2}
+\frac{\gamma_{{}_P}^2+2\cdot\gamma_{{}_P}}{(\gamma_{{}_P}+1)^2} \rbrack
   \cdot (\vec{s}\wtimes\vec{v}_{{}_P})^\dagger\cdot\vec{B}_{{}_P}
  \cdot (\vec{v}_{{}_P} \wtimes\vec{E}_{{}_P})
\nonumber\\&&\qquad
%P
+   \frac{e^2}{m^2} \cdot \lbrack
 \frac{g^2}{4}\cdot
 \frac{\gamma_{{}_P}^2-1}{\gamma_{{}_P}^2}
 -  g\cdot
 \frac{\gamma_{{}_P}^2+\gamma_{{}_P}-1}
    {\gamma_{{}_P}\cdot(\gamma_{{}_P}+1)}
+\frac{\gamma_{{}_P}^2+2\cdot\gamma_{{}_P}}{(\gamma_{{}_P}+1)^2}\rbrack
   \cdot \vec{E}_{{}_P}^\dagger\cdot\vec{s}\cdot
    (\vec{v}_{{}_P} \wtimes\vec{E}_{{}_P})
\nonumber\\&&\qquad
%Q
-   \frac{e^2}{m^2} \cdot
 \frac{(g-2)^2}{4}\cdot
      \vec{B}_{{}_P}^\dagger\cdot(\vec{s} \wtimes
                                     \vec{v}_{{}_P})\cdot \vec{B}_{{}_P}
\nonumber\\ &&\qquad
%R
    +\frac{2\cdot e^2}{m^2}\cdot
  \frac{\gamma_{{}_P}}{(\gamma_{{}_P}+1)^3} \cdot
       \vec{v}_{{}_P}^\dagger\cdot\vec{E}_{{}_P}\cdot
  \vec{v}_{{}_P}^\dagger \cdot\vec{E}_{{}_P}
                              \cdot (\vec{s}\wtimes\vec{v}_{{}_P})
\nonumber\\ &&\qquad
%S
-   \frac{e^2}{m^2} \cdot \frac{1}{(\gamma_{{}_P}+1)^2}
\cdot\vec{E}_{{}_P}^\dagger\cdot(\vec{v}_{{}_P}\wtimes\vec{B}_{{}_P})
  \cdot(\vec{s}\wtimes\vec{v}_{{}_P})
\nonumber\\ &&\qquad
%T
+   \frac{e^2}{m^2} \cdot \lbrack
   \frac{g}{2}\cdot
 \frac{1}{(\gamma_{{}_P}+1)^2}
+\frac{1}{(\gamma_{{}_P}+1)^2} \rbrack
\cdot  \vec{v}_{{}_P}^\dagger\cdot\vec{E}_{{}_P}\cdot
   \vec{v}_{{}_P}^\dagger \cdot\vec{B}_{{}_P}
                              \cdot \vec{s}
\nonumber\\ &&\qquad
%U
-   \frac{e^2\cdot g}{2\cdot m^2} \cdot
 \frac{1}{\gamma_{{}_P}\cdot(\gamma_{{}_P}+1)}
\cdot  \vec{v}_{{}_P}^\dagger\cdot\vec{B}_{{}_P}\cdot
                                          (\vec{s}\wtimes \vec{B}_{{}_P})
\nonumber\\&&\qquad
%V
  - \frac{e^2\cdot (g-2)}{2\cdot m^2}\cdot
 \frac{\gamma_{{}_P}}{\gamma_{{}_P}+1}\cdot
\vec{v}_{{}_P}^\dagger\cdot\vec{B}_{{}_P}\cdot
\vec{v}_{{}_P}^\dagger\cdot\vec{s}\cdot
  \lbrack \vec{E}_{{}_P}+\vec{v}_{{}_P}\wtimes\vec{B}_{{}_P}
\rbrack   \; .
 \label{B.43}
\end{eqnarray}
\subsection*{B.9}
In this subsection I simplify $\vec{\Delta}_{17}$ and $\vec{\Delta}_{23}$
and use:
\begin{eqnarray*}
&&  \lbrace \pi_{{}_{M,j}} ,  \pi_{{}_{M,m}}\rbrace_{{}_M} =
                                 e\cdot
   \sum_{n=1}^3 \varepsilon_{jmn}\cdot B_{{}_{M,n}}
            \; , \nonumber\\
&& \sigma_k \cdot \lbrace \pi_{{}_{M,j}} ,
                                      \pi_{{}_{M,m}}\rbrace_{{}_M} =
                                 e\cdot \sigma_k\cdot
   \sum_{n=1}^3 \varepsilon_{jmn}\cdot B_{{}_{P,n}}
            \; , \qquad (j,k,m=1,2,3)
\end{eqnarray*}
which follows from (1.1),(1.2),(B.21). Thus I get by (1.3),(B.5),(B.25):
\begin{eqnarray}
&&\vec{\Delta}_{17} = \vec{\Delta}_4 - \vec{\Delta}_{14}
 = \lbrace\vec{\pi}_{{}_M},\vec{\sigma}^{\,\dagger}\cdot
                                         \vec{W}_{{}_M}\rbrace_{{}_M}-
                                                       \vec{\Delta}_{14}
\nonumber\\
&=& -\frac{e}{m}\cdot\biggl(
\lbrace\vec{\pi}_{{}_M},\frac{m}{J_{{}_M}}\rbrace_{{}_M}\cdot
   \vec{\sigma}^{\,\dagger}\cdot\vec{B}_{{}_P}
-\frac{g-2}{2}\cdot
  \lbrace\vec{\pi}_{{}_M},\frac{1}{J_{{}_M}\cdot(J_{{}_M}+m)}\cdot
  \vec{\pi}_{{}_M}^\dagger\cdot\vec{\sigma}\cdot
   \vec{\pi}_{{}_M}^\dagger\rbrace_{{}_M} \cdot\vec{B}_{{}_P}
\nonumber\\ &&\qquad
-\lbrace \vec{\pi}_{{}_M},\lbrack \frac{g}{2\cdot J_{{}_M}}
    - \frac{1}{J_{{}_M}+m}\rbrack\cdot
 (\vec{\sigma}\wtimes\vec{\pi}_{{}_M})^\dagger \rbrace_{{}_M} \cdot
                                              \vec{E}_{{}_P} \biggr)
\nonumber\\
&=&  \frac{e^2}{J_{{}_M}^2}\cdot\vec{\sigma}^{\,\dagger}\cdot
                                                    \vec{B}_{{}_P}\cdot
 (\vec{v}_{{}_P}\wtimes\vec{B}_{{}_P})
 -  \frac{e^2}{m}\cdot\frac{g-2}{2}\cdot
 \frac{2\cdot J_{{}_M}+m}{J_{{}_M}^2\cdot(J_{{}_M}+m)^2}\cdot
 \vec{\pi}_{{}_M}^\dagger\cdot\vec{\sigma}\cdot
       \vec{\pi}_{{}_M}^\dagger \cdot\vec{B}_{{}_P}\cdot
 (\vec{v}_{{}_P}\wtimes\vec{B}_{{}_P})
\nonumber\\ &&\qquad
 +  \frac{e^2}{m}\cdot\frac{g-2}{2}\cdot
 \frac{1}{J_{{}_M}\cdot(J_{{}_M}+m)}\cdot
    \vec{\pi}_{{}_M}^\dagger\cdot\vec{B}_{{}_P}\cdot
 (\vec{\sigma}\wtimes\vec{B}_{{}_P})
\nonumber\\ &&\qquad
 +  \frac{e^2}{m}\cdot
   \lbrack-\frac{g}{2\cdot J_{{}_M}^2} +
                                    \frac{1}{(J_{{}_M}+m)^2}\rbrack\cdot
 (\vec{\sigma}\wtimes\vec{\pi}_{{}_M})^\dagger \cdot \vec{E}_{{}_P}\cdot
 (\vec{v}_{{}_P}\wtimes\vec{B}_{{}_P})
\nonumber\\ &&\qquad
 +  \frac{e^2}{m}\cdot
   \lbrack \frac{g}{2\cdot J_{{}_M}} -
   \frac{1}{J_{{}_M}+m}\rbrack\cdot\lbrack
 (\vec{E}_{{}_P}\wtimes\vec{\sigma}) \wtimes  \vec{B}_{{}_P} \rbrack
\nonumber\\
&=&  \frac{e^2}{m^2}\cdot\frac{1}{\gamma_{{}_P}^2}
   \cdot(\vec{v}_{{}_P}\wtimes\vec{B}_{{}_P})\cdot
                         \vec{B}_{{}_P}^\dagger\cdot
\lbrack \frac{1}{\gamma_{{}_P}}\cdot
    \vec{s}+\frac{\gamma_{{}_P}}{\gamma_{{}_P}+1}
              \cdot \vec{s}^{\,\dagger}\cdot
  \vec{v}_{{}_P}\cdot\vec{v}_{{}_P}\rbrack
\nonumber\\ &&\qquad
 -  \frac{e^2}{m^2}\cdot\frac{g-2}{2}\cdot
 \frac{2\cdot \gamma_{{}_P}+1}{(\gamma_{{}_P}+1)^2}\cdot
 \vec{v}_{{}_P}^\dagger\cdot\vec{s}\cdot
    \vec{v}_{{}_P}^\dagger \cdot\vec{B}_{{}_P}\cdot
 (\vec{v}_{{}_P}\wtimes\vec{B}_{{}_P})
\nonumber\\ &&\qquad
 -  \frac{e^2}{m^2}\cdot\frac{g-2}{2}\cdot
 \frac{1}{\gamma_{{}_P}+1}\cdot\vec{v}_{{}_P}^\dagger\cdot
       \vec{B}_{{}_P}\cdot\vec{B}_{{}_P}\wtimes
\lbrack \frac{1}{\gamma_{{}_P}}\cdot
   \vec{s}+\frac{\gamma_{{}_P}}{\gamma_{{}_P}+1}
              \cdot \vec{s}^{\,\dagger}\cdot
   \vec{v}_{{}_P}\cdot\vec{v}_{{}_P}\rbrack
\nonumber\\ &&\qquad
 +  \frac{e^2}{m^2}\cdot
   \lbrack-\frac{g}{2\cdot \gamma_{{}_P}} +
    \frac{\gamma_{{}_P}}{(\gamma_{{}_P}+1)^2}\rbrack\cdot
 (\vec{\sigma}\wtimes\vec{v}_{{}_P})^\dagger \cdot \vec{E}_{{}_P}\cdot
 (\vec{v}_{{}_P}\wtimes\vec{B}_{{}_P})
\nonumber\\ &&\qquad
 +  \frac{e^2}{m^2}\cdot
\lbrack \frac{g}{2\cdot\gamma_{{}_P}}-\frac{1}{\gamma_{{}_P}+1}\rbrack
   \cdot\biggl(
 \vec{E}_{{}_P}^\dagger\cdot\vec{B}_{{}_P}\cdot
\lbrack \frac{1}{\gamma_{{}_P}}\cdot
   \vec{s}+\frac{\gamma_{{}_P}}{\gamma_{{}_P}+1}
              \cdot \vec{s}^{\,\dagger}\cdot
   \vec{v}_{{}_P}\cdot\vec{v}_{{}_P}\rbrack
\nonumber\\ &&\qquad
-\vec{E}_{{}_P}\cdot\vec{B}_{{}_P}^\dagger\cdot
\lbrack \frac{1}{\gamma_{{}_P}}\cdot
   \vec{s}+\frac{\gamma_{{}_P}}{\gamma_{{}_P}+1}
   \cdot \vec{s}^{\,\dagger}\cdot
                             \vec{v}_{{}_P}\cdot\vec{v}_{{}_P}\rbrack
\biggr)
\nonumber\\
&=&  \frac{e^2}{m^2}\cdot\frac{1}{\gamma_{{}_P}^3}
  \cdot\vec{B}_{{}_P}^\dagger\cdot\vec{s}\cdot
                                   (\vec{v}_{{}_P}\wtimes\vec{B}_{{}_P})
 +  \frac{e^2}{m^2}\cdot\lbrack \frac{1}{\gamma_{{}_P}}
 -  \frac{g}{2}\cdot\frac{1}{\gamma_{{}_P}+1} \rbrack\cdot
 \vec{v}_{{}_P}^\dagger\cdot\vec{s}\cdot
    \vec{v}_{{}_P}^\dagger \cdot\vec{B}_{{}_P}\cdot
                                 (\vec{v}_{{}_P}\wtimes\vec{B}_{{}_P})
\nonumber\\ &&\qquad
 -  \frac{e^2}{m^2}\cdot\frac{g-2}{2}\cdot
 \frac{1}{\gamma_{{}_P}\cdot(\gamma_{{}_P}+1)}\cdot
  \vec{v}_{{}_P}^\dagger \cdot\vec{B}_{{}_P}\cdot
                                        (\vec{B}_{{}_P}\wtimes\vec{s})
\nonumber\\ &&\qquad
 +  \frac{e^2}{m^2}\cdot
   \lbrack-\frac{g}{2\cdot \gamma_{{}_P}^2} +
    \frac{1}{(\gamma_{{}_P}+1)^2}\rbrack\cdot
 (\vec{s}\wtimes\vec{v}_{{}_P})^\dagger \cdot \vec{E}_{{}_P}\cdot
 (\vec{v}_{{}_P}\wtimes\vec{B}_{{}_P})
\nonumber\\ &&\qquad
 +  \frac{e^2}{m^2}\cdot
\lbrack \frac{g}{2\cdot\gamma_{{}_P}}-\frac{1}{\gamma_{{}_P}+1}\rbrack
   \cdot\biggl(
 \vec{E}_{{}_P}^\dagger\cdot\vec{B}_{{}_P}\cdot
\lbrack \frac{1}{\gamma_{{}_P}}\cdot
   \vec{s}+\frac{\gamma_{{}_P}}{\gamma_{{}_P}+1}
  \cdot \vec{s}^{\,\dagger}\cdot\vec{v}_{{}_P}\cdot\vec{v}_{{}_P}\rbrack
\nonumber\\ &&\qquad
-\vec{E}_{{}_P}\cdot\vec{B}_{{}_P}^\dagger\cdot
\lbrack \frac{1}{\gamma_{{}_P}}\cdot
     \vec{s}+\frac{\gamma_{{}_P}}{\gamma_{{}_P}+1}
  \cdot \vec{s}^{\,\dagger}\cdot\vec{v}_{{}_P}\cdot\vec{v}_{{}_P}\rbrack
\biggr) \; .
 \label{B.44}
\end{eqnarray}
Next I calculate by using (B.17),(B.38):
\begin{eqnarray}
&&  \vec{\Delta}_{23} =  \biggl(
   -\frac{e^2}{m^2} \cdot
\frac{\gamma^2_{{}_P}+\gamma_{{}_P}+1}{\gamma_{{}_P}^3\cdot
                                            (\gamma_{{}_P} + 1)}
         \cdot\vec{B}_{{}_P}^\dagger\cdot\vec{s}\cdot\vec{v}_{{}_P}
 \nonumber\\ &&\qquad
-\frac{e^2}{m^2}\cdot \lbrack \frac{1}{\gamma_{{}_P}\cdot(\gamma_{{}_P}
                   +1)}-\frac{g-2}{2} \rbrack
\cdot\vec{s}^{\,\dagger}\cdot\vec{v}_{{}_P}\cdot
   \vec{v}_{{}_P}^\dagger\cdot\vec{B}_{{}_P}\cdot\vec{v}_{{}_P}
 \nonumber\\ &&\qquad
-\frac{e^2}{m^2}\cdot \lbrack \frac{1}{(\gamma_{{}_P}+1)^2}
     -\frac{g}{2\cdot\gamma_{{}_P}^2}\rbrack
\cdot\vec{s}^{\,\dagger}\cdot (\vec{v}_{{}_P}\wtimes
  \vec{E}_{{}_P})\cdot\vec{v}_{{}_P}
 \nonumber\\ &&\qquad
- \frac{e^2}{m^2}\cdot \lbrack -
  \frac{2}{\gamma_{{}_P}\cdot(\gamma_{{}_P}+1)}
     +\frac{g}{2\cdot\gamma_{{}_P}^2}\rbrack
              \cdot (\vec{E}_{{}_P}\wtimes \vec{s})
 \nonumber\\ &&\qquad
-\frac{e^2}{m^2}\cdot \lbrack
  -\frac{\gamma_{{}_P}^2+2\cdot\gamma_{{}_P}}{(\gamma_{{}_P} +1)^2}+
                                                      \frac{g}{2}\rbrack
  \cdot \vec{s}^{\,\dagger}\cdot\vec{v}_{{}_P}\cdot
                    (\vec{E}_{{}_P}\wtimes \vec{v}_{{}_P})
-\frac{e^2\cdot(g-2)}{2\cdot m^2}
  \cdot\vec{s}^{\,\dagger}\cdot\vec{v}_{{}_P}\cdot\vec{B}_{{}_P}
\nonumber\\ &&\qquad
    +\frac{e^2}{m^2}\cdot \vec{E}_{{}_P}^\dagger\cdot\vec{v}_{{}_P}
  \cdot \frac{1}{(\gamma_{{}_P}+1)^2} \cdot
                                       (\vec{s}\wtimes\vec{v}_{{}_P})
 \nonumber\\ &&\qquad
+  \frac{e^2}{m^2}\cdot
   \frac{1}{\gamma_{{}_P}\cdot(\gamma_{{}_P}+1)}\cdot
             \vec{B}_{{}_P}^\dagger\cdot\vec{v}_{{}_P}\cdot
                                                \vec{s} \biggr) \wtimes
             \vec{B}_{{}_P}
\nonumber\\
&=& -\frac{e^2}{m^2} \cdot
\frac{\gamma^2_{{}_P}+\gamma_{{}_P}+1}{\gamma_{{}_P}^3\cdot
                                            (\gamma_{{}_P} + 1)}
         \cdot\vec{B}_{{}_P}^\dagger\cdot
 \vec{s}\cdot(\vec{v}_{{}_P}\wtimes\vec{B}_{{}_P})
 \nonumber\\ &&\qquad
-\frac{e^2}{m^2}\cdot
  \lbrack \frac{1}{\gamma_{{}_P}\cdot(\gamma_{{}_P}+1)}
   -\frac{g-2}{2} \rbrack \cdot\vec{s}^{\,\dagger}\cdot\vec{v}_{{}_P}
   \cdot\vec{v}_{{}_P}^\dagger\cdot\vec{B}_{{}_P}\cdot
     (\vec{v}_{{}_P}\wtimes\vec{B}_{{}_P})
 \nonumber\\ &&\qquad
-\frac{e^2}{m^2}\cdot \lbrack \frac{1}{(\gamma_{{}_P}+1)^2}
     -\frac{g}{2\cdot\gamma_{{}_P}^2}\rbrack\cdot
                      \vec{s}^{\,\dagger}\cdot
(\vec{v}_{{}_P}\wtimes\vec{E}_{{}_P})\cdot
    (\vec{v}_{{}_P}\wtimes\vec{B}_{{}_P})
 \nonumber\\ &&\qquad
- \frac{e^2}{m^2}\cdot \lbrack
   -\frac{2}{\gamma_{{}_P}\cdot(\gamma_{{}_P}+1)}
+\frac{g}{2\cdot\gamma_{{}_P}^2}\rbrack\cdot\lbrack
  \vec{E}_{{}_P}^\dagger\cdot\vec{B}_{{}_P}\cdot\vec{s} -
  \vec{B}_{{}_P}^\dagger\cdot\vec{s}\cdot\vec{E}_{{}_P} \rbrack
 \nonumber\\ &&\qquad
-\frac{e^2}{m^2}\cdot \lbrack
  -\frac{\gamma_{{}_P}^2+2\cdot\gamma_{{}_P}}{(\gamma_{{}_P} +1)^2}+
                                                      \frac{g}{2}\rbrack
  \cdot \vec{s}^{\,\dagger}\cdot\vec{v}_{{}_P}\cdot\lbrack
  \vec{E}_{{}_P}^\dagger\cdot\vec{B}_{{}_P}\cdot\vec{v}_{{}_P} -
  \vec{B}_{{}_P}^\dagger\cdot\vec{v}_{{}_P}\cdot\vec{E}_{{}_P} \rbrack
\nonumber\\ &&\qquad
    +\frac{e^2}{m^2}\cdot \frac{1}{(\gamma_{{}_P}+1)^2} \cdot
  \vec{E}_{{}_P}^\dagger\cdot\vec{v}_{{}_P}\cdot\lbrack
  \vec{s}^{\,\dagger}\cdot\vec{B}_{{}_P}\cdot\vec{v}_{{}_P} -
  \vec{B}_{{}_P}^\dagger\cdot\vec{v}_{{}_P}\cdot\vec{s} \rbrack
 \nonumber\\ &&\qquad
+  \frac{e^2}{m^2}\cdot \frac{1}{\gamma_{{}_P}\cdot
                                              (\gamma_{{}_P}+1)}\cdot
   \vec{B}_{{}_P}^\dagger\cdot\vec{v}_{{}_P}\cdot
                                      (\vec{s}\wtimes\vec{B}_{{}_P}) \; .
 \label{B.45}
\end{eqnarray}
\subsection*{B.10}
Combining (B.39),(B.43-45) the second order SG terms read as follows:
\begin{eqnarray}
&&\vec{\Delta}_{21} = \vec{\Delta}_{17}
                    + \vec{\Delta}_{22}
                    + \vec{\Delta}_{23}
\nonumber\\
&=&
%B,beta
    \frac{e^2}{m^2} \cdot
      \vec{s}^{\,\dagger}\cdot\vec{B}_{{}_P}\cdot
     \lbrack \vec{v}_{{}_P}\wtimes\vec{B}_{{}_P}
 + \vec{E}_{{}_P} \rbrack
%C
-   \frac{e^2\cdot g}{2\cdot m^2} \cdot
    \frac{1}{(\gamma_{{}_P}+1)^2}\cdot
  \vec{E}_{{}_P}^\dagger\cdot\vec{s}\cdot
  \vec{v}_{{}_P}^\dagger\cdot\vec{B}_{{}_P}\cdot\vec{v}_{{}_P}
\nonumber\\&&\qquad
%D
+   \frac{e^2}{m^2} \cdot\lbrack \frac{g^2}{4}
-\frac{g}{2}\cdot
    \frac{\gamma_{{}_P}^2+2\cdot\gamma_{{}_P}}{(\gamma_{{}_P}+1)^2}
                                                  \rbrack\cdot
 \vec{s}^{\,\dagger}\cdot\vec{v}_{{}_P}\cdot
  \vec{E}_{{}_P}^\dagger\cdot\vec{B}_{{}_P}\cdot\vec{v}_{{}_P}
\nonumber\\&&\qquad
%F
  - \frac{e^2}{m^2} \cdot
\frac{2\cdot\gamma_{{}_P}}{(\gamma_{{}_P}+1)^3}
 \cdot\vec{v}_{{}_P}^\dagger\cdot\vec{E}_{{}_P}\cdot
       (\vec{s}\wtimes\vec{v}_{{}_P})^\dagger\cdot
                 \vec{E}_{{}_P}\cdot\vec{v}_{{}_P}
\nonumber\\&&\qquad
%G
-   \frac{e^2}{m^2} \cdot
\frac{\gamma_{{}_P}^2+2\cdot\gamma_{{}_P}}{(\gamma_{{}_P}+1)^2}
 \cdot(\vec{s}\wtimes\vec{v}_{{}_P})^\dagger\cdot\vec{E}_{{}_P}\cdot
 (\vec{v}_{{}_P}\wtimes\vec{B}_{{}_P})
\nonumber\\&&\qquad
%H
+   \frac{e^2}{m^2} \cdot \lbrack
 \frac{g^2}{4}\cdot \frac{1}{\gamma_{{}_P}^2}
- g\cdot \frac{1}{\gamma_{{}_P}\cdot(\gamma_{{}_P}+1)}
-\frac{\gamma_{{}_P}^2+2\cdot\gamma_{{}_P}}{(\gamma_{{}_P}+1)^2}
                                                 \rbrack
\cdot \vec{E}_{{}_P}^\dagger\cdot(\vec{s}\wtimes
                                    \vec{v}_{{}_P})\cdot\vec{E}_{{}_P}
\nonumber\\&&\qquad
%I
+   \frac{e^2}{m^2} \cdot \lbrack
 \frac{g^2}{4}\cdot \frac{1}{\gamma_{{}_P}^2}
   - g\cdot\frac{1}{\gamma_{{}_P}\cdot(\gamma_{{}_P}+1)}
   -\frac{\gamma_{{}_P}-2}{\gamma_{{}_P}\cdot(\gamma_{{}_P}+1)^2}\rbrack
\cdot  \vec{v}_{{}_P}^\dagger\cdot\vec{E}_{{}_P}\cdot
   (\vec{s}\wtimes \vec{E}_{{}_P})
\nonumber\\&&\qquad
%J
+   \frac{e^2}{m^2} \cdot \lbrack
 \frac{g^2}{4}\cdot \frac{1}{\gamma_{{}_P}^2}
- g \cdot \frac{1}{\gamma_{{}_P}\cdot(\gamma_{{}_P}+1)}
                                        \rbrack
\cdot  \vec{E}_{{}_P}^\dagger\cdot\vec{B}_{{}_P}\cdot\vec{s}
\nonumber\\&&\qquad
%K
+   \frac{e^2}{m^2} \cdot \lbrack
-\frac{g^2}{4}
+\frac{g}{2} \cdot \frac{2\cdot\gamma_{{}_P}^3+
   2\cdot\gamma_{{}_P}^2+\gamma_{{}_P}-1}
 {\gamma_{{}_P}^2\cdot(\gamma_{{}_P}+1)}-1\rbrack
\cdot  \vec{s}^{\,\dagger}\cdot\vec{E}_{{}_P}\cdot\vec{B}_{{}_P}
\nonumber\\&&\qquad
%L
+   \frac{e^2}{m^2} \cdot \lbrack
 \frac{g^2}{4}\cdot \frac{1}{\gamma_{{}_P}^2}
- g\cdot \frac{1}{\gamma_{{}_P}\cdot(\gamma_{{}_P}+1)}
+\frac{1}{(\gamma_{{}_P}+1)^2}\rbrack
\cdot  \vec{E}_{{}_P}^\dagger\cdot\vec{E}_{{}_P}\cdot
                                     (\vec{v}_{{}_P}\wtimes\vec{s})
\nonumber\\&&\qquad
%M
-   \frac{e^2\cdot g}{2\cdot m^2} \cdot \frac{1}{(1+\gamma_{{}_P})^2}
\cdot\vec{s}^{\,\dagger}\cdot\vec{v}_{{}_P}\cdot
     \vec{E}_{{}_P}^\dagger\cdot\vec{v}_{{}_P}\cdot\vec{B}_{{}_P}
\nonumber\\&&\qquad
%N
-   \frac{2\cdot e^2}{m^2} \cdot
     \frac{\gamma_{{}_P}}{(\gamma_{{}_P}+1)^3}
\cdot \vec{s}^{\,\dagger}\cdot\vec{v}_{{}_P}\cdot
   \vec{E}_{{}_P}^\dagger\cdot\vec{v}_{{}_P}\cdot
    (\vec{E}_{{}_P}\wtimes \vec{v}_{{}_P})
\nonumber\\&&\qquad
%O
+   \frac{e^2}{m^2} \cdot \lbrack
 \frac{g^2}{4}
 - \frac{g}{2}\cdot
 \frac{2\cdot\gamma_{{}_P}^2+4\cdot\gamma_{{}_P}+1}{(\gamma_{{}_P}+1)^2}
+\frac{\gamma_{{}_P}^2+2\cdot\gamma_{{}_P}}{(\gamma_{{}_P}+1)^2} \rbrack
   \cdot (\vec{s}\wtimes\vec{v}_{{}_P})^\dagger\cdot\vec{B}_{{}_P}
  \cdot (\vec{v}_{{}_P} \wtimes\vec{E}_{{}_P})
\nonumber\\&&\qquad
%P
+   \frac{e^2}{m^2} \cdot \lbrack
 \frac{g^2}{4}\cdot
 \frac{\gamma_{{}_P}^2-1}{\gamma_{{}_P}^2}
 -  g\cdot
 \frac{\gamma_{{}_P}^2+\gamma_{{}_P}-1}
    {\gamma_{{}_P}\cdot(\gamma_{{}_P}+1)}
+\frac{\gamma_{{}_P}^2+2\cdot\gamma_{{}_P}}{(\gamma_{{}_P}+1)^2}\rbrack
 \cdot \vec{E}_{{}_P}^\dagger\cdot\vec{s}\cdot
   (\vec{v}_{{}_P} \wtimes\vec{E}_{{}_P})
\nonumber\\&&\qquad
%Q
-   \frac{e^2}{m^2} \cdot
 \frac{(g-2)^2}{4}\cdot
      \vec{B}_{{}_P}^\dagger\cdot(\vec{s} \wtimes
                                     \vec{v}_{{}_P})\cdot \vec{B}_{{}_P}
\nonumber\\ &&\qquad
%R
    +\frac{2\cdot e^2}{m^2}\cdot
    \frac{\gamma_{{}_P}}{(\gamma_{{}_P}+1)^3} \cdot
       \vec{v}_{{}_P}^\dagger\cdot\vec{E}_{{}_P}\cdot
   \vec{v}_{{}_P}^\dagger \cdot\vec{E}_{{}_P}
                              \cdot (\vec{s}\wtimes\vec{v}_{{}_P})
\nonumber\\ &&\qquad
%S
-   \frac{e^2}{m^2} \cdot \frac{1}{(\gamma_{{}_P}+1)^2}
\cdot\vec{E}_{{}_P}^\dagger\cdot(\vec{v}_{{}_P}\wtimes\vec{B}_{{}_P})
     \cdot(\vec{s}\wtimes\vec{v}_{{}_P})
\nonumber\\ &&\qquad
%T
+   \frac{e^2\cdot g}{2\cdot m^2} \cdot
 \frac{1}{(\gamma_{{}_P}+1)^2}
\cdot  \vec{v}_{{}_P}^\dagger\cdot\vec{E}_{{}_P}\cdot
           \vec{v}_{{}_P}^\dagger \cdot\vec{B}_{{}_P}
                              \cdot \vec{s} \; .
 \label{B.46}
\end{eqnarray}
With (B.46) I have an explicit form of the second order SG terms and my
remaining task of this Appendix is to reduce  them to  those in (2.10).
I abbreviate the second order SG terms of (2.10) by
\begin{eqnarray}
&&\vec{\Delta}_{24} \equiv
 \frac{e^2}{4\cdot m^2}\cdot\lbrack
   -(g-2)^2\cdot
   \vec{v}_{{}_P}^\dagger\cdot\vec{B}_{{}_P}\cdot
                                      (\vec{B}_{{}_P} \wtimes\vec{s})
   +(g-2)^2\cdot
   \vec{B}_{{}_P}^\dagger\cdot\vec{B}_{{}_P}\cdot
                                 (\vec{v}_{{}_P} \wtimes\vec{s})
\nonumber\\&&\qquad
   -(g-2)^2\cdot
   \vec{E}_{{}_P}^\dagger\cdot\vec{s}\cdot\vec{B}_{{}_P}
   +(-g^2\cdot \vec{v}_{{}_P}^\dagger\cdot\vec{v}_{{}_P}
   +2\cdot g\cdot \vec{v}_{{}_P}^\dagger\cdot\vec{v}_{{}_P} +4)\cdot
   \vec{s}^{\,\dagger}\cdot\vec{B}_{{}_P}\cdot\vec{E}_{{}_P}
\nonumber\\&&\qquad
   -(g-2)^2\cdot
   \vec{v}_{{}_P}^\dagger\cdot\vec{E}_{{}_P}\cdot
                                      (\vec{E}_{{}_P} \wtimes\vec{s})
   +(g-2)^2\cdot
       ( \vec{v}_{{}_P}\wtimes\vec{B}_{{}_P})^\dagger
 \cdot\vec{E}_{{}_P} \cdot(\vec{v}_{{}_P}\wtimes\vec{s})
\nonumber\\&&\qquad
   +(g-2)^2\cdot
       \vec{E}_{{}_P}^\dagger\cdot\vec{E}_{{}_P} \cdot
                                      (\vec{v}_{{}_P}\wtimes\vec{s})
   -(g^2-4\cdot g)\cdot
   \vec{s}^{\,\dagger}\cdot\vec{B}_{{}_P}\cdot
                               (\vec{v}_{{}_P} \wtimes\vec{B}_{{}_P})
\nonumber\\&&\qquad
   +(g-2)\cdot g\cdot
   \vec{E}_{{}_P}^\dagger\cdot\vec{B}_{{}_P}\cdot \vec{s}
   -(g-2)\cdot g\cdot
   \vec{v}_{{}_P}^\dagger\cdot\vec{E}_{{}_P}\cdot
     \vec{v}_{{}_P}^\dagger\cdot\vec{B}_{{}_P}\cdot\vec{s}
\nonumber\\&&\qquad
   +(g-2)\cdot g\cdot
   \vec{v}_{{}_P}^\dagger\cdot\vec{E}_{{}_P}\cdot
     \vec{s}^{\,\dagger}\cdot\vec{B}_{{}_P}\cdot\vec{v}_{{}_P}
   +(g-2)\cdot g\cdot
   \vec{v}_{{}_P}^\dagger\cdot\vec{s}\cdot
   \vec{v}_{{}_P}^\dagger\cdot\vec{B}_{{}_P}\cdot\vec{E}_{{}_P}
\nonumber\\&&\qquad
   -(g^2-4\cdot g)\cdot
  \vec{E}_{{}_P}^\dagger\cdot
                       ( \vec{v}_{{}_P}\wtimes\vec{s})\cdot\vec{E}_{{}_P}
   +2\cdot g\cdot
\vec{E}_{{}_P}^\dagger\cdot(\vec{v}_{{}_P}\wtimes
   \vec{s})\cdot(\vec{v}_{{}_P}\wtimes\vec{B}_{{}_P})
\rbrack   \; ,
\label{B.47}
\end{eqnarray}
                                                      Therefore the remaining
task of this Appendix is to show that the rhs of (B.46) equals
$\vec{\Delta}_{24}$.
\par Introducing the abbreviation
\begin{eqnarray}
&&\vec{\Delta}_{25} \equiv  \vec{\Delta}_{21} - \vec{\Delta}_{24}
= \frac{e^2}{4\cdot m^2}\cdot\biggl(
%A
   (g-2)^2\cdot
      \vec{s}^{\,\dagger}\cdot\vec{B}_{{}_P}\cdot
                                  (\vec{v}_{{}_P}\wtimes\vec{B}_{{}_P})
\nonumber\\&&
%B
+  (g^2-2\cdot g)\cdot
   \vec{v}_{{}_P}^\dagger\cdot\vec{v}_{{}_P}\cdot
      \vec{s}^{\,\dagger}\cdot\vec{B}_{{}_P}\cdot\vec{E}_{{}_P}
\nonumber\\&&
%C
-   2\cdot g\cdot \frac{1}{(\gamma_{{}_P}+1)^2}\cdot
  \vec{E}_{{}_P}^\dagger\cdot\vec{s}\cdot
    \vec{v}_{{}_P}^\dagger\cdot\vec{B}_{{}_P}\cdot\vec{v}_{{}_P}
%D
+  \lbrack g^2 -2\cdot g\cdot
\frac{\gamma_{{}_P}^2+2\cdot\gamma_{{}_P}}{(\gamma_{{}_P}+1)^2}
                                                 \rbrack\cdot
 \vec{s}^{\,\dagger}\cdot\vec{v}_{{}_P}\cdot
  \vec{E}_{{}_P}^\dagger\cdot\vec{B}_{{}_P}\cdot\vec{v}_{{}_P}
\nonumber\\&&
%E
  - \frac{8\cdot\gamma_{{}_P}}{(\gamma_{{}_P}+1)^3}
 \cdot\vec{v}_{{}_P}^\dagger\cdot\vec{E}_{{}_P}\cdot
              (\vec{s}\wtimes\vec{v}_{{}_P})^\dagger\cdot
                 \vec{E}_{{}_P}\cdot\vec{v}_{{}_P}
\nonumber\\&&
%F
+ \lbrack 2\cdot g  -
\frac{4\cdot\gamma_{{}_P}^2+8\cdot\gamma_{{}_P}}{(\gamma_{{}_P}+1)^2}
                                                       \rbrack
 \cdot(\vec{s}\wtimes\vec{v}_{{}_P})^\dagger\cdot\vec{E}_{{}_P}\cdot
 (\vec{v}_{{}_P}\wtimes\vec{B}_{{}_P})
\nonumber\\&&
%G
+   \lbrack g^2 \cdot \frac{1-\gamma_{{}_P}^2}{\gamma_{{}_P}^2}
+4\cdot g\cdot
\frac{\gamma_{{}_P}^2+\gamma_{{}_P}-1}{\gamma_{{}_P}\cdot
                                          (\gamma_{{}_P}+1)}
-\frac{4\cdot \gamma_{{}_P}^2+8\cdot\gamma_{{}_P}}{(\gamma_{{}_P}+1)^2}
                                                          \rbrack
\cdot \vec{E}_{{}_P}^\dagger\cdot(\vec{s}\wtimes
                                    \vec{v}_{{}_P})\cdot\vec{E}_{{}_P}
\nonumber\\&&
%H
+   \lbrack g^2 \cdot \frac{1-\gamma_{{}_P}^2}{\gamma_{{}_P}^2}
+4\cdot g\cdot
\frac{\gamma_{{}_P}^2+\gamma_{{}_P}-1}{\gamma_{{}_P}\cdot
                                          (\gamma_{{}_P}+1)}
-\frac{4\cdot\gamma_{{}_P}^3+8\cdot\gamma_{{}_P}^2+8\cdot
                                               \gamma_{{}_P}-8}
  {\gamma_{{}_P}\cdot(\gamma_{{}_P}+1)^2} \rbrack
\cdot  \vec{v}_{{}_P}^\dagger\cdot\vec{E}_{{}_P}\cdot
   (\vec{s}\wtimes \vec{E}_{{}_P})
\nonumber\\&&
%I
+   \lbrack g^2 \cdot \frac{1-\gamma_{{}_P}^2}{\gamma_{{}_P}^2}
+2\cdot g\cdot
 \frac{\gamma_{{}_P}^2+\gamma_{{}_P}-2}{\gamma_{{}_P}\cdot
                                           (\gamma_{{}_P}+1)}
   \rbrack
\cdot  \vec{E}_{{}_P}^\dagger\cdot\vec{B}_{{}_P}\cdot\vec{s}
%J
+2\cdot g\cdot
 \frac{\gamma_{{}_P}-1}{\gamma_{{}_P}^2\cdot(\gamma_{{}_P}+1)}
\cdot  \vec{s}^{\,\dagger}\cdot\vec{E}_{{}_P}\cdot\vec{B}_{{}_P}
\nonumber\\&&
%K
+   \lbrack g^2 \cdot \frac{1-\gamma_{{}_P}^2}{\gamma_{{}_P}^2}
+4\cdot g\cdot
 \frac{\gamma_{{}_P}^2+\gamma_{{}_P}-1}{\gamma_{{}_P}\cdot
                                           (\gamma_{{}_P}+1)}
 -\frac{4\cdot \gamma_{{}_P}^2+8\cdot\gamma_{{}_P}}{(\gamma_{{}_P}+1)^2}
                                                         \rbrack
\cdot  \vec{E}_{{}_P}^\dagger\cdot\vec{E}_{{}_P}\cdot
                                     (\vec{v}_{{}_P}\wtimes\vec{s})
\nonumber\\&&
%L
-  2\cdot g \cdot \frac{1}{(1+\gamma_{{}_P})^2}
\cdot\vec{s}^{\,\dagger}\cdot\vec{v}_{{}_P}\cdot
   \vec{E}_{{}_P}^\dagger\cdot\vec{v}_{{}_P}\cdot\vec{B}_{{}_P}
%M
-   8\cdot  \frac{\gamma_{{}_P}}{(\gamma_{{}_P}+1)^3}
\cdot \vec{s}^{\,\dagger}\cdot\vec{v}_{{}_P}\cdot
  \vec{E}_{{}_P}^\dagger\cdot\vec{v}_{{}_P}\cdot
    (\vec{E}_{{}_P}\wtimes \vec{v}_{{}_P})
\nonumber\\&&
%N
+  \lbrack  g^2 - 2\cdot g\cdot
 \frac{2\cdot\gamma_{{}_P}^2+4\cdot\gamma_{{}_P}+1}{(\gamma_{{}_P}+1)^2}
+\frac{4\cdot\gamma_{{}_P}^2+8\cdot\gamma_{{}_P}}{(\gamma_{{}_P}+1)^2}
                                                        \rbrack
   \cdot (\vec{s}\wtimes\vec{v}_{{}_P})^\dagger\cdot\vec{B}_{{}_P}
  \cdot (\vec{v}_{{}_P} \wtimes\vec{E}_{{}_P})
\nonumber\\&&
%O
+   \lbrack  g^2 \cdot
 \frac{\gamma_{{}_P}^2-1}{\gamma_{{}_P}^2}
 -  4\cdot g\cdot
 \frac{\gamma_{{}_P}^2+\gamma_{{}_P}-1}
    {\gamma_{{}_P}\cdot(\gamma_{{}_P}+1)}
+\frac{4\cdot\gamma_{{}_P}^2+8\cdot\gamma_{{}_P}}{(\gamma_{{}_P}+1)^2}
                                                       \rbrack
   \cdot \vec{E}_{{}_P}^\dagger\cdot\vec{s}\cdot
   (\vec{v}_{{}_P} \wtimes\vec{E}_{{}_P})
\nonumber\\&&
%P
-  (g-2)^2\cdot
 \vec{B}_{{}_P}^\dagger\cdot(\vec{s} \wtimes
                                     \vec{v}_{{}_P})\cdot \vec{B}_{{}_P}
%Q
    + 8\cdot \frac{\gamma_{{}_P}}{(\gamma_{{}_P}+1)^3} \cdot
       \vec{v}_{{}_P}^\dagger\cdot\vec{E}_{{}_P}\cdot
  \vec{v}_{{}_P}^\dagger \cdot\vec{E}_{{}_P}
                              \cdot (\vec{s}\wtimes\vec{v}_{{}_P})
\nonumber\\ &&
%R
+   \lbrack g^2 -4\cdot g
+\frac{4\cdot \gamma_{{}_P}^2+8\cdot\gamma_{{}_P}}{(\gamma_{{}_P}+1)^2}
                                                          \rbrack
\cdot\vec{E}_{{}_P}^\dagger
    \cdot(\vec{v}_{{}_P}\wtimes\vec{B}_{{}_P})\cdot
                                        (\vec{s}\wtimes\vec{v}_{{}_P})
\nonumber\\ &&
%S
+   \lbrack g^2 -2\cdot g
\cdot\frac{\gamma_{{}_P}^2+2\cdot\gamma_{{}_P}}
          {(\gamma_{{}_P}+1)^2}     \rbrack
\cdot  \vec{v}_{{}_P}^\dagger\cdot\vec{E}_{{}_P}\cdot
    \vec{v}_{{}_P}^\dagger \cdot\vec{B}_{{}_P}
                              \cdot \vec{s}
\nonumber\\ &&
%T
   +(g-2)^2\cdot
   \vec{v}_{{}_P}^\dagger\cdot\vec{B}_{{}_P}\cdot
                                      (\vec{B}_{{}_P} \wtimes\vec{s})
%U
   -(g-2)^2\cdot
   \vec{B}_{{}_P}^\dagger\cdot\vec{B}_{{}_P}\cdot
                                 (\vec{v}_{{}_P} \wtimes\vec{s})
\nonumber\\ &&
%V
   -(g-2)\cdot g\cdot
   \vec{v}_{{}_P}^\dagger\cdot\vec{E}_{{}_P}\cdot
 \vec{s}^{\,\dagger}\cdot\vec{B}_{{}_P}\cdot\vec{v}_{{}_P}
%W
   -(g-2)\cdot g\cdot
   \vec{v}_{{}_P}^\dagger\cdot\vec{s}\cdot
  \vec{v}_{{}_P}^\dagger\cdot\vec{B}_{{}_P}\cdot\vec{E}_{{}_P}
\biggr) \; ,
\label{B.48}
\end{eqnarray}
I have to show that $\vec{\Delta}_{25}$ vanishes.
One can separate $\vec{\Delta}_{25}$ into a magnetic part
$\vec{\Delta}_{26}$ plus an electric part $\vec{\Delta}_{27}$ plus a
mixed part $\vec{\Delta}_{28}$. Therefore I abbreviate:
\begin{eqnarray}
&&\vec{\Delta}_{25} \equiv  \vec{\Delta}_{26} + \vec{\Delta}_{27}
                                     + \vec{\Delta}_{28} \; ,
\label{B.49}
\end{eqnarray}
where
\begin{eqnarray}
&&\vec{\Delta}_{26} \equiv \frac{e^2}{4\cdot m^2}\cdot\biggl(
%A
   (g-2)^2\cdot
      \vec{s}^{\,\dagger}\cdot\vec{B}_{{}_P}\cdot
                                  (\vec{v}_{{}_P}\wtimes\vec{B}_{{}_P})
%P
-  (g-2)^2\cdot
      \vec{B}_{{}_P}^\dagger\cdot(\vec{s} \wtimes
                                     \vec{v}_{{}_P})\cdot \vec{B}_{{}_P}
\nonumber\\ &&\qquad
%T
   +(g-2)^2\cdot
   \vec{v}_{{}_P}^\dagger\cdot\vec{B}_{{}_P}\cdot
                                      (\vec{B}_{{}_P} \wtimes\vec{s})
%U
   -(g-2)^2\cdot
   \vec{B}_{{}_P}^\dagger\cdot\vec{B}_{{}_P}\cdot
                                 (\vec{v}_{{}_P} \wtimes\vec{s}) \biggr)
\nonumber\\
&=&  \frac{e^2\cdot(g-2)^2}{4\cdot m^2}\cdot\biggl(
%A
      \vec{s}^{\,\dagger}\cdot\vec{B}_{{}_P}\cdot
                                  (\vec{v}_{{}_P}\wtimes\vec{B}_{{}_P})
%P
-  \vec{B}_{{}_P}^\dagger\cdot(\vec{s} \wtimes
                                  \vec{v}_{{}_P})\cdot \vec{B}_{{}_P}
%T
   + \vec{v}_{{}_P}^\dagger\cdot\vec{B}_{{}_P}\cdot
                                        (\vec{B}_{{}_P} \wtimes\vec{s})
\nonumber\\ &&\qquad
%U
-\vec{B}_{{}_P}^\dagger\cdot\vec{B}_{{}_P}\cdot
   (\vec{v}_{{}_P} \wtimes\vec{s}) \biggr) \; ,
\nonumber\\
&&\vec{\Delta}_{27} \equiv \frac{e^2}{4\cdot m^2}\cdot\biggl(
%E
  - \frac{8\cdot\gamma_{{}_P}}{(\gamma_{{}_P}+1)^3}
 \cdot\vec{v}_{{}_P}^\dagger\cdot\vec{E}_{{}_P}\cdot
        (\vec{s}\wtimes\vec{v}_{{}_P})^\dagger\cdot
                 \vec{E}_{{}_P}\cdot\vec{v}_{{}_P}
\nonumber\\&&\qquad
%G
+   \lbrack g^2 \cdot \frac{1-\gamma_{{}_P}^2}{\gamma_{{}_P}^2}
+4\cdot g\cdot \frac{\gamma_{{}_P}^2+\gamma_{{}_P}-1}{\gamma_{{}_P}
     \cdot(\gamma_{{}_P}+1)}
-\frac{4\cdot \gamma_{{}_P}^2+8\cdot\gamma_{{}_P}}{(\gamma_{{}_P}+1)^2}
                                                        \rbrack
\cdot \vec{E}_{{}_P}^\dagger\cdot(\vec{s}\wtimes
                                    \vec{v}_{{}_P})\cdot\vec{E}_{{}_P}
\nonumber\\&&\qquad
%H
+   \lbrack g^2 \cdot \frac{1-\gamma_{{}_P}^2}{\gamma_{{}_P}^2}
+4\cdot g\cdot
\frac{\gamma_{{}_P}^2+\gamma_{{}_P}-1}{\gamma_{{}_P}\cdot
                                          (\gamma_{{}_P}+1)}
-\frac{4\cdot\gamma_{{}_P}^3+8\cdot\gamma_{{}_P}^2+
                                         8\cdot\gamma_{{}_P}-8}
  {\gamma_{{}_P}\cdot(\gamma_{{}_P}+1)^2} \rbrack
\cdot  \vec{v}_{{}_P}^\dagger\cdot\vec{E}_{{}_P}\cdot
   (\vec{s}\wtimes \vec{E}_{{}_P})
\nonumber\\&&\qquad
%K
+   \lbrack g^2 \cdot \frac{1-\gamma_{{}_P}^2}{\gamma_{{}_P}^2}
+4\cdot g\cdot
\frac{\gamma_{{}_P}^2+\gamma_{{}_P}-1}{\gamma_{{}_P}\cdot
                                          (\gamma_{{}_P}+1)}
-\frac{4\cdot \gamma_{{}_P}^2+8\cdot\gamma_{{}_P}}{(\gamma_{{}_P}+1)^2}
                                                        \rbrack
\cdot  \vec{E}_{{}_P}^\dagger\cdot\vec{E}_{{}_P}\cdot
                                     (\vec{v}_{{}_P}\wtimes\vec{s})
\nonumber\\&&\qquad
%M
-   8\cdot  \frac{\gamma_{{}_P}}{(\gamma_{{}_P}+1)^3}
\cdot \vec{s}^{\,\dagger}\cdot\vec{v}_{{}_P}\cdot
      \vec{E}_{{}_P}^\dagger\cdot\vec{v}_{{}_P}\cdot
    (\vec{E}_{{}_P}\wtimes \vec{v}_{{}_P})
\nonumber\\&&\qquad
%O
+   \lbrack  g^2 \cdot
 \frac{\gamma_{{}_P}^2-1}{\gamma_{{}_P}^2}
 -  4\cdot g\cdot
 \frac{\gamma_{{}_P}^2+\gamma_{{}_P}-1}
    {\gamma_{{}_P}\cdot(\gamma_{{}_P}+1)}
+\frac{4\cdot\gamma_{{}_P}^2+8\cdot\gamma_{{}_P}}{(\gamma_{{}_P}+1)^2}
                                                       \rbrack
   \cdot \vec{E}_{{}_P}^\dagger\cdot\vec{s}\cdot
  (\vec{v}_{{}_P} \wtimes\vec{E}_{{}_P})
\nonumber\\&&\qquad
%Q
    + 8\cdot \frac{\gamma_{{}_P}}{(\gamma_{{}_P}+1)^3} \cdot
       \vec{v}_{{}_P}^\dagger\cdot\vec{E}_{{}_P}\cdot
 \vec{v}_{{}_P}^\dagger \cdot\vec{E}_{{}_P}
            \cdot (\vec{s}\wtimes\vec{v}_{{}_P})\biggr) \; ,
\nonumber\\
&&\vec{\Delta}_{28} \equiv \frac{e^2}{4\cdot m^2}\cdot\biggl(
%B
   (g^2-2\cdot g)\cdot
   \vec{v}_{{}_P}^\dagger\cdot\vec{v}_{{}_P}\cdot
      \vec{s}^{\,\dagger}\cdot\vec{B}_{{}_P}\cdot\vec{E}_{{}_P}
\nonumber\\&&
%C
-   2\cdot g\cdot \frac{1}{(\gamma_{{}_P}+1)^2}\cdot
  \vec{E}_{{}_P}^\dagger\cdot\vec{s}\cdot
   \vec{v}_{{}_P}^\dagger\cdot\vec{B}_{{}_P}\cdot\vec{v}_{{}_P}
%D
+  \lbrack g^2 -2\cdot g\cdot
\frac{\gamma_{{}_P}^2+2\cdot\gamma_{{}_P}}{(\gamma_{{}_P}+1)^2}
                                                 \rbrack\cdot
 \vec{s}^{\,\dagger}\cdot\vec{v}_{{}_P}\cdot
 \vec{E}_{{}_P}^\dagger\cdot\vec{B}_{{}_P}\cdot\vec{v}_{{}_P}
\nonumber\\&&\qquad
%F
+ \lbrack 2\cdot g  -
\frac{4\cdot\gamma_{{}_P}^2+8\cdot\gamma_{{}_P}}{(\gamma_{{}_P}+1)^2}
                                                       \rbrack
 \cdot(\vec{s}\wtimes\vec{v}_{{}_P})^\dagger\cdot\vec{E}_{{}_P}\cdot
 (\vec{v}_{{}_P}\wtimes\vec{B}_{{}_P})
\nonumber\\&&\qquad
%I
+   \lbrack g^2 \cdot \frac{1-\gamma_{{}_P}^2}{\gamma_{{}_P}^2}
+2\cdot g\cdot
  \frac{\gamma_{{}_P}^2+\gamma_{{}_P}-2}{\gamma_{{}_P}\cdot
                                            (\gamma_{{}_P}+1)}
   \rbrack
\cdot  \vec{E}_{{}_P}^\dagger\cdot\vec{B}_{{}_P}\cdot\vec{s}
%J
+2\cdot g\cdot
  \frac{\gamma_{{}_P}-1}{\gamma_{{}_P}^2\cdot(\gamma_{{}_P}+1)}
\cdot  \vec{s}^{\,\dagger}\cdot\vec{E}_{{}_P}\cdot\vec{B}_{{}_P}
\nonumber\\&&\qquad
%L
-  2\cdot g \cdot \frac{1}{(1+\gamma_{{}_P})^2}
\cdot\vec{s}^{\,\dagger}\cdot\vec{v}_{{}_P}\cdot
  \vec{E}_{{}_P}^\dagger\cdot\vec{v}_{{}_P}\cdot\vec{B}_{{}_P}
\nonumber\\&&\qquad
%N
+  \lbrack  g^2 - 2\cdot g\cdot
 \frac{2\cdot\gamma_{{}_P}^2+4\cdot\gamma_{{}_P}+1}{(\gamma_{{}_P}+1)^2}
+\frac{4\cdot\gamma_{{}_P}^2+8\cdot\gamma_{{}_P}}{(\gamma_{{}_P}+1)^2}
                                                        \rbrack
   \cdot (\vec{s}\wtimes\vec{v}_{{}_P})^\dagger\cdot\vec{B}_{{}_P}
  \cdot (\vec{v}_{{}_P} \wtimes\vec{E}_{{}_P})
\nonumber\\ &&\qquad
%R
+   \lbrack g^2 -4\cdot g
+\frac{4\cdot \gamma_{{}_P}^2+8\cdot\gamma_{{}_P}}{(\gamma_{{}_P}+1)^2}
                                                          \rbrack
\cdot\vec{E}_{{}_P}^\dagger\cdot(\vec{v}_{{}_P}\wtimes\vec{B}_{{}_P})
 \cdot(\vec{s}\wtimes\vec{v}_{{}_P})
\nonumber\\ &&\qquad
%S
+   \lbrack g^2 -2\cdot g
\cdot\frac{\gamma_{{}_P}^2+2\cdot\gamma_{{}_P}}
          {(\gamma_{{}_P}+1)^2}  \rbrack
\cdot  \vec{v}_{{}_P}^\dagger\cdot\vec{E}_{{}_P}\cdot
  \vec{v}_{{}_P}^\dagger \cdot\vec{B}_{{}_P}
                              \cdot \vec{s}
\nonumber\\ &&\qquad
%V
   -(g-2)\cdot g\cdot
   \vec{v}_{{}_P}^\dagger\cdot\vec{E}_{{}_P}\cdot
   \vec{s}^{\,\dagger}\cdot\vec{B}_{{}_P}\cdot\vec{v}_{{}_P}
%W
   -(g-2)\cdot g\cdot
   \vec{v}_{{}_P}^\dagger\cdot\vec{s}\cdot
   \vec{v}_{{}_P}^\dagger\cdot\vec{B}_{{}_P}\cdot\vec{E}_{{}_P}
\biggr) \; .
\label{B.50}
\end{eqnarray}
Hence the remaining task of this Appendix is to show that
$\vec{\Delta}_{26}, \vec{\Delta}_{27},\vec{\Delta}_{28}$ vanish.
\subsection*{B.11}
In this subsection I simplify $\vec{\Delta}_{26}$. I use the same
method as in section A.2. If
$\vec{v}_{{}_P},\vec{B}_{{}_P}$ are linearly independent,
I have the following 3 linearly independent vectors:
\begin{eqnarray*}
&& \vec{v}_{{}_P},\vec{B}_{{}_P},\vec{v}_{{}_P}\wtimes\vec{B}_{{}_P} \; .
\end{eqnarray*}
One concludes from (B.50):
\begin{eqnarray}
&& \vec{v}_{{}_P}^\dagger\cdot \vec{\Delta}_{26} = 0 \; ,
\nonumber\\
&& \vec{B}_{{}_P}^\dagger\cdot \vec{\Delta}_{26} = 0 \; ,
\nonumber\\
&& (\vec{v}_{{}_P}\wtimes\vec{B}_{{}_P})^\dagger\cdot \vec{\Delta}_{26} =
     \frac{e^2\cdot(g-2)^2}{4\cdot m^2}\cdot
   (\vec{v}_{{}_P}\wtimes\vec{B}_{{}_P})^\dagger\cdot
    \biggl(
%A
      \vec{s}^{\,\dagger}\cdot\vec{B}_{{}_P}\cdot
                                  (\vec{v}_{{}_P}\wtimes\vec{B}_{{}_P})
%P
-  \vec{B}_{{}_P}^\dagger\cdot(\vec{s} \wtimes
                                  \vec{v}_{{}_P})\cdot \vec{B}_{{}_P}
\nonumber\\ &&\qquad
%T
   + \vec{v}_{{}_P}^\dagger\cdot\vec{B}_{{}_P}\cdot
                                        (\vec{B}_{{}_P} \wtimes\vec{s})
%U
 - \vec{B}_{{}_P}^\dagger\cdot\vec{B}_{{}_P}\cdot
                                 (\vec{v}_{{}_P} \wtimes\vec{s}) \biggr)
 =   \frac{e^2\cdot(g-2)^2}{4\cdot m^2}\cdot\lbrack
\vec{s}^{\,\dagger}\cdot\vec{B}_{{}_P}\cdot\vec{B}_{{}_P}^\dagger\cdot
                                               \vec{B}_{{}_P}\cdot
    \vec{v}_{{}_P}^\dagger\cdot\vec{v}_{{}_P}
\nonumber\\ &&
-\vec{s}^{\,\dagger}\cdot\vec{B}_{{}_P}\cdot
    \vec{v}_{{}_P}^\dagger\cdot\vec{B}_{{}_P}\cdot
    \vec{v}_{{}_P}^\dagger\cdot\vec{B}_{{}_P}
+\vec{v}_{{}_P}^\dagger\cdot\vec{B}_{{}_P}\cdot
          \vec{v}_{{}_P}^\dagger\cdot\vec{B}_{{}_P}\cdot
          \vec{s}^{\,\dagger}\cdot\vec{B}_{{}_P}
-\vec{v}_{{}_P}^\dagger\cdot\vec{B}_{{}_P}\cdot
          \vec{B}_{{}_P}^\dagger\cdot\vec{B}_{{}_P}\cdot
          \vec{s}^{\,\dagger}\cdot\vec{v}_{{}_P}
\nonumber\\ &&\qquad
-\vec{B}_{{}_P}^\dagger\cdot\vec{B}_{{}_P}\cdot
       \vec{v}_{{}_P}^\dagger\cdot\vec{v}_{{}_P}\cdot
       \vec{B}_{{}_P}^\dagger\cdot\vec{s}
+\vec{B}_{{}_P}^\dagger\cdot\vec{B}_{{}_P}\cdot
                               \vec{v}_{{}_P}^\dagger\cdot\vec{s}\cdot
       \vec{v}_{{}_P}^\dagger\cdot\vec{B}_{{}_P} \rbrack = 0 \; ,
\label{B.51}
\end{eqnarray}
so that $\vec{\Delta}_{26}$ vanishes, if
$\vec{v}_{{}_P},\vec{B}_{{}_P}$ are linearly independent.
\par If $\vec{v}_{{}_P},\vec{B}_{{}_P}$ are linearly dependent, then
$\vec{v}_{{}_P} \wtimes\vec{B}_{{}_P}$ vanishes so that one gets from (B.50):
\begin{eqnarray}
&&\vec{\Delta}_{26}
 =   \frac{e^2\cdot(g-2)^2}{4\cdot m^2}\cdot\biggl(
     \vec{v}_{{}_P}^\dagger\cdot\vec{B}_{{}_P}\cdot
                                        (\vec{B}_{{}_P} \wtimes\vec{s})
-\vec{B}_{{}_P}^\dagger\cdot\vec{B}_{{}_P}\cdot
   (\vec{v}_{{}_P} \wtimes\vec{s}) \biggr) = 0 \; .
\label{B.52}
\end{eqnarray}
One observes by (B.52) that $\vec{\Delta}_{26}$
vanishes if $\vec{v}_{{}_P}=0$ or $\vec{B}_{{}_P}=0$.
It remains to consider
the subcase  with: $\vec{B}_{{}_P}=\lambda\cdot\vec{v}_{{}_P}$, where
$\lambda$ is a constant which balances the dimensions. Then the rhs of
(B.52)
vanishes. Therefore $\vec{\Delta}_{26}$ vanishes in any case.
\subsection*{B.12}
In this subsection I simplify $\vec{\Delta}_{27}$. If
$\vec{v}_{{}_P},\vec{E}_{{}_P}$ are linearly independent,
one has the following 3 linearly independent vectors:
\begin{eqnarray*}
&& \vec{v}_{{}_P},\vec{E}_{{}_P},\vec{v}_{{}_P}\wtimes\vec{E}_{{}_P} \; .
\end{eqnarray*}
One concludes from (B.50):
\begin{eqnarray}
&& \vec{v}_{{}_P}^\dagger\cdot \vec{\Delta}_{27} =
  \frac{e^2}{4\cdot m^2}\cdot \biggl(
%E
  - \frac{8\cdot\gamma_{{}_P}}{(\gamma_{{}_P}+1)^3}
 \cdot\vec{v}_{{}_P}^\dagger\cdot\vec{E}_{{}_P}\cdot
                  (\vec{s}\wtimes\vec{v}_{{}_P})^\dagger\cdot
  \vec{E}_{{}_P}\cdot\vec{v}_{{}_P}^\dagger\cdot\vec{v}_{{}_P}
\nonumber\\&&\qquad
%G
+   \lbrack g^2 \cdot \frac{1-\gamma_{{}_P}^2}{\gamma_{{}_P}^2}
+4\cdot g\cdot
\frac{\gamma_{{}_P}^2+\gamma_{{}_P}-1}{\gamma_{{}_P}\cdot
                                          (\gamma_{{}_P}+1)}
-\frac{4\cdot \gamma_{{}_P}^2+8\cdot\gamma_{{}_P}}{(\gamma_{{}_P}+1)^2}
                                                          \rbrack
\cdot \vec{E}_{{}_P}^\dagger\cdot(\vec{s}\wtimes
                  \vec{v}_{{}_P})\cdot\vec{E}_{{}_P}^\dagger
              \cdot\vec{v}_{{}_P}
\nonumber\\&&
%H
+   \lbrack g^2 \cdot \frac{1-\gamma_{{}_P}^2}{\gamma_{{}_P}^2}
+4\cdot g\cdot
\frac{\gamma_{{}_P}^2+\gamma_{{}_P}-1}{\gamma_{{}_P}\cdot
                                          (\gamma_{{}_P}+1)}
\nonumber\\&&\qquad
-\frac{4\cdot \gamma_{{}_P}^3+8\cdot\gamma_{{}_P}^2+
                                          8\cdot\gamma_{{}_P}-8}
  {\gamma_{{}_P}\cdot(\gamma_{{}_P}+1)^2} \rbrack
\cdot  \vec{v}_{{}_P}^\dagger\cdot\vec{E}_{{}_P}\cdot
  (\vec{s}\wtimes \vec{E}_{{}_P})^\dagger\cdot
       \vec{v}_{{}_P}  \biggr)
\nonumber\\
&=&  \frac{e^2}{4\cdot m^2}\cdot
\vec{v}_{{}_P}^\dagger\cdot\vec{E}_{{}_P}\cdot
                 (\vec{s}\wtimes\vec{v}_{{}_P})^\dagger\cdot
                 \vec{E}_{{}_P}\cdot
    \frac{1}{\gamma_{{}_P}\cdot(\gamma_{{}_P}+1)^3}\cdot\lbrack
 - 8\cdot \gamma_{{}_P}^2\cdot\vec{v}_{{}_P}^\dagger\cdot\vec{v}_{{}_P}
\nonumber\\&&\qquad
 - (4\cdot \gamma_{{}_P}^2+8\cdot\gamma_{{}_P})\cdot
    (\gamma_{{}_P}^2+\gamma_{{}_P})
+ (4\cdot \gamma_{{}_P}^3+8\cdot\gamma_{{}_P}^2+ 8\cdot\gamma_{{}_P}-8)
                                                        \cdot
   (\gamma_{{}_P}+1)\rbrack   = 0 \; ,
\nonumber\\
&& \vec{E}_{{}_P}^\dagger\cdot \vec{\Delta}_{27} =
  \frac{e^2}{4\cdot m^2}\cdot\biggl(
%E
  - \frac{8\cdot\gamma_{{}_P}}{(\gamma_{{}_P}+1)^3}
 \cdot\vec{v}_{{}_P}^\dagger\cdot\vec{E}_{{}_P}\cdot
        (\vec{s}\wtimes\vec{v}_{{}_P})^\dagger\cdot
                 \vec{E}_{{}_P}\cdot\vec{v}_{{}_P}^\dagger\cdot
                 \vec{E}_{{}_P}
\nonumber\\&&\qquad
%G
+   \lbrack g^2 \cdot \frac{1-\gamma_{{}_P}^2}{\gamma_{{}_P}^2}
+4\cdot g\cdot \frac{\gamma_{{}_P}^2+\gamma_{{}_P}-1}{\gamma_{{}_P}
     \cdot(\gamma_{{}_P}+1)}
-\frac{4\cdot \gamma_{{}_P}^2+8\cdot\gamma_{{}_P}}{(\gamma_{{}_P}+1)^2}
                                                        \rbrack
\cdot \vec{E}_{{}_P}^\dagger\cdot(\vec{s}\wtimes
    \vec{v}_{{}_P})\cdot \vec{E}_{{}_P}^\dagger\cdot \vec{E}_{{}_P}
\nonumber\\&&\qquad
%K
+   \lbrack g^2 \cdot \frac{1-\gamma_{{}_P}^2}{\gamma_{{}_P}^2}
+4\cdot g\cdot
\frac{\gamma_{{}_P}^2+\gamma_{{}_P}-1}{\gamma_{{}_P}\cdot
                                          (\gamma_{{}_P}+1)}
-\frac{4\cdot \gamma_{{}_P}^2+8\cdot\gamma_{{}_P}}{(\gamma_{{}_P}+1)^2}
                                                        \rbrack
\cdot  \vec{E}_{{}_P}^\dagger\cdot\vec{E}_{{}_P}\cdot
   (\vec{v}_{{}_P}\wtimes\vec{s})^\dagger\cdot\vec{E}_{{}_P}
\nonumber\\&&\qquad
%Q
    + 8\cdot \frac{\gamma_{{}_P}}{(\gamma_{{}_P}+1)^3} \cdot
       \vec{v}_{{}_P}^\dagger\cdot\vec{E}_{{}_P}\cdot
 \vec{v}_{{}_P}^\dagger \cdot\vec{E}_{{}_P}
   \cdot (\vec{s}\wtimes\vec{v}_{{}_P})^\dagger \cdot\vec{E}_{{}_P}
                  \biggr)   =  0 \; ,
\nonumber\\
&& (\vec{v}_{{}_P}\wtimes\vec{E}_{{}_P})^\dagger\cdot
\vec{\Delta}_{27} =
     \frac{e^2}{4\cdot m^2}\cdot
   (\vec{v}_{{}_P}\wtimes\vec{E}_{{}_P})^\dagger\cdot
   \biggl(
%H
    \lbrack g^2 \cdot \frac{1-\gamma_{{}_P}^2}{\gamma_{{}_P}^2}
+4\cdot g\cdot
\frac{\gamma_{{}_P}^2+\gamma_{{}_P}-1}{\gamma_{{}_P}\cdot
                                          (\gamma_{{}_P}+1)}
\nonumber\\&&\qquad
-\frac{4\cdot\gamma_{{}_P}^3+8\cdot\gamma_{{}_P}^2+
                                         8\cdot\gamma_{{}_P}-8}
  {\gamma_{{}_P}\cdot(\gamma_{{}_P}+1)^2} \rbrack
\cdot  \vec{v}_{{}_P}^\dagger\cdot\vec{E}_{{}_P}\cdot
   (\vec{s}\wtimes \vec{E}_{{}_P})
\nonumber\\&&\qquad
%K
+   \lbrack g^2 \cdot \frac{1-\gamma_{{}_P}^2}{\gamma_{{}_P}^2}
+4\cdot g\cdot
\frac{\gamma_{{}_P}^2+\gamma_{{}_P}-1}{\gamma_{{}_P}\cdot
                                          (\gamma_{{}_P}+1)}
-\frac{4\cdot \gamma_{{}_P}^2+8\cdot\gamma_{{}_P}}{(\gamma_{{}_P}+1)^2}
                                                        \rbrack
\cdot  \vec{E}_{{}_P}^\dagger\cdot\vec{E}_{{}_P}\cdot
                                     (\vec{v}_{{}_P}\wtimes\vec{s})
\nonumber\\&&\qquad
%M
-   8\cdot  \frac{\gamma_{{}_P}}{(\gamma_{{}_P}+1)^3}
\cdot \vec{s}^{\,\dagger}\cdot\vec{v}_{{}_P}\cdot
      \vec{E}_{{}_P}^\dagger\cdot\vec{v}_{{}_P}\cdot
    (\vec{E}_{{}_P}\wtimes \vec{v}_{{}_P})
\nonumber\\&&\qquad
%O
+   \lbrack  g^2 \cdot
 \frac{\gamma_{{}_P}^2-1}{\gamma_{{}_P}^2}
 -  4\cdot g\cdot
 \frac{\gamma_{{}_P}^2+\gamma_{{}_P}-1}
    {\gamma_{{}_P}\cdot(\gamma_{{}_P}+1)}
+\frac{4\cdot\gamma_{{}_P}^2+8\cdot\gamma_{{}_P}}{(\gamma_{{}_P}+1)^2}
                                                       \rbrack
   \cdot \vec{E}_{{}_P}^\dagger\cdot\vec{s}\cdot
  (\vec{v}_{{}_P} \wtimes\vec{E}_{{}_P})
\nonumber\\&&\qquad
%Q
    + 8\cdot \frac{\gamma_{{}_P}}{(\gamma_{{}_P}+1)^3} \cdot
       \vec{v}_{{}_P}^\dagger\cdot\vec{E}_{{}_P}\cdot
 \vec{v}_{{}_P}^\dagger \cdot\vec{E}_{{}_P}
            \cdot (\vec{s}\wtimes\vec{v}_{{}_P})\biggr)
\nonumber\\
&=&  \frac{e^2}{4\cdot m^2}\cdot
   \biggl(
    \lbrack -\frac{4\cdot\gamma_{{}_P}^3+8\cdot\gamma_{{}_P}^2+
       8\cdot\gamma_{{}_P}-8}{\gamma_{{}_P}\cdot(\gamma_{{}_P}+1)^2}
+\frac{4\cdot \gamma_{{}_P}^2+8\cdot\gamma_{{}_P}}{(\gamma_{{}_P}+1)^2}
\nonumber\\&&\qquad
+   8\cdot  \frac{\gamma_{{}_P}}{(\gamma_{{}_P}+1)^3}
\cdot \vec{v}_{{}_P}^\dagger\cdot\vec{v}_{{}_P} \rbrack
\cdot  \vec{E}_{{}_P}^\dagger\cdot\vec{E}_{{}_P}\cdot
       \vec{v}_{{}_P}^\dagger\cdot\vec{E}_{{}_P}\cdot
       \vec{v}_{{}_P}^\dagger\cdot\vec{s}
  + \lbrack  \frac{4\cdot\gamma_{{}_P}^3+8\cdot\gamma_{{}_P}^2+
       8\cdot\gamma_{{}_P}-8}{\gamma_{{}_P}\cdot(\gamma_{{}_P}+1)^2}
\nonumber\\&&\qquad
-\frac{4\cdot \gamma_{{}_P}^2+8\cdot\gamma_{{}_P}}{(\gamma_{{}_P}+1)^2}
-   8\cdot  \frac{\gamma_{{}_P}}{(\gamma_{{}_P}+1)^3}
\cdot \vec{v}_{{}_P}^\dagger\cdot\vec{v}_{{}_P} \rbrack
\cdot  \vec{v}_{{}_P}^\dagger\cdot\vec{E}_{{}_P}\cdot
       \vec{v}_{{}_P}^\dagger\cdot\vec{E}_{{}_P}\cdot
       \vec{E}_{{}_P}^\dagger\cdot\vec{s}
                                               \biggr) = 0 \; ,
\nonumber\\&&
\label{B.53}
\end{eqnarray}
so that $\vec{\Delta}_{27}$ vanishes, if
$\vec{v}_{{}_P},\vec{E}_{{}_P}$ are linearly independent.
\par If $\vec{v}_{{}_P},\vec{E}_{{}_P}$ are linearly dependent, then
$\vec{v}_{{}_P} \wtimes\vec{E}_{{}_P}$ vanishes so that one gets from (B.50):
\begin{eqnarray}
&&\vec{\Delta}_{27}  =     \frac{e^2}{4\cdot m^2}\cdot\biggl(
    \lbrack g^2 \cdot \frac{1-\gamma_{{}_P}^2}{\gamma_{{}_P}^2}
+4\cdot g\cdot
\frac{\gamma_{{}_P}^2+\gamma_{{}_P}-1}{\gamma_{{}_P}\cdot
                                          (\gamma_{{}_P}+1)}
\nonumber\\&&\qquad
-\frac{4\cdot\gamma_{{}_P}^3+8\cdot\gamma_{{}_P}^2+
                                         8\cdot\gamma_{{}_P}-8}
  {\gamma_{{}_P}\cdot(\gamma_{{}_P}+1)^2} \rbrack
\cdot  \vec{v}_{{}_P}^\dagger\cdot\vec{E}_{{}_P}\cdot
   (\vec{s}\wtimes \vec{E}_{{}_P})
+   \lbrack g^2 \cdot \frac{1-\gamma_{{}_P}^2}{\gamma_{{}_P}^2}
+4\cdot g\cdot
\frac{\gamma_{{}_P}^2+\gamma_{{}_P}-1}{\gamma_{{}_P}\cdot
                                          (\gamma_{{}_P}+1)}
\nonumber\\&&\qquad
-\frac{4\cdot \gamma_{{}_P}^2+8\cdot\gamma_{{}_P}}{(\gamma_{{}_P}+1)^2}
                                                        \rbrack
\cdot  \vec{E}_{{}_P}^\dagger\cdot\vec{E}_{{}_P}\cdot
                                     (\vec{v}_{{}_P}\wtimes\vec{s})
    + 8\cdot \frac{\gamma_{{}_P}}{(\gamma_{{}_P}+1)^3} \cdot
       \vec{v}_{{}_P}^\dagger\cdot\vec{E}_{{}_P}\cdot
 \vec{v}_{{}_P}^\dagger \cdot\vec{E}_{{}_P}
            \cdot (\vec{s}\wtimes\vec{v}_{{}_P})\biggr)  \; .
\nonumber\\&&
\label{B.54}
\end{eqnarray}
One observes by (B.54) that $\vec{\Delta}_{27}$
vanishes if $\vec{v}_{{}_P}=0$ or $\vec{E}_{{}_P}=0$.
It remains to consider
the subcase  with: $\vec{E}_{{}_P}=\lambda\cdot\vec{v}_{{}_P}$, where
$\lambda$ is a constant which balances
                                 the dimensions. Then (B.54) reads as:
\begin{eqnarray*}
&&\vec{\Delta}_{27}  =     \frac{e^2\cdot\lambda^2}{4\cdot m^2}\cdot
(\vec{s}\wtimes\vec{v}_{{}_P})\cdot\biggl(
\lbrack
-\frac{4\cdot\gamma_{{}_P}^3+8\cdot\gamma_{{}_P}^2+
                                         8\cdot\gamma_{{}_P}-8}
  {\gamma_{{}_P}\cdot(\gamma_{{}_P}+1)^2}
+\frac{4\cdot \gamma_{{}_P}^2+8\cdot\gamma_{{}_P}}{(\gamma_{{}_P}+1)^2}
                                                        \rbrack
\cdot  \vec{v}_{{}_P}^\dagger\cdot\vec{v}_{{}_P}
\nonumber\\&&\qquad
    + 8\cdot \frac{\gamma_{{}_P}}{(\gamma_{{}_P}+1)^3} \cdot
       \vec{v}_{{}_P}^\dagger\cdot\vec{v}_{{}_P}\cdot
 \vec{v}_{{}_P}^\dagger \cdot\vec{v}_{{}_P}\biggr) \nonumber\\
&=&  \frac{e^2\cdot\lambda^2}{4\cdot m^2}\cdot \vec{v}_{{}_P}^\dagger\cdot
\vec{v}_{{}_P}\cdot
(\vec{s}\wtimes\vec{v}_{{}_P})\cdot\biggl(
-\frac{8\cdot\gamma_{{}_P}-8}{\gamma_{{}_P}\cdot(\gamma_{{}_P}+1)^2}
    + 8\cdot \frac{\gamma_{{}_P}}{(\gamma_{{}_P}+1)^3} \cdot
 \vec{v}_{{}_P}^\dagger \cdot\vec{v}_{{}_P}\biggr) = 0 \; .
\end{eqnarray*}
Therefore $\vec{\Delta}_{27}$ vanishes in any case.
\subsection*{B.13}
In this subsection I simplify $\vec{\Delta}_{28}$. If
$\vec{v}_{{}_P},\vec{E}_{{}_P}$ are linearly independent,
one has the following 3 linearly independent vectors:
\begin{eqnarray*}
&& \vec{v}_{{}_P},\vec{E}_{{}_P},\vec{v}_{{}_P}\wtimes\vec{E}_{{}_P} \; .
\end{eqnarray*}
One concludes from (B.50):
\begin{eqnarray}
&&\vec{v}_{{}_P}^\dagger\cdot
  \vec{\Delta}_{28} = \frac{e^2}{4\cdot m^2}\cdot\biggl(
%B
   (g^2-2\cdot g)\cdot
   \vec{v}_{{}_P}^\dagger\cdot\vec{v}_{{}_P}\cdot
      \vec{s}^{\,\dagger}\cdot\vec{B}_{{}_P}\cdot
                    \vec{E}_{{}_P}^\dagger\cdot\vec{v}_{{}_P}
\nonumber\\&&\qquad
%C
-   2\cdot g\cdot \frac{1}{(\gamma_{{}_P}+1)^2}\cdot
 \vec{E}_{{}_P}^\dagger\cdot\vec{s}\cdot
    \vec{v}_{{}_P}^\dagger\cdot\vec{B}_{{}_P}\cdot
  \vec{v}_{{}_P}^\dagger\cdot\vec{v}_{{}_P}
\nonumber\\&&\qquad
%D
+  \lbrack g^2 -2\cdot g\cdot
\frac{\gamma_{{}_P}^2+2\cdot\gamma_{{}_P}}{(\gamma_{{}_P}+1)^2} \rbrack
                                                        \cdot
 \vec{s}^{\,\dagger}\cdot\vec{v}_{{}_P}\cdot
      \vec{E}_{{}_P}^\dagger\cdot\vec{B}_{{}_P}\cdot
         \vec{v}_{{}_P}^\dagger\cdot\vec{v}_{{}_P}
\nonumber\\&&\qquad
%I
+   \lbrack g^2 \cdot \frac{1-\gamma_{{}_P}^2}{\gamma_{{}_P}^2}
+2\cdot g\cdot
\frac{\gamma_{{}_P}^2+\gamma_{{}_P}-2}{\gamma_{{}_P}\cdot
                                          (\gamma_{{}_P}+1)}
   \rbrack
\cdot  \vec{E}_{{}_P}^\dagger\cdot\vec{B}_{{}_P}\cdot
                      \vec{s}^{\,\dagger}\cdot\vec{v}_{{}_P}
\nonumber\\&&\qquad
%J
+2\cdot g\cdot
\frac{\gamma_{{}_P}-1}{\gamma_{{}_P}^2\cdot(\gamma_{{}_P}+1)}
\cdot  \vec{s}^{\,\dagger}\cdot\vec{E}_{{}_P}\cdot
     \vec{B}_{{}_P}^\dagger\cdot\vec{v}_{{}_P}
%L
-  2\cdot g \cdot \frac{1}{(1+\gamma_{{}_P})^2}
\cdot \vec{s}^{\,\dagger}\cdot\vec{v}_{{}_P}\cdot
       \vec{E}_{{}_P}^\dagger\cdot\vec{v}_{{}_P}\cdot
      \vec{B}_{{}_P}^\dagger\cdot\vec{v}_{{}_P}
\nonumber\\&&\qquad
%S
+   \lbrack g^2 -2\cdot g
\cdot\frac{\gamma_{{}_P}^2+2\cdot\gamma_{{}_P}}
          {(\gamma_{{}_P}+1)^2}   \rbrack
\cdot  \vec{v}_{{}_P}^\dagger\cdot\vec{E}_{{}_P}\cdot
         \vec{v}_{{}_P}^\dagger \cdot\vec{B}_{{}_P}
             \cdot \vec{s}^{\,\dagger}\cdot\vec{v}_{{}_P}
\nonumber\\ &&\qquad
%V
   -(g-2)\cdot g\cdot
   \vec{v}_{{}_P}^\dagger\cdot\vec{E}_{{}_P}\cdot
      \vec{s}^{\,\dagger}\cdot\vec{B}_{{}_P}\cdot
   \vec{v}_{{}_P}^\dagger\cdot\vec{v}_{{}_P}
%W
   -(g-2)\cdot g\cdot
   \vec{v}_{{}_P}^\dagger\cdot\vec{s}\cdot
       \vec{v}_{{}_P}^\dagger\cdot\vec{B}_{{}_P}\cdot
   \vec{E}_{{}_P}^\dagger\cdot\vec{v}_{{}_P} \biggr)  = 0 \; ,
\nonumber\\
&&\vec{E}_{{}_P}^\dagger\cdot
  \vec{\Delta}_{28} = \frac{e^2}{4\cdot m^2}\cdot\biggl(
%B
   (g^2-2\cdot g)\cdot
   \vec{v}_{{}_P}^\dagger\cdot\vec{v}_{{}_P}\cdot
      \vec{s}^{\,\dagger}\cdot\vec{B}_{{}_P}\cdot
   \vec{E}_{{}_P}^\dagger\cdot\vec{E}_{{}_P}
\nonumber\\&&
%C
-   2\cdot g\cdot \frac{1}{(\gamma_{{}_P}+1)^2}\cdot
  \vec{E}_{{}_P}^\dagger\cdot\vec{s}\cdot
   \vec{v}_{{}_P}^\dagger\cdot\vec{B}_{{}_P}\cdot
   \vec{v}_{{}_P}^\dagger\cdot\vec{E}_{{}_P}
\nonumber\\&&
%D
+  \lbrack g^2 -2\cdot g\cdot
\frac{\gamma_{{}_P}^2+2\cdot\gamma_{{}_P}}{(\gamma_{{}_P}+1)^2}
                                                 \rbrack\cdot
 \vec{s}^{\,\dagger}\cdot\vec{v}_{{}_P}\cdot
 \vec{E}_{{}_P}^\dagger\cdot\vec{B}_{{}_P}\cdot
 \vec{v}_{{}_P}^\dagger\cdot\vec{E}_{{}_P}
\nonumber\\&&\qquad
%F
+ \lbrack 2\cdot g  -
\frac{4\cdot\gamma_{{}_P}^2+8\cdot\gamma_{{}_P}}{(\gamma_{{}_P}+1)^2}
                                                       \rbrack
 \cdot(\vec{s}\wtimes\vec{v}_{{}_P})^\dagger\cdot\vec{E}_{{}_P}\cdot
 (\vec{v}_{{}_P}\wtimes\vec{B}_{{}_P})^\dagger\cdot\vec{E}_{{}_P}
\nonumber\\&&\qquad
%I
+   \lbrack g^2 \cdot \frac{1-\gamma_{{}_P}^2}{\gamma_{{}_P}^2}
+2\cdot g\cdot
  \frac{\gamma_{{}_P}^2+\gamma_{{}_P}-2}{\gamma_{{}_P}\cdot
                                            (\gamma_{{}_P}+1)}
   \rbrack
\cdot  \vec{E}_{{}_P}^\dagger\cdot\vec{B}_{{}_P}\cdot
       \vec{E}_{{}_P}^\dagger\cdot\vec{s}
\nonumber\\&&
%J
+2\cdot g\cdot
  \frac{\gamma_{{}_P}-1}{\gamma_{{}_P}^2\cdot(\gamma_{{}_P}+1)}
\cdot  \vec{s}^{\,\dagger}\cdot\vec{E}_{{}_P}\cdot
       \vec{E}_{{}_P}^\dagger\cdot\vec{B}_{{}_P}
\nonumber\\&&\qquad
%L
-  2\cdot g \cdot \frac{1}{(1+\gamma_{{}_P})^2}
\cdot\vec{s}^{\,\dagger}\cdot\vec{v}_{{}_P}\cdot
  \vec{E}_{{}_P}^\dagger\cdot\vec{v}_{{}_P}\cdot
  \vec{E}_{{}_P}^\dagger\cdot\vec{B}_{{}_P}
\nonumber\\&&\qquad
%R
+   \lbrack g^2 -4\cdot g
+\frac{4\cdot \gamma_{{}_P}^2+8\cdot\gamma_{{}_P}}{(\gamma_{{}_P}+1)^2}
                                                          \rbrack
\cdot\vec{E}_{{}_P}^\dagger\cdot(\vec{v}_{{}_P}\wtimes\vec{B}_{{}_P})
 \cdot(\vec{s}\wtimes\vec{v}_{{}_P})^\dagger\cdot \vec{E}_{{}_P}
\nonumber\\ &&\qquad
%S
+   \lbrack g^2 -2\cdot g
\cdot\frac{\gamma_{{}_P}^2+2\cdot\gamma_{{}_P}}
          {(\gamma_{{}_P}+1)^2}  \rbrack
\cdot  \vec{v}_{{}_P}^\dagger\cdot\vec{E}_{{}_P}\cdot
  \vec{v}_{{}_P}^\dagger \cdot\vec{B}_{{}_P}\cdot
  \vec{E}_{{}_P}^\dagger \cdot\vec{s}
\nonumber\\ &&\qquad
%V
   -(g-2)\cdot g\cdot
   \vec{v}_{{}_P}^\dagger\cdot\vec{E}_{{}_P}\cdot
   \vec{s}^{\,\dagger}\cdot\vec{B}_{{}_P}\cdot
   \vec{v}_{{}_P}^\dagger\cdot\vec{E}_{{}_P}
%W
   -(g-2)\cdot g\cdot
   \vec{v}_{{}_P}^\dagger\cdot\vec{s}\cdot
   \vec{v}_{{}_P}^\dagger\cdot\vec{B}_{{}_P}\cdot
   \vec{E}_{{}_P}^\dagger\cdot\vec{E}_{{}_P}
\biggr)
\nonumber\\
&=&    \frac{e^2}{4\cdot m^2}\cdot\biggl(
%B
   (g^2-2\cdot g)\cdot
   \vec{v}_{{}_P}^\dagger\cdot\vec{v}_{{}_P}\cdot
      \vec{s}^{\,\dagger}\cdot\vec{B}_{{}_P}\cdot
   \vec{E}_{{}_P}^\dagger\cdot\vec{E}_{{}_P}
%D
+  \lbrack g^2 -2\cdot g \rbrack\cdot
 \vec{s}^{\,\dagger}\cdot\vec{v}_{{}_P}\cdot
 \vec{E}_{{}_P}^\dagger\cdot\vec{B}_{{}_P}\cdot
 \vec{v}_{{}_P}^\dagger\cdot\vec{E}_{{}_P}
\nonumber\\&&\qquad
%I
+   \lbrack g^2 - 2\cdot g \rbrack \cdot
                      \frac{1-\gamma_{{}_P}^2}{\gamma_{{}_P}^2}
\cdot  \vec{E}_{{}_P}^\dagger\cdot\vec{B}_{{}_P}\cdot
       \vec{E}_{{}_P}^\dagger\cdot\vec{s}
%R
+   \lbrack g^2 -2\cdot g \rbrack
\cdot\vec{E}_{{}_P}^\dagger\cdot(\vec{v}_{{}_P}\wtimes\vec{B}_{{}_P})
 \cdot(\vec{s}\wtimes\vec{v}_{{}_P})^\dagger\cdot \vec{E}_{{}_P}
\nonumber\\ &&\qquad
%S
+   \lbrack g^2 -2\cdot g  \rbrack
\cdot  \vec{v}_{{}_P}^\dagger\cdot\vec{E}_{{}_P}\cdot
  \vec{v}_{{}_P}^\dagger \cdot\vec{B}_{{}_P}\cdot
  \vec{E}_{{}_P}^\dagger \cdot\vec{s}
%V
   -(g-2)\cdot g\cdot
   \vec{v}_{{}_P}^\dagger\cdot\vec{E}_{{}_P}\cdot
   \vec{s}^{\,\dagger}\cdot\vec{B}_{{}_P}\cdot
   \vec{v}_{{}_P}^\dagger\cdot\vec{E}_{{}_P}
\nonumber\\ &&\qquad
%W
   -(g-2)\cdot g\cdot
   \vec{v}_{{}_P}^\dagger\cdot\vec{s}\cdot
   \vec{v}_{{}_P}^\dagger\cdot\vec{B}_{{}_P}\cdot
   \vec{E}_{{}_P}^\dagger\cdot\vec{E}_{{}_P}
\biggr)  \; .
\label{B.55}
\end{eqnarray}
To show that $\vec{E}_{{}_P}^\dagger\cdot\vec{\Delta}_{28}$ vanishes, I
introduce the abbreviation
\footnote{The nabla operator $\vec{\nabla}_B$ always acts on functions
depending on $t,\vec{B}_{{}_P},\vec{E}_{{}_P}, \vec{v}_{{}_P},\vec{s}$
and denotes the gradient w.r.t. to $\vec{B}_{{}_P}$.}
\begin{eqnarray}
&&\vec{\Delta}_{29} \equiv  \vec{\nabla}_B \lbrack
   \vec{E}_{{}_P}^\dagger\cdot \vec{\Delta}_{28} \rbrack
 =   \frac{e^2}{4\cdot m^2}\cdot  \biggl(
%B
   (g^2-2\cdot g)\cdot
   \vec{v}_{{}_P}^\dagger\cdot\vec{v}_{{}_P}\cdot
   \vec{E}_{{}_P}^\dagger\cdot\vec{E}_{{}_P}\cdot
      \vec{s}
\nonumber\\&&\qquad
%D
+  \lbrack g^2 -2\cdot g \rbrack\cdot
 \vec{s}^{\,\dagger}\cdot\vec{v}_{{}_P}\cdot
 \vec{v}_{{}_P}^\dagger\cdot\vec{E}_{{}_P}\cdot
 \vec{E}_{{}_P}
%I
+   \lbrack g^2 - 2\cdot g \rbrack \cdot
                      \frac{1-\gamma_{{}_P}^2}{\gamma_{{}_P}^2}
\cdot  \vec{E}_{{}_P}^\dagger\cdot\vec{s}
\cdot  \vec{E}_{{}_P}
\nonumber\\ &&\qquad
%R
+   \lbrack g^2 -2\cdot g \rbrack
 \cdot(\vec{s}\wtimes\vec{v}_{{}_P})^\dagger\cdot \vec{E}_{{}_P}
\cdot (\vec{E}_{{}_P}\wtimes\vec{v}_{{}_P})
%S
+   \lbrack g^2 -2\cdot g  \rbrack
\cdot  \vec{v}_{{}_P}^\dagger\cdot\vec{E}_{{}_P}\cdot
  \vec{E}_{{}_P}^\dagger \cdot\vec{s}\cdot
  \vec{v}_{{}_P}
\nonumber\\ &&\qquad
%V
   -(g-2)\cdot g\cdot
   \vec{v}_{{}_P}^\dagger\cdot\vec{E}_{{}_P}\cdot
   \vec{v}_{{}_P}^\dagger\cdot\vec{E}_{{}_P}\cdot
   \vec{s}
%W
   -(g-2)\cdot g\cdot
   \vec{v}_{{}_P}^\dagger\cdot\vec{s}\cdot
   \vec{E}_{{}_P}^\dagger\cdot\vec{E}_{{}_P}\cdot
   \vec{v}_{{}_P} \biggr)  \; ,
\label{B.56}
\end{eqnarray}
so that one gets
\begin{eqnarray}
&& \vec{v}_{{}_P}^\dagger\cdot \vec{\Delta}_{29}
 =   \frac{e^2}{4\cdot m^2}\cdot  \biggl(
%B
   (g^2-2\cdot g)\cdot
   \vec{v}_{{}_P}^\dagger\cdot\vec{v}_{{}_P}\cdot
   \vec{E}_{{}_P}^\dagger\cdot\vec{E}_{{}_P}\cdot
   \vec{v}_{{}_P}^\dagger\cdot\vec{s}
\nonumber\\&&\qquad
%D
+  \lbrack g^2 -2\cdot g \rbrack\cdot
 \vec{s}^{\,\dagger}\cdot\vec{v}_{{}_P}\cdot
 \vec{v}_{{}_P}^\dagger\cdot\vec{E}_{{}_P}\cdot
 \vec{v}_{{}_P}^\dagger\cdot\vec{E}_{{}_P}
%I
+   \lbrack g^2 - 2\cdot g \rbrack \cdot
                      \frac{1-\gamma_{{}_P}^2}{\gamma_{{}_P}^2}
\cdot  \vec{E}_{{}_P}^\dagger\cdot\vec{s}
\cdot  \vec{E}_{{}_P}^\dagger\cdot \vec{v}_{{}_P}
\nonumber\\ &&\qquad
%S
+   \lbrack g^2 -2\cdot g  \rbrack
\cdot  \vec{v}_{{}_P}^\dagger\cdot\vec{E}_{{}_P}\cdot
  \vec{E}_{{}_P}^\dagger \cdot\vec{s}\cdot
       \vec{v}_{{}_P}^\dagger\cdot\vec{v}_{{}_P}
%V
   -(g-2)\cdot g\cdot
   \vec{v}_{{}_P}^\dagger\cdot\vec{E}_{{}_P}\cdot
   \vec{v}_{{}_P}^\dagger\cdot\vec{E}_{{}_P}\cdot
   \vec{v}_{{}_P}^\dagger\cdot\vec{s}
\nonumber\\ &&\qquad
%W
   -(g-2)\cdot g\cdot
   \vec{v}_{{}_P}^\dagger\cdot\vec{s}\cdot
   \vec{E}_{{}_P}^\dagger\cdot\vec{E}_{{}_P}\cdot
   \vec{v}_{{}_P}^\dagger\cdot\vec{v}_{{}_P}
                  \biggr) = 0 \; ,
\nonumber\\
&& \vec{E}_{{}_P}^\dagger\cdot \vec{\Delta}_{29}
 =   \frac{e^2}{4\cdot m^2}\cdot  \biggl(
%B
   (g^2-2\cdot g)\cdot
   \vec{v}_{{}_P}^\dagger\cdot\vec{v}_{{}_P}\cdot
   \vec{E}_{{}_P}^\dagger\cdot\vec{E}_{{}_P}\cdot
   \vec{E}_{{}_P}^\dagger\cdot\vec{s}
\nonumber\\&&\qquad
%D
+  \lbrack g^2 -2\cdot g \rbrack\cdot
 \vec{s}^{\,\dagger}\cdot\vec{v}_{{}_P}\cdot
 \vec{v}_{{}_P}^\dagger\cdot\vec{E}_{{}_P}\cdot
 \vec{E}_{{}_P}^\dagger\cdot\vec{E}_{{}_P}
%I
+   \lbrack g^2 - 2\cdot g \rbrack \cdot
                      \frac{1-\gamma_{{}_P}^2}{\gamma_{{}_P}^2}
\cdot  \vec{E}_{{}_P}^\dagger\cdot\vec{s}\cdot
 \vec{E}_{{}_P}^\dagger\cdot\vec{E}_{{}_P}
\nonumber\\ &&\qquad
%S
+   \lbrack g^2 -2\cdot g  \rbrack
\cdot  \vec{v}_{{}_P}^\dagger\cdot\vec{E}_{{}_P}\cdot
  \vec{E}_{{}_P}^\dagger \cdot\vec{s}\cdot
   \vec{v}_{{}_P}^\dagger\cdot\vec{E}_{{}_P}
%V
   -(g-2)\cdot g\cdot
   \vec{v}_{{}_P}^\dagger\cdot\vec{E}_{{}_P}\cdot
   \vec{v}_{{}_P}^\dagger\cdot\vec{E}_{{}_P}\cdot
   \vec{E}_{{}_P}^\dagger\cdot\vec{s}
\nonumber\\ &&\qquad
%W
   -(g-2)\cdot g\cdot
   \vec{v}_{{}_P}^\dagger\cdot\vec{s}\cdot
   \vec{E}_{{}_P}^\dagger\cdot\vec{E}_{{}_P}\cdot
   \vec{E}_{{}_P}^\dagger\cdot\vec{v}_{{}_P} \biggr) = 0 \; ,
\nonumber\\
&& (\vec{v}_{{}_P}\wtimes\vec{E}_{{}_P})^\dagger\cdot
\vec{\Delta}_{29} =
     \frac{e^2}{4\cdot m^2}\cdot
   (\vec{v}_{{}_P}\wtimes\vec{E}_{{}_P})^\dagger\cdot
                                  \biggl(
%B
   (g^2-2\cdot g)\cdot
   \vec{v}_{{}_P}^\dagger\cdot\vec{v}_{{}_P}\cdot
   \vec{E}_{{}_P}^\dagger\cdot\vec{E}_{{}_P}\cdot
      \vec{s}
\nonumber\\&&
%R
+   \lbrack g^2 -2\cdot g \rbrack
 \cdot(\vec{s}\wtimes\vec{v}_{{}_P})^\dagger\cdot \vec{E}_{{}_P}
\cdot (\vec{E}_{{}_P}\wtimes\vec{v}_{{}_P})
%V
   -(g-2)\cdot g\cdot
   \vec{v}_{{}_P}^\dagger\cdot\vec{E}_{{}_P}\cdot
   \vec{v}_{{}_P}^\dagger\cdot\vec{E}_{{}_P}\cdot
   \vec{s}  \biggr)  = 0 \; . \qquad
\label{B.57}
\end{eqnarray}
Hence $\vec{\Delta}_{29}$ vanishes, if
$\vec{v}_{{}_P},\vec{E}_{{}_P}$ are linearly independent.
Combining this with (B.56) and using the fact that
$\vec{\Delta}_{28}$ is linear in $\vec{B}_{{}_P}$, one concludes:
\begin{eqnarray}
&& \vec{E}_{{}_P}^\dagger\cdot \vec{\Delta}_{28} = 0 \; ,
\label{B.58}
\end{eqnarray}
if $\vec{v}_{{}_P},\vec{E}_{{}_P}$ are linearly independent.
\par Next I calculate
\begin{eqnarray}
&& (\vec{v}_{{}_P}\wtimes\vec{E}_{{}_P})^\dagger\cdot
\vec{\Delta}_{28} =
     \frac{e^2}{4\cdot m^2}\cdot
   (\vec{v}_{{}_P}\wtimes\vec{E}_{{}_P})^\dagger\cdot \biggl(
%F
  \lbrack 2\cdot g
\nonumber\\&&\qquad
- \frac{4\cdot\gamma_{{}_P}^2+8\cdot\gamma_{{}_P}}{(\gamma_{{}_P}+1)^2}
 \rbrack
 \cdot(\vec{s}\wtimes\vec{v}_{{}_P})^\dagger\cdot\vec{E}_{{}_P}\cdot
 (\vec{v}_{{}_P}\wtimes\vec{B}_{{}_P})
\nonumber\\&&\qquad
%I
+   \lbrack g^2 \cdot \frac{1-\gamma_{{}_P}^2}{\gamma_{{}_P}^2}
+2\cdot g\cdot
  \frac{\gamma_{{}_P}^2+\gamma_{{}_P}-2}{\gamma_{{}_P}\cdot
                                            (\gamma_{{}_P}+1)}
   \rbrack
\cdot  \vec{E}_{{}_P}^\dagger\cdot\vec{B}_{{}_P}\cdot\vec{s}
%J
+2\cdot g\cdot
  \frac{\gamma_{{}_P}-1}{\gamma_{{}_P}^2\cdot(\gamma_{{}_P}+1)}
\cdot  \vec{s}^{\,\dagger}\cdot\vec{E}_{{}_P}\cdot\vec{B}_{{}_P}
\nonumber\\&&\qquad
%L
-  2\cdot g \cdot \frac{1}{(1+\gamma_{{}_P})^2}
\cdot\vec{s}^{\,\dagger}\cdot\vec{v}_{{}_P}\cdot
  \vec{E}_{{}_P}^\dagger\cdot\vec{v}_{{}_P}\cdot\vec{B}_{{}_P}
\nonumber\\&&\qquad
%N
+  \lbrack  g^2 - 2\cdot g\cdot
 \frac{2\cdot\gamma_{{}_P}^2+4\cdot\gamma_{{}_P}+1}{(\gamma_{{}_P}+1)^2}
+\frac{4\cdot\gamma_{{}_P}^2+8\cdot\gamma_{{}_P}}{(\gamma_{{}_P}+1)^2}
                                                        \rbrack
   \cdot (\vec{s}\wtimes\vec{v}_{{}_P})^\dagger\cdot\vec{B}_{{}_P}
  \cdot (\vec{v}_{{}_P} \wtimes\vec{E}_{{}_P})
\nonumber\\ &&\qquad
%R
+   \lbrack g^2 -4\cdot g
+\frac{4\cdot \gamma_{{}_P}^2+8\cdot\gamma_{{}_P}}{(\gamma_{{}_P}+1)^2}
                                                          \rbrack
\cdot\vec{E}_{{}_P}^\dagger\cdot(\vec{v}_{{}_P}\wtimes\vec{B}_{{}_P})
 \cdot(\vec{s}\wtimes\vec{v}_{{}_P})
\nonumber\\ &&\qquad
%S
+   \lbrack g^2 -2\cdot g
\cdot\frac{\gamma_{{}_P}^2+2\cdot\gamma_{{}_P}}
          {(\gamma_{{}_P}+1)^2}  \rbrack
\cdot  \vec{v}_{{}_P}^\dagger\cdot\vec{E}_{{}_P}\cdot
  \vec{v}_{{}_P}^\dagger \cdot\vec{B}_{{}_P}
                              \cdot \vec{s}
\biggr)
\nonumber\\
&=&  \frac{e^2}{4\cdot m^2}\cdot \biggl(
    \lbrack g^2 \cdot \frac{1-\gamma_{{}_P}^2}{\gamma_{{}_P}^2}
       +  2\cdot g \cdot \vec{v}_{{}_P}^\dagger\cdot\vec{v}_{{}_P}
+2\cdot g\cdot
  \frac{\gamma_{{}_P}^2+\gamma_{{}_P}-2}{\gamma_{{}_P}\cdot
                                            (\gamma_{{}_P}+1)}
\nonumber\\&&\qquad
 -   \vec{v}_{{}_P}^\dagger\cdot\vec{v}_{{}_P}\cdot
\frac{4\cdot\gamma_{{}_P}^2+8\cdot\gamma_{{}_P}}{(\gamma_{{}_P}+1)^2}
   \rbrack \cdot  \vec{E}_{{}_P}^\dagger\cdot\vec{B}_{{}_P}\cdot
   (\vec{v}_{{}_P}\wtimes\vec{E}_{{}_P})^\dagger\cdot \vec{s}
\nonumber\\ &&\qquad
+   \lbrack g^2 -2\cdot g \cdot
 \frac{2\cdot\gamma_{{}_P}^2+4\cdot\gamma_{{}_P}+1}{(\gamma_{{}_P}+1)^2}
+\frac{4\cdot\gamma_{{}_P}^2+8\cdot\gamma_{{}_P}}{(\gamma_{{}_P}+1)^2}
  \rbrack
\cdot  \vec{v}_{{}_P}^\dagger\cdot\vec{E}_{{}_P}\cdot
  \vec{v}_{{}_P}^\dagger \cdot\vec{B}_{{}_P}\cdot
   (\vec{v}_{{}_P}\wtimes\vec{E}_{{}_P})^\dagger\cdot \vec{s}
\nonumber\\ &&\qquad
+   \lbrack g^2\cdot \vec{v}_{{}_P}^\dagger \cdot\vec{v}_{{}_P}
+2\cdot g\cdot
  \frac{\gamma_{{}_P}-1}{\gamma_{{}_P}^2\cdot(\gamma_{{}_P}+1)}
 -4\cdot g\cdot \vec{v}_{{}_P}^\dagger \cdot\vec{v}_{{}_P}
\nonumber\\ &&\qquad
 +   \vec{v}_{{}_P}^\dagger\cdot\vec{v}_{{}_P}\cdot
\frac{4\cdot\gamma_{{}_P}^2+8\cdot\gamma_{{}_P}}{(\gamma_{{}_P}+1)^2}
\rbrack \cdot  \vec{E}_{{}_P}^\dagger\cdot\vec{s}\cdot
   (\vec{v}_{{}_P}\wtimes\vec{E}_{{}_P})^\dagger\cdot\vec{B}_{{}_P}
\nonumber\\ &&\qquad
+   \lbrack -g^2 +4\cdot g
-  2\cdot g \cdot \frac{1}{(1+\gamma_{{}_P})^2}
-\frac{4\cdot \gamma_{{}_P}^2+8\cdot\gamma_{{}_P}}{(\gamma_{{}_P}+1)^2}
                                                          \rbrack
        \cdot  \vec{E}_{{}_P}^\dagger\cdot\vec{v}_{{}_P}\cdot
               \vec{v}_{{}_P}^\dagger\cdot\vec{s}\cdot
     \vec{B}_{{}_P}^\dagger\cdot(\vec{v}_{{}_P}\wtimes\vec{E}_{{}_P})
\nonumber\\&&\qquad
+  \lbrack  g^2 - 2\cdot g\cdot
 \frac{2\cdot\gamma_{{}_P}^2+4\cdot\gamma_{{}_P}+1}{(\gamma_{{}_P}+1)^2}
+\frac{4\cdot\gamma_{{}_P}^2+8\cdot\gamma_{{}_P}}{(\gamma_{{}_P}+1)^2}
                                                        \rbrack
   \cdot (\vec{s}\wtimes\vec{v}_{{}_P})^\dagger\cdot\vec{B}_{{}_P}
  \cdot \lbrack \vec{v}_{{}_P}^\dagger\cdot\vec{v}_{{}_P}\cdot
               \vec{E}_{{}_P}^\dagger\cdot\vec{E}_{{}_P}
\nonumber\\&&\qquad
            -   \vec{v}_{{}_P}^\dagger\cdot\vec{E}_{{}_P}\cdot
   \vec{v}_{{}_P}^\dagger\cdot\vec{E}_{{}_P} \rbrack \biggr) \; .
\label{B.59}
\end{eqnarray}
To show that $(\vec{v}_{{}_P}\wtimes\vec{E}_{{}_P})^\dagger\cdot
\vec{\Delta}_{28}$ vanishes, I introduce the abbreviation
\begin{eqnarray}
&&\vec{\Delta}_{30} \equiv  \vec{\nabla}_B \lbrack
   (\vec{v}_{{}_P}\wtimes\vec{E}_{{}_P})^\dagger\cdot
   \vec{\Delta}_{28} \rbrack
 =   \frac{e^2}{4\cdot m^2}\cdot  \biggl(
    \lbrack g^2 \cdot \frac{1-\gamma_{{}_P}^2}{\gamma_{{}_P}^2}
       +  2\cdot g \cdot \vec{v}_{{}_P}^\dagger\cdot\vec{v}_{{}_P}
+2\cdot g\cdot
  \frac{\gamma_{{}_P}^2+\gamma_{{}_P}-2}{\gamma_{{}_P}\cdot
                                            (\gamma_{{}_P}+1)}
\nonumber\\&&\qquad
 -   \vec{v}_{{}_P}^\dagger\cdot\vec{v}_{{}_P}\cdot
\frac{4\cdot\gamma_{{}_P}^2+8\cdot\gamma_{{}_P}}{(\gamma_{{}_P}+1)^2}
   \rbrack \cdot  (\vec{v}_{{}_P}\wtimes\vec{E}_{{}_P})^\dagger\cdot
   \vec{s} \cdot  \vec{E}_{{}_P}
\nonumber\\ &&\qquad
+   \lbrack g^2 -2\cdot g \cdot
 \frac{2\cdot\gamma_{{}_P}^2+4\cdot\gamma_{{}_P}+1}{(\gamma_{{}_P}+1)^2}
+\frac{4\cdot\gamma_{{}_P}^2+8\cdot\gamma_{{}_P}}{(\gamma_{{}_P}+1)^2}
  \rbrack
\cdot  \vec{v}_{{}_P}^\dagger\cdot\vec{E}_{{}_P}\cdot
(\vec{v}_{{}_P}\wtimes\vec{E}_{{}_P})^\dagger\cdot\vec{s}\cdot
  \vec{v}_{{}_P}
\nonumber\\ &&\qquad
+   \lbrack g^2\cdot \vec{v}_{{}_P}^\dagger \cdot\vec{v}_{{}_P}
+2\cdot g\cdot
  \frac{\gamma_{{}_P}-1}{\gamma_{{}_P}^2\cdot(\gamma_{{}_P}+1)}
 -4\cdot g\cdot \vec{v}_{{}_P}^\dagger \cdot\vec{v}_{{}_P}
\nonumber\\ &&\qquad
 +   \vec{v}_{{}_P}^\dagger\cdot\vec{v}_{{}_P}\cdot
\frac{4\cdot\gamma_{{}_P}^2+8\cdot\gamma_{{}_P}}{(\gamma_{{}_P}+1)^2}
\rbrack \cdot  \vec{E}_{{}_P}^\dagger\cdot\vec{s}\cdot
   (\vec{v}_{{}_P}\wtimes\vec{E}_{{}_P})
\nonumber\\ &&\qquad
+   \lbrack -g^2 +4\cdot g
-  2\cdot g \cdot \frac{1}{(1+\gamma_{{}_P})^2}
-\frac{4\cdot \gamma_{{}_P}^2+8\cdot\gamma_{{}_P}}{(\gamma_{{}_P}+1)^2}
                                                          \rbrack
        \cdot  \vec{E}_{{}_P}^\dagger\cdot\vec{v}_{{}_P}\cdot
               \vec{v}_{{}_P}^\dagger\cdot\vec{s}\cdot
                    (\vec{v}_{{}_P}\wtimes\vec{E}_{{}_P})
\nonumber\\&&\qquad
+  \lbrack  g^2 - 2\cdot g\cdot
 \frac{2\cdot\gamma_{{}_P}^2+4\cdot\gamma_{{}_P}+1}{(\gamma_{{}_P}+1)^2}
+\frac{4\cdot\gamma_{{}_P}^2+8\cdot\gamma_{{}_P}}{(\gamma_{{}_P}+1)^2}
                                                        \rbrack
  \cdot \lbrack \vec{v}_{{}_P}^\dagger\cdot\vec{v}_{{}_P}\cdot
               \vec{E}_{{}_P}^\dagger\cdot\vec{E}_{{}_P}
\nonumber\\&&\qquad
            -   \vec{v}_{{}_P}^\dagger\cdot\vec{E}_{{}_P}\cdot
   \vec{v}_{{}_P}^\dagger\cdot\vec{E}_{{}_P} \rbrack
   \cdot (\vec{s}\wtimes\vec{v}_{{}_P}) \biggr) \; ,
\label{B.60}
\end{eqnarray}
so that one gets
\begin{eqnarray}
&& \vec{v}_{{}_P}^\dagger\cdot \vec{\Delta}_{30}
 =   \frac{e^2}{4\cdot m^2}\cdot  \biggl(
    \lbrack g^2 \cdot \frac{1-\gamma_{{}_P}^2}{\gamma_{{}_P}^2}
       +  2\cdot g \cdot \vec{v}_{{}_P}^\dagger\cdot\vec{v}_{{}_P}
+2\cdot g\cdot
  \frac{\gamma_{{}_P}^2+\gamma_{{}_P}-2}{\gamma_{{}_P}\cdot
                                            (\gamma_{{}_P}+1)}
\nonumber\\&&\qquad
 -   \vec{v}_{{}_P}^\dagger\cdot\vec{v}_{{}_P}\cdot
\frac{4\cdot\gamma_{{}_P}^2+8\cdot\gamma_{{}_P}}{(\gamma_{{}_P}+1)^2}
   \rbrack \cdot  (\vec{v}_{{}_P}\wtimes\vec{E}_{{}_P})^\dagger\cdot
   \vec{s} \cdot   \vec{v}_{{}_P}^\dagger\cdot\vec{E}_{{}_P}
\nonumber\\ &&
+   \lbrack g^2 -2\cdot g \cdot
 \frac{2\cdot\gamma_{{}_P}^2+4\cdot\gamma_{{}_P}+1}{(\gamma_{{}_P}+1)^2}
+\frac{4\cdot\gamma_{{}_P}^2+8\cdot\gamma_{{}_P}}{(\gamma_{{}_P}+1)^2}
  \rbrack
\cdot  \vec{v}_{{}_P}^\dagger\cdot\vec{E}_{{}_P}\cdot
(\vec{v}_{{}_P}\wtimes\vec{E}_{{}_P})^\dagger\cdot\vec{s}\cdot
       \vec{v}_{{}_P}^\dagger\cdot\vec{v}_{{}_P} \biggr) = 0 \; ,
\nonumber\\
&& \vec{E}_{{}_P}^\dagger\cdot \vec{\Delta}_{30}
 =   \frac{e^2}{4\cdot m^2}\cdot  \biggl(
    \lbrack g^2 \cdot \frac{1-\gamma_{{}_P}^2}{\gamma_{{}_P}^2}
       +  2\cdot g \cdot \vec{v}_{{}_P}^\dagger\cdot\vec{v}_{{}_P}
+2\cdot g\cdot
  \frac{\gamma_{{}_P}^2+\gamma_{{}_P}-2}{\gamma_{{}_P}\cdot
                                            (\gamma_{{}_P}+1)}
\nonumber\\&&\qquad
 -   \vec{v}_{{}_P}^\dagger\cdot\vec{v}_{{}_P}\cdot
\frac{4\cdot\gamma_{{}_P}^2+8\cdot\gamma_{{}_P}}{(\gamma_{{}_P}+1)^2}
   \rbrack \cdot  (\vec{v}_{{}_P}\wtimes\vec{E}_{{}_P})^\dagger\cdot
   \vec{s} \cdot  \vec{E}_{{}_P}^\dagger\cdot\vec{E}_{{}_P}
\nonumber\\ &&\qquad
+   \lbrack g^2 -2\cdot g \cdot
 \frac{2\cdot\gamma_{{}_P}^2+4\cdot\gamma_{{}_P}+1}{(\gamma_{{}_P}+1)^2}
+\frac{4\cdot\gamma_{{}_P}^2+8\cdot\gamma_{{}_P}}{(\gamma_{{}_P}+1)^2}
  \rbrack
\cdot  \vec{v}_{{}_P}^\dagger\cdot\vec{E}_{{}_P}\cdot
(\vec{v}_{{}_P}\wtimes\vec{E}_{{}_P})^\dagger\cdot\vec{s}\cdot
       \vec{v}_{{}_P}^\dagger\cdot\vec{E}_{{}_P}
\nonumber\\ &&\qquad
+  \lbrack  g^2 - 2\cdot g\cdot
 \frac{2\cdot\gamma_{{}_P}^2+4\cdot\gamma_{{}_P}+1}{(\gamma_{{}_P}+1)^2}
+\frac{4\cdot\gamma_{{}_P}^2+8\cdot\gamma_{{}_P}}{(\gamma_{{}_P}+1)^2}
                                                        \rbrack
  \cdot \lbrack \vec{v}_{{}_P}^\dagger\cdot\vec{v}_{{}_P}\cdot
               \vec{E}_{{}_P}^\dagger\cdot\vec{E}_{{}_P}
\nonumber\\&&\qquad
            -   \vec{v}_{{}_P}^\dagger\cdot\vec{E}_{{}_P}\cdot
   \vec{v}_{{}_P}^\dagger\cdot\vec{E}_{{}_P} \rbrack
   \cdot (\vec{s}\wtimes\vec{v}_{{}_P})^\dagger\cdot\vec{E}_{{}_P}
                                       \biggr) = 0 \; ,
\nonumber\\
&& (\vec{v}_{{}_P}\wtimes\vec{E}_{{}_P})^\dagger\cdot \vec{\Delta}_{30}
 =   \frac{e^2}{4\cdot m^2}\cdot
   (\vec{v}_{{}_P}\wtimes\vec{E}_{{}_P})^\dagger\cdot \biggl(
    \lbrack g^2\cdot \vec{v}_{{}_P}^\dagger \cdot\vec{v}_{{}_P}
+2\cdot g\cdot
  \frac{\gamma_{{}_P}-1}{\gamma_{{}_P}^2\cdot(\gamma_{{}_P}+1)}
\nonumber\\ &&\qquad
 -4\cdot g\cdot \vec{v}_{{}_P}^\dagger \cdot\vec{v}_{{}_P}
 +   \vec{v}_{{}_P}^\dagger\cdot\vec{v}_{{}_P}\cdot
\frac{4\cdot\gamma_{{}_P}^2+8\cdot\gamma_{{}_P}}{(\gamma_{{}_P}+1)^2}
\rbrack \cdot  \vec{E}_{{}_P}^\dagger\cdot\vec{s}\cdot
   (\vec{v}_{{}_P}\wtimes\vec{E}_{{}_P})
\nonumber\\ &&\qquad
+   \lbrack -g^2
                 +4\cdot g
-  2\cdot g \cdot \frac{1}{(1+\gamma_{{}_P})^2}
-\frac{4\cdot \gamma_{{}_P}^2+8\cdot\gamma_{{}_P}}{(\gamma_{{}_P}+1)^2}
                                                          \rbrack
        \cdot  \vec{E}_{{}_P}^\dagger\cdot\vec{v}_{{}_P}\cdot
               \vec{v}_{{}_P}^\dagger\cdot\vec{s}\cdot
                    (\vec{v}_{{}_P}\wtimes\vec{E}_{{}_P})
\nonumber\\&&\qquad
+  \lbrack  g^2
                - 2\cdot g\cdot
 \frac{2\cdot\gamma_{{}_P}^2+4\cdot\gamma_{{}_P}+1}{(\gamma_{{}_P}+1)^2}
+\frac{4\cdot\gamma_{{}_P}^2+8\cdot\gamma_{{}_P}}{(\gamma_{{}_P}+1)^2}
                                                        \rbrack
  \cdot \lbrack \vec{v}_{{}_P}^\dagger\cdot\vec{v}_{{}_P}\cdot
               \vec{E}_{{}_P}^\dagger\cdot\vec{E}_{{}_P}
\nonumber\\&&\qquad
            -   \vec{v}_{{}_P}^\dagger\cdot\vec{E}_{{}_P}\cdot
   \vec{v}_{{}_P}^\dagger\cdot\vec{E}_{{}_P} \rbrack
   \cdot (\vec{s}\wtimes\vec{v}_{{}_P}) \biggr)
\nonumber\\
&=&   \frac{e^2}{4\cdot m^2}\cdot \biggl(
    \lbrack 2\cdot g\cdot
  \frac{\gamma_{{}_P}-1}{\gamma_{{}_P}^2\cdot(\gamma_{{}_P}+1)}
 -4\cdot g\cdot \vec{v}_{{}_P}^\dagger \cdot\vec{v}_{{}_P}
\nonumber\\&&
                + 2\cdot g\cdot
                \vec{v}_{{}_P}^\dagger \cdot\vec{v}_{{}_P}\cdot
 \frac{2\cdot\gamma_{{}_P}^2+4\cdot\gamma_{{}_P}+1}{(\gamma_{{}_P}+1)^2}
  \rbrack\cdot
     \vec{v}_{{}_P}^\dagger\cdot\vec{v}_{{}_P}\cdot
               \vec{E}_{{}_P}^\dagger\cdot\vec{s}\cdot
     \vec{E}_{{}_P}^\dagger\cdot\vec{E}_{{}_P}
\nonumber\\&&\qquad
+\lbrack -2\cdot g\cdot
  \frac{\gamma_{{}_P}-1}{\gamma_{{}_P}^2\cdot(\gamma_{{}_P}+1)}
 +4\cdot g\cdot \vec{v}_{{}_P}^\dagger \cdot\vec{v}_{{}_P}
\nonumber\\&&\qquad
   - 2\cdot g\cdot \vec{v}_{{}_P}^\dagger \cdot\vec{v}_{{}_P}\cdot
 \frac{2\cdot\gamma_{{}_P}^2+4\cdot\gamma_{{}_P}+1}{(\gamma_{{}_P}+1)^2}
\rbrack \cdot
     \vec{v}_{{}_P}^\dagger\cdot\vec{E}_{{}_P}\cdot
     \vec{v}_{{}_P}^\dagger\cdot\vec{E}_{{}_P}\cdot
               \vec{E}_{{}_P}^\dagger\cdot\vec{s}
\nonumber\\&&
+ \lbrack  4\cdot g
-  2\cdot g \cdot \frac{1}{(1+\gamma_{{}_P})^2}
                - 2\cdot g\cdot
 \frac{2\cdot\gamma_{{}_P}^2+4\cdot\gamma_{{}_P}+1}{(\gamma_{{}_P}+1)^2}
\rbrack\cdot
     \vec{v}_{{}_P}^\dagger\cdot\vec{v}_{{}_P}\cdot
     \vec{v}_{{}_P}^\dagger\cdot\vec{E}_{{}_P}\cdot
     \vec{E}_{{}_P}^\dagger\cdot\vec{E}_{{}_P}\cdot
               \vec{v}_{{}_P}^\dagger\cdot\vec{s}
\nonumber\\&&
+ \lbrack -4\cdot g
+  2\cdot g \cdot \frac{1}{(1+\gamma_{{}_P})^2}
                + 2\cdot g\cdot
 \frac{2\cdot\gamma_{{}_P}^2+4\cdot\gamma_{{}_P}+1}{(\gamma_{{}_P}+1)^2}
\rbrack\cdot
     \vec{v}_{{}_P}^\dagger\cdot\vec{E}_{{}_P}\cdot
     \vec{v}_{{}_P}^\dagger\cdot\vec{E}_{{}_P}\cdot
     \vec{v}_{{}_P}^\dagger\cdot\vec{E}_{{}_P}\cdot
               \vec{v}_{{}_P}^\dagger\cdot\vec{s} \biggr)
\nonumber\\
&=&   0 \; .
\label{B.61}
\end{eqnarray}
Hence $\vec{\Delta}_{30}$ vanishes, if
$\vec{v}_{{}_P},\vec{E}_{{}_P}$ are linearly independent.
Combining this with (B.60) and using the fact that
$\vec{\Delta}_{28}$ is linear in $\vec{B}_{{}_P}$, one concludes:
\begin{eqnarray}
&& (\vec{v}_{{}_P}\wtimes\vec{E}_{{}_P})^\dagger\cdot
\vec{\Delta}_{28} = 0 \; ,
\label{B.62}
\end{eqnarray}
if $\vec{v}_{{}_P},\vec{E}_{{}_P}$ are linearly independent.
Collecting (B.55),(B.58),(B.62) one concludes that
$\vec{\Delta}_{28}$ vanishes, if $\vec{v}_{{}_P},\vec{E}_{{}_P}$ are
linearly independent.
\par To discuss the case  where $\vec{v}_{{}_P},\vec{E}_{{}_P}$ are
linearly dependent, one first observes by (B.50) that $\vec{\Delta}_{28}$
vanishes if $\vec{v}_{{}_P}=0$ or $\vec{E}_{{}_P}=0$.
It remains to consider
the subcase with: $\vec{E}_{{}_P}=\lambda\cdot\vec{v}_{{}_P}$, where $\lambda$
is a constant which balances
                       the dimensions. Then one concludes from (B.50):
\begin{eqnarray}
&&\vec{\Delta}_{28}  =     \frac{e^2\cdot\lambda}{4\cdot m^2}\cdot\biggl(
%B
   (g^2-2\cdot g)\cdot
   \vec{v}_{{}_P}^\dagger\cdot\vec{v}_{{}_P}\cdot
      \vec{s}^{\,\dagger}\cdot\vec{B}_{{}_P}\cdot\vec{v}_{{}_P}
\nonumber\\&&
%C
-   2\cdot g\cdot \frac{1}{(\gamma_{{}_P}+1)^2}\cdot
  \vec{v}_{{}_P}^\dagger\cdot\vec{s}\cdot
   \vec{v}_{{}_P}^\dagger\cdot\vec{B}_{{}_P}\cdot\vec{v}_{{}_P}
%D
+  \lbrack g^2 -2\cdot g\cdot
\frac{\gamma_{{}_P}^2+2\cdot\gamma_{{}_P}}{(\gamma_{{}_P}+1)^2}
                                                 \rbrack\cdot
 \vec{s}^{\,\dagger}\cdot\vec{v}_{{}_P}\cdot
 \vec{v}_{{}_P}^\dagger\cdot\vec{B}_{{}_P}\cdot\vec{v}_{{}_P}
\nonumber\\&&\qquad
%I
+   \lbrack g^2 \cdot \frac{1-\gamma_{{}_P}^2}{\gamma_{{}_P}^2}
+2\cdot g\cdot
  \frac{\gamma_{{}_P}^2+\gamma_{{}_P}-2}{\gamma_{{}_P}\cdot
                                            (\gamma_{{}_P}+1)}
   \rbrack
\cdot  \vec{v}_{{}_P}^\dagger\cdot\vec{B}_{{}_P}\cdot\vec{s}
%J
+2\cdot g\cdot
  \frac{\gamma_{{}_P}-1}{\gamma_{{}_P}^2\cdot(\gamma_{{}_P}+1)}
\cdot  \vec{s}^{\,\dagger}\cdot\vec{v}_{{}_P}\cdot\vec{B}_{{}_P}
\nonumber\\&&\qquad
%L
-  2\cdot g \cdot \frac{1}{(1+\gamma_{{}_P})^2}
\cdot\vec{s}^{\,\dagger}\cdot\vec{v}_{{}_P}\cdot
  \vec{v}_{{}_P}^\dagger\cdot\vec{v}_{{}_P}\cdot\vec{B}_{{}_P}
\nonumber\\&&\qquad
%S
+   \lbrack g^2 -2\cdot g
\cdot\frac{\gamma_{{}_P}^2+2\cdot\gamma_{{}_P}}
          {(\gamma_{{}_P}+1)^2}  \rbrack
\cdot  \vec{v}_{{}_P}^\dagger\cdot\vec{v}_{{}_P}\cdot
  \vec{v}_{{}_P}^\dagger \cdot\vec{B}_{{}_P}
                              \cdot \vec{s}
\nonumber\\ &&\qquad
%V
   -(g-2)\cdot g\cdot
   \vec{v}_{{}_P}^\dagger\cdot\vec{v}_{{}_P}\cdot
   \vec{s}^{\,\dagger}\cdot\vec{B}_{{}_P}\cdot\vec{v}_{{}_P}
%W
   -(g-2)\cdot g\cdot
   \vec{v}_{{}_P}^\dagger\cdot\vec{s}\cdot
   \vec{v}_{{}_P}^\dagger\cdot\vec{B}_{{}_P}\cdot\vec{v}_{{}_P}
\biggr) = 0  \; .
\label{B.63}
\end{eqnarray}
Hence $\vec{\Delta}_{28}$ vanishes, if $\vec{v}_{{}_P},\vec{E}_{{}_P}$
are linearly dependent. From this it follows that $\vec{\Delta}_{28}$
vanishes in any case.
\subsection*{B.14}
In subsections B.11-13 I have shown that
$\vec{\Delta}_{26},\vec{\Delta}_{27},\vec{\Delta}_{28}$ vanish so that
by (B.49) it follows that:
\begin{eqnarray}
&& \vec{\Delta}_{25} = 0 \; .
\label{B.64}
\end{eqnarray}
Thus one has by (B.48):
\begin{eqnarray}
&& \vec{\Delta}_{21} = \vec{\Delta}_{24} \; .
\label{B.65}
\end{eqnarray}
Inserting (B.47),(B.65) into (B.35) results in
\begin{eqnarray}
&& m\cdot (\gamma_{{}_P}\cdot\vec{v}_{{}_P})' =
    e\cdot (\vec{v}_{{}_P}\wtimes\vec{B}_{{}_P}) + e\cdot\vec{E}_{{}_P} +
\frac{e\cdot g}{2\cdot m\cdot\gamma_{{}_P}}\cdot \vec{\nabla}_{{}_P}
                                                               \biggl(
\vec{s}^{\,\dagger}\cdot\vec{B}_{{}_P}-\vec{E}_{{}_P}^\dagger
         \cdot(\vec{s}\wtimes\vec{v}_{{}_P}) \biggr)
\nonumber\\&&
+\frac{e\cdot \gamma_{{}_P}}{2\cdot m}\cdot \lbrack
 2\cdot\vec{s}^{\,\dagger}\cdot\vec{B}_{{}_P}'\cdot\vec{v}_{{}_P}
-g\cdot(\vec{s}\wtimes\vec{v}_{{}_P})^\dagger\cdot
  \vec{E}_{{}_P}'\cdot\vec{v}_{{}_P}
+(g-2)\cdot(\vec{E}_{{}_P}'\wtimes\vec{s})
\nonumber\\&&\qquad
+(g-2)\cdot\vec{v}_{{}_P}^\dagger\cdot\vec{s}\cdot\vec{B}_{{}_P}'
-(g-2)\cdot\vec{v}_{{}_P}^\dagger\cdot\vec{E}_{{}_P}'
     \cdot(\vec{v}_{{}_P}\wtimes\vec{s})\rbrack
\nonumber\\&&\qquad
+\frac{e^2}{4\cdot m^2}\cdot\lbrack
   -(g-2)^2\cdot
   \vec{v}_{{}_P}^\dagger\cdot\vec{B}_{{}_P}\cdot
                                      (\vec{B}_{{}_P} \wtimes\vec{s})
   +(g-2)^2\cdot
   \vec{B}_{{}_P}^\dagger\cdot\vec{B}_{{}_P}\cdot
                                 (\vec{v}_{{}_P} \wtimes\vec{s})
\nonumber\\&&\qquad
   -(g-2)^2\cdot
   \vec{E}_{{}_P}^\dagger\cdot\vec{s}\cdot\vec{B}_{{}_P}
   +(-g^2\cdot \vec{v}_{{}_P}^\dagger\cdot\vec{v}_{{}_P}
   +2\cdot g\cdot \vec{v}_{{}_P}^\dagger\cdot\vec{v}_{{}_P} +4)\cdot
   \vec{s}^{\,\dagger}\cdot\vec{B}_{{}_P}\cdot\vec{E}_{{}_P}
\nonumber\\&&\qquad
   -(g-2)^2\cdot
   \vec{v}_{{}_P}^\dagger\cdot\vec{E}_{{}_P}\cdot
                                      (\vec{E}_{{}_P} \wtimes\vec{s})
   +(g-2)^2\cdot
       ( \vec{v}_{{}_P}\wtimes\vec{B}_{{}_P})^\dagger
 \cdot\vec{E}_{{}_P} \cdot(\vec{v}_{{}_P}\wtimes\vec{s})
\nonumber\\&&\qquad
   +(g-2)^2\cdot
       \vec{E}_{{}_P}^\dagger\cdot\vec{E}_{{}_P} \cdot
                                      (\vec{v}_{{}_P}\wtimes\vec{s})
   -(g^2-4\cdot g)\cdot
   \vec{s}^{\,\dagger}\cdot\vec{B}_{{}_P}\cdot
                               (\vec{v}_{{}_P} \wtimes\vec{B}_{{}_P})
\nonumber\\&&\qquad
   +(g-2)\cdot g\cdot
   \vec{E}_{{}_P}^\dagger\cdot\vec{B}_{{}_P}\cdot \vec{s}
   -(g-2)\cdot g\cdot
   \vec{v}_{{}_P}^\dagger\cdot\vec{E}_{{}_P}\cdot
          \vec{v}_{{}_P}^\dagger\cdot\vec{B}_{{}_P}\cdot\vec{s}
\nonumber\\&&\qquad
   +(g-2)\cdot g\cdot
   \vec{v}_{{}_P}^\dagger\cdot\vec{E}_{{}_P}\cdot
           \vec{s}^{\,\dagger}\cdot\vec{B}_{{}_P}\cdot\vec{v}_{{}_P}
   +(g-2)\cdot g\cdot
   \vec{v}_{{}_P}^\dagger\cdot\vec{s}\cdot
            \vec{v}_{{}_P}^\dagger\cdot\vec{B}_{{}_P}\cdot\vec{E}_{{}_P}
\nonumber\\&&\qquad
   -(g^2-4\cdot g)\cdot
  \vec{E}_{{}_P}^\dagger\cdot( \vec{v}_{{}_P}\wtimes
                                             \vec{s})\cdot\vec{E}_{{}_P}
   +2\cdot g\cdot
\vec{E}_{{}_P}^\dagger\cdot( \vec{v}_{{}_P}\wtimes
      \vec{s})\cdot(\vec{v}_{{}_P}\wtimes\vec{B}_{{}_P})
\rbrack   \; .
\label{B.66}
\end{eqnarray}
This completes the proof of (2.10).
\setcounter{section}{3}
\setcounter{subsection}{0}
\setcounter{equation}{0}
\section*{Appendix C}
\addcontentsline{toc}{section}{Appendix C}
In this Appendix I derive the equation of motion (8.15) for
$m\cdot\gamma_{{}_M}\cdot\vec{v}_{{}_M}$ by using those equations in sections
1,2,5 and Appendix B which are valid for arbitrary values of $c_1,...,c_5$.
I only consider the case of static (i.e. time independent) magnetic fields
and vanishing electric fields. I do this for the
general case, i.e. for arbitrary values of $c_1,...,c_5$.
First of all one concludes from (B.12),(B.15-16):
\begin{eqnarray}
 && m\cdot \gamma_{{}_M}\cdot\vec{v}_{{}_M} =
    m\cdot \gamma_{{}_P}\cdot\vec{v}_{{}_P}
- K_{{}_P}\cdot( \vec{\Delta}_7 + \gamma_{{}_P}^2\cdot\vec{v}_{{}_P}\cdot
\vec{v}_{{}_P}^\dagger\cdot\vec{\Delta}_7) \; .
\label{C.1}
\end{eqnarray}
By (B.13) one has
\begin{eqnarray*}
 && \vec{\Delta}_7 + \gamma_{{}_P}^2\cdot\vec{v}_{{}_P}\cdot
\vec{v}_{{}_P}^\dagger\cdot\vec{\Delta}_7
 = - \frac{e\cdot g}{2\cdot m^2}\cdot
   \frac{1}{\gamma_{{}_P}\cdot(\gamma_{{}_P}+1)}
\cdot \vec{B}_{{}_P}^\dagger\cdot\vec{v}_{{}_P} \cdot \vec{s}
\nonumber\\ &&\qquad
   +\frac{e\cdot(g-2)}{2\cdot m^2}\cdot
\frac{\gamma_{{}_P}}{\gamma_{{}_P}+1}\cdot
    \vec{s}^{\,\dagger}\cdot\vec{v}_{{}_P}\cdot\vec{B}_{{}_P}
+  \frac{e}{m^2}\cdot \frac{\gamma_{{}_P}}{\gamma_{{}_P}+1}\cdot
             \vec{s}^{\,\dagger}\cdot\vec{B}_{{}_P}\cdot\vec{v}_{{}_P}
\nonumber\\ &&\qquad
- \frac{e\cdot g}{2\cdot m^2}\cdot
    \frac{\gamma_{{}_P}}{\gamma_{{}_P}+1}
\cdot \vec{B}_{{}_P}^\dagger\cdot\vec{v}_{{}_P} \cdot
       \vec{s}^{\,\dagger}\cdot\vec{v}_{{}_P}\cdot\vec{v}_{{}_P}
 \; ,
\end{eqnarray*}
from which follows by (2.9),(B.42):
\begin{eqnarray*}
 && (\vec{\Delta}_7 + \gamma_{{}_P}^2\cdot\vec{v}_{{}_P}\cdot
\vec{v}_{{}_P}^\dagger\cdot\vec{\Delta}_7)'
 = - \frac{e\cdot g}{2\cdot m^2}\cdot
   \frac{1}{\gamma_{{}_P}\cdot(\gamma_{{}_P}+1)}
\cdot (\vec{B}_{{}_P}^\dagger\cdot\vec{v}_{{}_P} \cdot \vec{s})'
\nonumber\\ &&\qquad
   +\frac{e\cdot(g-2)}{2\cdot m^2}\cdot
\frac{\gamma_{{}_P}}{\gamma_{{}_P}+1}\cdot
    (\vec{s}^{\,\dagger}\cdot\vec{v}_{{}_P}\cdot\vec{B}_{{}_P})'
+  \frac{e}{m^2}\cdot \frac{\gamma_{{}_P}}{\gamma_{{}_P}+1}\cdot
             (\vec{s}^{\,\dagger}\cdot\vec{B}_{{}_P}\cdot\vec{v}_{{}_P})'
\nonumber\\ &&\qquad
- \frac{e\cdot g}{2\cdot m^2}\cdot
    \frac{\gamma_{{}_P}}{\gamma_{{}_P}+1}
\cdot (\vec{B}_{{}_P}^\dagger\cdot\vec{v}_{{}_P} \cdot
       \vec{s}^{\,\dagger}\cdot\vec{v}_{{}_P}\cdot\vec{v}_{{}_P})'
\nonumber\\
 &=& - \frac{e\cdot g}{2\cdot m^2}\cdot
   \frac{1}{\gamma_{{}_P}\cdot(\gamma_{{}_P}+1)}
\cdot (\vec{v}_{{}_P}^\dagger\cdot\vec{B}_{{}_P}'\cdot \vec{s} +
\vec{B}_{{}_P}^\dagger\cdot\vec{v}_{{}_P} \cdot \vec{s}\ ')
\nonumber\\ &&\qquad
   +\frac{e\cdot(g-2)}{2\cdot m^2}\cdot
\frac{\gamma_{{}_P}}{\gamma_{{}_P}+1}\cdot\biggl(
    (\vec{s}^{\,\dagger}\cdot\vec{v}_{{}_P})'\cdot\vec{B}_{{}_P} +
\vec{s}^{\,\dagger}\cdot\vec{v}_{{}_P}\cdot\vec{B}_{{}_P}'\biggr)
\nonumber\\ &&\qquad
+  \frac{e}{m^2}\cdot \frac{\gamma_{{}_P}}{\gamma_{{}_P}+1}\cdot
( \vec{s}^{\,\dagger}\cdot\vec{B}_{{}_P}'\cdot\vec{v}_{{}_P}
+ \vec{B}_{{}_P}^\dagger\cdot \vec{s}\ '\cdot\vec{v}_{{}_P}
+ \vec{s}^{\,\dagger}\cdot\vec{B}_{{}_P}\cdot\vec{v}_{{}_P}')
\nonumber\\ &&
- \frac{e\cdot g}{2\cdot m^2}\cdot
    \frac{\gamma_{{}_P}}{\gamma_{{}_P}+1}
\cdot \biggl(
\vec{v}_{{}_P}^\dagger\cdot\vec{B}_{{}_P}' \cdot
       \vec{s}^{\,\dagger}\cdot\vec{v}_{{}_P}\cdot\vec{v}_{{}_P}
+ \vec{B}_{{}_P}^\dagger\cdot\vec{v}_{{}_P} \cdot
       (\vec{s}^{\,\dagger}\cdot\vec{v}_{{}_P})'\cdot\vec{v}_{{}_P}
+ \vec{B}_{{}_P}^\dagger\cdot\vec{v}_{{}_P} \cdot
       \vec{s}^{\,\dagger}\cdot\vec{v}_{{}_P}\cdot\vec{v}_{{}_P}'\biggr)
\nonumber\\
 &=& - \frac{e\cdot g}{2\cdot m^2}\cdot
   \frac{1}{\gamma_{{}_P}\cdot(\gamma_{{}_P}+1)}
\cdot \vec{v}_{{}_P}^\dagger\cdot\vec{B}_{{}_P}'\cdot \vec{s}
   +\frac{e\cdot(g-2)}{2\cdot m^2}\cdot
\frac{\gamma_{{}_P}}{\gamma_{{}_P}+1}\cdot
    (\vec{s}^{\,\dagger}\cdot\vec{v}_{{}_P})'\cdot\vec{B}_{{}_P} \nonumber\\
&&\qquad +  \frac{e\cdot(g-2)}{2\cdot m^2}\cdot
\frac{\gamma_{{}_P}}{\gamma_{{}_P}+1}\cdot
\vec{s}^{\,\dagger}\cdot\vec{v}_{{}_P}\cdot\vec{B}_{{}_P}'
+  \frac{e}{m^2}\cdot \frac{\gamma_{{}_P}}{\gamma_{{}_P}+1}\cdot
( \vec{s}^{\,\dagger}\cdot\vec{B}_{{}_P}'\cdot\vec{v}_{{}_P}
+ \vec{s}^{\,\dagger}\cdot\vec{B}_{{}_P}\cdot\vec{v}_{{}_P}')
\nonumber\\ &&
- \frac{e\cdot g}{2\cdot m^2}\cdot
    \frac{\gamma_{{}_P}}{\gamma_{{}_P}+1}
\cdot \biggl(
\vec{v}_{{}_P}^\dagger\cdot\vec{B}_{{}_P}' \cdot
       \vec{s}^{\,\dagger}\cdot\vec{v}_{{}_P}\cdot\vec{v}_{{}_P}
+ \vec{B}_{{}_P}^\dagger\cdot\vec{v}_{{}_P} \cdot
       (\vec{s}^{\,\dagger}\cdot\vec{v}_{{}_P})'\cdot\vec{v}_{{}_P}\biggr)
\nonumber\\ &&
- \frac{e^2\cdot g^2}{4\cdot m^3}\cdot\frac{1}{\gamma_{{}_P}+1}
\cdot \vec{v}_{{}_P}^\dagger\cdot\vec{B}_{{}_P} \cdot
       \vec{s}^{\,\dagger}\cdot\vec{v}_{{}_P}\cdot
(\vec{v}_{{}_P}\wtimes\vec{B}_{{}_P})
- \frac{e^2\cdot g^2}{4\cdot m^3}\cdot
    \frac{1}{\gamma_{{}_P}^2\cdot(\gamma_{{}_P}+1)}
\cdot \vec{v}_{{}_P}^\dagger\cdot\vec{B}_{{}_P} \cdot
       (\vec{s}\wtimes\vec{B}_{{}_P})
\nonumber\\
 &=& - \frac{e\cdot g}{2\cdot m^2}\cdot
   \frac{1}{\gamma_{{}_P}\cdot(\gamma_{{}_P}+1)}
\cdot \vec{v}_{{}_P}^\dagger\cdot\vec{B}_{{}_P}'\cdot \vec{s}
   -\frac{e^2\cdot(g-2)^2}{4\cdot m^3}\cdot
\frac{1}{\gamma_{{}_P}+1}\cdot
(\vec{s}\wtimes\vec{v}_{{}_P})^\dagger\cdot\vec{B}_{{}_P}\cdot\vec{B}_{{}_P}
 \nonumber\\
&&\qquad +  \frac{e\cdot(g-2)}{2\cdot m^2}\cdot
\frac{\gamma_{{}_P}}{\gamma_{{}_P}+1}\cdot
\vec{s}^{\,\dagger}\cdot\vec{v}_{{}_P}\cdot\vec{B}_{{}_P}'
+  \frac{e}{m^2}\cdot \frac{\gamma_{{}_P}}{\gamma_{{}_P}+1}\cdot
\vec{s}^{\,\dagger}\cdot\vec{B}_{{}_P}'\cdot\vec{v}_{{}_P} \nonumber\\
&&\qquad+  \frac{e^2}{m^3}\cdot \frac{1}{\gamma_{{}_P}+1}\cdot
\vec{s}^{\,\dagger}\cdot\vec{B}_{{}_P}\cdot(\vec{v}_{{}_P}\wtimes
\vec{B}_{{}_P})
- \frac{e\cdot g}{2\cdot m^2}\cdot
    \frac{\gamma_{{}_P}}{\gamma_{{}_P}+1}\cdot
\vec{v}_{{}_P}^\dagger\cdot\vec{B}_{{}_P}' \cdot
       \vec{s}^{\,\dagger}\cdot\vec{v}_{{}_P}\cdot\vec{v}_{{}_P}
\nonumber\\&&\qquad
+\frac{e^2\cdot g\cdot(g-2)}{4\cdot m^3}\cdot
    \frac{1}{\gamma_{{}_P}+1}
\cdot \vec{v}_{{}_P}^\dagger\cdot\vec{B}_{{}_P}\cdot
(\vec{s}\wtimes\vec{v}_{{}_P})^\dagger\cdot\vec{B}_{{}_P}\cdot\vec{v}_{{}_P}
\nonumber\\ &&
- \frac{e^2\cdot g^2}{4\cdot m^3}\cdot\frac{1}{\gamma_{{}_P}+1}
\cdot \vec{v}_{{}_P}^\dagger\cdot\vec{B}_{{}_P} \cdot
       \vec{s}^{\,\dagger}\cdot\vec{v}_{{}_P}\cdot
(\vec{v}_{{}_P}\wtimes\vec{B}_{{}_P})
- \frac{e^2\cdot g^2}{4\cdot m^3}\cdot
    \frac{1}{\gamma_{{}_P}^2\cdot(\gamma_{{}_P}+1)}
\cdot \vec{v}_{{}_P}^\dagger\cdot\vec{B}_{{}_P} \cdot
       (\vec{s}\wtimes\vec{B}_{{}_P})
 \; .
\end{eqnarray*}
One thus has:
\begin{eqnarray}
 && \biggl(K_{{}_P}\cdot
(\vec{\Delta}_7 + \gamma_{{}_P}^2\cdot\vec{v}_{{}_P}\cdot
\vec{v}_{{}_P}^\dagger\cdot\vec{\Delta}_7)\biggr)'
 = K_{{}_P}\cdot \biggl(\vec{\Delta}_7 +
\gamma_{{}_P}^2\cdot\vec{v}_{{}_P}\cdot
\vec{v}_{{}_P}^\dagger\cdot\vec{\Delta}_7\biggr)' \nonumber\\
&=& - \frac{e\cdot g}{2\cdot m}\cdot
   \frac{1}{\gamma_{{}_P}+1}
\cdot \vec{v}_{{}_P}^\dagger\cdot\vec{B}_{{}_P}'\cdot \vec{s}
   -\frac{e^2\cdot(g-2)^2}{4\cdot m^2}\cdot
\frac{\gamma_{{}_P}}{\gamma_{{}_P}+1}\cdot
(\vec{s}\wtimes\vec{v}_{{}_P})^\dagger\cdot\vec{B}_{{}_P}\cdot\vec{B}_{{}_P}
 \nonumber\\
&&\qquad +  \frac{e\cdot(g-2)}{2\cdot m}\cdot
\frac{\gamma_{{}_P}^2}{\gamma_{{}_P}+1}\cdot
\vec{s}^{\,\dagger}\cdot\vec{v}_{{}_P}\cdot\vec{B}_{{}_P}'
+  \frac{e}{m}\cdot \frac{\gamma_{{}_P}^2}{\gamma_{{}_P}+1}\cdot
\vec{s}^{\,\dagger}\cdot\vec{B}_{{}_P}'\cdot\vec{v}_{{}_P} \nonumber\\
&&\qquad+  \frac{e^2}{m^2}\cdot \frac{\gamma_{{}_P}}{\gamma_{{}_P}+1}\cdot
\vec{s}^{\,\dagger}\cdot\vec{B}_{{}_P}\cdot(\vec{v}_{{}_P}\wtimes
\vec{B}_{{}_P})
- \frac{e\cdot g}{2\cdot m}\cdot
    \frac{\gamma_{{}_P}^2}{\gamma_{{}_P}+1}\cdot
\vec{v}_{{}_P}^\dagger\cdot\vec{B}_{{}_P}' \cdot
       \vec{s}^{\,\dagger}\cdot\vec{v}_{{}_P}\cdot\vec{v}_{{}_P}
\nonumber\\&&\qquad
+\frac{e^2\cdot g\cdot(g-2)}{4\cdot m^2}\cdot
    \frac{\gamma_{{}_P}}{\gamma_{{}_P}+1}
\cdot \vec{v}_{{}_P}^\dagger\cdot\vec{B}_{{}_P}\cdot
(\vec{s}\wtimes\vec{v}_{{}_P})^\dagger\cdot\vec{B}_{{}_P}\cdot\vec{v}_{{}_P}
\nonumber\\ &&
- \frac{e^2\cdot g^2}{4\cdot m^2}\cdot\frac{\gamma_{{}_P}}{\gamma_{{}_P}+1}
\cdot \vec{v}_{{}_P}^\dagger\cdot\vec{B}_{{}_P} \cdot
       \vec{s}^{\,\dagger}\cdot\vec{v}_{{}_P}\cdot
(\vec{v}_{{}_P}\wtimes\vec{B}_{{}_P})
- \frac{e^2\cdot g^2}{4\cdot m^2}\cdot
    \frac{1}{\gamma_{{}_P}\cdot(\gamma_{{}_P}+1)}
\cdot \vec{v}_{{}_P}^\dagger\cdot\vec{B}_{{}_P} \cdot
       (\vec{s}\wtimes\vec{B}_{{}_P}) \; . \nonumber\\&&
\label{C.2}
\end{eqnarray}
Combining this with (5.12b) yields by (C.1):
\begin{eqnarray}
 && m\cdot (\gamma_{{}_M}\cdot\vec{v}_{{}_M})' =
    m\cdot (\gamma_{{}_P}\cdot\vec{v}_{{}_P})'
-\biggl(K_{{}_P}\cdot( \vec{\Delta}_7 + \gamma_{{}_P}^2\cdot\vec{v}_{{}_P}\cdot
\vec{v}_{{}_P}^\dagger\cdot\vec{\Delta}_7)\biggr)' \nonumber\\
&=&    e\cdot (\vec{v}_{{}_P}\wtimes\vec{B}_{{}_P})
+\frac{e\cdot c_2}{2\cdot m\cdot\gamma_{{}_P}}\cdot \vec{\nabla}_{{}_P}
(\vec{s}^{\,\dagger}\cdot\vec{B}_{{}_P})
\nonumber\\&& \qquad
+\frac{e\cdot \gamma_{{}_P}}{2\cdot m}\cdot \lbrack
 (c_1+2)\cdot\vec{s}^{\,\dagger}\cdot\vec{B}_{{}_P}'\cdot\vec{v}_{{}_P}
+(c_2-c_1-2)\cdot\vec{v}_{{}_P}^\dagger\cdot\vec{s}\cdot\vec{B}_{{}_P}'\rbrack
\nonumber\\&&\qquad
+\frac{e^2}{4\cdot m^2}\cdot\lbrack
    c_5\cdot
   \vec{v}_{{}_P}^\dagger\cdot\vec{B}_{{}_P}\cdot
                                      (\vec{B}_{{}_P} \wtimes\vec{s})
   -c_5\cdot
   \vec{B}_{{}_P}^\dagger\cdot\vec{B}_{{}_P}\cdot
                                 (\vec{v}_{{}_P} \wtimes\vec{s})
\nonumber\\&&\qquad
   +(2\cdot c_4-c_3)\cdot
   \vec{s}^{\,\dagger}\cdot\vec{B}_{{}_P}\cdot
                               (\vec{v}_{{}_P} \wtimes\vec{B}_{{}_P})
\rbrack
 + \frac{e\cdot g}{2\cdot m}\cdot
   \frac{1}{\gamma_{{}_P}+1}
\cdot \vec{v}_{{}_P}^\dagger\cdot\vec{B}_{{}_P}'\cdot \vec{s}
 \nonumber\\&&\qquad
   +\frac{e^2\cdot(g-2)^2}{4\cdot m^2}\cdot
\frac{\gamma_{{}_P}}{\gamma_{{}_P}+1}\cdot
(\vec{s}\wtimes\vec{v}_{{}_P})^\dagger\cdot\vec{B}_{{}_P}\cdot\vec{B}_{{}_P}
 \nonumber\\
&&\qquad -  \frac{e\cdot(g-2)}{2\cdot m}\cdot
\frac{\gamma_{{}_P}^2}{\gamma_{{}_P}+1}\cdot
\vec{s}^{\,\dagger}\cdot\vec{v}_{{}_P}\cdot\vec{B}_{{}_P}'
-  \frac{e}{m}\cdot \frac{\gamma_{{}_P}^2}{\gamma_{{}_P}+1}\cdot
\vec{s}^{\,\dagger}\cdot\vec{B}_{{}_P}'\cdot\vec{v}_{{}_P} \nonumber\\
&&\qquad -  \frac{e^2}{m^2}\cdot \frac{\gamma_{{}_P}}{\gamma_{{}_P}+1}\cdot
\vec{s}^{\,\dagger}\cdot\vec{B}_{{}_P}\cdot(\vec{v}_{{}_P}\wtimes
\vec{B}_{{}_P})
+ \frac{e\cdot g}{2\cdot m}\cdot
    \frac{\gamma_{{}_P}^2}{\gamma_{{}_P}+1}\cdot
\vec{v}_{{}_P}^\dagger\cdot\vec{B}_{{}_P}' \cdot
       \vec{s}^{\,\dagger}\cdot\vec{v}_{{}_P}\cdot\vec{v}_{{}_P}
\nonumber\\&&\qquad
-\frac{e^2\cdot g\cdot(g-2)}{4\cdot m^2}\cdot
    \frac{\gamma_{{}_P}}{\gamma_{{}_P}+1}
\cdot \vec{v}_{{}_P}^\dagger\cdot\vec{B}_{{}_P}\cdot
(\vec{s}\wtimes\vec{v}_{{}_P})^\dagger\cdot\vec{B}_{{}_P}\cdot\vec{v}_{{}_P}
\nonumber\\ &&
+ \frac{e^2\cdot g^2}{4\cdot m^2}\cdot\frac{\gamma_{{}_P}}{\gamma_{{}_P}+1}
\cdot \vec{v}_{{}_P}^\dagger\cdot\vec{B}_{{}_P} \cdot
       \vec{s}^{\,\dagger}\cdot\vec{v}_{{}_P}\cdot
(\vec{v}_{{}_P}\wtimes\vec{B}_{{}_P})
+ \frac{e^2\cdot g^2}{4\cdot m^2}\cdot
    \frac{1}{\gamma_{{}_P}\cdot(\gamma_{{}_P}+1)}
\cdot \vec{v}_{{}_P}^\dagger\cdot\vec{B}_{{}_P} \cdot
       (\vec{s}\wtimes\vec{B}_{{}_P})
 \nonumber\\
&=&    e\cdot (\vec{v}_{{}_P}\wtimes\vec{B}_{{}_P})
+\frac{e\cdot c_2}{2\cdot m\cdot\gamma_{{}_P}}\cdot \vec{\nabla}_{{}_P}
(\vec{s}^{\,\dagger}\cdot\vec{B}_{{}_P})
\nonumber\\&& \qquad
+\frac{e}{2\cdot m}\cdot
\vec{s}^{\,\dagger}\cdot\vec{B}_{{}_P}'\cdot\vec{v}_{{}_P}\cdot
(\frac{2\cdot \gamma_{{}_P}}{\gamma_{{}_P}+1} + c_1\cdot\gamma_{{}_P})
\nonumber\\&&\qquad
+\frac{e}{2\cdot m}\cdot
\vec{v}_{{}_P}^\dagger\cdot\vec{s}\cdot\vec{B}_{{}_P}'\cdot \gamma_{{}_P}
\cdot (c_2-c_1-\frac{2}{\gamma_{{}_P}+1}-g\cdot\frac{\gamma_{{}_P}}
{\gamma_{{}_P}+1})
 + \frac{e\cdot g}{2\cdot m}\cdot
   \frac{1}{\gamma_{{}_P}+1}
\cdot \vec{v}_{{}_P}^\dagger\cdot\vec{B}_{{}_P}'\cdot \vec{s}
\nonumber\\&&\qquad
+ \frac{e\cdot g}{2\cdot m}\cdot
    \frac{\gamma_{{}_P}^2}{\gamma_{{}_P}+1}\cdot
\vec{v}_{{}_P}^\dagger\cdot\vec{B}_{{}_P}' \cdot
       \vec{s}^{\,\dagger}\cdot\vec{v}_{{}_P}\cdot\vec{v}_{{}_P}
\nonumber\\&&\qquad
+\frac{e^2}{4\cdot m^2}\cdot
  \vec{v}_{{}_P}^\dagger\cdot\vec{B}_{{}_P}\cdot
  (\vec{B}_{{}_P} \wtimes\vec{s})\cdot (c_5
   - g^2\cdot\frac{1}{\gamma_{{}_P}\cdot(\gamma_{{}_P}+1)})
\nonumber\\&&\qquad
-\frac{e^2}{4\cdot m^2}\cdot c_5\cdot
  \vec{B}_{{}_P}^\dagger\cdot\vec{B}_{{}_P}\cdot
  (\vec{v}_{{}_P} \wtimes\vec{s})
\nonumber\\&&\qquad
+\frac{e^2}{4\cdot m^2}\cdot
   \vec{s}^{\,\dagger}\cdot\vec{B}_{{}_P}\cdot
                               (\vec{v}_{{}_P} \wtimes\vec{B}_{{}_P})\cdot
(2\cdot c_4-c_3 -\frac{4\cdot\gamma_{{}_P}}{\gamma_{{}_P}+1})
 \nonumber\\&&\qquad
   +\frac{e^2\cdot(g-2)^2}{4\cdot m^2}\cdot
\frac{\gamma_{{}_P}}{\gamma_{{}_P}+1}\cdot
(\vec{s}\wtimes\vec{v}_{{}_P})^\dagger\cdot\vec{B}_{{}_P}\cdot\vec{B}_{{}_P}
\nonumber\\&&\qquad
-\frac{e^2\cdot g\cdot(g-2)}{4\cdot m^2}\cdot
    \frac{\gamma_{{}_P}}{\gamma_{{}_P}+1}
\cdot \vec{v}_{{}_P}^\dagger\cdot\vec{B}_{{}_P}\cdot
(\vec{s}\wtimes\vec{v}_{{}_P})^\dagger\cdot\vec{B}_{{}_P}\cdot\vec{v}_{{}_P}
\nonumber\\ &&
+ \frac{e^2\cdot g^2}{4\cdot m^2}\cdot\frac{\gamma_{{}_P}}{\gamma_{{}_P}+1}
\cdot \vec{v}_{{}_P}^\dagger\cdot\vec{B}_{{}_P} \cdot
       \vec{s}^{\,\dagger}\cdot\vec{v}_{{}_P}\cdot
(\vec{v}_{{}_P}\wtimes\vec{B}_{{}_P})
\; .
\label{C.3}
\end{eqnarray}
From (2.3),(C.3) follows:
\begin{eqnarray}
 && m\cdot (\gamma_{{}_M}\cdot\vec{v}_{{}_M})' =
    e\cdot (\vec{v}_{{}_P}\wtimes\vec{B}_{{}_P}) \nonumber\\&&\qquad
+\frac{e\cdot c_2}{2\cdot m\cdot\gamma_{{}_P}}\cdot \vec{\nabla}_{{}_P}(
\lbrack \gamma_{{}_P} \cdot \vec{\sigma} -
\frac{\gamma_{{}_P}^2}{\gamma_{{}_P}+1}\cdot\vec{\sigma}^{\,\dagger}
\cdot\vec{v}_{{}_P}\cdot \vec{v}_{{}_P}\rbrack
^\dagger\cdot\vec{B}_{{}_P})
\nonumber\\&& \qquad
+\frac{e}{2\cdot m}\cdot
\lbrack \gamma_{{}_P} \cdot \vec{\sigma} -
\frac{\gamma_{{}_P}^2}{\gamma_{{}_P}+1}\cdot\vec{\sigma}^{\,\dagger}
\cdot\vec{v}_{{}_P}\cdot \vec{v}_{{}_P}\rbrack
^\dagger\cdot\vec{B}_{{}_P}'\cdot\vec{v}_{{}_P}\cdot
(\frac{2\cdot \gamma_{{}_P}}{\gamma_{{}_P}+1} + c_1\cdot\gamma_{{}_P})
\nonumber\\&&\qquad
+\frac{e}{2\cdot m}\cdot
\vec{v}_{{}_P}^\dagger\cdot\vec{\sigma}\cdot\vec{B}_{{}_P}'\cdot \gamma_{{}_P}
\cdot (c_2-c_1-\frac{2}{\gamma_{{}_P}+1}-g\cdot\frac{\gamma_{{}_P}}
{\gamma_{{}_P}+1})\nonumber\\&&\qquad
 + \frac{e\cdot g}{2\cdot m}\cdot
   \frac{1}{\gamma_{{}_P}+1}
\cdot \vec{v}_{{}_P}^\dagger\cdot\vec{B}_{{}_P}'\cdot
\lbrack \gamma_{{}_P} \cdot \vec{\sigma} -
\frac{\gamma_{{}_P}^2}{\gamma_{{}_P}+1}\cdot\vec{\sigma}^{\,\dagger}
\cdot\vec{v}_{{}_P}\cdot \vec{v}_{{}_P}\rbrack
\nonumber\\&&\qquad
+ \frac{e\cdot g}{2\cdot m}\cdot
    \frac{\gamma_{{}_P}^2}{\gamma_{{}_P}+1}\cdot
\vec{v}_{{}_P}^\dagger\cdot\vec{B}_{{}_P}' \cdot
       \vec{\sigma}^{\,\dagger}\cdot\vec{v}_{{}_P}\cdot\vec{v}_{{}_P}
\nonumber\\&&\qquad
+\frac{e^2}{4\cdot m^2}\cdot
  \vec{v}_{{}_P}^\dagger\cdot\vec{B}_{{}_P}\cdot
(c_5  - g^2\cdot\frac{1}{\gamma_{{}_P}\cdot(\gamma_{{}_P}+1)})
  \cdot\vec{B}_{{}_P} \wtimes
  \lbrack \gamma_{{}_P} \cdot \vec{\sigma} -
\frac{\gamma_{{}_P}^2}{\gamma_{{}_P}+1}\cdot\vec{\sigma}^{\,\dagger}
\cdot\vec{v}_{{}_P}\cdot \vec{v}_{{}_P}\rbrack
 \nonumber\\&&\qquad
-\frac{e^2}{4\cdot m^2}\cdot c_5\cdot
  \vec{B}_{{}_P}^\dagger\cdot\vec{B}_{{}_P}\cdot\gamma_{{}_P}\cdot
 (\vec{v}_{{}_P} \wtimes\vec{\sigma})
\nonumber\\&&\qquad
+\frac{e^2}{4\cdot m^2}\cdot
  \lbrack \gamma_{{}_P} \cdot \vec{\sigma} -
\frac{\gamma_{{}_P}^2}{\gamma_{{}_P}+1}\cdot\vec{\sigma}^{\,\dagger}
\cdot\vec{v}_{{}_P}\cdot \vec{v}_{{}_P}\rbrack
^\dagger\cdot\vec{B}_{{}_P}\cdot
                               (\vec{v}_{{}_P} \wtimes\vec{B}_{{}_P})\cdot
(2\cdot c_4-c_3 -\frac{4\cdot\gamma_{{}_P}}{\gamma_{{}_P}+1})
 \nonumber\\&&\qquad
   +\frac{e^2\cdot(g-2)^2}{4\cdot m^2}\cdot
\frac{\gamma_{{}_P}^2}{\gamma_{{}_P}+1}\cdot
(\vec{\sigma}\wtimes\vec{v}_{{}_P})^\dagger\cdot\vec{B}_{{}_P}\cdot
\vec{B}_{{}_P}
\nonumber\\&&\qquad
-\frac{e^2\cdot g\cdot(g-2)}{4\cdot m^2}\cdot
    \frac{\gamma_{{}_P}^2}{\gamma_{{}_P}+1}
\cdot \vec{v}_{{}_P}^\dagger\cdot\vec{B}_{{}_P}\cdot(\vec{\sigma}\wtimes
\vec{v}_{{}_P})^\dagger\cdot\vec{B}_{{}_P}\cdot\vec{v}_{{}_P}
\nonumber\\ &&
+ \frac{e^2\cdot g^2}{4\cdot m^2}\cdot\frac{\gamma_{{}_P}}{\gamma_{{}_P}+1}
\cdot \vec{v}_{{}_P}^\dagger\cdot\vec{B}_{{}_P} \cdot
       \vec{\sigma}^{\,\dagger}\cdot\vec{v}_{{}_P}\cdot
(\vec{v}_{{}_P}\wtimes\vec{B}_{{}_P}) \nonumber\\
&=&  e\cdot (\vec{v}_{{}_P}\wtimes\vec{B}_{{}_P})
+\frac{e\cdot c_2}{2\cdot m}\cdot\vec{\nabla}_{{}_M}(
\vec{\sigma}^{\,\dagger}\cdot\vec{B}_{{}_M})
-\frac{e\cdot c_2}{2\cdot m}\cdot\frac{\gamma_{{}_M}}{\gamma_{{}_M}+1}\cdot
\vec{\sigma}^{\,\dagger}\cdot\vec{v}_{{}_M}\cdot \vec{B}_{{}_M}'
\nonumber\\&& \qquad
+\frac{e}{2\cdot m}\cdot
\lbrack \gamma_{{}_M} \cdot \vec{\sigma} -
\frac{\gamma_{{}_M}^2}{\gamma_{{}_M}+1}\cdot\vec{\sigma}^{\,\dagger}
\cdot\vec{v}_{{}_M}\cdot \vec{v}_{{}_M}\rbrack
^\dagger\cdot\vec{B}_{{}_M}'\cdot\vec{v}_{{}_M}\cdot
(\frac{2\cdot \gamma_{{}_M}}{\gamma_{{}_M}+1} + c_1\cdot\gamma_{{}_M})
\nonumber\\&&\qquad
+\frac{e}{2\cdot m}\cdot
\vec{v}_{{}_M}^\dagger\cdot\vec{\sigma}\cdot\vec{B}_{{}_M}'\cdot \gamma_{{}_M}
\cdot (c_2-c_1-\frac{2}{\gamma_{{}_M}+1}-g\cdot\frac{\gamma_{{}_M}}
{\gamma_{{}_M}+1})\nonumber\\&&\qquad
 + \frac{e\cdot g}{2\cdot m}\cdot
   \frac{1}{\gamma_{{}_M}+1}
\cdot \vec{v}_{{}_M}^\dagger\cdot\vec{B}_{{}_M}'\cdot
\lbrack \gamma_{{}_M} \cdot \vec{\sigma} -
\frac{\gamma_{{}_M}^2}{\gamma_{{}_M}+1}\cdot\vec{\sigma}^{\,\dagger}
\cdot\vec{v}_{{}_M}\cdot \vec{v}_{{}_M}\rbrack
\nonumber\\&&\qquad
+ \frac{e\cdot g}{2\cdot m}\cdot
    \frac{\gamma_{{}_M}^2}{\gamma_{{}_M}+1}\cdot
\vec{v}_{{}_M}^\dagger\cdot\vec{B}_{{}_M}' \cdot
       \vec{\sigma}^{\,\dagger}\cdot\vec{v}_{{}_M}\cdot\vec{v}_{{}_M}
\nonumber\\&&\qquad
+\frac{e^2}{4\cdot m^2}\cdot
  \vec{v}_{{}_M}^\dagger\cdot\vec{B}_{{}_M}\cdot
(c_5  - g^2\cdot\frac{1}{\gamma_{{}_M}\cdot(\gamma_{{}_M}+1)})
  \cdot\vec{B}_{{}_M} \wtimes
  \lbrack \gamma_{{}_M} \cdot \vec{\sigma} -
\frac{\gamma_{{}_M}^2}{\gamma_{{}_M}+1}\cdot\vec{\sigma}^{\,\dagger}
\cdot\vec{v}_{{}_M}\cdot \vec{v}_{{}_M}\rbrack
 \nonumber\\&&\qquad
-\frac{e^2}{4\cdot m^2}\cdot c_5\cdot
  \vec{B}_{{}_M}^\dagger\cdot\vec{B}_{{}_M}\cdot\gamma_{{}_M}\cdot
 (\vec{v}_{{}_M} \wtimes\vec{\sigma})
\nonumber\\&&\qquad
+\frac{e^2}{4\cdot m^2}\cdot
  \lbrack \gamma_{{}_M} \cdot \vec{\sigma} -
\frac{\gamma_{{}_M}^2}{\gamma_{{}_M}+1}\cdot\vec{\sigma}^{\,\dagger}
\cdot\vec{v}_{{}_M}\cdot \vec{v}_{{}_M}\rbrack
^\dagger\cdot\vec{B}_{{}_M}\cdot
                               (\vec{v}_{{}_M} \wtimes\vec{B}_{{}_M})\cdot
(2\cdot c_4-c_3 -\frac{4\cdot\gamma_{{}_M}}{\gamma_{{}_M}+1})
 \nonumber\\&&\qquad
   +\frac{e^2\cdot(g-2)^2}{4\cdot m^2}\cdot
\frac{\gamma_{{}_M}^2}{\gamma_{{}_M}+1}\cdot
(\vec{\sigma}\wtimes\vec{v}_{{}_M})^\dagger\cdot\vec{B}_{{}_M}\cdot
\vec{B}_{{}_M}
\nonumber\\&&\qquad
-\frac{e^2\cdot g\cdot(g-2)}{4\cdot m^2}\cdot
    \frac{\gamma_{{}_M}^2}{\gamma_{{}_M}+1}
\cdot \vec{v}_{{}_M}^\dagger\cdot\vec{B}_{{}_M}\cdot(\vec{\sigma}\wtimes
\vec{v}_{{}_M})^\dagger\cdot\vec{B}_{{}_M}\cdot\vec{v}_{{}_M}
\nonumber\\ &&
+ \frac{e^2\cdot g^2}{4\cdot m^2}\cdot\frac{\gamma_{{}_M}}{\gamma_{{}_M}+1}
\cdot \vec{v}_{{}_M}^\dagger\cdot\vec{B}_{{}_M} \cdot
       \vec{\sigma}^{\,\dagger}\cdot\vec{v}_{{}_M}\cdot
(\vec{v}_{{}_M}\wtimes\vec{B}_{{}_M})
\nonumber\\
&=&  e\cdot (\vec{v}_{{}_P}\wtimes\vec{B}_{{}_P})
+\frac{e\cdot c_2}{2\cdot m}\cdot\vec{\nabla}_{{}_M}(
\vec{\sigma}^{\,\dagger}\cdot\vec{B}_{{}_M})\nonumber\\&&\qquad
+\frac{e}{2\cdot m}\cdot\vec{\sigma}^{\,\dagger}\cdot\vec{v}_{{}_M}\cdot
\vec{B}_{{}_M}'\cdot\gamma_{{}_M}\cdot \lbrack
\frac{\gamma_{{}_M}}{\gamma_{{}_M}+1}\cdot(c_2-g)-c_1-
\frac{2}{\gamma_{{}_M}+1}\rbrack
\nonumber\\&& \qquad
+\frac{e}{2\cdot m}\cdot\vec{\sigma}^\dagger\cdot\vec{B}_{{}_M}'\cdot
\vec{v}_{{}_M}\cdot
(\frac{2\cdot \gamma_{{}_M}^2}{\gamma_{{}_M}+1} + c_1\cdot\gamma_{{}_M}^2)
\nonumber\\&&\qquad
+\frac{e}{2\cdot m}\cdot\vec{\sigma}^{\,\dagger}\cdot\vec{v}_{{}_M}\cdot
\vec{v}_{{}_M}^\dagger\cdot\vec{B}_{{}_M}'\cdot\vec{v}_{{}_M}\cdot
\frac{\gamma_{{}_M}^2}{\gamma_{{}_M}+1}\cdot\lbrack
\frac{\gamma_{{}_M}}{\gamma_{{}_M}+1}\cdot(g-2)
- c_1\cdot\gamma_{{}_M}\rbrack
\nonumber\\&&\qquad
+ \frac{e\cdot g}{2\cdot m}\cdot
    \frac{\gamma_{{}_M}}{\gamma_{{}_M}+1}\cdot
\vec{v}_{{}_M}^\dagger\cdot\vec{B}_{{}_M}' \cdot\vec{\sigma}
\nonumber\\&&\qquad
+\frac{e^2}{4\cdot m^2}\cdot
  \vec{v}_{{}_M}^\dagger\cdot\vec{B}_{{}_M}\cdot
 (\vec{B}_{{}_M} \wtimes\vec{\sigma})\cdot
\gamma_{{}_M}\cdot
(c_5  - g^2\cdot\frac{1}{\gamma_{{}_M}\cdot(\gamma_{{}_M}+1)})
 \nonumber\\&&\qquad
+\frac{e^2}{4\cdot m^2}\cdot \vec{v}_{{}_M}^\dagger\cdot\vec{B}_{{}_M}\cdot
       \vec{\sigma}^{\,\dagger}\cdot\vec{v}_{{}_M}\cdot
 (\vec{v}_{{}_M} \wtimes\vec{B}_{{}_M})\cdot\lbrack
\frac{\gamma_{{}_M}^2}{\gamma_{{}_M}+1}\cdot(c_3-2\cdot c_4+c_5)
+g^2\cdot\frac{\gamma_{{}_M}^2}{(\gamma_{{}_M}+1)^2}
 \nonumber\\&&\qquad
+4\cdot\frac{\gamma_{{}_M}^3}{(\gamma_{{}_M}+1)^2}\rbrack
-\frac{e^2}{4\cdot m^2}\cdot c_5\cdot
  \vec{B}_{{}_M}^\dagger\cdot\vec{B}_{{}_M}\cdot\gamma_{{}_M}\cdot
 (\vec{v}_{{}_M} \wtimes\vec{\sigma})
\nonumber\\&&\qquad
+\frac{e^2}{4\cdot m^2}\cdot\vec{\sigma}^\dagger
\cdot\vec{B}_{{}_M}\cdot(\vec{v}_{{}_M} \wtimes\vec{B}_{{}_M})\cdot
\gamma_{{}_M}\cdot
(2\cdot c_4-c_3 -\frac{4\cdot\gamma_{{}_M}}{\gamma_{{}_M}+1})
 \nonumber\\&&\qquad
   +\frac{e^2\cdot(g-2)^2}{4\cdot m^2}\cdot
\frac{\gamma_{{}_M}^2}{\gamma_{{}_M}+1}\cdot
(\vec{\sigma}\wtimes\vec{v}_{{}_M})^\dagger\cdot\vec{B}_{{}_M}\cdot
\vec{B}_{{}_M}
\nonumber\\&&\qquad
-\frac{e^2\cdot g\cdot(g-2)}{4\cdot m^2}\cdot
    \frac{\gamma_{{}_M}^2}{\gamma_{{}_M}+1}
\cdot \vec{v}_{{}_M}^\dagger\cdot\vec{B}_{{}_M}\cdot(\vec{\sigma}\wtimes
\vec{v}_{{}_M})^\dagger\cdot\vec{B}_{{}_M}\cdot\vec{v}_{{}_M} \; .
\label{C.4}
\end{eqnarray}
This can be simplified by calculating via (1.6-7),(B.4),(B.7),(B.12-13):
\begin{eqnarray}
&& \vec{v}_{{}_P} =  \vec{v}_{{}_M} + \vec{\Delta}_7 =
\vec{v}_{{}_M} +  \frac{1}{m}\cdot\frac{\gamma_{{}_M}}{\gamma_{{}_M}+1}
\cdot (\vec{\sigma}\wtimes\vec{v}_{{}_M})'
= \frac{1}{m}\cdot\frac{\gamma_{{}_M}}{\gamma_{{}_M}+1}
\cdot \biggl( (\vec{\Omega}_{{}_M}\wtimes \vec{\sigma})
                                                 \wtimes\vec{v}_{{}_M}
 + \vec{\sigma}\wtimes\vec{v}_{{}_M}' \biggr)
\nonumber\\
 &=& \vec{v}_{{}_M} +
\frac{1}{m}\cdot\frac{\gamma_{{}_M}}{\gamma_{{}_M}+1}
\cdot\biggl(\vec{v}_{{}_M}^{\,\dagger}\cdot\vec{\Omega}_{{}_M}\cdot\vec{\sigma}
 -\vec{v}_{{}_M}^{\,\dagger}\cdot\vec{\sigma}\cdot\vec{\Omega}_{{}_M}
+\frac{e}{K_{{}_M}}\cdot\vec{B}_{{}_M}^{\,\dagger}\cdot\vec{\sigma}\cdot
\vec{v}_{{}_M}
-\frac{e}{K_{{}_M}}\cdot\vec{v}_{{}_M}^{\,\dagger}\cdot\vec{\sigma}\cdot
\vec{B}_{{}_M} \biggr) \nonumber\\
&=& \frac{e}{2\cdot m^2}\cdot\frac{\gamma_{{}_M}}{\gamma_{{}_M}+1}
\cdot\biggl(-\frac{g}{\gamma_{{}_M}}\cdot
\vec{v}_{{}_M}^{\,\dagger}\cdot\vec{B}_{{}_M}\cdot\vec{\sigma}
+ \lbrack \frac{2}{\gamma_{{}_M}}+g-2 \rbrack \cdot
\vec{v}_{{}_M}^{\,\dagger}\cdot\vec{\sigma}\cdot\vec{B}_{{}_M} \nonumber\\
&&\qquad -(g-2)\cdot\frac{\gamma_{{}_M}}{\gamma_{{}_M}+1}\cdot
\vec{\sigma}^{\,\dagger}\cdot\vec{v}_{{}_M}\cdot
 \vec{v}_{{}_M}^{\,\dagger}\cdot\vec{B}_{{}_M}\cdot\vec{v}_{{}_M}
+\frac{2}{\gamma_{{}_M}}\cdot\vec{B}_{{}_M}^{\,\dagger}\cdot\vec{\sigma}\cdot
\vec{v}_{{}_M}
-\frac{2}{\gamma_{{}_M}}\cdot\vec{v}_{{}_M}^{\,\dagger}\cdot\vec{\sigma}\cdot
\vec{B}_{{}_M}  \biggr) \nonumber\\
&=& \vec{v}_{{}_M} +
\frac{e}{2\cdot m^2}\cdot\frac{1}{\gamma_{{}_M}+1}
\cdot\biggl( -g\cdot
\vec{v}_{{}_M}^{\,\dagger}\cdot\vec{B}_{{}_M}\cdot\vec{\sigma}
+(g-2)\cdot\gamma_{{}_M}\cdot
\vec{v}_{{}_M}^{\,\dagger}\cdot\vec{\sigma}\cdot\vec{B}_{{}_M}
\nonumber\\
&&\qquad -(g-2)\cdot\frac{\gamma_{{}_M}^2}{\gamma_{{}_M}+1}\cdot
\vec{\sigma}^{\,\dagger}\cdot\vec{v}_{{}_M}\cdot
 \vec{v}_{{}_M}^{\,\dagger}\cdot\vec{B}_{{}_M}\cdot\vec{v}_{{}_M}
 +2\cdot\vec{B}_{{}_M}^{\,\dagger}\cdot\vec{\sigma}\cdot
\vec{v}_{{}_M} \biggr) \; ,
\label{C.5}
\end{eqnarray}
from which follows by (B.21)
\begin{eqnarray}
&& e\cdot(\vec{v}_{{}_P}\wtimes\vec{B}_{{}_P})  =
  e\cdot(\vec{v}_{{}_M}\wtimes\vec{B}_{{}_P})
 +     \frac{e^2}{m^2}\cdot\frac{1}{\gamma_{{}_M}+1}
\cdot\biggl( -\frac{g}{2}\cdot \vec{v}_{{}_M}^{\,\dagger}\cdot
       \vec{B}_{{}_M}\cdot\vec{\sigma}
 \nonumber\\&&\qquad
 -\frac{g-2}{2}\cdot\frac{\gamma_{{}_M}^2}{\gamma_{{}_M}+1}\cdot
    \vec{v}_{{}_M}^{\,\dagger}\cdot\vec{\sigma}\cdot
  \vec{v}_{{}_M}^{\,\dagger}\cdot\vec{B}_{{}_M}\cdot\vec{v}_{{}_M}
  +  \vec{\sigma}^{\,\dagger}\cdot \vec{B}_{{}_M}\cdot\vec{v}_{{}_M}
              \biggr)
\wtimes\vec{B}_{{}_M} \nonumber\\
&=&    e\cdot(\vec{v}_{{}_M}\wtimes\vec{B}_{{}_M})
  - e\cdot(\vec{v}_{{}_M}\wtimes\vec{\Delta}_{12})
 +     \frac{e^2}{m^2}\cdot\frac{1}{\gamma_{{}_M}+1}
\cdot\biggl( -\frac{g}{2}\cdot \vec{v}_{{}_M}^{\,\dagger}\cdot
       \vec{B}_{{}_M}\cdot\vec{\sigma}
 \nonumber\\&&\qquad
 -\frac{g-2}{2}\cdot\frac{\gamma_{{}_M}^2}{\gamma_{{}_M}+1}\cdot
    \vec{v}_{{}_M}^{\,\dagger}\cdot\vec{\sigma}\cdot
  \vec{v}_{{}_M}^{\,\dagger}\cdot\vec{B}_{{}_M}\cdot\vec{v}_{{}_M}
  +  \vec{\sigma}^{\,\dagger}\cdot \vec{B}_{{}_M}\cdot\vec{v}_{{}_M}
              \biggr)
\wtimes\vec{B}_{{}_M} \; .
\label{C.6}
\end{eqnarray}
This can be further simplified by calculating via (B.20),(B.30):
\begin{eqnarray*}
&&  - e\cdot(\vec{v}_{{}_M}\wtimes\vec{\Delta}_{12})
 =  - e\cdot\lbrack \vec{v}_{{}_M}\wtimes
 \vec{\Delta}_{10}^\dagger\cdot\vec{\nabla}_{{}_M}\vec{B}_{{}_M}
 \rbrack
 =  - e\cdot\vec{\Delta}_{10}^\dagger\cdot\vec{\nabla}_{{}_M}
            (\vec{v}_{{}_M}\wtimes \vec{B}_{{}_M})
 \nonumber\\ &&
 = - e\cdot\vec{\nabla}_{{}_M}\biggl(\vec{\Delta}_{10}^\dagger\cdot
            (\vec{v}_{{}_M}\wtimes \vec{B}_{{}_M})\biggr)
 + e\cdot\vec{\Delta}_{10}\wtimes\biggl(
\vec{\nabla}_{{}_M}\wtimes(\vec{v}_{{}_M}\wtimes\vec{B}_{{}_M})\biggr)
\nonumber\\
 &=& \frac{e}{m}\cdot \frac{\gamma_{{}_M}}{\gamma_{{}_M}+1}\cdot
\vec{\nabla}_{{}_M}
   \biggl((\vec{\sigma}\wtimes\vec{v}_{{}_M})^{\dagger}\cdot
    (\vec{v}_{{}_M}\wtimes\vec{B}_{{}_M})\biggr)
 + \frac{e}{m}\cdot \frac{\gamma_{{}_M}}{\gamma_{{}_M}+1}\cdot
   (\vec{\sigma}\wtimes\vec{v}_{{}_M})\wtimes\vec{B}_{{}_M}'
 \nonumber\\
&=&\frac{e}{m}\cdot\frac{\gamma_{{}_M}}{\gamma_{{}_M}+1}\cdot
\vec{\nabla}_{{}_M}
   (\vec{\sigma}^{\,\dagger}\cdot \vec{v}_{{}_M}\cdot
   \vec{v}_{{}_M}^{\,\dagger}\cdot\vec{B}_{{}_M}
    - \vec{\sigma}^{\,\dagger}\cdot \vec{B}_{{}_M}\cdot
   \vec{v}_{{}_M}^{\,\dagger}\cdot\vec{v}_{{}_M})
 \nonumber\\
&+&\frac{e}{m}\cdot\frac{\gamma_{{}_M}}{\gamma_{{}_M}+1}\cdot
   \lbrack \vec{\sigma}^{\,\dagger}\cdot \vec{B}_{{}_M}'\cdot
   \vec{v}_{{}_M}
    -  \vec{v}_{{}_M}^{\,\dagger}\cdot\vec{B}_{{}_M}'\cdot\vec{\sigma}\rbrack
 \nonumber\\
&=&\frac{e}{m}\cdot\frac{\gamma_{{}_M}}{\gamma_{{}_M}+1}\cdot\lbrack
  -\vec{v}_{{}_M}^{\,\dagger}\cdot\vec{v}_{{}_M}\cdot
\vec{\nabla}_{{}_M}(\vec{\sigma}^{\,\dagger}\cdot\vec{B}_{{}_M})
 + \vec{\sigma}^{\,\dagger}\cdot \vec{v}_{{}_M}\cdot\vec{B}_{{}_M}'
 +\vec{\sigma}^{\,\dagger}\cdot \vec{B}_{{}_M}'\cdot\vec{v}_{{}_M}
-  \vec{v}_{{}_M}^{\,\dagger}\cdot\vec{B}_{{}_M}'\cdot\vec{\sigma}\rbrack \; ,
\end{eqnarray*}
so that (C.6) reads as:
\begin{eqnarray*}
&& e\cdot(\vec{v}_{{}_P}\wtimes\vec{B}_{{}_P})  =
        e\cdot(\vec{v}_{{}_M}\wtimes\vec{B}_{{}_M})
 + \frac{e}{m}\cdot\frac{\gamma_{{}_M}}{\gamma_{{}_M}+1}\cdot\lbrack
    \vec{\sigma}^{\,\dagger}\cdot \vec{v}_{{}_M}\cdot\vec{B}_{{}_M}'
 +\vec{\sigma}^{\,\dagger}\cdot \vec{B}_{{}_M}'\cdot\vec{v}_{{}_M}
\nonumber\\ &&
-  \vec{v}_{{}_M}^{\,\dagger}\cdot\vec{B}_{{}_M}'\cdot\vec{\sigma}
  -\vec{v}_{{}_M}^{\,\dagger}\cdot\vec{v}_{{}_M}\cdot
  \vec{\nabla}_{{}_M}(\vec{\sigma}^{\,\dagger}\cdot
                                               \vec{B}_{{}_M})\rbrack
 +     \frac{e^2}{m^2}\cdot\frac{1}{\gamma_{{}_M}+1}
\cdot\biggl( -\frac{g}{2}\cdot \vec{v}_{{}_M}^{\,\dagger}\cdot
       \vec{B}_{{}_M}\cdot\vec{\sigma}
 \nonumber\\&&\qquad
 -\frac{g-2}{2}\cdot\frac{\gamma_{{}_M}^2}{\gamma_{{}_M}+1}\cdot
    \vec{v}_{{}_M}^{\,\dagger}\cdot\vec{\sigma}\cdot
  \vec{v}_{{}_M}^{\,\dagger}\cdot\vec{B}_{{}_M}\cdot\vec{v}_{{}_M}
  +  \vec{\sigma}^{\,\dagger}\cdot \vec{B}_{{}_M}\cdot\vec{v}_{{}_M}
              \biggr)
\wtimes\vec{B}_{{}_M} \; .
\end{eqnarray*}
Inserting this into (C.4) yields:
\begin{eqnarray}
 && m\cdot (\gamma_{{}_M}\cdot\vec{v}_{{}_M})' =
         e\cdot(\vec{v}_{{}_M}\wtimes\vec{B}_{{}_M})
 + \frac{e}{m}\cdot\frac{\gamma_{{}_M}}{\gamma_{{}_M}+1}\cdot\lbrack
    \vec{\sigma}^{\,\dagger}\cdot \vec{v}_{{}_M}\cdot\vec{B}_{{}_M}'
 +\vec{\sigma}^{\,\dagger}\cdot \vec{B}_{{}_M}'\cdot\vec{v}_{{}_M}
\nonumber\\ &&
-  \vec{v}_{{}_M}^{\,\dagger}\cdot\vec{B}_{{}_M}'\cdot\vec{\sigma}
  -\vec{v}_{{}_M}^{\,\dagger}\cdot\vec{v}_{{}_M}\cdot
  \vec{\nabla}_{{}_M}(\vec{\sigma}^{\,\dagger}\cdot
                                               \vec{B}_{{}_M})\rbrack
 +     \frac{e^2}{m^2}\cdot\frac{1}{\gamma_{{}_M}+1}
\cdot\biggl( -\frac{g}{2}\cdot \vec{v}_{{}_M}^{\,\dagger}\cdot
       \vec{B}_{{}_M}\cdot\vec{\sigma}
 \nonumber\\&&
 -\frac{g-2}{2}\cdot\frac{\gamma_{{}_M}^2}{\gamma_{{}_M}+1}\cdot
    \vec{v}_{{}_M}^{\,\dagger}\cdot\vec{\sigma}\cdot
  \vec{v}_{{}_M}^{\,\dagger}\cdot\vec{B}_{{}_M}\cdot\vec{v}_{{}_M}
  +  \vec{\sigma}^{\,\dagger}\cdot \vec{B}_{{}_M}\cdot\vec{v}_{{}_M}
              \biggr)
\wtimes\vec{B}_{{}_M}
+\frac{e\cdot c_2}{2\cdot m}\cdot\vec{\nabla}_{{}_M}(
\vec{\sigma}^{\,\dagger}\cdot\vec{B}_{{}_M})\nonumber\\&&\qquad
+\frac{e}{2\cdot m}\cdot\vec{\sigma}^{\,\dagger}\cdot\vec{v}_{{}_M}\cdot
\vec{B}_{{}_M}'\cdot\gamma_{{}_M}\cdot \lbrack
\frac{\gamma_{{}_M}}{\gamma_{{}_M}+1}\cdot(c_2-g)-c_1-
\frac{2}{\gamma_{{}_M}+1}\rbrack
\nonumber\\&& \qquad
+\frac{e}{2\cdot m}\cdot\vec{\sigma}^\dagger\cdot\vec{B}_{{}_M}'\cdot
\vec{v}_{{}_M}\cdot
(\frac{2\cdot \gamma_{{}_M}^2}{\gamma_{{}_M}+1} + c_1\cdot\gamma_{{}_M}^2)
\nonumber\\&&\qquad
+\frac{e}{2\cdot m}\cdot\vec{\sigma}^{\,\dagger}\cdot\vec{v}_{{}_M}\cdot
\vec{v}_{{}_M}^\dagger\cdot\vec{B}_{{}_M}'\cdot\vec{v}_{{}_M}\cdot
\frac{\gamma_{{}_M}^2}{\gamma_{{}_M}+1}\cdot\lbrack
\frac{\gamma_{{}_M}}{\gamma_{{}_M}+1}\cdot(g-2)
- c_1\cdot\gamma_{{}_M}\rbrack
\nonumber\\&&\qquad
+ \frac{e\cdot g}{2\cdot m}\cdot
    \frac{\gamma_{{}_M}}{\gamma_{{}_M}+1}\cdot
\vec{v}_{{}_M}^\dagger\cdot\vec{B}_{{}_M}' \cdot\vec{\sigma}
\nonumber\\&&\qquad
+\frac{e^2}{4\cdot m^2}\cdot
  \vec{v}_{{}_M}^\dagger\cdot\vec{B}_{{}_M}\cdot
 (\vec{B}_{{}_M} \wtimes\vec{\sigma})\cdot
\gamma_{{}_M}\cdot
(c_5  - g^2\cdot\frac{1}{\gamma_{{}_M}\cdot(\gamma_{{}_M}+1)})
 \nonumber\\&&\qquad
+\frac{e^2}{4\cdot m^2}\cdot \vec{v}_{{}_M}^\dagger\cdot\vec{B}_{{}_M}\cdot
       \vec{\sigma}^{\,\dagger}\cdot\vec{v}_{{}_M}\cdot
 (\vec{v}_{{}_M} \wtimes\vec{B}_{{}_M})\cdot\lbrack
\frac{\gamma_{{}_M}^2}{\gamma_{{}_M}+1}\cdot(c_3-2\cdot c_4+c_5)
+g^2\cdot\frac{\gamma_{{}_M}^2}{(\gamma_{{}_M}+1)^2}
 \nonumber\\&&\qquad
+4\cdot\frac{\gamma_{{}_M}^3}{(\gamma_{{}_M}+1)^2}\rbrack
-\frac{e^2}{4\cdot m^2}\cdot c_5\cdot
  \vec{B}_{{}_M}^\dagger\cdot\vec{B}_{{}_M}\cdot\gamma_{{}_M}\cdot
 (\vec{v}_{{}_M} \wtimes\vec{\sigma})
\nonumber\\&&\qquad
+\frac{e^2}{4\cdot m^2}\cdot\vec{\sigma}^\dagger
\cdot\vec{B}_{{}_M}\cdot(\vec{v}_{{}_M} \wtimes\vec{B}_{{}_M})\cdot
\gamma_{{}_M}\cdot
(2\cdot c_4-c_3 -\frac{4\cdot\gamma_{{}_M}}{\gamma_{{}_M}+1})
 \nonumber\\&&\qquad
   +\frac{e^2\cdot(g-2)^2}{4\cdot m^2}\cdot
\frac{\gamma_{{}_M}^2}{\gamma_{{}_M}+1}\cdot
(\vec{\sigma}\wtimes\vec{v}_{{}_M})^\dagger\cdot\vec{B}_{{}_M}\cdot
\vec{B}_{{}_M}
\nonumber\\&&\qquad
-\frac{e^2\cdot g\cdot(g-2)}{4\cdot m^2}\cdot
    \frac{\gamma_{{}_M}^2}{\gamma_{{}_M}+1}
\cdot \vec{v}_{{}_M}^\dagger\cdot\vec{B}_{{}_M}\cdot(\vec{\sigma}\wtimes
\vec{v}_{{}_M})^\dagger\cdot\vec{B}_{{}_M}\cdot\vec{v}_{{}_M}
\nonumber\\
&=& e\cdot (\vec{v}_{{}_M}\wtimes\vec{B}_{{}_M})
 + \frac{e}{2\cdot m}\cdot
  (-\frac{2\cdot\gamma_{{}_M}}{\gamma_{{}_M}+1}\cdot
   \vec{v}_{{}_M}^{\,\dagger}\cdot\vec{v}_{{}_M}+c_2)\cdot
  \vec{\nabla}_{{}_M}(\vec{\sigma}^{\,\dagger}\cdot \vec{B}_{{}_M})
\nonumber\\&&
 + \frac{e}{2\cdot m}\cdot\biggl(
 \lbrack(c_2-g)\cdot\frac{\gamma_{{}_M}^2}{\gamma_{{}_M}+1}-c_1\cdot
\gamma_{{}_M}\rbrack
\cdot    \vec{\sigma}^{\,\dagger}\cdot \vec{v}_{{}_M}\cdot\vec{B}_{{}_M}'
\nonumber\\&&
+ (c_1\cdot\gamma_{{}_M}^2+2\cdot\gamma_{{}_M})\cdot
  \vec{\sigma}^{\,\dagger}\cdot \vec{B}_{{}_M}'\cdot\vec{v}_{{}_M}
\nonumber\\&&
+ \frac{\gamma_{{}_M}^2}{\gamma_{{}_M}+1}\cdot\lbrack
\frac{\gamma_{{}_M}}{\gamma_{{}_M}+1}\cdot(g-2)
- c_1\cdot\gamma_{{}_M}\rbrack\cdot
      \vec{\sigma}^{\,\dagger}\cdot
      \vec{v}_{{}_M}\cdot\vec{v}_{{}_M}^{\,\dagger}\cdot
      \vec{B}_{{}_M}'\cdot\vec{v}_{{}_M}
\nonumber\\&&
   + (g-2)\cdot\frac{\gamma_{{}_M}}{\gamma_{{}_M}+1}\cdot
   \vec{v}_{{}_M}^{\,\dagger}\cdot\vec{B}_{{}_M}'\cdot\vec{\sigma}\biggr)
\nonumber\\&&
+\frac{e^2}{4\cdot m^2}\cdot\biggl(
\lbrack 4-4\cdot\gamma_{{}_M}+\gamma_{{}_M}\cdot
   (2\cdot c_4-c_3)\rbrack\cdot
   \vec{B}_{{}_M}^\dagger\cdot  \vec{\sigma}
      \cdot  (\vec{v}_{{}_M} \wtimes\vec{B}_{{}_M})
\nonumber\\&&
 + \frac{\gamma_{{}_M}^2}{\gamma_{{}_M}+1}\cdot\lbrack
   -2\cdot c_4+c_3+c_5+4+(g^2-2\cdot g)\cdot
 \frac{1}{\gamma_{{}_M}+1}\rbrack\cdot
 \vec{v}_{{}_M}^{\,\dagger}\cdot\vec{\sigma}\cdot
 \vec{v}_{{}_M}^{\,\dagger}\cdot\vec{B}_{{}_M}
       \cdot(\vec{v}_{{}_M}\wtimes \vec{B}_{{}_M})
\nonumber\\&&
+\lbrack(g-2)\cdot g\cdot\frac{1}{\gamma_{{}_M}+1}-\gamma_{{}_M}\cdot c_5
\rbrack\cdot \vec{v}_{{}_M}^{\,\dagger}\cdot\vec{B}_{{}_M}\cdot
  (\vec{\sigma}\wtimes\vec{B}_{{}_M})
   -c_5\cdot\gamma_{{}_M}\cdot
   \vec{B}_{{}_M}^\dagger\cdot\vec{B}_{{}_M}\cdot
                                 (\vec{v}_{{}_M} \wtimes\vec{\sigma})
\nonumber\\&&\qquad
+(g-2)^2\cdot\frac{\gamma_{{}_M}^2}{\gamma_{{}_M}+1}\cdot
(\vec{\sigma}\wtimes\vec{v}_{{}_M})^\dagger\cdot\vec{B}_{{}_M}\cdot
\vec{B}_{{}_M}
\nonumber\\&&
-(g^2-2\cdot g)\cdot
   \frac{\gamma_{{}_M}^2}{\gamma_{{}_M}+1}\cdot
 (\vec{\sigma}\wtimes\vec{v}_{{}_M})^\dagger\cdot\vec{B}_{{}_M}
   \cdot\vec{v}_{{}_M}^{\,\dagger}\cdot\vec{B}_{{}_M}
                                      \cdot\vec{v}_{{}_M}\biggr) \; .
\label{C.7}
\end{eqnarray}
This completes the derivation of (8.15).
\setcounter{section}{4}
\setcounter{subsection}{0}
\setcounter{equation}{0}
\section*{Appendix D}
\addcontentsline{toc}{section}{Appendix D}
\subsection*{D.1}
In this Appendix I show that (5.12) is equivalent to (5.5).
First of all one observes by (3.1-2) that (5.12a) is equivalent to
(5.5a).
\subsection*{D.2}
In this subsection I show that (5.12b) is equivalent to (5.5b)
and to do that I only have to show that the spatial part of (5.5b)
is equivalent to (5.12b).
\footnote{The temporal part of equation (5.5b) follows from the spatial part
of equation (5.5b) by using the constraints (3.8).}
To come to that I have to calculate the spatial parts of several
4-vectors. Given arbitrary antisymmetric tensors $N,\hat{N}$ of rank 2
and a 4-vector $a$ with the following notation:
\begin{eqnarray*}
 && N  \leftrightarrow (\vec{b},\vec{d}) \; , \qquad
    \hat{N}  \leftrightarrow (\hat{\vec{b}},\hat{\vec{d}}) \; ,
\qquad
    a_{\mu} = (\vec{a}^\dagger, a_4)_{\mu} \; ,
 \qquad (\mu =1,...,4)
\end{eqnarray*}
one gets:
\begin{eqnarray}
&& N_{\mu\nu}\cdot a_{\nu}  =  \biggl( \lbrack \vec{a} \wtimes\vec{b}
 + a_4 \cdot \vec{d}\rbrack^\dagger ,  - \vec{a}^\dagger\cdot
                                                 \vec{d} \biggr)_{\mu}
                       \; , \;
   N_{\nu\omega}\cdot \hat{N}_{\omega\nu} =
   -2\cdot\vec{b}^\dagger\cdot\hat{\vec{b}}
   -2\cdot\vec{d}^{\,\dagger}\cdot\hat{\vec{d}} \; .
 \qquad (\mu =1,...,4)
\nonumber\\
\label{D.1}
\end{eqnarray}
One thus gets
\footnote{The partial derivative $\partial/\partial t$ in this equation
acts on functions depending on $\vec{r}_{{}_P},t$.}
\begin{eqnarray*}
 && F^{{}^P}_{\mu\nu}\cdot U^{{}^P}_{\nu} =
\gamma_{{}_P}\cdot\biggl(\lbrack \vec{v}_{{}_P}
  \wtimes\vec{B}_{{}_P}+\vec{E}_{{}_P}\rbrack^\dagger,
   i\cdot\vec{v}_{{}_P}^\dagger\cdot\vec{E}_{{}_P}\biggr)_{\mu} \; ,
\nonumber\\
&& F^{{}^P}_{\mu\nu}\cdot F^{{}^P}_{\nu\rho}\cdot U^{{}^P}_{\rho}=
 \gamma_{{}_P}\cdot
\biggl( \lbrack (\vec{v}_{{}_P}\wtimes
  \vec{B}_{{}_P}+\vec{E}_{{}_P} )\wtimes\vec{B}_{{}_P} +
 \vec{v}_{{}_P}^\dagger\cdot\vec{E}_{{}_P}\cdot
                                         \vec{E}_{{}_P}\rbrack^\dagger ,
i\cdot (\vec{v}_{{}_P}\wtimes\vec{B}_{{}_P}+\vec{E}_{{}_P} )^\dagger
  \cdot\vec{E}_{{}_P} \biggr)_{\mu}\; ,
\nonumber\\
&& S^{{}^P}_{\mu\nu}\cdot F^{{}^P}_{\nu\rho}\cdot U^{{}^P}_{\rho}
\nonumber\\&&
=  \gamma_{{}_P}\cdot
\biggl( \lbrack (\vec{v}_{{}_P}\wtimes\vec{B}_{{}_P}+\vec{E}_{{}_P} )
\wtimes
\vec{s}-\vec{v}_{{}_P}^\dagger\cdot\vec{E}_{{}_P}
     \cdot(\vec{v}_{{}_P}\wtimes\vec{s})\rbrack^\dagger,
-i\cdot (\vec{v}_{{}_P}\wtimes\vec{B}_{{}_P}+\vec{E}_{{}_P} )^\dagger
                         \cdot(\vec{v}_{{}_P}\wtimes\vec{s})
               \biggr)_{\mu}\; ,
\nonumber\\
&& U^{{}^P}_{\lambda}\cdot\partial^{{}^P}_{\lambda} =
  \gamma_{{}_P}\cdot (\vec{v}_{{}_P}^\dagger\cdot\vec{\nabla}_{{}_P}
 + \partial/\partial t ) \; ,
 \qquad (\mu =1,...,4)
\end{eqnarray*}
and
\begin{eqnarray}
 && U^{{}^P}_{\mu} \rightarrow  \gamma_{{}_P}\cdot\vec{v}_{{}_P}  \; ,
\nonumber\\
&&\dot{U}^{{}^P}_{\mu} \rightarrow  \gamma_{{}_P}\cdot
             (\gamma_{{}_P}\cdot\vec{v}_{{}_P} )' \; ,
\nonumber\\
 && F^{{}^P}_{\mu\nu}\cdot U^{{}^P}_{\nu}
   \rightarrow                 \gamma_{{}_P}
\cdot \lbrack \vec{v}_{{}_P}\wtimes
 \vec{B}_{{}_P}  + \vec{E}_{{}_P}  \rbrack \; ,
\nonumber\\
&& S^{{}^P}_{\nu\omega}\cdot \partial^{{}^P}_{\mu}
               F^{{}^P}_{\omega\nu}\rightarrow
-  2\cdot         \vec{\nabla}_{{}_P}\biggl(
\vec{s}^{\,\dagger}\cdot\vec{B}_{{}_P} -\vec{E}_{{}_P}^\dagger
 \cdot(\vec{s}\wtimes\vec{v}_{{}_P})\biggr)
  \; ,
\nonumber\\
&& U^{{}^P}_{\mu}\cdot S^{{}^P}_{\nu\omega}\cdot
                                     F^{{}^P}_{\omega\nu}  \rightarrow
-2\cdot\gamma_{{}_P}\cdot
       \lbrack \vec{s}^{\,\dagger}\cdot\vec{B}_{{}_P}
     -\vec{E}_{{}_P}^\dagger\cdot(\vec{s}\wtimes
                                         \vec{v}_{{}_P}) \rbrack \cdot
      \vec{v}_{{}_P} \; ,
\nonumber\\
&& U^{{}^P}_{\mu}\cdot S^{{}^P}_{\nu\omega}\cdot U^{{}^P}_{\lambda}
                                      \cdot\partial^{{}^P}_{\lambda}
               F^{{}^P}_{\omega\nu}  \rightarrow
-2\cdot\gamma_{{}_P}^2\cdot
       \lbrack \vec{s}^{\,\dagger}\cdot\vec{B}_{{}_P}'
     -(\vec{s}\wtimes\vec{v}_{{}_P})^\dagger\cdot
                                          \vec{E}_{{}_P}' \rbrack \cdot
      \vec{v}_{{}_P} \; ,
\nonumber\\
&& S^{{}^P}_{\mu\nu}\cdot U^{{}^P}_{\omega}\cdot
   U^{{}^P}_{\lambda}\cdot\partial^{{}^P}_{\lambda}
                            F^{{}^P}_{\nu\omega}
   \rightarrow
\gamma_{{}_P}^2\cdot \lbrack
\vec{E}_{{}_P}'+\vec{v}_{{}_P}\wtimes\vec{B}_{{}_P}'-
  \vec{v}_{{}_P}^\dagger\cdot\vec{E}_{{}_P}'\cdot\vec{v}_{{}_P} \rbrack
  \wtimes\vec{s}
\nonumber\\&&\qquad
 = \gamma_{{}_P}^2\cdot \lbrack
\vec{E}_{{}_P}' \wtimes\vec{s}
+ \vec{v}_{{}_P}^\dagger\cdot\vec{s}\cdot \vec{B}_{{}_P}'
- \vec{s}^{\,\dagger}\cdot\vec{B}_{{}_P}'\cdot \vec{v}_{{}_P}
- \vec{v}_{{}_P}^\dagger\cdot\vec{E}_{{}_P}'\cdot
(\vec{v}_{{}_P} \wtimes\vec{s})\rbrack
                \; ,
\nonumber\\
&& S^{{}^P}_{\mu\nu}\cdot F^{{}^P}_{\nu\omega}\cdot
               F^{{}^P}_{\omega\rho}\cdot U^{{}^P}_{\rho}
   \rightarrow
 \gamma_{{}_P}\cdot
\biggl( ( \vec{v}_{{}_P}\wtimes\vec{B}_{{}_P}+\vec{E}_{{}_P} )
\wtimes
\vec{B}_{{}_P} + \vec{v}_{{}_P}^\dagger\cdot\vec{E}_{{}_P}
  \cdot\vec{E}_{{}_P}\biggr) \wtimes\vec{s}
\nonumber\\&&
- \gamma_{{}_P}\cdot
       ( \vec{v}_{{}_P}\wtimes\vec{B}_{{}_P}+\vec{E}_{{}_P} )^\dagger
 \cdot\vec{E}_{{}_P} \cdot(\vec{v}_{{}_P}\wtimes\vec{s})
 = \gamma_{{}_P}\cdot \lbrack
   \vec{v}_{{}_P}^\dagger\cdot\vec{B}_{{}_P}\cdot
                                      (\vec{B}_{{}_P} \wtimes\vec{s})
 - \vec{B}_{{}_P}^\dagger\cdot\vec{B}_{{}_P}\cdot
                                 (\vec{v}_{{}_P} \wtimes\vec{s})
\nonumber\\&&\qquad
 + \vec{E}_{{}_P}^\dagger\cdot\vec{s}\cdot\vec{B}_{{}_P}
 - \vec{B}_{{}_P}^\dagger\cdot\vec{s}\cdot\vec{E}_{{}_P}
 + \vec{v}_{{}_P}^\dagger\cdot\vec{E}_{{}_P}\cdot
                                      (\vec{E}_{{}_P} \wtimes\vec{s})
-
       ( \vec{v}_{{}_P}\wtimes\vec{B}_{{}_P}+\vec{E}_{{}_P} )^\dagger
 \cdot\vec{E}_{{}_P} \cdot(\vec{v}_{{}_P}\wtimes\vec{s})\rbrack
                                     \; ,
\nonumber\\
&& F^{{}^P}_{\mu\nu}\cdot S^{{}^P}_{\nu\omega}\cdot
             F^{{}^P}_{\omega\rho}\cdot U^{{}^P}_{\rho}
                             \rightarrow
   \gamma_{{}_P}\cdot
        \lbrack (\vec{v}_{{}_P}\wtimes\vec{B}_{{}_P}+\vec{E}_{{}_P} )
\wtimes
\vec{s}-\vec{v}_{{}_P}^\dagger\cdot\vec{E}_{{}_P}
     \cdot(\vec{v}_{{}_P}\wtimes\vec{s})\rbrack \wtimes \vec{B}_{{}_P}
\nonumber\\&&\qquad
- \gamma_{{}_P}\cdot
    (\vec{v}_{{}_P}\wtimes\vec{B}_{{}_P}+\vec{E}_{{}_P} )^\dagger
         \cdot(\vec{v}_{{}_P}\wtimes\vec{s})\cdot\vec{E}_{{}_P}
 = \gamma_{{}_P}\cdot \lbrack
-  \vec{s}^{\,\dagger}\cdot\vec{B}_{{}_P}\cdot
                               (\vec{v}_{{}_P} \wtimes\vec{B}_{{}_P})
 + \vec{E}_{{}_P}^\dagger\cdot\vec{B}_{{}_P}\cdot \vec{s}
\nonumber\\&&\qquad
 - (1+ \vec{v}_{{}_P}^\dagger\cdot\vec{v}_{{}_P} )\cdot
   \vec{B}_{{}_P}^\dagger\cdot\vec{s}\cdot\vec{E}_{{}_P}
 - \vec{v}_{{}_P}^\dagger\cdot\vec{E}_{{}_P}
    \cdot\vec{v}_{{}_P}^\dagger\cdot\vec{B}_{{}_P}\cdot\vec{s}
 + \vec{v}_{{}_P}^\dagger\cdot\vec{E}_{{}_P}
   \cdot\vec{B}_{{}_P}^\dagger\cdot\vec{s}\cdot\vec{v}_{{}_P}
\nonumber\\&&\qquad
 + \vec{v}_{{}_P}^\dagger\cdot\vec{s}
 \cdot\vec{v}_{{}_P}^\dagger\cdot\vec{B}_{{}_P}\cdot\vec{E}_{{}_P}
- \vec{E}_{{}_P}^\dagger\cdot( \vec{v}_{{}_P}\wtimes
 \vec{s})\cdot\vec{E}_{{}_P} \rbrack
                                     \; ,
\nonumber\\
 && F^{{}^P}_{\mu\nu}\cdot U^{{}^P}_{\nu} \cdot
   S^{{}^P}_{\nu\omega}\cdot F^{{}^P}_{\omega\nu}\rightarrow
-2\cdot \gamma_{{}_P}\cdot
       \lbrack \vec{s}^{\,\dagger}\cdot\vec{B}_{{}_P}
     -\vec{E}_{{}_P}^\dagger\cdot(\vec{s}\wtimes
                                         \vec{v}_{{}_P}) \rbrack \cdot
       \lbrack \vec{v}_{{}_P}\wtimes
 \vec{B}_{{}_P}  + \vec{E}_{{}_P}  \rbrack \; ,
\label{D.2}
\end{eqnarray}
where the expressions on the rhs of the arrows denote the corresponding
spatial parts. With (D.2) the spatial part of the rhs of (5.5b) reads
as:
\begin{eqnarray}
&& \frac{e}{m}\cdot\gamma_{{}_P} \cdot
          (\vec{v}_{{}_P}\wtimes\vec{B}_{{}_P}) +
   \frac{e}{m}\cdot\gamma_{{}_P} \cdot \vec{E}_{{}_P}  +
\frac{e\cdot c_2}{2\cdot m^2}\cdot \vec{\nabla}_{{}_P}\biggl(
\vec{s}^{\,\dagger}\cdot\vec{B}_{{}_P}-\vec{E}_{{}_P}^\dagger
   \cdot(\vec{s}\wtimes\vec{v}_{{}_P}) \biggr)
\nonumber\\&& \qquad
+\frac{e\cdot \gamma_{{}_P}^2}{2\cdot m^2}\cdot \lbrack
 (c_1+2)\cdot\vec{s}^{\,\dagger}\cdot\vec{B}_{{}_P}'\cdot\vec{v}_{{}_P}
-c_2\cdot(\vec{s}\wtimes\vec{v}_{{}_P})^\dagger\cdot
  \vec{E}_{{}_P}'\cdot\vec{v}_{{}_P}
+(c_2-c_1-2)\cdot(\vec{E}_{{}_P}'\wtimes\vec{s})
\nonumber\\&&\qquad
+(c_2-c_1-2)\cdot\vec{v}_{{}_P}^\dagger\cdot\vec{s}\cdot\vec{B}_{{}_P}'
-(c_2-c_1-2)\cdot\vec{v}_{{}_P}^\dagger\cdot\vec{E}_{{}_P}'
     \cdot(\vec{v}_{{}_P}\wtimes\vec{s})\rbrack
\nonumber\\&&\qquad
+\frac{e^2}{4\cdot m^3}\cdot\gamma_{{}_P} \cdot\lbrack
    c_5\cdot
   \vec{v}_{{}_P}^\dagger\cdot\vec{B}_{{}_P}\cdot
                                      (\vec{B}_{{}_P} \wtimes\vec{s})
   -c_5\cdot
   \vec{B}_{{}_P}^\dagger\cdot\vec{B}_{{}_P}\cdot
                                 (\vec{v}_{{}_P} \wtimes\vec{s})
   +c_5\cdot
   \vec{E}_{{}_P}^\dagger\cdot\vec{s}\cdot\vec{B}_{{}_P}
\nonumber\\&&\qquad
+(-c_3\cdot \vec{v}_{{}_P}^\dagger\cdot\vec{v}_{{}_P}
-c_5+2\cdot c_4 -c_3   )\cdot
   \vec{s}^{\,\dagger}\cdot\vec{B}_{{}_P}\cdot\vec{E}_{{}_P}
   +c_5\cdot
   \vec{v}_{{}_P}^\dagger\cdot\vec{E}_{{}_P}\cdot
                                      (\vec{E}_{{}_P} \wtimes\vec{s})
\nonumber\\&&\qquad
   -c_5\cdot
       ( \vec{v}_{{}_P}\wtimes\vec{B}_{{}_P})^\dagger
 \cdot\vec{E}_{{}_P} \cdot(\vec{v}_{{}_P}\wtimes\vec{s})
   -c_5\cdot
       \vec{E}_{{}_P}^\dagger\cdot\vec{E}_{{}_P} \cdot
                                      (\vec{v}_{{}_P}\wtimes\vec{s})
\nonumber\\&&\qquad
   +(2\cdot c_4-c_3)\cdot
   \vec{s}^{\,\dagger}\cdot\vec{B}_{{}_P}\cdot
                               (\vec{v}_{{}_P} \wtimes\vec{B}_{{}_P})
   +c_3\cdot
   \vec{E}_{{}_P}^\dagger\cdot\vec{B}_{{}_P}\cdot \vec{s}
   -c_3\cdot
   \vec{v}_{{}_P}^\dagger\cdot\vec{E}_{{}_P}\cdot
          \vec{v}_{{}_P}^\dagger\cdot\vec{B}_{{}_P}\cdot\vec{s}
\nonumber\\&&\qquad
   +c_3\cdot
   \vec{v}_{{}_P}^\dagger\cdot\vec{E}_{{}_P}\cdot
           \vec{s}^{\,\dagger}\cdot\vec{B}_{{}_P}\cdot\vec{v}_{{}_P}
   +c_3\cdot
   \vec{v}_{{}_P}^\dagger\cdot\vec{s}\cdot
            \vec{v}_{{}_P}^\dagger\cdot\vec{B}_{{}_P}\cdot\vec{E}_{{}_P}
\nonumber\\&&\qquad
   +(2\cdot c_4-c_3)\cdot
  \vec{E}_{{}_P}^\dagger\cdot( \vec{v}_{{}_P}\wtimes
                                             \vec{s})\cdot\vec{E}_{{}_P}
   +2\cdot c_4\cdot
\vec{E}_{{}_P}^\dagger\cdot( \vec{v}_{{}_P}\wtimes
      \vec{s})\cdot(\vec{v}_{{}_P}\wtimes\vec{B}_{{}_P})
\rbrack   \; ,
\label{D.3}
\end{eqnarray}
and the spatial part of the lhs of (5.5b) reads as:
\begin{eqnarray}
 &&   \gamma_{{}_P}\cdot (\gamma_{{}_P}\cdot\vec{v}_{{}_P})' \; .
\label{D.4}
\end{eqnarray}
By (D.3-4),(5.5b),(5.12b) one sees that multiplying the spatial part of
(5.5b) by $m/\gamma_{{}_P}$ results in (5.12b) so that the spatial
part of (5.5b) is equivalent to (5.12b). Thus I have shown that
(5.5b) is equivalent to (5.12b).
\subsection*{D.3}
In this subsection I show that the spatial part of (5.5c) is
equivalent to (5.12c). To come to that I have to calculate the spatial
parts of several antisymmetric tensors of rank 2. Using again the
notation
\begin{eqnarray*}
 && a_{\mu} = (\vec{a}^\dagger, a_4)_{\mu} \; ,
 \qquad (\mu =1,...,4)
\end{eqnarray*}
I calculate first of all:
\footnote{Note that: $
    \sum_{k=1}^3 \varepsilon_{jpk}\cdot\varepsilon_{nmk}=
     \delta_{jn}\cdot\delta_{pm} - \delta_{jm}\cdot\delta_{pn}$ for
$j,p,m,n=1,2,3$.}
\begin{eqnarray}
&& \frac{1}{2}\cdot \sum_{k,i=1}^3 \varepsilon_{jki} \cdot\lbrack
    S^{{}^P}_{k\omega}\cdot F^{{}^P}_{\omega\lambda}\cdot a_{\lambda}\cdot a_i
-S^{{}^P}_{i\omega}\cdot F^{{}^P}_{\omega\lambda}\cdot
                                           a_{\lambda}\cdot a_k\rbrack
 =   \sum_{k,i=1}^3 \varepsilon_{jki} \cdot
    S^{{}^P}_{k\omega}\cdot F^{{}^P}_{\omega\lambda}\cdot a_{\lambda}\cdot a_i
\nonumber\\
&=&\sum_{k,i=1}^3 \varepsilon_{jki}\cdot\lbrack
   \sum_{m=1}^3 S^{{}^P}_{km}\cdot F^{{}^P}_{m\lambda}\cdot
a_{\lambda}\cdot a_i
 +  S^{{}^P}_{k4}\cdot F^{{}^P}_{4\lambda}\cdot a_{\lambda}\cdot a_i \rbrack
\nonumber\\
&=&\sum_{k,i=1}^3 \varepsilon_{jki}\cdot\lbrack
   \sum_{m,n=1}^3 \varepsilon_{kmn}\cdot
            s_n\cdot F^{{}^P}_{m\lambda}\cdot
a_{\lambda}\cdot a_i
 +  S^{{}^P}_{k4}\cdot F^{{}^P}_{4\lambda}\cdot a_{\lambda}\cdot a_i \rbrack
\nonumber\\
&=&\sum_{k,i=1}^3 \varepsilon_{jki}\cdot\biggl(
   \sum_{m,n=1}^3 \varepsilon_{kmn}\cdot s_n\cdot\lbrack
   \sum_{r=1}^3 F^{{}^P}_{mr}\cdot a_r\cdot a_i
    +  F^{{}^P}_{m4}\cdot a_4\cdot a_i \rbrack
 +  S^{{}^P}_{k4}\cdot F^{{}^P}_{4\lambda}\cdot a_{\lambda}\cdot a_i
\biggr)
\nonumber\\
&=& \sum_{k,i=1}^3 \varepsilon_{jik}\cdot\biggl(
   \sum_{m,n=1}^3 \varepsilon_{nmk}\cdot s_n\cdot\lbrack
\sum_{q,r=1}^3 \varepsilon_{mrq}\cdot B_{{}_{P,q}}\cdot a_r\cdot a_i
    +  F^{{}^P}_{m4}\cdot a_4\cdot a_i \rbrack
 - \sum_{m=1}^3 S^{{}^P}_{k4}\cdot F^{{}^P}_{4m}\cdot a_m\cdot a_i
\biggr)
\nonumber\\
&=& \sum_{k,i=1}^3 \varepsilon_{jik}\cdot\biggl(
   \sum_{m,n=1}^3 \varepsilon_{nmk}\cdot s_n\cdot\lbrack
\sum_{q,r=1}^3 \varepsilon_{mrq}\cdot B_{{}_{P,q}}\cdot a_r\cdot a_i
 - i\cdot E_{{}_{P,m}}\cdot a_4\cdot a_i \rbrack
\nonumber\\&&\qquad
 - \sum_{m=1}^3 q_k\cdot E_{{}_{P,m}}\cdot a_m\cdot a_i
\biggr)
\nonumber\\
&=& \sum_{i,m,n=1}^3 \lbrack \delta_{jn}\cdot\delta_{im} -
  \delta_{jm}\cdot\delta_{in} \rbrack\cdot s_n\cdot\lbrack
\sum_{q,r=1}^3 \varepsilon_{mrq}\cdot B_{{}_{P,q}}\cdot a_r\cdot a_i
 - i\cdot E_{{}_{P,m}}\cdot a_4\cdot a_i \rbrack
\nonumber\\&&\qquad
 + \vec{a}^\dagger\cdot\vec{E}_{{}_P}\cdot (\vec{q}\wtimes\vec{a})_j
\nonumber\\
&=& \sum_{i,q,r=1}^3
\varepsilon_{irq}\cdot B_{{}_{P,q}}\cdot a_r\cdot a_i\cdot s_j
  - \sum_{n,q,r=1}^3
\varepsilon_{jrq}\cdot B_{{}_{P,q}}\cdot a_r\cdot a_n\cdot s_n
\nonumber\\&&\qquad
 - i\cdot a_4\cdot \sum_{i=1}^3  \lbrack
E_{{}_{P,i}}\cdot a_i\cdot s_j-E_{{}_{P,j}}\cdot a_i\cdot s_i \rbrack
 + \vec{a}^\dagger\cdot\vec{E}_{{}_P}\cdot (\vec{q}\wtimes\vec{a})_j
\nonumber\\
&=& \vec{a}^\dagger\cdot\vec{s}\cdot (\vec{B}_{{}_P}\wtimes\vec{a})_j
 - i\cdot a_4\cdot \vec{a}^\dagger\cdot\vec{E}_{{}_P}\cdot s_j
 + i\cdot a_4\cdot \vec{a}^\dagger\cdot\vec{s}\cdot E_{{}_{P,j}}
 + \vec{a}^\dagger\cdot\vec{E}_{{}_P}\cdot
                                     (\vec{q}\wtimes\vec{a})_j \; ,
\nonumber\\&&
\qquad\qquad\qquad\qquad\qquad\qquad
\qquad\qquad\qquad\qquad\qquad\qquad
\qquad
(j=1,2,3)
\label{D.5}
\end{eqnarray}
from which follows
\begin{eqnarray}
&&S^{{}^P}_{\mu\omega}\cdot F^{{}^P}_{\omega\lambda}\cdot
                                              a_{\lambda}\cdot a_{\nu}-
    S^{{}^P}_{\nu\omega}\cdot F^{{}^P}_{\omega\lambda}\cdot
                                                a_{\lambda}\cdot a_{\mu}
   \rightarrow
\nonumber\\&&\qquad
    \vec{a}^\dagger\cdot\vec{s}\cdot (\vec{B}_{{}_P}\wtimes\vec{a})
 - i\cdot a_4\cdot \vec{a}^\dagger\cdot\vec{E}_{{}_P}\cdot \vec{s}
 + i\cdot a_4\cdot \vec{a}^\dagger\cdot\vec{s}\cdot \vec{E}_{{}_P}
 + \vec{a}^\dagger\cdot\vec{E}_{{}_P}\cdot (\vec{q}\wtimes\vec{a})\; ,
\qquad
\label{D.6}
\end{eqnarray}
where the expression on the rhs of the arrow denotes the corresponding
spatial part. For the special case: $a=U^{{}^P}$ one gets from (D.6):
\begin{eqnarray}
&&S^{{}^P}_{\mu\omega}\cdot F^{{}^P}_{\omega\lambda}\cdot U^{{}^P}_{\lambda}
                                                   \cdot U^{{}^P}_{\nu}-
S^{{}^P}_{\nu\omega}\cdot F^{{}^P}_{\omega\lambda}\cdot
     U^{{}^P}_{\lambda}\cdot U^{{}^P}_{\mu}
   \rightarrow
\nonumber\\&&\qquad
           -\gamma_{{}_P}^2 \cdot
\vec{s}^{\,\dagger}\cdot\vec{v}_{{}_P}\cdot
     \lbrack \vec{E}_{{}_P}+\vec{v}_{{}_P}\wtimes\vec{B}_{{}_P}
\rbrack + \vec{v}_{{}_P}^\dagger\cdot
 \vec{E}_{{}_P}\cdot\vec{s} + \gamma_{{}_P}^2 \cdot
\vec{s}^{\,\dagger}\cdot\vec{v}_{{}_P}\cdot
     \vec{v}_{{}_P}^\dagger\cdot\vec{E}_{{}_P}\cdot \vec{v}_{{}_P}
           \; .
\label{D.7}
\end{eqnarray}
Also I calculate:
\begin{eqnarray}
&& \frac{1}{2}\cdot \sum_{k,l=1}^3 \varepsilon_{jkl} \cdot\lbrack
S^{{}^P}_{k\omega}\cdot F^{{}^P}_{\omega l}-S^{{}^P}_{l\omega}\cdot
            F^{{}^P}_{\omega k}\rbrack
 =   \sum_{k,l=1}^3 \varepsilon_{jkl} \cdot
    S^{{}^P}_{k\omega}\cdot F^{{}^P}_{\omega l}
\nonumber\\
&=&\sum_{k,l=1}^3 \varepsilon_{jkl}\cdot\lbrack
   \sum_{m=1}^3 S^{{}^P}_{km}\cdot F^{{}^P}_{ml} +
           S^{{}^P}_{k4}\cdot F^{{}^P}_{4l} \rbrack
 = \sum_{k,l=1}^3 \varepsilon_{jkl}\cdot\lbrack
   \sum_{m,n=1}^3 \varepsilon_{kmn}\cdot s_n\cdot F^{{}^P}_{ml}
 +  S^{{}^P}_{k4}\cdot F^{{}^P}_{4l} \rbrack
\nonumber\\
&=& \sum_{k,l=1}^3 \varepsilon_{jlk}\cdot\biggl(
   \sum_{m,n,r=1}^3 \varepsilon_{nmk}\cdot s_n\cdot
    \varepsilon_{mlr}\cdot B_r
 - S^{{}^P}_{k4}\cdot F^{{}^P}_{4l} \biggr)
\nonumber\\
&=& \sum_{k,l=1}^3 \varepsilon_{jlk}\cdot\biggl(
   \sum_{m,n,r=1}^3 \varepsilon_{nmk}\cdot s_n\cdot
       \varepsilon_{mlr}\cdot B_{{}_{P,r}}
 -  q_k\cdot E_{{}_{P,l}} \biggr)
\nonumber\\
&=& \sum_{l,m,n,r=1}^3 \lbrack \delta_{jn}\cdot\delta_{lm} -
  \delta_{jm}\cdot\delta_{ln} \rbrack\cdot s_n\cdot
     \varepsilon_{mlr}\cdot B_{{}_{P,r}}
 +  (\vec{q}\wtimes\vec{E}_{{}_P})_j
\nonumber\\
&=& - \sum_{n,r=1}^3
  \varepsilon_{jnr}\cdot s_n\cdot B_{{}_{P,r}}
 +  (\vec{q}\wtimes\vec{E}_{{}_P})_j
 =    (\vec{B}_{{}_P}\wtimes\vec{s})_j
 +  (\vec{q}\wtimes\vec{E}_{{}_P})_j  \; ,
\qquad
(j=1,2,3)
\label{D.8}
\end{eqnarray}
from which follows
\begin{eqnarray}
&&F^{{}^P}_{\mu\omega}\cdot S^{{}^P}_{\omega\nu}-
           S^{{}^P}_{\mu\omega}\cdot F^{{}^P}_{\omega\nu}
   \rightarrow
 \vec{s}\wtimes\vec{B}_{{}_P} +
    (\vec{v}_{{}_P}\wtimes\vec{s})\wtimes\vec{E}_{{}_P}  \; ,
\label{D.9}
\end{eqnarray}
where the expression on the rhs of the arrow denotes the corresponding
spatial part. With (D.7),(D.9) the spatial part of the rhs of (5.5c)
reads as:
\begin{eqnarray}
 && \frac{e}{m}\cdot \lbrack
 \frac{g-2}{2}\cdot\gamma_{{}_P}^2\cdot
             \vec{s}^{\,\dagger}\cdot\vec{v}_{{}_P}\cdot
    (\vec{E}_{{}_P}+\vec{v}_{{}_P}\wtimes\vec{B}_{{}_P})
+\vec{v}_{{}_P}^\dagger\cdot\vec{E}_{{}_P}\cdot\vec{s}
\nonumber\\&&
-\frac{g-2}{2}\cdot\gamma_{{}_P}^2\cdot \vec{s}^{\,\dagger}\cdot
                                                    \vec{v}_{{}_P}\cdot
        \vec{v}_{{}_P}^\dagger\cdot\vec{E}_{{}_P}\cdot \vec{v}_{{}_P}
+\frac{g}{2}\cdot (\vec{s}\wtimes\vec{B}_{{}_P})
-\frac{g}{2}\cdot
\vec{s}^{\,\dagger}\cdot\vec{E}_{{}_P}\cdot \vec{v}_{{}_P} \rbrack \; ,
\label{D.10}
\end{eqnarray}
and the spatial part of the lhs of (5.5c) reads as:
\begin{eqnarray}
 &&   \gamma_{{}_P}\cdot \vec{s}\ ' \; .
\label{D.11}
\end{eqnarray}
Thus I have shown that the spatial part of equation (5.5c) is equivalent to
equation (5.12c). Using (3.6) one also finds that the temporal part of
equation (5.5c) follows from equation (5.12c) so that equation (5.5c) is
equivalent to (5.12c).
\setcounter{section}{5}
\setcounter{subsection}{0}
\setcounter{equation}{0}
\section*{Appendix E}
\addcontentsline{toc}{section}{Appendix E}
In this Appendix I show that (6.8) is equivalent to (5.5). From
(6.2),(6.7) follows
\footnote{Note that: $
   \varepsilon_{\alpha\nu\beta\gamma}\cdot
   \varepsilon_{\alpha\rho\lambda\sigma} =
 \delta_{\nu\rho}\cdot\delta_{\beta\lambda}\cdot\delta_{\gamma\sigma}
+\delta_{\nu\sigma}\cdot\delta_{\beta\rho}\cdot\delta_{\gamma\lambda}
+\delta_{\nu\lambda}\cdot\delta_{\beta\sigma}\cdot\delta_{\gamma\rho}
\\
-\delta_{\nu\rho}\cdot\delta_{\beta\sigma}\cdot\delta_{\gamma\lambda}
-\delta_{\nu\sigma}\cdot\delta_{\beta\lambda}\cdot\delta_{\gamma\rho}
-\delta_{\nu\lambda}\cdot\delta_{\beta\rho}\cdot\delta_{\gamma\sigma}
$ for $\nu,\beta,\gamma,\rho,\lambda,\sigma=1,...,4$.}
\begin{eqnarray}
&&    S^{{}^P}_{\nu\alpha}\cdot F^{{}^P}_{\alpha\rho} = \frac{1}{2}\cdot
   \varepsilon_{\nu\alpha\beta\gamma}\cdot
   \varepsilon_{\alpha\rho\lambda\sigma} \cdot
                       T^{{}^P}_{\beta}\cdot U^{{}^P}_{\gamma}\cdot
               \tilde{F}^{{}^P}_{\lambda\sigma}
 =  - \frac{1}{2}\cdot
   \varepsilon_{\alpha\nu\beta\gamma}\cdot
   \varepsilon_{\alpha\rho\lambda\sigma} \cdot
                       T^{{}^P}_{\beta}\cdot U^{{}^P}_{\gamma}\cdot
               \tilde{F}^{{}^P}_{\lambda\sigma}
\nonumber\\
&=& - \frac{1}{2}\cdot\lbrack
 \delta_{\nu\rho}\cdot\delta_{\beta\lambda}\cdot\delta_{\gamma\sigma}
+\delta_{\nu\sigma}\cdot\delta_{\beta\rho}\cdot\delta_{\gamma\lambda}
+\delta_{\nu\lambda}\cdot\delta_{\beta\sigma}\cdot\delta_{\gamma\rho}
-\delta_{\nu\rho}\cdot\delta_{\beta\sigma}\cdot\delta_{\gamma\lambda}
-\delta_{\nu\sigma}\cdot\delta_{\beta\lambda}\cdot\delta_{\gamma\rho}
\nonumber\\&&
-\delta_{\nu\lambda}\cdot\delta_{\beta\rho}\cdot\delta_{\gamma\sigma}
\rbrack\cdot  T^{{}^P}_{\beta}\cdot U^{{}^P}_{\gamma}\cdot
               \tilde{F}^{{}^P}_{\lambda\sigma}
 = \delta_{\nu\rho} \cdot
U^{{}^P}_{\alpha}\cdot T^{{}^P}_{\beta}\cdot \tilde{F}^{{}^P}_{\alpha\beta}
- \tilde{F}^{{}^P}_{\nu\alpha}\cdot T^{{}^P}_{\alpha}\cdot U^{{}^P}_{\rho}
+ \tilde{F}^{{}^P}_{\nu\alpha}\cdot U^{{}^P}_{\alpha}\cdot T^{{}^P}_{\rho} \; ,
\nonumber\\
&&    F^{{}^P}_{\nu\alpha}\cdot  S^{{}^P}_{\alpha\rho} =
      \delta_{\nu\rho} \cdot
U^{{}^P}_{\alpha}\cdot T^{{}^P}_{\beta}\cdot \tilde{F}^{{}^P}_{\alpha\beta}
- \tilde{F}^{{}^P}_{\rho\alpha}\cdot T^{{}^P}_{\alpha}\cdot U^{{}^P}_{\nu}
+ \tilde{F}^{{}^P}_{\rho\alpha}\cdot U^{{}^P}_{\alpha}\cdot T^{{}^P}_{\nu} \; ,
\nonumber\\
&&    S^{{}^P}_{\mu\nu}\cdot F^{{}^P}_{\nu\omega}\cdot U^{{}^P}_{\omega} =
U^{{}^P}_{\mu}\cdot U^{{}^P}_{\alpha}\cdot
    T^{{}^P}_{\beta}\cdot\tilde{F}^{{}^P}_{\alpha\beta}
+ \tilde{F}^{{}^P}_{\mu\nu}\cdot T^{{}^P}_{\nu} \; ,
\nonumber\\
&&    S^{{}^P}_{\lambda\omega}\cdot F^{{}^P}_{\omega\lambda}  =
2\cdot U^{{}^P}_{\alpha}\cdot T^{{}^P}_{\beta}\cdot
      \tilde{F}^{{}^P}_{\alpha\beta} \; ,
 \qquad (\mu,\nu,\rho =1,...,4)
\label{E.1}
\end{eqnarray}
so that one gets by (5.5c),(6.1-2):
\footnote{Note that: $
\varepsilon_{\nu\rho\mu\omega}\cdot \varepsilon_{\nu\rho\beta\gamma}
 =  \delta_{\mu\beta}\cdot\delta_{\omega\gamma}
-\delta_{\mu\gamma}\cdot\delta_{\omega\beta}$ for
$\mu,\omega,\beta,\gamma=1,...,4$.}
\begin{eqnarray}
&&\dot{T}^{{}^P}_{\mu} = -\frac{i}{2}\cdot\varepsilon_{\mu\nu\rho\omega}\cdot
\frac{d}{d\tau}\; (S^{{}^P}_{\nu\rho}\cdot U^{{}^P}_{\omega})
                = -\frac{i}{2}\cdot\varepsilon_{\mu\nu\rho\omega}\cdot
\lbrack \dot{S}^{{}^P}_{\nu\rho}\cdot U^{{}^P}_{\omega}
 +  S^{{}^P}_{\nu\rho}\cdot \dot{U}^{{}^P}_{\omega}   \rbrack
\nonumber\\
&=& -\frac{i}{2}\cdot\varepsilon_{\mu\nu\rho\omega}\cdot
\lbrack \frac{e\cdot g}{2\cdot m}\cdot
       ( F^{{}^P}_{\nu\alpha}\cdot S^{{}^P}_{\alpha\rho}\cdot U^{{}^P}_{\omega}
 - S^{{}^P}_{\nu\alpha}\cdot F^{{}^P}_{\alpha\rho}\cdot U^{{}^P}_{\omega})
+\frac{e}{m}\cdot S^{{}^P}_{\nu\rho}\cdot
 F^{{}^P}_{\omega\alpha}\cdot U^{{}^P}_{\alpha}
          \rbrack
\nonumber\\
&=& -\frac{i}{2}\cdot\varepsilon_{\mu\nu\rho\omega}\cdot
\lbrack \frac{e\cdot g}{2\cdot m}\cdot
(-\tilde{F}^{{}^P}_{\rho\alpha}\cdot T^{{}^P}_{\alpha}\cdot U^{{}^P}_{\nu}
                                              \cdot U^{{}^P}_{\omega}
+ \tilde{F}^{{}^P}_{\rho\alpha}\cdot U^{{}^P}_{\alpha}\cdot T^{{}^P}_{\nu}
                                              \cdot U^{{}^P}_{\omega}
  +\tilde{F}^{{}^P}_{\nu\alpha}\cdot T^{{}^P}_{\alpha}\cdot U^{{}^P}_{\rho}
                                              \cdot U^{{}^P}_{\omega}
\nonumber\\&&\qquad
- \tilde{F}^{{}^P}_{\nu\alpha}\cdot U^{{}^P}_{\alpha}\cdot T^{{}^P}_{\rho}
                                              \cdot U^{{}^P}_{\omega})
 -  \frac{i\cdot e}{m}\cdot
\varepsilon_{\nu\rho\beta\gamma}\cdot T^{{}^P}_{\beta}\cdot U^{{}^P}_{\gamma}
        \cdot F^{{}^P}_{\omega\alpha}\cdot U^{{}^P}_{\alpha} \rbrack
\nonumber\\
&=& -\frac{i}{2}\cdot\varepsilon_{\mu\nu\rho\omega}\cdot
\lbrack -\frac{i\cdot e\cdot g}{4\cdot m}\cdot
 ( -\varepsilon_{\rho\alpha\beta\gamma}\cdot
  F^{{}^P}_{\beta\gamma}\cdot T^{{}^P}_{\alpha}\cdot
 U^{{}^P}_{\nu}\cdot U^{{}^P}_{\omega}
   +\varepsilon_{\rho\alpha\beta\gamma}\cdot
  F^{{}^P}_{\beta\gamma}\cdot U^{{}^P}_{\alpha}\cdot
  T^{{}^P}_{\nu}\cdot U^{{}^P}_{\omega}
\nonumber\\&&
   +\varepsilon_{\nu\alpha\beta\gamma}\cdot
  F^{{}^P}_{\beta\gamma}\cdot T^{{}^P}_{\alpha}\cdot
   U^{{}^P}_{\rho}\cdot U^{{}^P}_{\omega}
   -\varepsilon_{\nu\alpha\beta\gamma}\cdot
  F^{{}^P}_{\beta\gamma}\cdot U^{{}^P}_{\alpha}\cdot
   T^{{}^P}_{\rho}\cdot U^{{}^P}_{\omega} )
 -  \frac{i\cdot e}{m}\cdot
\varepsilon_{\nu\rho\beta\gamma}\cdot T^{{}^P}_{\beta}\cdot U^{{}^P}_{\gamma}
        \cdot F^{{}^P}_{\omega\alpha}\cdot U^{{}^P}_{\alpha} \rbrack
\nonumber\\
&=&  \frac{e\cdot g}{8\cdot m}\cdot
\varepsilon_{\rho\nu\mu\omega}\cdot \varepsilon_{\rho\alpha\beta\gamma}
  \cdot \lbrack
- F^{{}^P}_{\beta\gamma}\cdot T^{{}^P}_{\alpha}\cdot
  U^{{}^P}_{\nu}\cdot U^{{}^P}_{\omega}
+ F^{{}^P}_{\beta\gamma}\cdot U^{{}^P}_{\alpha}\cdot T^{{}^P}_{\nu}\cdot
                                               U^{{}^P}_{\omega}\rbrack
\nonumber\\
&& - \frac{e\cdot g}{8\cdot m}\cdot
\varepsilon_{\nu\rho\mu\omega}\cdot \varepsilon_{\nu\alpha\beta\gamma}
  \cdot \lbrack
  F^{{}^P}_{\beta\gamma}\cdot T^{{}^P}_{\alpha}\cdot
  U^{{}^P}_{\rho}\cdot U^{{}^P}_{\omega}
-F^{{}^P}_{\beta\gamma}\cdot U^{{}^P}_{\alpha}\cdot T^{{}^P}_{\rho}\cdot
                                               U^{{}^P}_{\omega}\rbrack
\nonumber\\&&
 -   \frac{e}{2\cdot m}\cdot
\varepsilon_{\nu\rho\mu\omega}\cdot \varepsilon_{\nu\rho\beta\gamma}
\cdot T^{{}^P}_{\beta}\cdot U^{{}^P}_{\gamma}\cdot F^{{}^P}_{\omega\alpha}\cdot
        U^{{}^P}_{\alpha}
\nonumber\\
&=&  \frac{e\cdot g}{4\cdot m}\cdot
\varepsilon_{\rho\nu\mu\omega}\cdot \varepsilon_{\rho\alpha\beta\gamma}
  \cdot \lbrack
- F^{{}^P}_{\beta\gamma}\cdot T^{{}^P}_{\alpha}\cdot
      U^{{}^P}_{\nu}\cdot U^{{}^P}_{\omega}
+ F^{{}^P}_{\beta\gamma}\cdot U^{{}^P}_{\alpha}\cdot
    T^{{}^P}_{\nu}\cdot U^{{}^P}_{\omega}
                                                                \rbrack
\nonumber\\&&
 -   \frac{e}{2\cdot m}\cdot
\varepsilon_{\nu\rho\mu\omega}\cdot \varepsilon_{\nu\rho\beta\gamma}
\cdot T^{{}^P}_{\beta}\cdot U^{{}^P}_{\gamma}\cdot F^{{}^P}_{\omega\alpha}\cdot
        U^{{}^P}_{\alpha}
\nonumber\\
&=&  \frac{e\cdot g}{4\cdot m}\cdot
\varepsilon_{\rho\nu\mu\omega}\cdot \varepsilon_{\rho\alpha\beta\gamma}
\cdot F^{{}^P}_{\beta\gamma}\cdot U^{{}^P}_{\alpha}\cdot T^{{}^P}_{\nu}
                                            \cdot U^{{}^P}_{\omega}
 -   \frac{e}{2\cdot m}\cdot
\varepsilon_{\nu\rho\mu\omega}\cdot \varepsilon_{\nu\rho\beta\gamma}
\cdot T^{{}^P}_{\beta}\cdot U^{{}^P}_{\gamma}\cdot F^{{}^P}_{\omega\alpha}\cdot
        U^{{}^P}_{\alpha}
\nonumber\\
&=&  \frac{e\cdot g}{4\cdot m}\cdot\lbrack
(\delta_{\nu\alpha}\cdot\delta_{\mu\beta}\cdot\delta_{\omega\gamma}
+\delta_{\nu\gamma}\cdot\delta_{\mu\alpha}\cdot\delta_{\omega\beta}
+\delta_{\nu\beta}\cdot\delta_{\mu\gamma}\cdot\delta_{\omega\alpha}
-\delta_{\nu\alpha}\cdot\delta_{\mu\gamma}\cdot\delta_{\omega\beta}
-\delta_{\nu\gamma}\cdot\delta_{\mu\beta}\cdot\delta_{\omega\alpha}
\nonumber\\&&
-\delta_{\nu\beta}\cdot\delta_{\mu\alpha}\cdot\delta_{\omega\gamma})
\cdot F^{{}^P}_{\beta\gamma}\cdot U^{{}^P}_{\alpha}\cdot
      T^{{}^P}_{\nu}\cdot U^{{}^P}_{\omega}
\rbrack
- \frac{e}{m}\cdot\lbrack \delta_{\mu\beta}\cdot\delta_{\omega\gamma}
-\delta_{\mu\gamma}\cdot\delta_{\omega\beta} \rbrack
\cdot T^{{}^P}_{\beta}\cdot U^{{}^P}_{\gamma}\cdot F^{{}^P}_{\omega\alpha}
 \cdot U^{{}^P}_{\alpha}
\nonumber\\
&=&  \frac{e\cdot g}{2\cdot m}\cdot\lbrack
F^{{}^P}_{\alpha\beta}\cdot
 U^{{}^P}_{\alpha}\cdot T^{{}^P}_{\beta}\cdot U^{{}^P}_{\mu}
+F^{{}^P}_{\mu\alpha}\cdot T^{{}^P}_{\alpha}  \rbrack
+ \frac{e}{m}\cdot
F^{{}^P}_{\beta\alpha}\cdot
 T^{{}^P}_{\beta}\cdot U^{{}^P}_{\alpha}\cdot U^{{}^P}_{\mu}
\nonumber\\
&=&  \frac{e\cdot g}{2\cdot m}\cdot F^{{}^P}_{\mu\nu}\cdot T^{{}^P}_{\nu}
  +   \frac{e\cdot (g-2)}{2\cdot m}\cdot
    U^{{}^P}_{\mu}\cdot
    T^{{}^P}_{\omega}\cdot F^{{}^P}_{\nu\omega}\cdot U^{{}^P}_{\nu} \; .
 \qquad (\mu =1,...,4)
\label{E.2}
\end{eqnarray}
Also one gets from (E.1):
\begin{eqnarray}
&&S^{{}^P}_{\nu\omega}\cdot \partial^{{}^P}_{\mu}
                                           F^{{}^P}_{\omega\nu}
+U^{{}^P}_{\mu}\cdot S^{{}^P}_{\nu\omega}\cdot U^{{}^P}_{\lambda}\cdot
                                            \partial^{{}^P}_{\lambda}
                            F^{{}^P}_{\omega\nu}  =
2\cdot U^{{}^P}_{\alpha}\cdot T^{{}^P}_{\beta}\cdot\partial^{{}^P}_{\mu}
                                      \tilde{F}^{{}^P}_{\alpha\beta}
+2\cdot U^{{}^P}_{\mu}\cdot U^{{}^P}_{\alpha}\cdot T^{{}^P}_{\beta}\cdot
U^{{}^P}_{\lambda}\cdot\partial^{{}^P}_{\lambda}
     \tilde{F}^{{}^P}_{\alpha\beta} \; ,
\nonumber\\
&& S^{{}^P}_{\mu\nu}\cdot U^{{}^P}_{\omega}\cdot U^{{}^P}_{\lambda}\cdot
                           \partial^{{}^P}_{\lambda}
                            F^{{}^P}_{\nu\omega} =
    U^{{}^P}_{\mu}\cdot U^{{}^P}_{\alpha}\cdot T^{{}^P}_{\beta}\cdot
U^{{}^P}_{\lambda}\cdot\partial^{{}^P}_{\lambda}
         \tilde{F}^{{}^P}_{\alpha\beta}
+T^{{}^P}_{\nu}\cdot U^{{}^P}_{\lambda}\cdot\partial^{{}^P}_{\lambda}
                                      \tilde{F}^{{}^P}_{\mu\nu} \; ,
\nonumber\\
&&F^{{}^P}_{\mu\nu}\cdot S^{{}^P}_{\nu\omega}\cdot
  F^{{}^P}_{\omega\rho}\cdot U^{{}^P}_{\rho}=
F^{{}^P}_{\mu\nu}\cdot U^{{}^P}_{\nu}\cdot
  U^{{}^P}_{\alpha}\cdot T^{{}^P}_{\beta}\cdot
  \tilde{F}^{{}^P}_{\alpha\beta}
+ F^{{}^P}_{\mu\nu}\cdot \tilde{F}^{{}^P}_{\nu\alpha} \cdot
                                                       T^{{}^P}_{\alpha} \; ,
\nonumber\\
&&F^{{}^P}_{\mu\nu}\cdot U^{{}^P}_{\nu}\cdot
  S^{{}^P}_{\omega\rho}\cdot F^{{}^P}_{\rho\omega}=
2\cdot F^{{}^P}_{\mu\nu}\cdot U^{{}^P}_{\nu}\cdot
  U^{{}^P}_{\alpha}\cdot T^{{}^P}_{\beta}\cdot
  \tilde{F}^{{}^P}_{\alpha\beta}  \; ,
\nonumber\\
&&S^{{}^P}_{\mu\nu}\cdot F^{{}^P}_{\nu\rho}\cdot
  F^{{}^P}_{\rho\omega}\cdot U^{{}^P}_{\omega}=
  F^{{}^P}_{\mu\nu}\cdot
  U^{{}^P}_{\nu}\cdot U^{{}^P}_{\alpha}\cdot T^{{}^P}_{\beta}\cdot
  \tilde{F}^{{}^P}_{\alpha\beta}
- \tilde{F}^{{}^P}_{\mu\nu}\cdot
 U^{{}^P}_{\nu}\cdot U^{{}^P}_{\alpha}\cdot T^{{}^P}_{\beta}\cdot
  F^{{}^P}_{\alpha\beta} \; .
 \qquad (\mu =1,...,4)
\nonumber\\
\label{E.3}
\end{eqnarray}
Combining (5.5),(6.8),(E.2-3) one observes that (6.8) is equivalent to
(5.5).
\section*{Remarks accompanying the text}
\addcontentsline{toc}{section}{Remarks accompanying the text}
$\lbrack a \rbrack$:
For a particle with `normal' intrinsic magnetic dipole moment one has by
definition: $g=2$.
\\\par\noindent
$\lbrack b \rbrack$:
%As far as the author is aware this transformation has not been
%published. However,
                    Transformations have been given in \cite{Blo62,DS70}
and in \cite{CaM55} for special cases and Jackson mentions the `Pauli
reduction' \cite{Jac76}. The general transformation is straightforward
and has been carried out by the author. It will be published
separately \cite{Hei}. Note that it is consistent with the 
(semi-relativistic) Foldy-Wouthuysen transformation
\cite{BD64,CoM95,FW50,Fol62}, because expanding the Hamiltonian
$H_{{}_{M,op}}$ up to second order in $1/c$ yields by (0.1) for $g=2$:
\begin{eqnarray*}
&& H_{{}_{M,op}} = \frac{m}{2}\cdot\beta + \frac{1}{4\cdot m}\cdot\beta\cdot
  \vec{\pi}_{{}_{M,op}}^\dagger\cdot \vec{\pi}_{{}_{M,op}}
                     - \frac{1}{16\cdot m^3}\cdot\beta\cdot
 (\vec{\pi}_{{}_{M,op}}^\dagger\cdot \vec{\pi}_{{}_{M,op}})^2
                           + \frac{e}{2}\cdot \phi_{{}_{M,op}}
\nonumber\\&&\qquad
 - \frac{e}{2\cdot m}\cdot \beta\cdot\vec{\sigma}_{{}_{op}}^{\,\dagger}\cdot
         \vec{B}_{{}_{M,op}}
%\nonumber\\&&\qquad
 + \frac{e}{4\cdot m^2}\cdot \vec{\sigma}_{{}_{op}}^{\,\dagger}\cdot
(\vec{\pi}_{{}_{M,op}}\wtimes\vec{E}_{{}_{M,op}}) + {\rm hermitian\;conjugate}
 \; .
\end{eqnarray*}
The Darwin term does not
appear here because of the {\it vacuum} Maxwell equations (see section 2).
Expanding $H_{{}_{M,op}}$ only up to first order in $1/c$ yields for $g=2$:
\begin{eqnarray*}
&& H_{{}_{M,op}} = m\cdot\beta + \frac{1}{2\cdot m}\cdot\beta\cdot
  \vec{\pi}_{{}_{M,op}}^\dagger\cdot \vec{\pi}_{{}_{M,op}}
                           + e\cdot \phi_{{}_{M,op}}
 - \frac{e}{m}\cdot \beta\cdot\vec{\sigma}_{{}_{op}}^{\,\dagger}\cdot
         \vec{B}_{{}_{M,op}}  \; ,
\end{eqnarray*}
which has the form of the Schroedinger-Pauli Hamiltonian.
\\\par\noindent
$\lbrack c \rbrack$:
Thus the DK equations turn out to be
$\rm{Poincar\acute{e}}$ covariant but not {\it manifestly}
$\rm{Poincar\acute{e}}$ covariant whereas the Frenkel equations
are manifestly covariant.
\\\par\noindent
$\lbrack d \rbrack$:
The particle described in this work has arbitrary but nonvanishing charge,
arbitrary intrinsic magnetic dipole moment and vanishing intrinsic electric
dipole moment. For a neutral particle the equations can be easily
modified.
\\\par\noindent
$\lbrack e \rbrack$:
Here $\varepsilon_{ijk}$ is the antisymmetric symbol with
$\varepsilon_{123}=1$ and $\delta$ denotes the Kronecker delta. All
three-component quantities $\vec{r}_{{}_M}, \vec{p}_{{}_M},\vec{\sigma},
\dots$ denoted by an arrow are {\it column} vectors. The components
$a_j$ of any $\vec{a}$ are defined by:
\begin{eqnarray*}
 && \vec{a} = (a_1,a_2,a_3)^\dagger \; .
\end{eqnarray*}
The transpose of a three-component quantity is denoted by `$\dagger$'.
Therefore $\vec{r}_{{}_M}^{\,\dagger},
\vec{p}_{{}_M}^{\,\dagger},\vec{\sigma}^{\,\dagger},\dots$ are {\it row}
vectors.
\\\par\noindent
$\lbrack f \rbrack$:
In this paper the nabla operator $\vec{\nabla}_{{}_M}$ always acts on
functions depending on $\vec{r}_{{}_M},t, \vec{p}_{{}_M},\vec{\sigma}$
and it is the gradient w.r.t. $\vec{r}_{{}_M}$.
\\\par\noindent
$\lbrack g \rbrack$:
Note that in this paper the multiplication symbol `$\cdot$' always
denotes {\it matrix} multiplication and that a single number is a
$1\times 1$ matrix.
\par To avoid the mushrooming of the bracket symbol I avoid its use
even in places where it would usually help to find the correct order of
matrix multiplications to be performed. If for example a matrix product
like
\begin{eqnarray*}
&& \vec{\pi}_{{}_M}^{\,\dagger}\cdot\vec{B}_{{}_M}\cdot\vec{\pi}_{{}_M}
\end{eqnarray*}
occurs (in this example associativity of multiplication does not hold!),
then the matrix structures of the factors suggest the correct order of
the matrix multiplications. In the present example one has:
\begin{eqnarray*}
&& \vec{\pi}_{{}_M}^{\,\dagger}\cdot\vec{B}_{{}_M}\cdot\vec{\pi}_{{}_M}
 = (\vec{\pi}_{{}_M}^{\,\dagger}\cdot\vec{B}_{{}_M})\cdot\vec{\pi}_{{}_M}
\end{eqnarray*}
because
\begin{eqnarray*}
&& \vec{\pi}_{{}_M}^{\,\dagger}\cdot(\vec{B}_{{}_M}\cdot\vec{\pi}_{{}_M})
\end{eqnarray*}
is a meaningless expression. Note also that the multiplication symbol
`$\wtimes$' always the denotes the vector product of three-component 
quantities.
\\\par\noindent
$\lbrack h \rbrack$:
The nabla operator $\vec{\nabla}_{{}_P}$ in this paper always acts on
functions depending on $\vec{r}_{{}_P},t,\vec{v}_{{}_P},\vec{s}$ and it
is the gradient w.r.t. $\vec{r}_{{}_P}$. My notation is chosen so as to
indicate that the functional dependence of $\vec{E}_{{}_P}$ on
$\vec{r}_{{}_P},t$ is the same as the dependence of $\vec{E}_{{}_M}$ on
$\vec{r}_{{}_M},t$ and likewise for $\vec{B}_{{}_M}$. The explicit way
in which the `P' fields are derived from the `M' fields is shown in
(B.21).
\\\par\noindent
$\lbrack i \rbrack$:
I am dealing with special relativistic space-time positions, tensors,
pseudotensors, tensor fields and pseudotensor fields
and I do so by using the {\it complex} notation where the fourth (=temporal)
component is imaginary. I use this convention following the usage in
much of the literature on the relativistic SG force
\cite{Cor68,Fre26,Goo62,Nyb62,Nyb64,Pla66a,Pla66b,Raf70}. Note also that
Einstein's summation conventions are applied to Greek indices. The Greek
indices assume the values 1,2,3,4. One of the advantages of the complex
convention is that only lower components occur. For textbooks using this
convention, see for example \cite{Moe72,Syn58}. Of course one could use
the {\it real} convention where, however, one has to distinguish between
covariant and contravariant components. Most textbooks use the real
convention, e.g. \cite{Jac75}.
\\\par\noindent
$\lbrack j \rbrack$:
Any antisymmetric tensor $N$ of rank 2 can be characterized by
two three-component quantities $\vec{b}$ resp. $\vec{d}$, called the
spatial resp. temporal part of the tensor, and they are defined by:
\begin{eqnarray*}
&& b_j = \frac{1}{2}\cdot
         \sum_{k,l=1}^3 \varepsilon_{jkl}\cdot N_{kl}  \; , \qquad
   d_j =   N_{j4}    \; . \qquad (j=1,2,3)
\end{eqnarray*}
I denote this correspondence as follows:
\begin{eqnarray*}
 && N  \leftrightarrow (\vec{b},\vec{d}) \; .
\end{eqnarray*}
Thus the spatial resp. temporal part of $M$ is given by
$\vec{\mu}_{{}_P}$ resp. $i\cdot\vec{\varepsilon}_{{}_P}$, i.e.
\begin{eqnarray*}
 && M  \leftrightarrow (\vec{\mu}_{{}_P},i\cdot\vec{\varepsilon}_{{}_P}) \; .
\end{eqnarray*}
\par\noindent
$\lbrack k \rbrack$:
The constraint (3.8b) follows from (3.2),(3.6) and states that the
particle has no intrinsic electric dipole moment \cite{Cor68,Fre26,Nyb64}.
\\\par\noindent
$\lbrack l \rbrack$:
In this paper the partial derivatives $\partial^{{}^P}_1,...,
\partial^{{}^P}_4$ always act on functions depending on
$X^{{}^P}_1,X^{{}^P}_2,X^{{}^P}_3,X^{{}^P}_4$.
\\\par\noindent
$\lbrack m \rbrack$:
One also gets this Hamiltonian if one neglects terms of second order
in spin in
the Hamiltonian given by Corben. For details, see \cite{Cor68},
especially Chapter 7 thereof. If one neglects all spin terms then
$H^{{}^P}$ reduces to the following well known expression:
\begin{eqnarray*}
 && H^{{}^P} = \frac{1}{2\cdot m}\cdot
 \Pi^{{}^P}_{\mu}\cdot \Pi^{{}^P}_{\mu}  +  \frac{m}{2} \; .
\end{eqnarray*}
See for example \cite{Bar64,Gol80,Jac75}.
\\\par\noindent
$\lbrack n \rbrack$:
The assumptions made on $Y^{{}^P}$ are as follows. The proper time dependence
in the SG term on the rhs of (5.1b) is assumed to come in only via the
proper time dependence of $X^{{}^P},U^{{}^P},S^{{}^P}$.
To get a useful class of allowed $Y^{{}^P}$, I assume in addition that the
$X^{{}^P}$ dependence only comes in via $F^{{}^P}$ and its space-time
derivatives. Specifically I assume that
$Y^{{}^P}_1,...,Y^{{}^P}_4$ are functions of the following arguments:
$m,U^{{}^P}_1,...,U^{{}^P}_4$, the six components of $S^{{}^P}$ and
all space time derivatives of the six components of
$F^{{}^P}$. This function is supposed to be a polynomial in all its arguments
(except $m$) and first order in spin. Thus the coefficients of this
polynomial
are functions of $m$ and it  turns out that they are just powers of $m$
times dimensionless numbers.
\\\par\noindent
$\lbrack o \rbrack$:
Another remark on the quantum mechanical aspect is in order. In deriving
the classical equations from the Dirac equation (plus the Pauli term)
one does not get a unique answer because one depends on the choice
of the proper operators. For example the
Frenkel equations can be obtained by using a certain special
relativistic generalization of the Foldy-Wouthuysen transformation
\cite{Blo62,DS70,DS72,Hei} with the emphasis on the Newton-Wigner
position operator whereas the GNR equations can be derived by
the `Gordon decomposition' \cite{Raf70}. Note again that in the present
work I am only interested in the classical aspect. Further details on
the quantum mechanics including the opinions of Pauli and Bohr are
discussed for example in \cite{DSV86,Goo62,Roh72}.
\\\par\noindent
$\lbrack p \rbrack$:
The Lorentz group consists of the homogeneous part of the
$\rm{Poincar\acute{e}}$ group, i.e. it does not contain the space-time
translations. Note that the $\rm{Poincar\acute{e}}$ group is also called
the inhomogeneous Lorentz group. A $\rm{Poincar\acute{e}}$
transformation is composed of a translation $a$ and a Lorentz
transformation $L$ so that the space-time position $X^{{}^P}$
transforms as:
\begin{eqnarray*}
&& X^{{}^P}_{\mu} \rightarrow
   L_{\mu\nu}\cdot X^{{}^P}_{\nu} + a_{\mu} \; .
 \qquad (\mu =1,...,4)
\end{eqnarray*}
Since $L$ belongs to the Lorentz group it satisfies
\begin{eqnarray*}
 && L_{\mu\nu} \cdot L_{\mu\rho} = \delta_{\nu\rho} \; .
 \qquad (\nu,\rho =1,...,4)
\end{eqnarray*}
The restricted $\rm{Poincar\acute{e}}$ group
(=proper orthochronous $\rm{Poincar\acute{e}}$ group)
contains those $\rm{Poincar\acute{e}}$ transformations, where:
\begin{eqnarray*}
 && \det(L) =1 \; , \qquad L_{44}> 0 \; .
\end{eqnarray*}
Hence the restricted $\rm{Poincar\acute{e}}$ group contains
neither the parity transformation nor the time
reversal transformation. For more details on the subgroups of the
Lorentz group resp. ${\rm Poincar\acute{e}}$ group, see for example
\cite{BLT75,SW89}.
\\\par\noindent
$\lbrack q \rbrack$:
This follows because the relations (3.1),(3.6),(3.9) between the variables
$X^{{}^P},S^{{}^P},F^{{}^P}$ and the variables
$\vec{r}_{{}_P},t,\vec{s},\vec{B}_{{}_P},\vec{E}_{{}_P}$
are the same for every ${\rm Poincar\acute{e}}$ frame.
\\\par\noindent
$\lbrack r \rbrack$:
If one performs a ${\rm Poincar\acute{e}}$ transformation, then (as shown in
subsection 5.3) the equations of motion (2.11) for the `P' variables
remain the
same (except that the `P'
fields have transformed in a specified way). Under the same
${\rm Poincar\acute{e}}$ transformation the equations of motion (1.5)
for the `M' variables also
                      transform in a definite way. The transformed equations
of motion for the `M' variables can be derived from the transformed equations
of motion for the `P' variables in the same way as the
original equations of motion (1.5) for the `M' variables were derived in
section 2 from the original equations of motion (2.11) for the `P' variables
because the relations (2.1),(2.3a),(2.4-5),(2.7) between the `M' variables 
and the `P'
variables are the same for every ${\rm Poincar\acute{e}}$ frame. The result is
that the transformed equations of motion for the M' variables are the same
as the original equations of motion (1.5) for the `M' variables
(except that the `M' fields have transformed in a specified way). Therefore
the DK equations (1.5) are $\rm{Poincar\acute{e}}$ covariant.
\section*{Acknowledgements}
\addcontentsline{toc}{section}{Acknowledgements}
I wish to thank Desmond P.\ Barber for careful reading of and valuable
remarks on the manuscript and Ya.S. Derbenev, G.H. Hoffst\"atter and G. Ripken
for useful discussions. Thanks also go to J.P. Costella and
R. Jagannathan for their valuable comments.
\end{document}